\documentclass{aa}    
\usepackage{graphicx,url,twoopt,natbib}
\usepackage[varg]{txfonts}           
\usepackage[usenames]{color}
\usepackage{pagecolor}
\usepackage{acronym}      
\usepackage[titles]{tocloft}
\usepackage{stdclsdv} 
\usepackage{hyperref} 
\definecolor{rrmaroon}{rgb}{0.3, 0.05, 0.01}
\hypersetup{
  colorlinks=true,  
  urlcolor=rrblue,    
  citecolor=rrblue,   
  linkcolor=rrmaroon    
}
\usepackage[all]{hypcap} 

\usepackage{eurosym}

\usepackage{fancyhdr}
\definecolor{bllsht}{rgb}{0.40, 0.05, 0.01} 
\long\def\TODO#1TODO{\par\vspace{0.5ex}\noindent
  {\Large \textcolor{red}{@~}}\color{bllsht}{#1}\color{black}\\[0.5ex]}

\definecolor{rrblue}{rgb}{0.15,0.0,0.8} 

\def\rrref#1#2{{\em \hyperref[#2]{#1\,}\ref{#2}}\/}

\makeatletter
\newcommand{\bibnote}[2]{\global\@namedef{#1note}{#2}}
\newcommand{\adslink}[2]{\global\@namedef{#1adslink}{#2}}
\newcommand{\urllink}[2]{\global\@namedef{#1urllink}{#2}}
\makeatother

\definecolor{amber}{rgb}{1.0, 0.49, 0.0}

\bibpunct{(}{)}{;}{a}{}{,}    
\makeatletter
  \protected\def\sppresslink{\def\hyper@linkstart##1##2{}\let\hyper@linkend\@empty}
  \newcommandtwoopt{\citeads}[3][][]{%
   \href{http://ui.adsabs.harvard.edu/abs/#3/abstract}%
        {\sppresslink \citealp[#1][#2]{#3}}
   \adslink{#3}{\href{http://ui.adsabs.harvard.edu/abs/#3/abstract}{~ADS}}}
 \newcommandtwoopt{\citepads}[3][][]{%
   \href{http://ui.adsabs.harvard.edu/abs/#3/abstract}%
        {\sppresslink \citep[#1][#2]{#3}}
   \adslink{#3}{\href{http://ui.adsabs.harvard.edu/abs/#3/abstract}{~ADS}}}
 \newcommandtwoopt{\citetads}[3][][]{%
   \href{http://ui.adsabs.harvard.edu/abs/#3/abstract}%
        {\sppresslink \citet[#1][#2]{#3}}
  \adslink{#3}{\href{http://ui.adsabs.harvard.edu/abs/#3/abstract}{~ADS}}}
 \newcommandtwoopt{\citeyearads}[3][][]{%
   \href{http://ui.adsabs.harvard.edu/abs/#3/abstract}%
        {\sppresslink \citeyear[#1][#2]{#3}}
   \adslink{#3}{\href{http://ui.adsabs.harvard.edu/abs/#3/abstract}{~ADS}}}
 \newcommandtwoopt{\citeweb}[4][][]{
   \href{#3}{\sppresslink \citealp[#1][#2]{#4}}%
   \urllink{#4}{\href{#3}{~URL}}}
 \newcommandtwoopt{\citepweb}[4][][]{
   \href{#3}{\sppresslink \citep[#1][#2]{#4}}%
   \urllink{#4}{\href{#3}{~URL}}}
 \newcommandtwoopt{\citetweb}[4][][]{
   \href{#3}{\sppresslink \citet[#1][#2]{#4}}%
   \urllink{#4}{\href{#3}{~URL}}}
 \newcommandtwoopt{\citeyearweb}[4][][]{
   \href{#3}{\sppresslink \citeyear[#1][#2]{#4}}%
   \urllink{#4}{\href{#3}{~URL}}}
\makeatother

\newcommand{\citenoweb}[1]{\citealp{#1}}
\newcommand{\citetnoweb}[1]{\citet{#1}}

\def\linkadspage#1#2#3{\href{http://adsabs.harvard.edu/cgi-bin/nph-data_query?bibcode=#1\&link_type=ARTICLE\&db_key=AST\#page=#2}{#3 (pdf\,#2)}}

\def\linkpdfpage#1#2#3{\href{#1\#page=#2}{#3 (pdf\,#2)}}

\def\linkrjrpage#1#2#3{%
  \href{https://robrutten.nl/rrweb/rjr-pubs/#1.pdf\#page=#2}{#3 (pdf\,#2)}}

\def\ISSFpage#1#2{\href{https://robrutten.nl/rrweb/rjr-pubs/1993-issf-notes.pdf\#page=#1}{[ISSF #2 (pdf\,#1)]}}

\def\IARTpage#1#2{\href{https://robrutten.nl/rrweb/rjr-pubs/2015-iart-notes.pdf\#page=#1}{[IART #2 (pdf\,#1)]}}

\def\RTSApage#1#2{\href{https://robrutten.nl/rrweb/rjr-pubs/2003rtsa.book.....R.pdf\#page=#1}{[RTSA #2 (pdf\,#1)]}}

\newcount\longrefs
\def\adv{\ifnum\longrefs=1 {Adv.\ Space Res.} \else 
                           {Adv.\ Sp'\ Res.}\fi}
\def\aap{\ifnum\longrefs=1 {Astron.\ Astrophys.}\else 
                           {A\hbox{\rm \&}A}\fi}
\def\aapr{\ifnum\longrefs=1 {Astron.\ Astrophys.\ Rev.}\else 
                            {A\hbox{\rm \&}AR}\fi}
\def\aaps{\ifnum\longrefs=1 {Astron.\ Astrophys.\ Suppl.}\else 
                            {A\hbox{\rm \&}A Suppl.}\fi}
\def\actaa{\ifnum\longrefs=1 {Acta Astronomica}\else
                            {Acta Astron.}\fi}
\def\aipcs{\ifnum\longrefs=1 {Am.\ Inst.\ Phys.\ Conf.\ Series}\else
                             {AIP Conf.\ Ser.}\fi}
\def\aj{\ifnum\longrefs=1 {Astron.\ J.}\else 
                          {AJ}\fi} 
\def\ao{\ifnum\longrefs=1 {Applied Optics}\else 
                           {Appl.\ Opt.}\fi} 
\def\aspcs{\ifnum\longrefs=1 {Astron.\ Soc.\ Pacific Conf.\ Series}\else 
                           {ASP Conf.\ Ser.}\fi} 
\def\apj{\ifnum\longrefs=1 {Astrophys.\ J.}\else 
                           {ApJ}\fi} 
\def\apjl{\ifnum\longrefs=1 {Astrophys.\ J. Lett.}\else 
                            {ApJL}\fi} 
\def\aplett{\ifnum\longrefs=1 {Astrophys.\ J. Lett.}\else 
                            {ApJ}\fi} 
\def\apjs{\ifnum\longrefs=1 {Astrophys.\ J. Suppl.}\else 
                            {ApJS}\fi}
\def\apss{\ifnum\longrefs=1 {Astrophys.\ Space Sci.}\else 
                            {Astrophys.\ Space Sci.}\fi}
\def\araa{\ifnum\longrefs=1 {Ann.\ Rev.\ Astron.\ Astrophys.}\else 
                            {ARA\hbox{\rm \&}A}\fi}
\def\azh{\ifnum\longrefs=1 {Astronomicheskii Zhurnal}\else 
                            {Astron.\ Zhur.}\fi}
\def\baas{\ifnum\longrefs=1 {Bull.\ Am.\ Astron.\ Soc.}\else 
                            {BAAS}\fi}
\def\bain{\ifnum\longrefs=1 {Bull.\ Astron.\ Inst. Neth.}\else
                            {Bull.\ Astr.\ Inst.\ Neth.}\fi}
\def\cjaa{\ifnum\longrefs=1 {Chin.\ J.\ Astron.\ Astrophys.}\else 
                            {Chin.\ J.\ A\&A}\fi}
\def\gca{\ifnum\longrefs=1 {Geochim.\ Cosmochim.\ Acta}\else 
                           {Geochim.\ Cosmochim.\ Acta}\fi}
\def\grl{\ifnum\longrefs=1 {Geophys.\ Res.\ Lett.}\else 
                           {Geoph.\ Res.\ Lett.}\fi}
\def\iaucirc{\ifnum\longrefs=1 {IAU Circulars}\else 
                          {IAU Circ.}\fi}
\def\icarus{\ifnum\longrefs=1 {Icarus}\else 
                          {Icarus}\fi}
\def\ip{\ifnum\longrefs=1 {in press}\else 
                          {in press}\fi}
\def\jcap{\ifnum\longrefs=1 {Jour.\ Cosmology Astropart.\ Phys.}\else 
                          {JCAP}\fi}
\def\jgr{\ifnum\longrefs=1 {J.\ Geophys.\ Res.}\else 
                           {J.\ Geophys.\ Res.}\fi}  
\def\jrasc{\ifnum\longrefs=1 {J.\ Royal Astron.\ Soc.\ Canada}\else 
                             {JRAS Can.}\fi}  
\def\memsai{\ifnum\longrefs=1 {Mem.~Soc.~Astron.~Italiana}\else
                              {MmSAI}\fi}
\def\mnras{\ifnum\longrefs=1 {Mon.\ Not.\ Roy.\ Astron.\ Soc.}\else 
                             {MNRAS}\fi} 
\def\na{\ifnum\longrefs=1 {New Astronomy}\else 
                          {New Astron.}\fi}
\def\nar{\ifnum\longrefs=1 {New Astronomy rev.}\else 
                           {New Astron.\ Rev.}\fi}
\def\nat{\ifnum\longrefs=1 {Nature}\else 
                           {Nat}\fi}
\def\pasa{\ifnum\longrefs=1 {Pub.\ Astron.\ Soc.\ Australia}\else 
                            {PASA}\fi} 
\def\pasj{\ifnum\longrefs=1 {Pub.\ Astron.\ Soc.\ Japan}\else 
                            {PASJ}\fi} 
\def\pasp{\ifnum\longrefs=1 {Pub.\ Astron.\ Soc.\ Pacific}\else 
                            {PASP}\fi} 
\def\physscr{\ifnum\longrefs=1 {Physica Scripta}\else 
                               {Phys.\ Scrip.}\fi} 
\def\planss{\ifnum\longrefs=1 {Planetary \& Space Science}\else 
                              {Plan. \& Space Sci.}\fi} 
\def\pre{\ifnum\longrefs=1 {Phys.\ Rev.\ E}\else
                           {Phys.\ Rev.\ E}\fi}
\def\procspie{\ifnum\longrefs=1 {Proc.\ SPIE}\else 
                                {Proc.\ SPIE}\fi} 
\def\qjras{\ifnum\longrefs=1 {Quarterly J.\ Royal Astron.\ Soc.}\else 
                             {QJRAS}\fi} 
\def\rmxaa{\ifnum\longrefs=1 {Revista Mexicana de Astron.\ y Astrofys.}\else 
                             {RMxAA}\fi} 
\def\sa{\ifnum\longrefs=1 {Soviet Astron..}\else 
                          {Sov.\ Astron.}\fi}
\def\skytel{\ifnum\longrefs=1 {Sky \& Telescope}\else 
                              {Sky \& Tel.}\fi} 
\def\solphys{\ifnum\longrefs=1 {Solar Phys.}\else 
                               {SoPh}\fi}
\def\sovast{\ifnum\longrefs=1 {Soviet Astron.}\else 
                              {Sov.\ Ast.}\fi}
\def\ssr{\ifnum\longrefs=1 {Space Sci.\ Rev.}\else 
                           {Space Sci.\ Rev.}\fi}
\def\zap{\ifnum\longrefs=1 {Zeit.\ f.\ Astrophys.}\else
                               {Z.\ Astrophys.}\fi}

\def\acp#1{#1} 
\newacro{AA}{Astronomy \& Astrophysics}  
\newacro{ADS}{Astrophysics Data System}
\newacro{AIA}{Atmospheric Imaging Assembly}
\newacro{ALMA}{Atacama Large Millimeter/submillimeter Array}
\newacro{AO}{adaptive optics}
\newacro{ApJ}{Astrophysical Journal}
\newacro{AR}{active region}
\newacro{bb}{bound-bound}
\newacro{bf}{bound-free}
\newacro{BFI}{Broad-band Filter Imager}
\newacro{CE}{coronal equilibrium}
\newacro{CfA}{Center for Astrophysics}
\newacro{CME}{coronal mass ejection}
\newacro{CRD}{complete redistribution}
\newacro{CRISP}{CRisp Imaging SpectroPolarimeter}
\newacro{CRISPEX}{CRisp SPectral EXplorer}
\newacro{CS}{coherent scattering}
\newacro{DEM}{Differential Emission Measure}
\newacro{DF}{dynamic fibril}
\newacro{DKIST}{Daniel K. Inouye Solar Telescope}
\newacro{DLR}{Deutsches Zentrum f\"ur Luft- und Raumfahrt}
\newacro{DOT}{Dutch Open Telescope}
\newacro{DST}{Richard B. Dunn Solar Telescope}   
\newacro{EB}{Ellerman bomb}
\newacro{EDP}{\'{E}dition Diffusion Presse}  
\newacro{EIT}{Extreme ultraviolet Imaging Telescope}
\newacro{EPIC}{European participation in Solar-C}
\newacro{ERC}{European Research Council}
\newacro{ESA}{European Space Agency}
\newacro{EST}{European Solar Telescope}
\newacro{EUV}{extreme ultraviolet}
\newacro{FAF}{flaring active-region fibril}
\newacro{ff}{free-free}
\newacro{FITS}{Flexible Image Transport System}
\newacro{FOV}{field of view}
\newacro{fov}{field of view}
\newacro{FWHM}{full width at half maximum}
\newacro{HAO}{High Altitude Observatory}
\newacro{HD}{hydrodynamics}
\newacro{Hi-C}{High Resolution Coronal Imager Sounding Rocket}
\newacro{HMI}{Helioseismic and Magnetic Imager}
\newacro{IAA}{Instituto de Astrof\'{i}sica de Andaluc\'{i}a}
\newacro{IAC}{Instituto de Astrof\'{i}sica de Canarias}
\newacro{IAS}{Institut d'Astrophysique Spatiale}
\newacro{IAU}{International Astronomical Union}
\newacro{IBIS}{Interferometric Bi-dimensional Spectrometer}
\newacro{IDL}{Interactive Data Language}
\newacro{IMaX}{Imaging Magnetograph eXperiment}
\newacro{INAF}{Istituto Nazionale di Astrofisica}
\newacro{IB}{IRIS bomb}
\newacro{IR}{infrared}
\newacro{IRIS}{Interface Region Imaging Spectrograph}
\newacro{ISAS}{Institute of Space and Astronautical Science}
\newacro{ISP}{Institute for Solar Physics}
\newacro{ISS}{International Space Station}
\newacro{ISSI}{International Space Science Institute}
\newacro{ITA}{Institute for Theoretical Astrophysics}
\newacro{JAXA}{Japan Aerospace Exploration Agency}
\newacro{JSOC}{Joint Science Operations Center}
\newacro{KIS}{Kiepenheuer--Institut f\"{u}r Sonnenphysik}
\newacro{KPNO}{Kitt Peak National Observatory}
\newacro{LASP}{Laboratory for Atmospheric and Space Physics}
\newacro{LC}{liquid cristal}
\newacro{LMSAL}{Lockheed Martin Solar and Astrophysics Labratory}
\newacro{LOS}{line of sight}
\newacro{LTE}{local thermodynamic equilibrium}
\newacro{MC}{magnetic concentration}
\newacro{MCAO}{multi-conjugate adaptive optics} 
\newacro{MDI}{Michelson Doppler Imager}
\newacro{ME}{Milne-Eddington} 
\newacro{MHD}{magnetohydrodynamics}
\newacro{MOMFBD}{Multi-Object Multi-Frame Blind Deconvolution}
\newacro{MPE}{Max--Planck--Institut f\"ur extraterrestrische Physik}
\newacro{MPG}{Max--Planck--Gesellschaft}
\newacro{MPS}{Max Planck Institute for Solar System Research}
\newacro{MSSL}{Mullard Space Science Laboratory}
\newacro{MTF}{modulation transfer function}
\newacro{NAOJ}{National Astronomical Observatory of Japan}
\newacro{NASA}{National Aeronautics and Space Administration}
\newacro{NIST}{National Institute of Standards and Technology}
\newacro{NLTE}{non-local thermodynamic equilibrium}
\newacro{NLFFF}{non-linear force-free field}
\newacro{NOAA}{National Oceanic and Atmospheric Administration}
\newacro{non-E}{non-equilibrium}
\newacro{NSO}{National Solar Observatory}
\newacro{NWO}{Netherlands Organisation for Scientific Research}
\newacro{PHE}{propagating heating event}
\newacro{PRD}{partial redistribution}
\newacro{PROBA2}{PRoject for OnBoard Autonomy}
\newacro{PSBE}{post Saha-Boltzmann extinction}
\newacro{PSF}{point spread function}
\newacro{QS}{quiet Sun}
\newacro{QSEB}{quiet-Sun Ellerman-like brightening} 
\newacro{RAL}{Rutherford Appleton Laboratory}
\newacro{RBE}{rapid blue-shifted excursion}
\newacro{R-MHD}{radiation hydrodynamics}
\newacro{rms}{root mean square}
\newacro{RMS}{root mean square}
\newacro{ROB}{Royal Observatory of Belgium}
\newacro{ROI}{region of interest}
\newacro{RRE}{rapid red-shifted excursion}
\newacro{RTE}{radiative transfer equation}
\newacro{RTSA}{Radiative Transfer in Stellar Atmospheres}
\newacro{SCF}{slender \CaIIH\ fibril}
\newacro{SE}{statistical equilibrium}
\newacro{SB}{Saha Boltzmann}
\newacro{SDO}{Solar Dynamics Observatory}
\newacro{SJI}{slit-jaw image}
\newacro{SLI}{slit image}
\newacro{SNR}{signal-to-noise ratio}
\newacro{SO}{Solar Orbiter}
\newacro{SoHO}{Solar and Heliospheric Observatory}
\newacro{SP}{Spectropolarimeter}
\newacro{SST}{Swedish 1-m Solar Telescope}
\newacro{SUMER}{Solar Ultraviolet Measurements of Emitted Radiation}
\newacro{SUFI}{Sunrise Filter Imager}
\newacro{SVD}{singular value decomposition}
\newacro{SVST}{Swedish Vacuum Solar Telescope}
\newacro{STX}{Solar Telescope X}
\newacro{THEMIS}{T\'{e}lescope H\'{e}liographique pour l'Etude du 
   Magn\'{e}tisme et des Instabilit\'{e} Solaires}     
\newacro{TR}{transition region}
\newacro{TRACE}{Transition Region and Coronal Explorer}
\newacro{TSI}{total solar irradiance}
\newacro{UT}{Universal Time}
\newacro{UV}{ultraviolet}
\newacro{VAULT}{Very high angular resolution ultraviolet telescope}
\newacro{VIRGO}{Variability of solar IRradiance and Gravity Oscillations}
\newacro{VTT}{Vacuum Tower Telescope}    
\newacro{XRT}{X-Ray Telescope}

\long\def\startignore #1\stopignore{}   

\def\rmit#1{{\it #1}}              
\def\etal{\rmit{et al.}}           
           
\def\ie{\rmit{i.e.,}}              
\def\eg{\rmit{e.g.,}}              
\def\cf{cf.}                       

\def\specchar#1{\uppercase{#1}}    
\def\specand{ and }                
\def\specand{\,\&\,}               
\def\AlI{\mbox{Al\,\specchar{i}}}  

\def\BaII{\mbox{Ba\,\specchar{ii}}} 
 
\def\CII{\mbox{C\,\specchar{ii}}} 
 
\def\CIV{\mbox{C\,\specchar{iv}}} 
 
\def\CaI{\mbox{Ca\,\specchar{i}}} 
\def\CaII{\mbox{Ca\,\specchar{ii}}} 
\def\CaIII{\mbox{Ca\,\specchar{iii}}} 
\def\CeII{\mbox{Ce\,\specchar{ii}}}

\def\FeI{\mbox{Fe\,\specchar{i}}} 
\def\FeII{\mbox{Fe\,\specchar{ii}}} 
 
\def\FeVII{\mbox{Fe\,\specchar{vii}}} 

\def\FeX{\mbox{Fe\,\specchar{x}}}

\def\FeXIII{\mbox{Fe\,\specchar{xiii}}}
\def\FeXIV{\mbox{Fe\,\specchar{xiv}}}

\def\HI{\mbox{H\,\specchar{i}}} 
 
\def\Hmin{\hbox{{\rm H}$^{^{_-}}$}}      
\def\HeI{\mbox{He\,\specchar{i}}} 
\def\HeII{\mbox{He\,\specchar{ii}}} 
 
\def\KI{\mbox{K\,\specchar{i}}}

\def\LaII{\mbox{La\,\specchar{ii}}}

\def\MgI{\mbox{Mg\,\specchar{i}}} 
\def\MgII{\mbox{Mg\,\specchar{ii}}}

\def\MnI{\mbox{Mn\,\specchar{i}}}

\def\NaI{\mbox{Na\,\specchar{i}}}

\def\NiI{\mbox{Ni\,\specchar{i}}} 
\def\NiII{\mbox{Ni\,\specchar{ii}}}

\def\OVII{\mbox{O\,\specchar{vii}}}
\def\OVIII{\mbox{O\,\specchar{viii}}}
 
\def\SiI{\mbox{Si\,\specchar{i}}}

\def\SiIV{\mbox{Si\,\specchar{iv}}}

\def\SmII{\mbox{Sm\,\specchar{ii}}} 
\def\SrI{\mbox{Sr\,\specchar{i}}}

\def\TiII{\mbox{Ti\,\specchar{ii}}}

\def\Halpha{\mbox{H\hspace{0.1ex}$\alpha$}} 

\def\Hbeta{\mbox{H\hspace{0.2ex}$\beta$}}
\def\Hgamma{\mbox{H\hspace{0.2ex}$\gamma$}}

\def\Hepsilon{\mbox{H\hspace{0.2ex}$\epsilon$}}
\def\Lyalpha{\mbox{Ly$\hspace{0.2ex}\alpha$}}
\def\Lybeta{\mbox{Ly$\hspace{0.2ex}\beta$}}

\def\Baalpha{\mbox{Ba$\hspace{0.2ex}\alpha$}}

\def\HeIDthree{\mbox{He\,\specchar{i}\,\,D$_3$}}

\def\NaDone{\mbox{Na\,\specchar{i}\,\,D$_1$}}

\def\NaID{\mbox{Na\,\specchar{i}\,\,D}}
\def\NaIDone{\mbox{Na\,\specchar{i}\,\,D$_1$}}
\def\NaIDtwo{\mbox{Na\,\specchar{i}\,\,D$_2$}}

\def\MgIb{\mbox{Mg\,\specchar{i}\,b}}

\def\MgIbtwo{\mbox{Mg\,\specchar{i}\,b$_2$}}

\def\CaIIK{\mbox{Ca\,\specchar{ii}\,\,K}}       
\def\CaIIH{\mbox{Ca\,\specchar{ii}\,\,H}}
\def\CaIIHK{\mbox{Ca\,\specchar{ii}\,\,H{\specand}K}}
\def\HK{\mbox{H{\specand}K}}
\def\Kthree{\mbox{K$_3$}}      

\def\Ktwo{\mbox{K$_2$}}
\def\Htwo{\mbox{H$_2$}}

\def\KtwoV{\mbox{K$_{2V}$}}

\def\HtwoV{\mbox{H$_{2V}$}}
\def\HtwoR{\mbox{H$_{2R}$}}

\def\CaIR{\mbox{Ca\,\specchar{ii}\,8542\,\AA}} 

\def\MgIIk{\mbox{Mg\,\specchar{ii}\,\,k}}

\def\MgIIhk{\mbox{Mg\,\specchar{ii}{\,h\specand}k}}
\def\hk{\mbox{h{\specand}k}}

\def\level #1 #2#3#4{$#1 \; ^{#2} \mbox{#3} ^{#4}$}   

\def\rma{{\rm a}}     
\def\rmb{{\rm b}} \def\rmB{{\rm B}}    
  
\def\rmd{{\rm d}} \def\rmD{{\rm D}} 
\def\rme{{\rm e}} \def\rmE{{\rm E}}

 \def\rmH{{\rm H}}

\def\rmp{{\rm p}}

\def\rms{{\rm s}}

\def\kms{\hbox{km$\;$s$^{-1}$}}

\def\tis{\!=\!\!}                          
\def\tapprox{\!\approx\!\!}                
\def\tsim{\!\sim\!\!}                      
\def\dif{\: {\rm d}}                       
\def\ep{\:{\rm e}^}                        
\def\lambdop{\hbox{$\bf \Lambda$}}         

\def\komega{($k, \omega$)}                 

\spacing{0.998}   

\def\rmit#1{#1}    
\def\paragraphrr#1{\paragraph*{#1.~~} \addcontentsline{toc}{subsection}{#1}}

\usepackage{endnotes}

\usepackage{mfirstuc}
\long\def\rrendnote#1#2#3{%
\phantomsection\label{#2main}
\stepcounter{endnote}%
\hyperref[#2]{[{\color{rrmaroon}{\bf {\em \theendnote\ #1\/}}}]}%
{\endnotetext{\phantomsection\label{#2}
\parindent=3ex
{\bf {\em \nopagebreak\xmakefirstuc{#1}\/}}\\[0.5ex]
{~\,#3}
~~[\Acrobatmenu{GoBack}{\color{rrmaroon}\em Back\/}]%
~~\hyperref[#2main]{[{\em Main call\/}]}%
}}}

\makeatletter
\renewcommand{\@makefnmark}{\hbox{\textsuperscript{\bf\scriptsize{\@thefnmark}}}}
\makeatother

\usepackage{footnotebackref}

\usepackage{xcolor,eso-pic}
\AddToShipoutPictureBG{
  \AtPageLowerLeft{%
    \ifnum\value{page}>18   
      \ifnum\value{page}<55  
        \textcolor{yellow!8!white}{\rule{\paperwidth}{\paperheight}}%
      \fi
    \fi
  }%
}

\bibnote{2003rtsa.book.....R}{~(RTSA)}
\bibnote{2020arXiv200900376R}{~(LAR-1)}
\bibnote{2021LingAstRep...3R}{~(LAR-3)}
\bibnote{1974SoPh...39...19H}{~(HOLMUL)}
\bibnote{1976ApJS...30....1V}{~(VALII)}
\bibnote{1981ApJS...45..635V}{~(VALIII)}
\bibnote{1986ApJ...306..284M}{~(MACKKL)}
\bibnote{1993ApJ...406..319F}{~(FAL)}
\bibnote{2008ApJS..175..229A}{~(ALC7)}

\def\SSIpage#1{\href{https://robrutten.nl/rrweb/rjr-edu/lectures/rutten_ssi_2020.pdf\#page=#1}{[SSI (pdf\,#1)]}}

\def\SSFpage#1{\href{https://robrutten.nl/rrweb/rjr-edu/lectures/rutten_ssf_2020.pdf\#page=#1}{[SSF (pdf\,#1)]}}

\def\SSXpage#1{\href{https://robrutten.nl/rrweb/rjr-edu/lectures/rutten_ssx_2020.pdf\#page=#1}{[SSX (pdf\,#1)]}}

\begin{document}  
\selectlanguage{english}

\twocolumn[{%
\mbox{}\vspace*{-8ex}
\includegraphics[width=27mm]{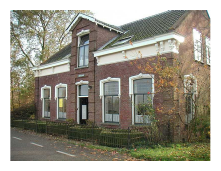}\\[-7mm]
\hspace*{30mm} {\sf Lingezicht Astrophysics Reports} 2 
~~--~~\href{https://arxiv.org/abs/2103.02369}{arXiv}/%
\href{https://ui.adsabs.harvard.edu/abs/2021arXiv210302369R/abstract}{ADS}
~~--~~
\today 
~~--~~
\href{https://robrutten.nl/rrweb/rjr-pubs/2021LingAstRep...2R.pdf}
{newest}\\[2ex]
\hspace*{30mm} {\large \em Solar Activities and their influences in
the Heliosphere and Planetary Atmospheres}\\[0.5ex]
\hspace*{30mm} {Workshop NIT Calicut, India, March 8-14 2021}

\begin{center}
\vspace*{4ex} {\Huge \bf Compendium solar spectrum
formation}\\[2ex]   
{\large \bf Robert J. Rutten$^{1, 2, 3}$}\\[2ex]
\begin{minipage}[t]{15cm} \small 
\mbox{}\hspace{20ex}$^1$ Lingezicht Astrophysics, Deil, The
Netherlands ~~\href{mailto:robenrietjerutten@gmail.com}{email}
~~\href{https://robrutten.nl}{website}\\
\mbox{}\hspace{20ex}$^2$ \href{http://www.mn.uio.no/astro/english}
{Institute of Theoretical Astrophysics},
University in Oslo, Oslo, Norway\\
\mbox{}\hspace{20ex}$^3$ \href{https://www.mn.uio.no/rocs/english}
{Rosseland Centre for Solar Physics},
University in Oslo, Oslo, Norway\\

\mbox{} \hspace{90mm}
\begin{parbox}[t]{100mm}
{\sl 
Myself when young did eagerly frequent\\
doctor and saint, and heard great argument\\
about it and about; but evermore\\
came out by the same door as in I went.}\\[0.7ex]
\end{parbox}
\mbox{} \hspace{110mm}  {\footnotesize Omar Khayyam (1048--1131)}\\[1ex]

{\bf Abstract.~} \small The solar spectrum conveys most of our
diagnostics to find out how our star works. 
They must be understood for utilization but solar spectrum formation
is complex because the interaction of matter and radiation within the
solar atmosphere suffers non-local control in space, wavelength, and
time. 
These complexities are summarized and illustrated with classic
literature. 
They combine in chromospheric spectrum formation.
\end{minipage}
\end{center}
}] 

\noindent
\begin{minipage}[t]{0.9\columnwidth}  
  {\small \tableofcontents}
  \label{sec:contents}
\end{minipage}


\section*{{\rm \bf Yellow-page detailing notes, roughly grouped}}
\begin{small} 
\label{tab:notes} 
\begin{itemize}
\item[\ref{note:books}] other texts
\item[\ref{note:silly}] ADS page opening, silly publishers
\vspace{1.4ex}
\item[\ref{note:bosons}] quanta, bosons, photons
\item[\ref{note:bandwidth}] spectral bandwidth and vision
\item[\ref{note:Tb}] equivalent temperatures
\item[\ref{note:absint}] intensity calibrations
\item[\ref{note:Lockyer}] Schuster-Schwarzschild approximation
\item[\ref{note:ME}] Milne-Eddington approximation
\item[\ref{note:response}] contribution, response, sensitivity
\item[\ref{note:extcont}] extinction diagrams, H-minus
\item[\ref{note:deepest}] deepest wavelengths
\vspace{1.4ex}
\item[\ref{note:wlcorona}] white-light corona
\item[\ref{note:forbidden}] optical coronal lines
\item[\ref{note:dielectronic}] dielectronic ionization and recombination
\item[\ref{note:EM}] emission measure, DEM
\item[\ref{note:DEM}] multi-DEM analysis 
\item[\ref{note:cloudmodeling}] cloud modeling
\item[\ref{note:CEcloud}] coronal clouds
\vspace{1.4ex}
\item[\ref{note:sqrteps}] \!$\sqrt{\varepsilon}$ law
\item[\ref{note:flatSl}] \!$\sqrt{\varepsilon}$ validity
\item[\ref{note:epssmall}] epsilon is small: ALI, acceleration, 
  net radiative bracket
\item[\ref{note:aureoles}] scatter aureoles 
\item[\ref{note:photoelectric}] no Thomas photoelectric control 
\item[\ref{note:bdef}] beware of Menzel departure coefficients
\item[\ref{note:pumping}] photon pumping, lasering, Wien transcription 
\item[\ref{note:suction}] photon loss, no nanoflame heating, photon suction
\item[\ref{note:interlocking}] spectral interlocking
\item[\ref{note:haze}] line haze modeling 
\item[\ref{note:detourdist}] detour and cross redistribution
\vspace{1.4ex}
\item[\ref{note:interlocking}] interlocking \FeI, CeII, \FeII\ lines
\item[\ref{note:6302-4571}] \NiI\,6768, \MgI\,4571, \FeI\,6302, 
   \FeI\,6173\AA
\item[\ref{note:4554}] \BaII\,4554\,\AA
\item[\ref{note:4554}] \MnI\ lines as activity monitor
\item[\ref{note:suction}] \NaI\ and \KI\ resonance lines
\item[\ref{note:NaD}] \NaID, \MgIb, \CaIR\ heights
\item[\ref{note:Halpha-high}] \Halpha\ photosphere and chromosphere 
                              visibilities
\item[\ref{note:HKreversals}] \CaII\ \HK\ core reversals
\item[\ref{note:gillespie}] \CaII\ \KtwoV\ and CN spectroheliograms
\item[\ref{note:MgI12micron}] \MgI\ 12\,micron emission lines
\item[\ref{note:Lockyer}] \HeIDthree\ off-limb
\item[\ref{note:HeI}] optical \HeI\ lines
\item[\ref{note:Rydberg}]  Rydberg \HI\ lines
\item[\ref{note:ALMA-line}] Rydberg \HI\ candidate for ALMA
\vspace{1.4ex}
\item[\ref{note:REphot}] RE upper photosphere
\item[\ref{note:abundance}] classic abundance determination
\item[\ref{note:masking}] NLTE masking 
\item[\ref{note:VAL-FeII}] Fe ionization in VALIIIC
\item[\ref{note:ALC7plots}] triptychs ALC7 line formation
\item[\ref{note:Ha-RE}] Ha--Ha scattering
\item[\ref{note:CO}] cool COmosphere
\item[\ref{note:runaway}] runaway cooling
\vspace{1.4ex} 
\item[\ref{note:CSshocks}] Carlsson-Stein shocks
\item[\ref{note:4pdiagrams}] Carlsson four-panel breakdown
\item[\ref{note:clapotis}] internetwork clapotisphere
\item[\ref{note:INshocksHa}] internetwork shocks in \CaIIH\ and \Halpha\
\item[\ref{note:ALMA-waves}] internetwork acoustics with ALMA
\vspace{1.4ex}
\item[\ref{note:MCs}] MCs, network, plage, faculae
\item[\ref{note:1600-1700}] bright-point shifts 1600--1700\,\AA\
\item[\ref{note:basal}] internetwork fields 
\vspace{1.4ex} 
\item[\ref{note:Lockyer}] Lockyer chromosphere, \HeIDthree, flash color
\item[\ref{note:spicules-II}] spicules-II
\item[\ref{note:nonEchrom}] RBE--RRE--fibril ionization-recombination
\item[\ref{note:quietHa}] quiet-Sun \Halpha\ scenes
\item[\ref{note:triples}] quiet-Sun SDO images
\item[\ref{note:ionization}] photosphere, chromosphere, corona: 
  hydrogen ionization 
\item[\ref{note:noTR}] no transition region 
\item[\ref{note:sheaths}] transition radiation, sheath ionization, 
coronal contact cooling
\item[\ref{note:campfires}] SolO campfires
\item[\ref{note:CBPfeet}] CBP foot visibility
\vspace{1.4ex} 
\item[\ref{note:IBIS}] IBIS--SDO active-region blinker
\item[\ref{note:rain}] dark EUV features
\item[\ref{note:DFs}] dynamic fibrils
\item[\ref{note:moss}] active-region moss
\item[\ref{note:EBs}] EBs and FAFs
\item[\ref{note:around-upward}] around and upward heating
\item[\ref{note:FIP}] FIP effect
\item[\ref{note:Lyalpha}] \Lyalpha\ dream
\item[\ref{note:Lya-features}] \Lyalpha\ features
\item[\ref{note:blobs}] filament blobs
\vspace{1.4ex} 
\item[\ref{note:ALMA-temp}] ALMA as thermometer
\item[\ref{note:ALMA-waves}] internetwork acoustics with ALMA
\item[\ref{note:ALMA-chrom}] quiet chromosphere with ALMA
\item[\ref{note:ALMA-line}] Rydberg \HI\ candidate for ALMA
\item[\ref{note:ALMA-Ha}] ALMA\,--\,\Halpha\,--\,SDO alignment
\item[\ref{note:ALMA-GONG}] ALMA\,--\,GONG\,--\,SDO alignment
\item[\ref{note:ALMA-SOLIS}] ALMA\,--\,SOLIS\,--\,SDO alignment
\vspace{1.4ex} 
\item[\ref{note:atlases}] optical spectrum atlases and line lists
\item[\ref{note:ppmodels}] plane-parallel atmosphere files
\item[\ref{note:abundance}] Kieler program for 1D LTE spectrum synthesis
\item[\ref{note:Pandora}] Pandora program for 1D NLTE spectrum synthesis
\item[\ref{note:MULTI}] MULTI program for 1D NLTE spectrum synthesis
\item[\ref{note:MULTI3D}] MULTI3D program for 3D NLTE spectrum synthesis
\item[\ref{note:RH}] RH program for 1D-2D-3D NLTE--PRD spectrum synthesis
\item[\ref{note:Bifrost}] Bifrost program for 3D radiative MHD simulation
\item[\ref{note:shortcuts}] chromospheric radiation shortcuts
\item[\ref{note:Bifroststar}] public Bifrost star
\item[\ref{note:CHIANTI}] CHIANTI package for CE analysis
\item[\ref{note:IDL}] my IDL programs
\item[\ref{note:SDO-STX}] SDO\,--\,STX alignment programs
\vspace{1.4ex}
\item[\ref{note:telescopes}] solar telescopes 
\item[\ref{note:restoration}] solar image restoration
\item[\ref{note:neutrinos}] solar neutrinos
\item[\ref{note:p-modes}] seismology success story
\item[\ref{note:g-modes}] no gravity-modes success story
\item[\ref{note:waves}] no coronal heating by acoustic or gravity waves
\item[\ref{note:HKreversals}] Wilson-Bappu effect
\item[\ref{note:basal}] basal flux
\vspace{1.4ex} 
\item[\ref{note:unsold}] LTE diehard
\item[\ref{note:grancontrast}] illustrious quartet
\vspace{1.4ex} 
\item[\ref{note:personal}] personal background
\item[\ref{note:posting}] posting course notes
\item[\ref{note:projection}] projection technology
\item[\ref{note:mainframes}] mainframe computing
\item[\ref{note:PC}] personal computing
\item[\ref{note:Bilderberg}] Bilderberg study week on the quiet photosphere
\end{itemize}
\end{small}

\section{Preamble: on-line format and on-line material} 
\label{sec:materials}
This is a compact graduate-level course on solar spectrum
formation.\footnote{Not treated:
\label{note:except}
high-energy spectra, radio spectra, polarization spectra, particle
spectra.
For high-energy radiation start with
\citetads{1986rpa..book.....R}. 
For radio diagnostics start with
\citetads{2004ASSL..314...71G}; 
they were treated by P.K.\,Manoharan in this school.
For polarization and spectropolarimetry (treated by Rohan Eugene Louis
and Debi Prasad Choudhary)
\citetads{2007insp.book.....D} 
is a clear introduction,
\citetads{2004ASSL..307.....L} 
the bible, C.U.\,Keller's lecture at the
\href{https://robrutten.nl/uso-dwingeloo-2009/Home.html}{2009
Dwingeloo school} a summary.
For high-energy particles see
\citetads{2021LNP...978.....R}. 
Neutrinos are largely done (\rrref{endnote}{note:neutrinos}).}
It is display-oriented adding many figures to basic equations but for
brevity and to avoid repetition this text contains only few figures
and equations,
instead page-linking to webposted
\href{https://robrutten.nl/Astronomy_course.html}{teaching
material} and \href{https://ui.adsabs.harvard.edu}{open-access
publications}.\footnote{Link colors: weblinks are blue, internal
crosslinks maroon. 
The suggestion is to open blue items till you are blue in the face --
meaning lack of sub-skin red scattering ('tis all scattering here
\SSFpage{116}) because your blood drained too far from your brain
\SSFpage{120}.}
I therefore recommend computer-screen
pdf reading with web access.\footnote{Page opening: weblinks
\label{note:openpage} to specific pdf pages (as the SSF ones in the
footnote above) should serve the specified page, best in a browser
window beside your pdf reader. 
Adobe Acrobat/Reader may need web access in Preferences > Trust
Manager. 
The built-in pdf readers of Chrome and Firefox for PCs handle pdf page
opening properly but macOS readers and pdf readers for tablet computers
may instead hang at the first page. 
With such poor pdf readers you must manually go to the pdf page specified
in each page link. 
At slow connection it may be better to first download my files in
\rrref{Table}{tab:abbreviations} and open them in parallel.}

The links make this compendium Wikipedia-style multilayer: you may
delve deeper by opening linked graphs, equations, references that suit
your interest.
In addition this is a two-tier text with a relatively straightforward
presentation in the main text
(p.\,\pageref{sec:basics}-\pageref{sec:conclusion}, contents above)
but detailing specific topics in yellow-page
elaborations\footnote{Navigation: endnotes are linked in the
main text with maroon bold-italics topic definers.
Each ends with a return link [Main call] to avoid getting marooned
there and a link [Back] that also returns there or back to another
link that you used -- working in some pdf readers but not all, then
try the back button.
Footnotes are just that (plenty, my trademark I'm told).
Their number jumps back to their call.}
(p.\,\pageref{sec:notes}-\pageref{sec:endendnotes}, list
above).\footnote{Selection: personal, favoring topics that I
have been involved in and references and figures which I knew best and came
first to mind in this 
writeup for the 
{Calicut school}. 
Plus reminiscences from my half century in solar physics.}
There is no subject or name index because your pdf reader offers search.
Items easily found in Wikipedia are not detailed here.

The links and two-tier format invite study at different levels:
\begin{itemize}  \vspace{-0.5ex} \itemsep=1ex

\item read only the six pages on radiative transfer basics 
(\rrref{Sect.}{sec:basics})
and go through the corresponding
\href{https://robrutten.nl/rrweb/rjr-edu/lectures/rutten_ssf_2020.pdf}{SSF}
equation displays for a theory overview (as I did in my Calicut zoom
lecturing). 
Readers only interested in optically thin spectrum formation may stop
already at the separation point on page \pageref{sec:SP}.
For thicker readers the triptychs of line formation in the didactic
ALC7 star (\rrref{Sect.}{sec:Avrett}) starting at \SSXpage{78} and
described in \rrref{endnote}{note:ALC7plots} are suited exercise
material, with exam in \SSFpage{113};

\item also read the main text of
\rrref{Sects.}{sec:Holweger}\rrref{\,--}{sec:Oslo}
for classic developments
LTE$\,\rightarrow\,$NLS$\,\rightarrow\,$NLW$\,\rightarrow\,$NLT.
For photospheric abundancers, irradiancers, inverters the
breakpoint is on page \pageref{sec:SP2};

\item also study detailing yellow-page endnotes and their sources on
topics of interest to you;

\item brainwash your brain by draining mine in reading
everything and opening each linked page and studying all references
for a full-fledged full-semester comprehensive course on many
intricacies of solar spectrum formation and some underlying solar
physics, with this text serving as convenient pointer and page and
reference opener;

\item when no longer a student but a researcher (doctor?) and/or teacher
(saint?) this text may serve as page- and reference-opening resource in
taking your pick in Khayyam-wise refreshing arguments you heard before.
\end{itemize} 
My \href{https://robrutten.nl/Astronomy\_course.html}{online teaching
material} used here amounts to nine files listed at the top
of \rrref{Table}{tab:abbreviations}
\rrendnote{personal background}{note:personal}{%
I developed this material in decades of teaching at Utrecht University
and abroad. 
In the 1960s I learned solar spectrum formation at Utrecht from
C. (Kees) de
Jager\footnote{Remembrance:
\label{note:CdeJ}
\href{https://robrutten.nl/astronomershots/album1967/kees-de-jager.jpg}
{Cornelis (Kees) de Jager} (1921\,--\,2021) achieved fast expansion of
Utrecht astronomy after M.G.J.\,Minnaert (including my job), started
Flemish astrophysics at Brussels, started Dutch space research, was
instrumental in starting ESRO and its transition to ESA, founded {\em
Space Science Reviews\/} and co-founded {\em Solar Physics}, was IAU
General Secretary, twice president of COSPAR, president of ICSU, and
much more -- including running marathons and giving public lectures up
to record age.
I honored his 80th birthday with a review of solar atmosphere modeling
(\citeads{2002JAD.....8....8R}) 
and his 100th birthday with an SDO triptych album in LAR-3
(\citeweb{https://robrutten.nl/rrweb/rjr-pubs/2021LingAstRep...3R.pdf}
{2021LingAstRep...3R}).
He passed away four weeks later.
I then put his thesis on solar hydrogen lines
(\citeads{1952RAOU....1.....D}) 
on ADS with help from Oslo and co-authored an ``In Memoriam'' for {\em
Solar Physics\/}
(\citeads{2022SoPh..297...15R}), 
a misedited affair ({\tt cdejsp\_log.txt} in the tarred
\href{https://arxiv.org/e-print/2201.11496} {arXiv source}).}.
Our bible was Uns\"old's book (\rrref{endnote}{note:unsold}) but NLTE
theory had started and was loudly advocated to us by A.B.\,Underhill,
then also professor at Utrecht. 
The guru was R.N.\,Thomas but I found his writings 
incomprehensible and Kees much clearer\footnote{Just as R.M.\,Bonnet
who wrote {\em ``As seen through the writings of Thomas, radiative
transfer for me would have remained an unintelligible theoretical
exercise forever''\/} and found De Jager much clearer, acknowledging
{\em ``the clarity of Kees' views''\/}
(\linkadspage{1996SoPh..169..233B}{2}{page~2} of
\citeads{1996SoPh..169..233B}). 
More Kees versus Thomas in \rrref{footnote}{note:Thomas+Athay}.}.
Jefferies' (\citeyearads{1968slf..book.....J}) 
book\footnote{\citetads{1968slf..book.....J} 
was scanned by A.V.\,Sukhorukov and put on ADS by me with Jefferies'
consent, enabling page openers here.}
was a welcome improvement on Thomas-speak\footnote{Telltale specimen:
NASA ``orange book'' {\em Stellar Atmospheric Structural Patterns\/}
(\citeads{1983NASSP.471.....T}). 
I called it \linkrjrpage{1985SSRv...41..394T}{2}{Nebraskan Franglais}
(\citeads{1985SSRv...41..394T}); 
\citetads{1996BAAS...28.1465J} 
called it ``hard-to-read vintage Thomas''; L.E.\,Cram suggested to
read it a dozen times to get its meaning.
ADS lists 26 citations versus 250 for
\citetads{1968slf..book.....J}, 
408 for \citetads{1970stat.book.....M}, 
2097 for \citetads{1978stat.book.....M}. 
However, Thomas (1921\,--\,1996) was truly a game-changer in
stellar-atmosphere RT.~ JILA (which he co-founded) celebrated his
centennial while I wrote this which made me reread my
\href{https://robrutten.nl/rrweb/rjr-pubs/2003-thomas-epsilon.pdf}
{1997 Thomas memorial} and realize that in publishing this link-strewn
compendium I finally fulfill my
\linkrjrpage{2003-thomas-epsilon}{9}{``On publishing''} promise
there.}, with the \linkadspage{1968slf..book.....J}{6}{preface}
stating that the then literature ``can well be criticized as being
overly esoteric''.

I had started on solar radio diagnostics
(\citeads{1967BAN....19..254F}) 
but switched to optical spectrum formation thanks to De Jager's
lecturing and then concentrated on the chromosphere thanks to
J.\,Houtgast who took me as junior help to both 1966 eclipses to
record the flash spectrum.  
Half a century later the chromospheric spectrum still fascinates me,
as evident in \rrref{Sect.}{sec:chromosphere} which is more status
report than course. 

C. (Kees) Zwaan\footnote{Remembrance:
\href{https://robrutten.nl/robshots/Kees-Zwaan.jpg}{Cornelis (Kees)
Zwaan} (1928--1999) was my thesis adviser and a close friend.  
He was a phenomenal educator with main interest in solar and stellar
magnetism wrapped up in Schrijver and Zwaan
(\citeyearads{2000ssma.book.....S}, 
\citeyearads{2008ssma.book.....S}). 
I not only took over his teaching but also his
\href{https://robrutten.nl/dot/DOT_home.html}{DOT} stewardship.
The revolutionary ``open principle'' of the DOT (no vacuum but rely on
wind flushing to avoid convective turbulence from heating in the
convergent beam to focus) was his idea and proposal to
\href{https://robrutten.nl/dot/dotweb/dot-albums/photographs/2005-dt-dot-rh.jpg}{R.H.\,Hammerschlag}. 
DKIST is sort of open.}
took over the Utrecht radiative transfer courses in the 1970s,
modernized RTSA and set up IART. 
I copied much of his (neatly hand-written Dutch) material when I took
over IART in 1985, RTSA in 1994 with English-version webposting in
1995 (\rrref{endnote}{note:posting}).
I then also started
\href{https://robrutten.nl/rrweb/rjr-elsewherecourses/dircontent.html}{teaching
abroad}, appreciating getting to know an appreciable fraction of later
colleagues worldwide.\footnote{I much missed face-to-face contact with
the students in this zoom school.
Just talking to your own screen is no good. 
I also missed who-what-where student introductions and joint meal and
pub sessions.}

I quit teaching at Utrecht at my mandatory retirement in 2007. 
In 2011 Utrecht University
\href{https://robrutten.nl/Closure_Utrecht.html}{abruptly quit its
astronomy} (even its A\&A and ApJ subscriptions) and effectively
killed \href{https://robrutten.nl/Utrecht_solar.html}{Dutch solar
physics}.
The \href{https://robrutten.nl/dot/DOT_home.html}{DOT} is mothballed
since. 
I still \href{https://robrutten.nl/My_courses.html}{teach abroad} when
invited but without Utrecht affiliation.
So it goes.}. 

The first three are lecture notes.
\href{https://robrutten.nl/rrweb/rjr-pubs/1993-issf-notes.pdf}{ISSF}
was a bachelor-level introduction to
\href{https://robrutten.nl/rrweb/rjr-pubs/2018arXiv180408709R.pdf}{Sac
Peak} summer students in 1993.
\href{https://robrutten.nl/rrweb/rjr-pubs/2015-iart-notes.pdf}{IART}
was my Utrecht course for second-year bachelors students, written in
Dutch in the 1980s, translated by R.C. Peterson in 1992 and revived by
L.H.M.~Rouppe van der Voort in 2015.
It follows \citetads{1986rpa..book.....R} 
with most emphasis on explaining their first chapter summarizing basic
RT up to NLS treatment.
\href{https://robrutten.nl/rrweb/rjr-pubs/2003rtsa.book.....R.pdf}{RTSA}
was initially written in 1995 for my Utrecht masters course, intended
as an easier-to-read rehash of
\citetads{1970stat.book.....M} 
\rrendnote{other texts}{note:books}{%
A consumer-report inventory of relevant textbooks is in the
\RTSApage{17}{bibliography}.
I found \citetads{1970stat.book.....M} 
easier to read then
\citetads{1978stat.book.....M}. 
Newer books are Oslo-school-triggered
\citetads{2007rahy.book.....C} 
and comprehensive
\citetads{2014tsa..book.....H}, 
both more advanced than this course.
The textbook on heliophysics of
\citetads{2019arXiv191014022S} 
is complementary: no radiative transfer, barely photosphere
and chromosphere but much about what the Sun gives us beyond
spectrum-formation puzzles.  The lectures on solar magnetism 
at the \href{https://robrutten.nl/uso-dwingeloo-2009/Home.html}
{2009 Dwingeloo school} are also complementary.}
with emphasis on solar NLS $\sqrt{\varepsilon}$ scattering and adding
NLW in particular for solar bound-free continua.
I immediately made it web-available 
\rrendnote{posting course notes}{note:posting}{%
At my RTSA completion in the spring of 1995 Mats Carlsson suggested to
ftp the postscript files to Oslo for printing as handout in his first
Oslo RT school. 
I did and realized I might spread them likewise for printing by
anybody, asked our software engineer how to make a website (``You? 
In latex because that's all you know'') and started
\href{https://robrutten.nl}{my website}.\footnote{Still made per
latex. 
After starting this compendium I migrated from serving by Utrecht
University to AWS S3 for higher speed (\rrref{footnote}{note:speeds})
and reliability.
Indeed a year later Utrecht University suddenly killed my account.}
This was my start in self-publishing; RTSA should have been
LAR-1.\footnote{Except that part was written away from {\em
Lingezicht\/} when a quarter-million Dutch evacuated central Holland
in fear of the Rhine.} 
Initially I offered it in 10-page chunks for piecemeal ftp and
printing and announced it by email to colleagues, then put an ad in
SolarNews in 1997 and finally asked ADS to link to it as single pdf
file in 2003 (\citeads{2003rtsa.book.....R}). 
It still says ``Not yet'' in enough places to forego publisher offers
to print it. 
This compendium (also self-published but arXived and hence on ADS)
represents an overdue update.

This RTSA experience taught me that the effort of formal publishing
isn't needed to spread course notes beyond the course; later I also
posted
\href{https://robrutten.nl/rrweb/rjr-pubs/1993-issf-notes.pdf}{ISSF}
and
\href{https://robrutten.nl/rrweb/rjr-pubs/2015-iart-notes.pdf}{IART}
(yet to be ADSed).
However, at the time I didn't yet think of reading on-screen whereas
now I work paper-free\footnote{If you still say ``paper'' for a
publication or even for a presentation or if you read this text on
non-linking paper after printing it you are likely older 
than me, maybe in spirit.} (except when browsing classic
monographs from the {\em Lingezicht\/} library) and expect active links
in what I read\footnote{Publishers
\label{note:vulture}
reselling printed works anew as electronic files should earn their
predator/vulture profit by activating crossreferences and hyperlinking
citations and likewise retroactively update all files that they sell.
Textbooks that aim at on-line study should add pdf page openers
wherever they refer to earlier results.
Research publications discussing other's results should also do
this.}.
With the take-or-leave link-rich format of this text I wikipedia-style my
course offerings.\footnote{It has been suggested that {\tt html5},
{\tt reveal.js} and {\tt SVG} suit better than slow pdf opening but
while ADS and arXiv serve pdfs that's my way (comforted by
\rrref{footnote}{note:slow}).
Full wikistyling would let you correct and expand -- next?}}.
The 2003 update (still incomplete) is linked at
\href{https://ui.adsabs.harvard.edu/abs/2003rtsa.book.....R/abstract}{ADS}
and therefore most frequently used but at Oslo and elsewhere IART
is preferred as broader course material.
None of these courses includes NLT nor multi-level detours while these
are now appreciated as key ingredients of chromospheric spectrum
formation.
I included them in more recent summary tutorials
(Rutten \citeyearads{2017IAUS..327....1R}, 
\citeyearads{2019SoPh..294..165R}) 
and do so here: this text represents an update.

\begin{table}
\caption[]{Acronyms and abbreviations, roughly grouped.
\label{tab:abbreviations} \vspace*{-2ex}}
\fbox{
\begin{small}
\begin{tabular}[t]{l}
\href{https://robrutten.nl/rrweb/rjr-pubs/1993-issf-notes.pdf}{ISSF} = introduction to solar spectrum formation\\
\href{https://robrutten.nl/rrweb/rjr-pubs/2015-iart-notes.pdf}{IART} = introduction to astrophysical radiative transfer \\
\href{https://robrutten.nl/rrweb/rjr-pubs/2003rtsa.book.....R.pdf}{RTSA} = radiative transfer in stellar atmospheres\\[1ex]

\href{https://robrutten.nl/rrweb/rjr-edu/lectures/rutten_ssi_2020.pdf}{SSI} = introduction to solar spectrum formation\\
\href{https://robrutten.nl/rrweb/rjr-edu/lectures/rutten_ssf_2020.pdf}{SSF} = solar spectrum formation theory\\
\href{https://robrutten.nl/rrweb/rjr-edu/lectures/rutten_ssx_2020.pdf}{SSX} = solar spectrum formation examples\\[1ex]

\href{https://robrutten.nl/rrweb/rjr-edu/exercises/ssa/ssa.pdf}{SSA} = stellar spectra A (Cannon\,--\,Payne\,--\,Minnaert)\\
\href{https://robrutten.nl/rrweb/rjr-edu/exercises/ssb/ssb.pdf}{SSB} = stellar spectra B (Avrett \,--\,Chandrasekhar\,--\,Uns\"old)\\
\href{https://robrutten.nl/rrweb/rjr-edu/exercises/rtsa-practical-2003/exercises.pdf}{SSC} = stellar spectra C 
[Mihalas\,--\,Judge\,--\,Feautrier]\\[1ex]

\href{https://robrutten.nl/rrweb/rjr-pubs/2020LingAstRep...1R.pdf}{LAR-1}
= {\em Lingezicht Astrophysics Report 1\/} 
(\citeads{2020arXiv200900376R})\\[1ex]  

RT = radiative transfer $\sim$ summed fermion-boson changes of $I_\nu$\\[1ex]

TE = thermodynamic equilibrium $\sim$ detailed balance in all\\
SB = Saha-Boltzmann population equilibrium ($\rightarrow$ $S\!_\nu = B_\nu$)\\
SE = statistical equilibrium $\equiv$ sum net population rates zero\\
LTE = local thermodynamic equilibrium $\equiv$ SB populations\\
NLTE = no LTE = non-LTE $\sim$ no SB but still SE\\
NSE = no SE = non-E = NEQ $\sim$ populations with memory\\[1ex]

RE = radiative equilibrium $\sim \int \alpha_\nu\,[S\!_\nu-J_\nu]\dif\nu=0$\\
CE = coronal equilibrium $\sim$ SE + thin collisional creation\\[1ex]

DEM = differential emission measure = CE density part $j_\nu$\\
EM = emission measure $\equiv \int \mbox{DEM}(T) \dif T$\\
FIP = first ionization potential = neutral-atom ionization energy\\[1ex]
 
NLS = non-local in space $\sim$ $J_\nu$ important\\
NLW = non-local in wavelength $\sim$ other $J_\nu$ important\\
NLT = non-local in time $\sim$ NSE with some memory\\[1ex]

CS = coherent scattering $\sim$  monofrequent, monochromatic\\
CRD = complete redistribution $\sim$ resample extinction profile\\
PRD = partial redistribution $\sim$ mix CS and CRD\\[1ex]

HD = hydrodynamics $\sim$ no magnetism\\
MHD = magnetohydrodynamics $\sim$ single-fluid HD + B\\[1ex]

UV = ultraviolet \eg\ AIA 1600 \& 1700\,\AA\\
EUV = extreme ultraviolet \eg\ shorter AIA wavelengths\\[1ex]

DOT = Dutch Open Telescope @ La Palma, 0.45\,m\\
SST = Swedish 1-m Solar Telescope @ La Palma, 1\,m\\
DKIST = Daniel K. Inouye Solar Telescope @ Maui, 4\,m\\

SDO = Solar Dynamics Observatory @ space\\
AIA = Atmospheric Imaging Assembly @ SDO\\
HMI = Helioseismic \& Magnetic Imager @ SDO\\
IRIS = Interface Region Imaging Spectrograph @ space\\
ALMA = Atacama Large Millimeter Array @ Chajnantor\\[1ex]

MC = small kilogauss magnetic concentration (``fluxtube'')\\
RBE = rapid blue-shifted excursion in blue wing \Halpha\\
RRE = rapid red-shifted excursion in red wing \Halpha\\
CBP = coronal bright point $\sim$ small hot quiet-Sun loop group\\
EB = Ellerman bomb (or burst, heed \rrref{footnote}{note:bomber})\\
FAF = flaring active-region fibril \eg\ in AIA\,1600\,\AA
\end{tabular} 
\end{small}  } 
\end{table}

The next three files in \rrref{Table}{tab:abbreviations} are sets of
projection displays that I use in teaching since the passing of the
viewgraph era\footnote{The
\href{https://robrutten.nl/rrweb/rjr-pubs/2003rtsa_equations.pdf}{RTSA
equation compendium} still supplies 90 viewgraph pages copying all
RTSA's 545 labeled equations.
The first few are empty. 
I used to overhead-project (\rrref{endnote}{note:projection})
them next to the blackboard.}
\rrendnote{projection technology}{note:projection}{%
When I was a student boxes of slides were brought by teachers to
class, by speakers to meetings.
A ``slide'' was a black-and-white or color transparency made from a
negative (emulsion on a glass plate, later 35-mm film) exposed in a
camera and chemically reversed (``developed'') into positive and
mounted in a frame holder to slide it into the focal plane of the
enlarging projector lens.\footnote{Kodak carousel projectors didn't
slide but dropped the transparency from a circular ``tray'' permitting
repeating sequence showing.}
Hence the funny name.\footnote{Dutch ``lantaarnplaatje'', later
``diapositief'' or ``dia'' was better. 
If you say ``next slide'' for your next presentation display you are a
worse mastodon than me.}

In the late 1970s Fresnel-lens overhead projectors for
page-size\footnote{Europeans using A4 had to take care not to exceed
letter-size projecting in the US.}
transparencies (viewgraphs, viewfoils) took over. 
They were better than the fixed order of both antique slide and modern
viewscreen projection\footnote{Dutch called viewgraphs ``sheets'' and
call a computer projector ``beamer'' but in the US that is a BMW
motorcycle or car.} in easy shuffling, skipping, overlap showing,
and the bodily obviousness of the speaker passing to the next
one.\footnote{Some speakers had the obnoxious habit of covering the
not yet discussed part of the viewgraph (called ``strip-teasing'' by
\href{https://robrutten.nl/stuff/vdkruit-hints.pdf}{P.\,van\,der\,Kruit}),
unfortunately sometimes aped by modern powerpointers.
Some boring teachers put their whole course on a continuous polyester
roll without page-selection liveliness in repeating the same course
every year.}

The menu openers of
\href{https://robrutten.nl/rrweb/rjr-edu/lectures/rutten_ssi_2020.pdf}{SSI},
\href{https://robrutten.nl/rrweb/rjr-edu/lectures/rutten_ssf_2020.pdf}{SSF},
\href{https://robrutten.nl/rrweb/rjr-edu/lectures/rutten_ssx_2020.pdf}{SSX}
restore selection flexibility in my teaching.
Their selection and pointing per on-screen cursor made me forego using
a bamboo stick or laser pen 
before covid-19 zoom practice.

Perhaps the covid-19 transition from 3D reality to 2D video and also the
hyperjump from 3D printed-text reading to 2D page linking
exploited here will in future transform to 3D augmented-virtuality
attending and many-D avatar observation and simulation exploring.}.
\rrref{Figure}{fig:quantities} shows an example.
\href{https://robrutten.nl/rrweb/rjr-edu/lectures/rutten_ssi_2020.pdf}{SSI}
is bachelors level,
\href{https://robrutten.nl/rrweb/rjr-edu/lectures/rutten_ssf_2020.pdf}{SSF}
and
\href{https://robrutten.nl/rrweb/rjr-edu/lectures/rutten_ssx_2020.pdf}{SSX}
are masters--graduate level.
These also contain newer aspects.
I showed a selection in my Calicut lectures.

The final three linked files in \rrref{Table}{tab:abbreviations} are
practicals. 
They date back to the 1990s\footnote{Secondary teaching goals were to
start students on IDL
(\href{https://robrutten.nl/rrweb/rjr-edu/manuals/idl-simple-manual.html}
{beginner instruction}),
\href{https://robrutten.nl/rrweb/rjr-edu/exercises/image-fourier-2006/dircontent.html}{image
and Fourier processing},
\href{https://robrutten.nl/rrweb/rjr-edu/manuals/idl-cube-manual.html}
{time-sequence analysis}, article-like
\href{https://robrutten.nl/Report_recipe.html}{report writing} using
latex and bibtex, giving
\href{https://robrutten.nl/stuff/vdkruit-hints.pdf}{presentations}.} 
and therefore use IDL but are easily coded in Python
\rrendnote{mainframe computing}{note:mainframes}{%
Utrecht University first had Dutch-made Electrologica X1 and X8
mainframes, then modernized to a CDC 6600 from lobbying by later
Nobelist M.J.G.\, Veltman
 who knew it from CERN.
It was the first supercomputer but far less powerful than the one in
your smartphone. 

For the Electrologica machines one delivered Algol programs punched
into papertape rolls in brown bags, for the CDC Fortran pograms
punched into Hollerith cards in brown boxes, to the job-submission
desk at the computer center, a large climate-controlled building with
whitecoated staff tending the single (but many-cabinet) holy-cow
machine. 
The output from the chain printer (per line slamming a chain of
upfront-spinned characters on fanfold paper at 1000 lines/minute)
was then collected from the computer center; towering stacks of
printout\footnote{With only uppercase letters so not suited for
manuscript printing. 
Slower daisywheel printers enabled that later and made me switch
wavelengths from {\AA}ngstrom to nm until latex and laserprinters
permitted return to {\AA}ngstrom (and classic cgs, also used here
following Mats Carlsson's didactic principle to expose students to
unit variety).}
adorned many offices.

Multilevel NLTE runs as my 15-level \FeII\ for
\citetads{1980ApJ...241..374C} 
took a full week or more from getting low priority in queued time
sharing by booking an overestimate of the required processor time --
the job was killed at request overrun.
Two-minute jobs had guaranteed return within the day, eight-minute
jobs overnight, mine waited on those (mostly black-box runs of SPSS by
sociology students) and primarily got done in weekends.\footnote{I
then controlled my jobs from home with an expensive
\href{https://robrutten.nl/robshots/digilog.jpg}{Digi-Log gadget}
combining an acoustic phone receptacle for 110~baud (then equal to
bps) modulated whistling feeding a modem, a keyboard for input and
40 character/line video output to our analog TV as monitor.
During weekends I so occupied both our telephone and television.} 
My inch-thick NLTE printout had graphs printplotted at the printer
line and character resolution to be traced on overlay transparent
paper for the observatory draftsman to draw figures as
\linkrjrpage{1980ApJ...241..374C}{6}{Fig.\,5}. 
All labeling is also handdrawn! 
These large drawings were then reduced photographically for 
journal offset printing.
I wrote the manuscript in pencil, editing snail-mail coauthor
iterations per eraser, and finally read it aloud into a dictaphone
audio cassette tape for an observatory typist.

The CDC 6600 was succeeded by various CDC CYBER models.
We then started jobs from monitors at Sonnenborgh Observatory 
that were modem-connected via permanent
phone lines (``terminals'') -- a fisticuffs-fought-over
dozen. 
Finally we used a self-written CDC UT-200 emulator with cardreader
input and lineprinter output on our HP\,2114A prior to the 1987
astronomy relocation to the campus outside Utrecht.

In March 1986 Mats Carlsson visited to skate frozen Holland and
suggested to use Emacs, IDL and (La)TeX -- still my staple pastime. 
The CDC machine was replaced by a cheaper Data General Eclipse
computer (scoring a Pulitzer with Tracy Kidder's {\em The soul of a
new machine\/})
and after leaving Sonnenborgh we got our own MicroVax, then switched to
distributed workstations serving X-terminals (Sun, Silicon Graphics,
DEC).}
\rrendnote{personal computing}{note:PC}{%
Utrecht's Observatory first single-user computer was a 1970
\href{https://www.hpmuseum.net/upload_htmlFile/PrintAds/Ad1970-Jan_2114B-2752A_scientificamerican.pdf}{HP
2114A} ``mini-computer''.
For the price of a luxury car a large heavy box with 4K core memory
(actual magnetic cores) running Algol, Fortran and Basic with a
papertape operating system, a papertape-typing Teletype console and a
fancy handheld electric papertape winder. 
After papertape (first 5-hole telex tape, then 1-inch for 8-hole
bytes) came 80-character IBM Hollerith cards for FORTRAN ``records''
(still quaintly defining fits file header format), magnetic tape on
reels and in cassettes, bendable magnetic diskettes, optical video
platters permitting random image access, non-bendable floppy disks up
to a full megabyte.
My Atari ST ``home computer'', the first with megabyte memory, needed
a multi-hour multi-disk sequence for its floppy-disk operating system
to latex my lecture notes but my later ``subnotebook'' ST-Book clone
had a 40\,MB hard disk. 
Result inspection needed laserprinting the postscript output.
Then came zip disks of 100\,MB and even 250\,MB, read-write CDs and DVDs --
all museum stuff now. 

I transited from Ultrix on
\href{https://robrutten.nl/robshots/kop-rob-pict0002.jpg}
{DECstations} at work, the Atari at home (ten years!) and the ST-Book
for travel (five years) via a
\href{https://robrutten.nl/robshots/schlepptopf.jpg}{heavy Dell} and a
\href{https://robrutten.nl/robshots/rob-toaster.jpg} {dysfunctional
Mac} to \href{https://robrutten.nl/Recipes_Ubuntu.html}{Ubuntu on
Toshiba laptops}.
Help went from thick vendor manuals in large binders to ICT personnel
to graduate students to Stackexchange.
FORTRAN still exists but IDL (my start on it was for teaching SSA and
SSB) presently museum-ripens in the Python take-over.
Me too. 
So it goes.}.
I recommend
\href{https://robrutten.nl/rrweb/rjr-edu/exercises/ssa/ssa.pdf}{SSA}
and
\href{https://robrutten.nl/rrweb/rjr-edu/exercises/ssb/ssb.pdf}{SSB},
designed to accompany IART, for fresh RT students to gain hands-on
insights. 
The third is a practical for RTSA developed by former graduate
students that still awaits conversion into proper SSC format.

Here I refer and link to these nine files for equations and graphs
that otherwise would swamp this text. 
In oral 3D bodily-on-the-spot classroom teaching I project and discuss
these; here you should open and scrutinize them yourself. 

Page links to these nine files (square-bracketed blue) should
open the particular page in your browser 
(\rrref{footnote}{note:openpage}).
Page links to ADS-accessible literature should open the
cited page 
\rrendnote{ADS page opening, silly publishers}{note:silly}{%
I describe linking citations to the ADS abstract page and opening
particular pages in ADS-accessible pdf files on
\href{https://robrutten.nl/Turning_citations.html}{this webpage}.
The latex macros and usage examples are given in my
\href{https://robrutten.nl/Report_recipe.html}{report recipe for
astronomy students}.
ADS pdf opening does not work for publications at what I call ``silly
publishers'' who do not furnish direct pdf access on the ADS abstract
page; I avoid citing these altogether.
More detail in my
\href{https://robrutten.nl/Recipes_publications.html}{recipes for
publications}. 

Until 2022 ApJ and A\&A furnished ADS pdf openers after one
year\footnote{For license-limited articles ADS links to the publisher
html access. 
\href{https://www.leanlibrary.com}{Lean Library} then silently uses
your remote-institute license to pass directly to the pdf but
presently opens the first page, making proper-page arXiv pdf opening
preferable. 
ADS often does this automatically.} while permitting arXiving
immediately; now they are open-access immediately.
Thus, the immense top-quality astrophysics literature in ApJ and A\&A
from their start to the present is directly accessible: any particular
page with a figure or equation or paragraph worth showing is only a
single click/tap away in the very sentence where it is
cited.\footnote{Unfortunately \label{note:slow} your browser may
reload the full pdf to show another page, taking a while
(\rrref{footnote}{note:speeds}).
But during these long seconds you might comfortingly contemplate the
much slower pre-internet/ADS chore: go to your institute library or
some dark storage for older publications, locate the journal rack and
volume shelf, find a ladder, climb it, get the volume, descend, find
the paper (literally), find the page, find a table and chair and make
notes or locate the nearest Xerox, write the full reference neatly on
an index card (for alphabetic reshuffle by the manuscript typist),
return the volume to its place, return the ladder, 
return to your desk.
More modern on-line procedure: click/tap the citation in the text to
jump to the reference at the end (or find it), open the ADS abstract
page or DOI-linked publisher page (or find it), find a download button
and load the pdf, find the desired page.
Still too many actions and prone to silly-publisher mazes if not
payment; you also may have trouble to get back to where you were
reading.
Much better to open the desired page per single click/tap parallel to
your reading!
What you miss from old-style library browsing is serendipitous finding
adjacent articles or volumes of unexpected interest to you, but this
is well-compensated by ADS listing which publications cited the one
you are looking up.}
This text has many such openers.  
You should use them too (\rrref{footnote}{note:vulture}).

In contrast, Springer promises ``documents\footnote{Springerese for
publications, not passports. 
``Reviewer'' means referee, ``ticket'' means into a black hole unless
``escalated''.} at your fingertips'' but requires many clicks/taps
that may impose hefty payment and do not bring you to the right page.
{\em Solar Physics\/}\footnote{History: unfortunately C.\,de Jager and
Z.\,{\v{S}}vestka started {\em Solar Physics\/} in 1967 as proprietary
for-profit business of a commercial publisher (Reidel > Kluwer >
Springer), similar to De Jager's start of {\em Space Science
Reviews\/} in 1962 also at Reidel. 
Actually A.\,Reidel pushed De Jager into starting {\em Solar
Physics\/} because he liked the commercial success of the earlier
journal and wanted more.
A year later J.H.\,Oort got upset about the commercial nature of
Z.\,Kopal's new journal {\em Astrophysics and Space Science\/}, also
at Reidel, and urged S.R.\,Pottasch and J.-L.\,Steinberg to found A\&A
as a non-profit astronomy organization
(\citeads{2011EAS....49...23P}) 
that contracts production to a hired publisher (Springer > EDP),
similarly to the non-profit AAS contracting ApJ production out (Univ.\
Chicago > IoP).
More recently some editorial boards of well-established Elsevier
journals have quit the
\href{https://www.theguardian.com/science/2017/jun/27/profitable-business-scientific-publishing-bad-for-science}{over-profitable
publisher} and started a fresh non-profit remake -- a repair the {\em
Solar Physics\/} editors might consider.} 
has no ADS pdf openers\footnote{Springer also forbids arXiving
accepted manuscripts forever unless open access is paid. 
Springer permits arXiving only for the initially submitted version,
not the peer-reviewed iteration -- an affront to referees whose
voluntary unpaid community effort in improving the manuscript is
blatantly discarded
(\href{https://robrutten.nl/Recipes_publications.html}{Springer
nasturtiums}).}
since Springer bought it in 1996 whereas ADS still page-opens scans
from the preceding Reidel-Kluwer era.\footnote{The hospitable and
well-browsable solar physics library of the SST which was my classroom
in a dozen intensive two-week
\href{https://robrutten.nl/Students_to_the_DOT_program.html}{solar-physics
courses} (my best teaching experience) conveniently offers all older
{\em Solar Physics} printed volumes including their often noteworthy
frontispieces, the pick of solar imaging at their time and worth
inspection.
Some would be great page openers here (spectroheliograms as the
Gillespie ones revived in \rrref{endnote}{note:gillespie}) but
Springer offers only a few, without index, costly, non-openable
whereas images in the pre-1996 ADS scans have too low scan quality.}
{\em Living Reviews in Solar Physics\/} is open access but also has no
ADS pdf openers since Springer bought it.\footnote{Example:
\label{note:LRSP}
opening some of the 62 informative figures in authoritative LRSP
\citetads{2009LRSP....6....2N} 
(\rrref{endnote}{note:grancontrast}) would suit well in this
display-oriented course. 
The ADS abstract page shows a pdf-promising green-dotted pdf button
but it serves the hassle of navigating the Springer maze. 
Among solar physics publications the LRSP reviews are top candidate to
page-link but notwithstanding their open access such showcasing is
sillily inhibited by Springer -- unless they are arXived: then ADS
directly page-opens that pdf instead. 
This one wasn't arXived hence remains silly-walled unusable. 
The LRSP editors should push authors to arXive, retroactively.}}.
The citations themselves
\href{https://robrutten.nl/Turning_citations.html} should open the ADS
abstract page.\footnote{
Using the {\tt \textbackslash citeads} commands in
\href{https://robrutten.nl/Turning_citations.html}{these recipes}.
They make a classic reference list superfluous except for the few
non-ADS non-WWW publications.
I yet add one (p.\,\pageref{sec:refs}-\pageref{sec:end}) giving
titles, bibcodes for copy-paste into your {\tt \textbackslash citeads} calls
and links as in my
\href{https://robrutten.nl/bibfiles/ads/publists/dircontent.html}
{solar physicist publication lists}, which with my more useful
\href{https://robrutten.nl/bibfiles/ads/abstracts/dircontent.html}
{solar abstract collection} are a byproduct of my
\href{https://robrutten.nl/Recipes_publications.html}{wholesome
bibtexing} with my entire 
\href{https://robrutten.nl/bibfiles/ads/bib/dircontent.html} {solar bibfile
collection}.}


\begin{figure*}[!p]
  \centering
  \includegraphics[width=15cm]{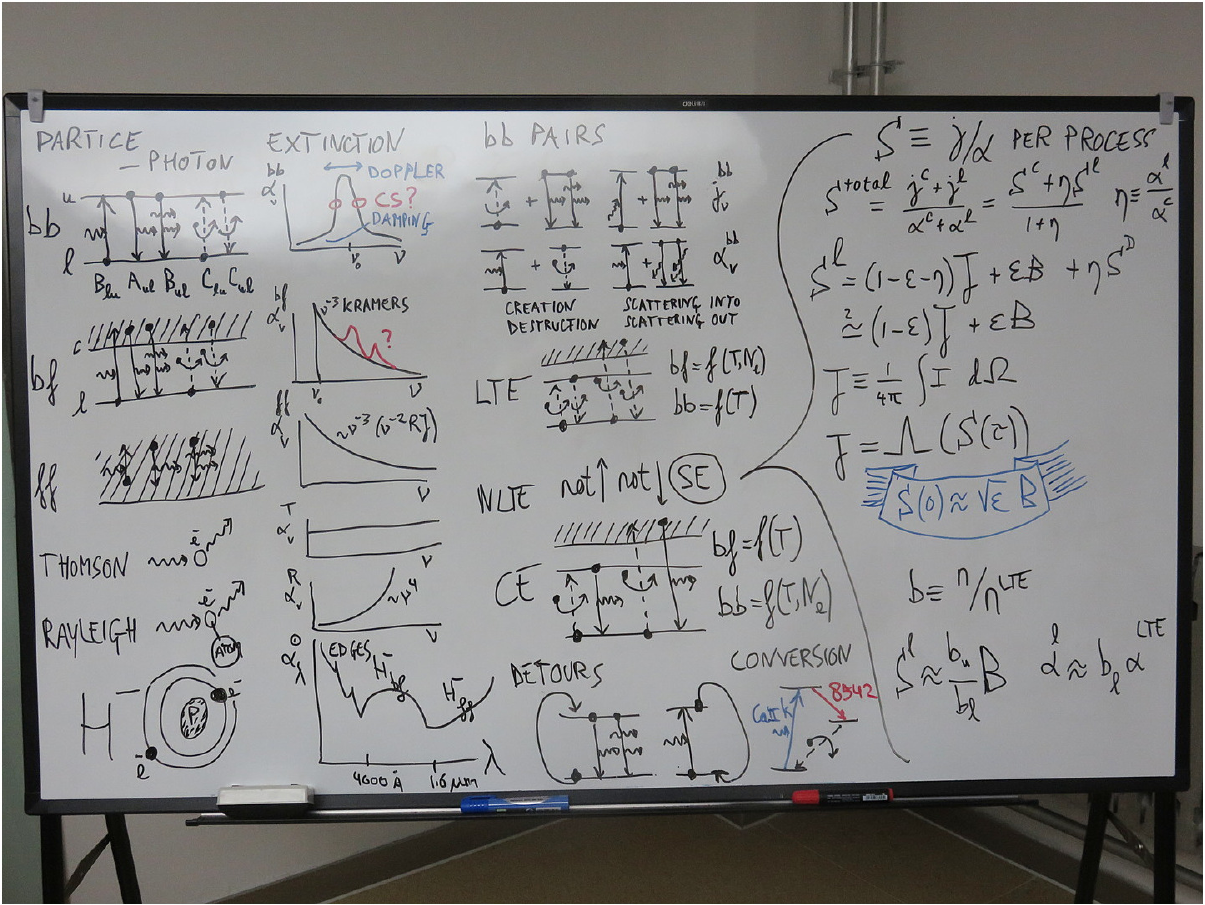}
  \caption[]{\label{fig:weihai}%
  Snapshot course summary, October 2018, Shandong University, Weihai.
  First column: particle--photon processes. 
  Second column: corresponding extinction profiles.
  Third column: process pairs.
  Last column: key equations.
  The first word is already wrong but nevertheless I
  \href{https://robrutten.nl/astronomershots/weihai2018_group.jpg}
  {publicly signed} this painting.\\[1ex]
  }
  \includegraphics[width=14cm]{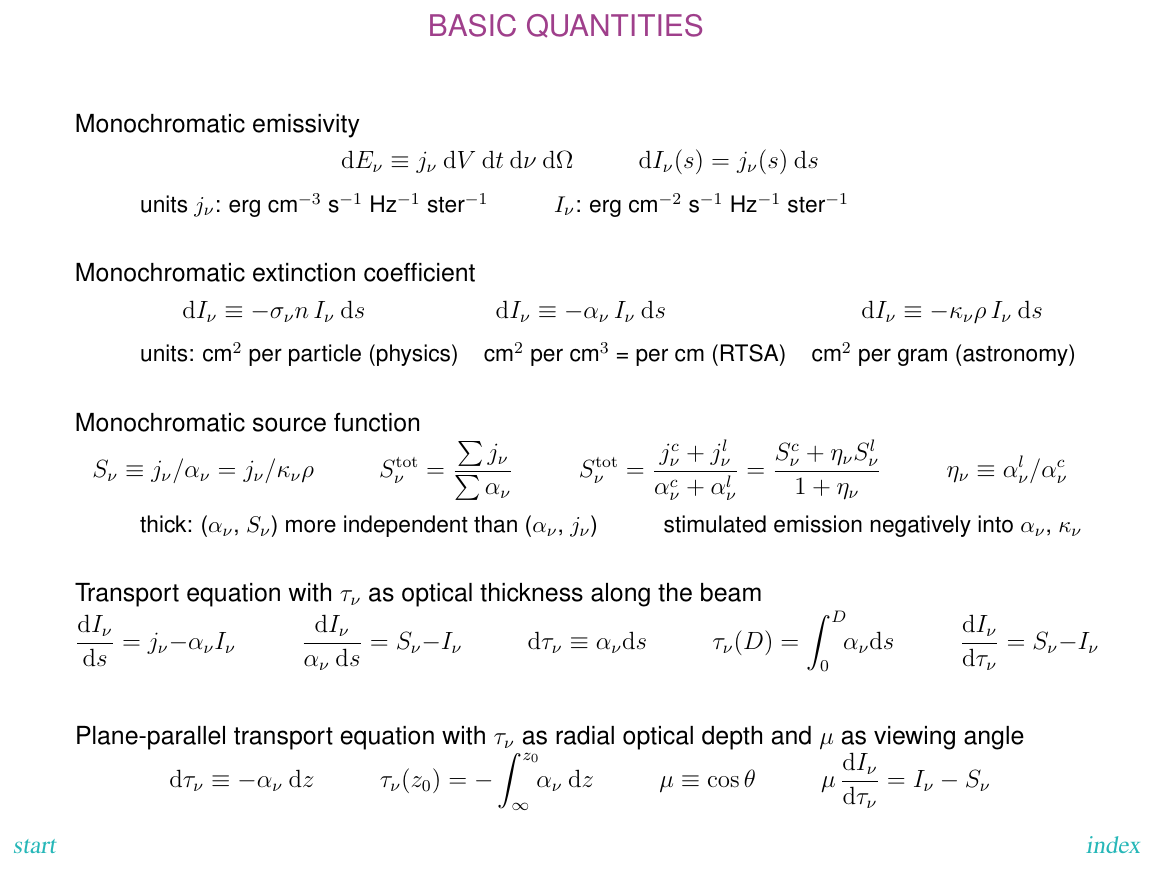}
  \caption[]{\label{fig:quantities}%
  Screenshot of \SSFpage{23}.
  \href{https://robrutten.nl/rrweb/rjr-edu/lectures/rutten_ssf_2020.pdf}{SSF}
  offers 119 similarly-styled theory displays (with a negative
  evaluation in \SSFpage{120}).
  \href{https://robrutten.nl/rrweb/rjr-edu/lectures/rutten_ssx_2020.pdf}{SSX}
  offers 185 example displays.
  \href{https://robrutten.nl/rrweb/rjr-edu/lectures/rutten_ssi_2020.pdf}{SSI}
  offers only 16 introductory displays -- you might start with these. 
  The light-blue button {\em start\/} at bottom left should open the
  contents overview, {\em index\/} at bottom right a thumbnail
  index.
  The display title may return to the previously shown one.
  For me most figures open full-page when I click on them and many
  blue linkers open figures, images, blinkers 
  and movies elsewhere in my laptop -- hence not for you.
  }
\end{figure*}

\section{Introduction}     \label{sec:introduction}
Understanding solar spectrum formation is necessary to understand our
star. 
The state of its matter is relatively simple: just gas, no fluids nor
solids, even close to ideal.
However, complexity is brought at wide-ranging spatiotemporal scales
by dynamics and electrodynamics with ionization and charging varying
from none to full. 
The resulting structures and features may vary from fully
transparent to extremely opaque between different spectral samples.
Using the emergent spectrum as diagnostic to solar structures and solar
happenings requires full elucidation of radiative interactions within
the solar atmosphere.
Observation, theory and simulation constitute the three-fold way to
enlightenment.
Shortcuts as 1D, 2D, LTE, CE, SE may be misleading into dead ends.

Brief history: the solar spectrum caused the dawn of astrophysics in
the nineteenth century in the hands of Fraunhofer, Herschel, Bunsen
and Kirchhoff \SSFpage{3} and others.  
Followed by stellar spectrum classification \SSFpage{6} at Harvard by
many ladies \SSFpage{5} but foremost A.\,Cannon \SSFpage{7} -- I
invite you to now do practical
\linkpdfpage{https://robrutten.nl/rrweb/rjr-edu/exercises/ssa/ssa.pdf}{5}{SSA\,1.1}
before reading further.\footnote{Cut and order
\linkpdfpage{https://robrutten.nl/rrweb/rjr-edu/exercises/ssa/ssa.pdf}
{7}{these spectral ribbons}.
Young kids often do this illuminating rediscovery exercise from Zwaan
better than
\href{https://robrutten.nl/astronomershots/SSA/album.html}{solar
colleagues} by sorting spectra without prejudice. 
Cannon's ``early -- late'' suggests counting lines as wrinkles or by
element number. [{\em Spoiler\/}:
\href{https://robrutten.nl/rrweb/rjr-edu/exercises/ssa/ssa1_answers.pdf}{answer}.]. 
In actual teaching I then present \SSFpage{2-16}. 
The eye-opening Payne part is treated in practical
\linkpdfpage{https://robrutten.nl/rrweb/rjr-edu/exercises/ssa/ssa.pdf}{9}{SSA2}.}
In the twentieth century stellar spectrum interpretation evolved from
basic concepts to understanding with Schwarzschild, Russell, Milne,
Eddington, Minnaert, Chandrasekhar and others.
Major breakthroughs were Payne's demonstration with the Saha law that
Cannon's classification represents temperature ordering \SSFpage{57},
the Pannekoek--Wildt--Chandrasekhar identification of \Hmin\ as major
continuous opacity provider (\rrref{endnote}{note:extcont}), and
Grotrian's establishing the outrageous high temperature of the corona
(\rrref{endnote}{note:wlcorona}).
Solar spectrum interpretation went from LTE (Uns\"old,
\rrref{endnote}{note:unsold}) to NLTE with Menzel, Thomas, Athay,
Hummer, Jefferies, Mihalas, Avrett and others. 
Then started (M)HD simulation modeling by Nordlund, Sch\"ussler,
Stein, Carlsson, Steiner and others with spectral synthesis programs
by Carlsson, Uitenbroek, Heinzel, Leenaarts, Pereira and others,
presently emphasizing the new frontier of NLT interpretation.
ADS serves nearly all their works.

This text gives an overview dividing radiative transfer (RT)
complexities between nonlocal in space (NLS), nonlocal in wavelength
(NLW), nonlocal in time (NLT).\footnote{As appetizer some nonlocal XXX
combinations detailed in this text: %
\begin{itemize}  \vspace{-2ex}
\item \mbox{\Hmin} bound-free (optical): near-LTE 
$S$ and near-LTE $\alpha$
\item \HI\ free-free (ALMA): LTE $S$ but NLS+NLW+NLT $\alpha$
\item \HI\ Lyman lines: dense gas LTE $\alpha$, hot dense gas LTE $S$, 
  cooling dense gas NLS+NLT $S$, tenuous gas NLS $\alpha$, NLS+PRD $S$
\item \HI\ Balmer lines: hot gas LTE $\alpha$, cooling gas
  NLS+NLW+NLT $\alpha$, NLS+NLW $S$
\item optical \HeI\ lines: likely NLS+NLW+NLT $\alpha$ and NLS+NLW $S$
\item \HeII\,304\,\AA: likely  NLS+NLW+NLT $\alpha$ and NLS+PRD $S$
\item \MgI\,4571\,\AA: near-LTE $S$ but NLS+NLW $\alpha$
\item \MgI\ 12\,$\mu$m lines: NLW $\alpha$ and NLW $S$
\item \MgIIhk, \BaII\,4554\,\AA: near-LTE $\alpha$ but NLS+PRD $S$
\item \CaIIHK: NLS $\alpha$ and NLS+PRD $S$
\item \CaIR:  NLS $\alpha$ and NLS $S$
\item neutral alkali resonance lines: NLW $\alpha$  and NLS $S$
\item strong neutral-metal lines: NLS $S$ and NLS+NLW $\alpha$
\item weak neutral-metal lines: near-LTE $S$ but NLS+NLW $\alpha$
\item optical coronal lines: CE $\alpha$, no RT, negligible $S$  but NLS $j$.
\end{itemize}  \vspace{-4ex} \mbox{} 
} 
Solar physicists generally appreciate NLS better than NLW and NLT least.

NLS means that radiation received from the location we are studying is
influenced by local radiation there that came from elsewhere.
It is daily familiar to us since our daytime outdoors is NLS
illuminated.
Whether sunshine or overcast, all photons we see were made in the Sun
and made it into our eyes by scattering, mostly multiple. 
As NLS as it can get! 
Very much out of LTE since the radiation temperature is about 6000~K,
higher than our local 300~K 
\rrendnote{spectral bandwidth and vision}{note:bandwidth}{%
Only our eyesight is adapted to solar radiation temperature but not
simply to the Planck function for the solar effective temperature. 
Stating that needs specification whether our retinas measure per
wavelength, frequency, wavenumber or count photons -- see \SSFpage{50}
and remember $\rmd\nu\!\tis\,\!-(c/\lambda^2)\,\rmd\lambda$.

The third panel shows why J.W.\,Brault said in 1978 that the intensity
at \CaII\ \HK\ was too low for his Kitt Peak Fourier Transform
Spectrometer: a photon-counting wavenumber device (but we got the only
center-limb \HK\ scans in the FTS archive). 
The last panel shows why high-energy astrophysicists plot
$\lambda\,\!B_\lambda$ or $\nu\,\!B_\nu$.

History; the spectral sensitivity of our eyesight was optimized in our
aquatic past for the penetration of sunlight in sea water, as for
giant squids whose giant eyes have similar but better design
(\href{https://www.feynmanlectures.caltech.edu/I\_36.html\#Ch36-F10}{Feynman
I Fig.\,36-10}) and might suit nighttime astronomers better.
As early mammals ashore we became dichromats being nocturnal to hide
from predating dinosaurs. 
After these were blasted into oblivion\footnote{The Chicxulub impact
(66\,Myr ago) was likely so devastating because the soft sulfur-rich and
feldspar-rich Yucatan limestone went pulverized up into the
stratosphere causing global acid rain, global re-entry firestorms and
long global winter 
(\cf\ \citeweb{https://pubs.geoscienceworld.org/jgs/article/doi/10.1144/jgs2021-055/609522/Meteorites-that-produce-K-feldspar-rich-ejecta}
{2021JGS..055.....P}).
So it goes.
The equal-size but 
\href{https://robrutten.nl/Kayak_Manicouagan.html}{kayakable 
Manicouagan impact}
(215\,Myr) without known mass extinction was into hard Precambrian
Grenville gabbro rock with energy dissipation in fluidizing shock
waves.}
early hominids restored red-detecting cones to spot low-hanging fruit,
giving us color-triangle vision
(\href{https://www.feynmanlectures.caltech.edu/I\_35.html}{Feynman I
Sect.\,35}) to appreciate Kodachrome Basin.}.  

NLW means that radiation received from the location we are studying is
influenced by radiation at other wavelengths at that location.
We do not suffer NLW because sunshine scattering around us is
monochromatic (monofrequent): every detected photon still has the
energy (frequency, color) with which it left the Sun.
All solar Fraunhofer lines also reach our retinas unmolested (but
disk-averaged and unnoticed\footnote{Not even when you look at your
beloveds although they get so richly adorned: 27500 colorful lines
(not wrinkles) resolvable on their face!}). 
We see the sky blue and the setting Sun red not because of underway
color change but because Rayleigh scattering off molecules has higher
probability at shorter wavelength.
The green flash\footnote{My most beautiful was from the Neemach Mata
temple in Udaipur.} combines that with larger refraction in the blue.

NLT means that radiation received from the location we are studying is
influenced by what happened there or thereabouts beforehand. 
We suffer outdoors NLT because our sunshine is eight minutes retarded
-- but the same for all photons so we don't care.

Solar NLS defines outward decline of line source functions\footnote{I
call this $\sqrt{\varepsilon}$ scattering after the blue-bannered
equation in \rrref{Fig.}{fig:weihai} which results for isothermal
constant-$\varepsilon$ atmospheres from the 2-level simplifications of
\rrref{Eq.}{eq:S_CS} and \rrref{Eq.}{eq:S_CRD} below.  
More in \rrref{endnote}{note:sqrteps}.}
that is well exemplified by the \NaID\ lines and most exemplified by
\Lyalpha, but it also brightens ultraviolet continua in and from the
solar atmosphere \SSFpage{109}, \SSXpage{42}. 
More below.

Solar NLW is exemplified by most atomic lines having opacity
deficiencies up to an order of magnitude from sensing the scattering
ultraviolet continua in ionization. 
This is less commonly appreciated, for example not in
profile-fitting ``inversion'' codes nor in spectral irradiance
modeling.  More below.

Solar NLT is exemplified by solar \Halpha\ which is a NLS
$\sqrt{\varepsilon}$ scattering line but it is also NLW because
\Lyalpha\ defines its extinction (opacity) and a loop ionizing per
scattering Balmer continuum from $n\tis\,2$ with cascade recombination
back to $n\tis\,2$ including \Halpha\ photon losses contributes to its
source function.
It is also NLT because this loop is controlled by the $n_2$
population which lags badly in gas that was first heated and then
cools, as happens continuously in the dynamic structures constituting
the solar chromosphere and even in long-lived filaments/prominences
(\rrref{endnote}{note:blobs}). 
Solar NLT can reach many orders of magnitude but is ignored in almost
all modeling by invoking statistical equilibrium (SE). 
More below.

I summarize the theory in \rrref{Sect.}{sec:basics}, followed by
classic developments in order of complexity in
\rrref{Sects.}{sec:Holweger}\rrref{\,--}{sec:Oslo}.
The dense and tenuous limits of solar spectrum formation are
easiest: LTE in the deep photosphere, CE in the corona. 
In between is the hardest nut to crack: the NLS+NLW+NLT spectrum of
the chromosphere discussed in \rrref{Sect.}{sec:chromosphere}.

\begin{figure*}[hbtp]
  \centering
  \includegraphics[width=15cm]{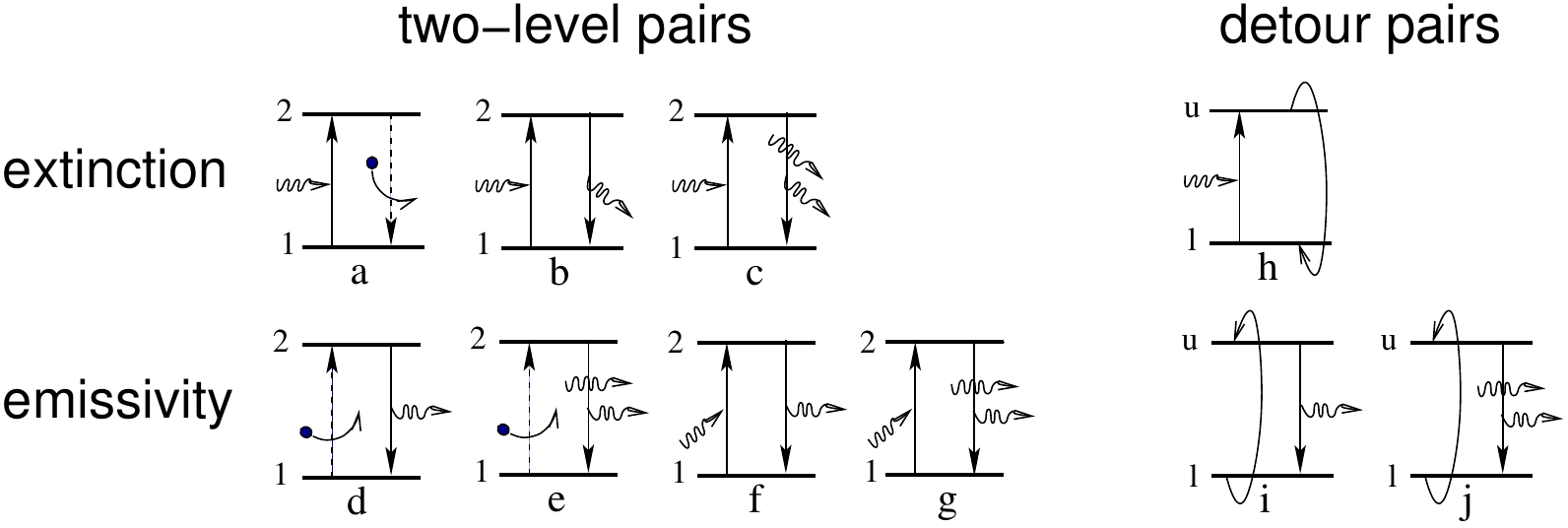}
  \caption[]{\label{fig:pairs}%
  Atomic transitions governing line formation arranged in
  photon-involving pairs. 
  The intensity in the beam of interest travels from left to right.
  All photons drawn here are at the line frequency = jump energy.
  The pairing categorizes what may happen to an atom after excitation
  (at left in each pair).
  It won't stay excited long, for a 2-level atom only between
  $1/A_{ul}\tapprox\,10^{-8}$\,s (resonance) and
  $1/A_{ul}\tapprox\,10^{-2}$\,s (forbidden) until {\bf b}, {\bf d} or
  {\bf f} occurs spontaneously, yet shorter if the atom while excited
  meets a deexciting collider {\bf a} or an inviting akin photon to
  join on its way ({\bf c}, {\bf e}, {\bf g}).  
  Multi-level detour paths (righthand cartoons) combine
  other-wavelength transitions involving other levels and may include
  analogous bound--free transitions. 
  There may be radiative or collisional deexcitation to another lower
  level or radiative or collisional excitation to a higher level
  including ionization and recombination. 
  Other transitions eventually return the atom to the start-off level 1.
  There may be endless up-down scattering in between so the detour
  possibilities are infinite.
  The upper row shows pair combinations contributing to line
  extinction: collisional photon destruction ({\bf a}), scattering out
  of the beam ({\bf b} and {\bf c}), photon conversion out of the beam
  ({\bf h}, into other-wavelength photons and/or kinetic energy).
  The lower row shows pairs contributing to line emissivity:
  collisional photon creation ({\bf d} and {\bf e}), scattering into
  the beam ({\bf f} and {\bf g}), detour photon production into the
  beam ({\bf i} and {\bf j}). 
  Pairs {\bf c} and {\bf g} have equal probability by requiring one
  photon in the beam and one with arbitrary direction.
  }
\end{figure*}

\section{RT basics} \label{sec:basics}
The figures on the preceding page show didactic
overviews of this section.
\rrref{Table}{tab:abbreviations} specifies acronyms I bombard you
with\footnote{It would be good to hover your cursor over an acronym to
Wikipedia-style pop up its meaning.
I have tried latex command pdftooltip in pdfcomment.sty for such
popups but they worked too different between pdf readers.}.
A summary of main RT quantities and equations is in the
\RTSApage{275}{RT Rap} and \SSFpage{107}. 

\paragraphrr{Intensity} 
The basic RT quantity is specific intensity $I_\nu$ describing energy
in a beam of photons at a given location at a given time at a given
frequency in a given direction, measured per units of area, time,
frequency bandwidth and beam spreading \ISSFpage{9}{Sect.\,3.1}
\IARTpage{22}{Sect.\,2.1} \RTSApage{29}{Eq.\,2.1}.
The last is hardest, \ISSFpage{8}{Fig.\,2} may help. 
It ensures that intensity does not vary with travel distance
\SSFpage{25}, the macroscopic way to express that photons do not decay
and may travel through the whole universe, not hindering one
another\footnote{Unlike fermionic you and Archimedes pushing bathwater
aside. 
The reflected beam from a mirror does not hinder the incoming beam,
they travel through another. 
As do ripples of incoming and outgoing waves on a beach, summing
height at their crossings.
In a sunlit swimming pool I like to piston-spread waves with
spread-out hands and admire the spreading dual-boson caustics on the
pool floor doubling radiance at their crossings.} being bosons
\rrendnote{quanta, bosons, photons}{note:bosons}{%
Since high-energy, polarization and radio spectra are not treated in
this compendium (\rrref{footnote}{note:except}) I ignore the
wave/particle duality of electromagnetic radiation by simply taking
photons as bullets of non-relativistic energy $h\nu$ without impulse
or wave properties beyond frequency: the ``quanta'' introduced by
Planck.
On p.\,134 of ``Inward Bound''
\citetads{1986ibmf.book.....P} 
calls his
\href{http://web.ihep.su/dbserv/compas/src/planck00b/eng.pdf} {quantum
explanation} 
(\citenoweb{planck1900b}) 
for the brilliant \mbox{-1} that he put a few weeks before in his
\href{http://web.ihep.su/dbserv/compas/src/planck00/eng.pdf}{empirical
radiation law}
(\citenoweb{planck1900a}) 
the most revolutionary physics discovery of the nineteenth century and
adds that Einstein was likely the first to recognize this ``as if the
ground was pulled from under him''.
Without the -1 the Wien simplification resembled the Boltzmann
distribution understood from probability theory but the \mbox{-1}
required energy quantization. 
Much later (1916) Einstein added that also for impulse and in 1924
attributed light particles to Bose-Einstein statistics.
See \citetads{1982sils.book.....P} 
``Subtle is the Lord'' Chapts.\,19 and 21 for more. 
On p.\,405 Pais attributes the name ``photon'' to paywalled
\citetads{1926Natur.118..874L}. 

Here the simplistic bosonic lightspeed bullet of energy suffices, as
does the Bohr model for fermionic atoms and ions. 
The bullets get created by excitable fermions and may be absorbed or
scattered by other fermions -- that's all!
Radiative transfer has a bad name of intransparency (literally
correct) but remains simple physics. 
As is solar physics: no dark matter or energy while the solar neutrino
problem went away (\rrref{endnote}{note:neutrinos}) and so did the
solar oblateness problem
(\citeads{1967PhRvL..18..313D}) 
with attendant Brans-Dicke scalar-tensor gravity.
Inside just gaseous loose but closely confined bosons and fermions
humming harmoniously in boxed-in resonances with tiny surface leaks
enabling our lives and also filling our careers with magnetic
complexities from the Navier-Stokes and Maxwell equations and
non-fluid properties converging outwardly into tenuous but stormy
extrasolar plasma outside -- that's all!
Solar physics is not reductionist inward-bound discovery of laws for
everything but Spinozist outward-bound discovery of what nature most
inventively makes from them.},
to most informatively convey the whole universe to us for inspection
with splendid spectral encoding to measure distance of and conditions
in the source wherever it is.\footnote{History: the stupendous insight
that spectroscopy permits measurement of local atomic properties in
faraway objects independent of distance came to R.\,Bunsen on his
evening stroll with G.\,Kirchhoff along the Philosopher's Path in
Heidelberg after they had pointed their spectroscope from their
laboratory window at a fire in nearby Mannheim and detected barium and
strontium in it, inspiring the idea that all Fraunhofer lines that do
not darken at sunset from being terrestrial describe the same elements
within the Sun as those giving bright lines at their specific
refrangibilities (wavelengths) when sprinkled in their Bunsen-burner
flame \href{https://books.google.nl/books?id=yDBep9y0SVsC&pg=PA138}
{(W.~Gratzer, Eurekas and Euphorias p.\,138)}.  
Kirchhoff later honoringly quoted Herschel (1827): 
``colours communicated by the different bases to a flame afford, in
many cases, a ready and neat way of detecting extremely minute
quantities of them'' but with his laws \SSFpage{3} founded
astrophysics elaborated here.}

Common sense suggests that the Sun is brighter than a distant star,
but that is irradiance which is flux \SSFpage{24}
\RTSApage{30}{Eq.\,2.4} not measured per unit of beam spreading (solid
angle \RTSApage{30}{Fig.\,2.1}).
If the distant star is solar-type it shows the same intensity, also
when your time-travel starship parks you just above its
surface.\footnote{If our universe were infinite, eternal and filled with
stars our night sky would be boringly solar-bright
everywhere, also in neutrinos.}
If your giant groundbased telescope resolves it the stellar intensity
in the focus is the same as the solar intensity in the focus of
photon-starved DKIST
\rrendnote{solar telescopes}{note:telescopes}{%
Solar telescopes record intensity and therefore are photon-starved,
worse so when larger (your nighttime colleagues won't believe either).
Reason: for pixels that resolve the diffraction limit the flux of
photons per pixel does not vary with aperture size while smaller
pixels sample smaller solar surface area (that's the game) for which
the solar change time (say crossing at solar sound speed) is shorter.
DKIST therefore needs four times faster data-taking cadences than the
SST unless it is used as (expensive) light bucket. 
DKIST thus has four times less time for multi-frame collection for
numerical post-detection restoration
(\rrref{endnote}{note:restoration}), \eg\ at \Halpha\ only 1\,\!s for
all desired wavelengths. 
A frequent design or set-up mistake is to overdo spectral resolution
and coverage at the cost of cadence, hence angular
resolution.
It is easier to sacrifice the latter so that larger aperture
light-buckets more photons, the nighttime ELT approach.

On the other hand, fortunately our resolvable daytime star kindly
offers wavefront-encoding granulation everywhere on her surface so
that there is no need to cover the target field with a \NaID\ laser
star in every isoplanatic patch (as ELT should have but won't). 
Thanks to adaptive optics and numerical reconstruction, both enabled
by this granular encoding, groundbased optical telescopes pointed
within the solar disk may reach their diffraction limit over their
full field of view when the seeing is good: modern telescopes that do
not reach double-digit percentage granular rms contrast
(\rrref{endnote}{note:grancontrast}) are not doing their job
(\citeads{2019A&A...626A..55S}).\footnote{To 
\label{note:Scharmer}
stop the hype in calling optical observations ``high resolution'' I
propose as qualifying ``Scharmer threshold'' criterion that the
continuum rms granulation contrast exceeds 80\% of the sky-limited SST
value in
\linkadspage{2019A&A...626A..55S}{8}{Fig.\,6}, 
even though that remains significantly below simulation estimates
reaching 20\% at 5000\,\AA\
(\citeads{2009A&A...503..225W}, 
\cf\ G-band comparison of
\citeads{2007ApJ...668..586U}).} 
However, high-altitude seeing spoils granular wavefront encoding 
by reducing isoplanatic patches to subgranular scales, requiring
multi-conjugate adaptive optics 
(Beckers \eg\
  \citeyearads{1989ASIC..263...43B}, 
  \citeyearads{2014SPIE.9148E..60B}).} 
\rrendnote{solar image restoration}{note:restoration}{%
Speckle and MOMFBD restoration exploit the lucky circumstance that the
terrestrial seeing-freezing time is only about 10\,ms, giving \eg\
DKIST time to collect a hundred independently disturbed frames to
obtain seeing statistics at \Halpha\ within its 1\,\!s solar change
time (\rrref{endnote}{note:telescopes}).
But for say five-wavelength \Halpha\ sampling 20 frames per wavelength
is not enough for good speckle reconstruction which needs 50-100
frames, whereas MOMFBD (\citeads{2005SoPh..228..191V}) 
needs only 5-10 frames, an important advantage. 
However, the latter technique is less robust and needs better seeing,
optics quality and adaptive optics -- so far performing best at the
SST.

At the \href{https://robrutten.nl/dot/DOT_home.html}{DOT} we used
two-channel speckle reconstruction
(\citeads{1992A&A...261..321K}, 
\citeads{1994ASIC..433...43K}; 
\href{https://robrutten.nl/dot/DOT_speckle.html}{demonstration}) to
reduce speckle burst sizes per wavelength to 20 in five-wavelength
\Halpha\ sampling by adding simultaneous 100-frame wide-band
registration as is also done for SST MOMFBD. 
Comparisons (the two telescopes share their seeing) showed that SST
MOMFBD yields much better quality than DOT two-channel speckle but
needs a higher seeing-quality Fried value to work, typically
$r_0\tapprox\,10-12$\,cm for the SST while only $r_0\tapprox\,5-7$\,cm for
the DOT.\footnote{This property and the DOT's easy mosaicing from its
parallactic configuration and superstable pointing (Hammerschlag tower
and telescope mount, no adaptive optics) make the DOT highly suited
for science programs with large-field five-wavelength \Halpha\
mosaicing by the DOT as context imager for other telescopes. 
After the DOT lost Utrecht funding I proposed such in 2012 for
co-pointing with IRIS (same resolution, field, cadence) but did not
succeed in getting money -- it did not help that I was officially
retired for years already and that all Utrecht astronomy was then
\href{https://robrutten.nl/Closure_Utrecht.html}{brutally closed
down}. 
I now suggest that moving the DOT to near ALMA is a good idea
(\rrref{endnote}{note:ALMA-Ha}).}

MOMFBD and two-channel speckle reconstruction both apply field
tesselation into isoplanatic patches and subsequent reassembly
with the important advantage of perfect co-registration of the
different wavelengths with the wide-band sequence, which independent
speckle reconstruction per wavelength does not give. 
Since optical wide-band registration shows granulation it is then easy
to cross-align that sequence precisely to SDO/HMI continuum -- or even
to use the latter within the reconstruction to improve the patch
reassembly -- and then to co-align all narrow-band sequences
precisely to all SDO diagnostics with my
\href{https://robrutten.nl/rridl/00-README/sdo-manual.html}
{co-alignment pipeline} (\rrref{endnote}{note:SDO-STX}).}.

At far-infrared, mm and radio wavelengths the Planck function
simplifies to the Rayleigh-Jeans limit \SSIpage{3} so that the
intensity of an optically thick object with an LTE source function
scales with its temperature and gives that when calibrated.
In the optical and shorter-wavelength domains the Wien non-linearity
implies that intensity contrast comparisons between different
wavelengths in terms of temperature need conversion into brightness
temperature
\rrendnote{equivalent temperatures}{note:Tb}{%
These serve for deWiening Planck function temperature sensitivity
in intensity-related and population-related quantities.
They are formal by being defined as the temperature value that entered
into the pertinent TE expression equals the quantity: brightness
temperature $T_\rmb$ for $I_\nu$ into the Planck function, radiation
temperature $T_{\rm rad}$ for $J_\nu$ into the Planck function,
excitation temperature $T_{\rm exc}$ for actual level population ratio
into the Boltzmann ratio making
$S\!_\nu^l \tis\,B_\nu(T_{\rm exc})$, ionization temperature
$T_{\rm ion}$ for actual level/continuum population ratio into the SB
expression making $S\!_\nu^c\tis\,B_\nu(T_{\rm ion})$
\SSFpage{72} \RTSApage{57}{Eqs.2.128\,--\,2/132} after
\citetads{1972SoPh...23..265W}. 
Thus, the difference $T_{\rm exc}-T_\rme$ is a deWiened alternative
to the $S\!_\nu^l-B_\nu$ measure of NLTE affecting a source function.
Scattering lines with small $\varepsilon$ likely have
$T_{\rm exc}\tapprox\,T_{\rm rad}$.

The required $[B]^{-1}$ operator
(\linkadspage{1979A&A....74..225S}{1}{Eq.\,1} of
\citeads{1979A&A....74..225S}) 
is straightforward but needs attention to units. 
My IDL \href{https://robrutten.nl/rridl/misclib/brighttemp.pro}{\tt
brighttemp.pro} handles it for $B_\nu$ and $B_\lambda$ and different
bandwidth units.

These conversions should always be applied when comparing
intensity-related quantities at different wavelengths. 
Examples:
\linkrjrpage{2011AAp...531A..17R}{5}{Fig.\,4} 
of \citetads{2011A&A...531A..17R} 
comparing brightness scenes in \NaIDone, \MgIbtwo\ and \CaIR, their
atlas profile comparisons in
\linkrjrpage{2011AAp...531A..17R}{10}{Figs.\,10\,--\,12} there, and
the intercomparable $S, B, J$ panels for many lines from the ALC7 star
in the displays starting at \SSXpage{78}.}
which requires absolute units
\rrendnote{intensity calibrations}{note:absint}{%
For the optical spectrum Neckel 
(\citeyearads{2003SoPh..212..239N}, 
\citeyearads{2005SoPh..229...13N}) 
provides limb-darkening expressions based on Brault's FTS observations
used by
\citetads{1984SoPh...90..205N} 
that \citetads{1999SoPh..184..421N} 
converted into a calibrated atlas (\rrref{endnote}{note:atlases}).

In the ultraviolet the SUMER atlas of
\citetads{2001A&A...375..591C} 
provides calibrated disk-center spectra for the 668\,--\,1611\,\AA\
range, with center-to-limb observations available in the
\href{https://sohowww.nascom.nasa.gov
/data/archive} {SUMER archive}.
Earlier 1400\,--\,2100\,\AA\ absolute-intensity and center-to-limb
estimates are in \citetads{1979A&A....74..225S}. 
Calibrated 1100\,--\,1900\,\AA\ flare spectra are shown and provided
by \citetads{2019ApJ...870..114S} 
in their inventory of spectral contributions to flare intensities
in the AIA UV passbands.} 
as in the optical spectrum atlas of
\citetads{1999SoPh..184..421N} 
\rrendnote{optical spectrum atlases and line lists}{note:atlases}{%
Solar spectrum atlases started with the beautiful engravings of
Fraunhofer and Kirchhoff, Rowland's photographs, and then the
graphical Utrecht Atlas of
\citetads{1940pass.book.....M}
\footnote{History: made by scanning photographic spectrum plates taken
at Mt.\,Wilson with an ingenious microdensitometer using analog
tandem-galvanometer nonlinear ``calibration curve'' conversion of
emulsion opacities into solar intensities, illustrated in Minnaert's
prefaces (English and Esperanto).}, all four sampled in 
\RTSApage{26}{Fig.\,3.1} from 
\citetads{1961hias.book.....P}. 
All lines in the Utrecht Atlas were laboriously
measured\footnote{History: manually by counting mm$^2$ area for each
line dip to get its equivalent width, for which the atlas was printed
on mm-grid paper.
While I was a student De Jager purchased a planimeter as
\href{https://en.wikipedia.org/wiki/Planimeter\#/media/File:Polarplanimeter\_01.JPG}{this
one} for faster measurement.
Later I co-developed electronic tracing readers until replacing the
Honeywell-Brown chart recorders of the Utrecht solar spectrograph and
Houtgast's microdensitometer with a self-built analog-to-digital
converter and an 8-bit papertape puncher
(\citeads{1973SoPh...28..347R}, 
\citeads{1975ASSL...54..261R}). 
Regrettably the impressively noisy tape-spewing (120 bytes/s!)
puncher got lost so that these gadgets are not on display in the De
Jager-started \href{https://www.sonnenborgh.nl}{Museum Sterrenwacht
Sonnenborgh}.}
to produce the MMH solar line table of
\citetads{1966sst..book.....M}. 
I got it as ascii file per magnetic tape from J.W.\,Harvey and
consult this often although I also have the book. 

While writing this note I realized that I might supply this file and
similar leftovers from the past and started a
\href{https://robrutten.nl/rrweb/rjr-archive/dircontent.html}{solar
file archive} supplying them.
Next to MMH the digital line tables include the legendary Revised
Multiplet Table of \citetads{1959mtai.book.....M} 
(RMT, \rrref{footnote}{note:Grotrian}) used as input to MMH (also a
tape from KPNO/Tucson), the more precise wavelength list of
\citetads{1974kptp.book.....P}, 
the yet more precise wavelength list of
\citetads{1998A&AS..131..431A}, 
the measurements for 750 clean optical lines in the Jungfraujoch
disk-center atlas of
\citetads{1973apds.book.....D} 
in \linkrjrpage{1984AApS...55..143R}{11\,ff}{Table\,IV} of
\citetads{1984A&AS...55..143R}, 
the corresponding measurements for 602 clean optical lines in the
Sacramento Peak irradiance atlas of
\citetads{1976hrsa.book.....B} 
in \linkrjrpage{1984AApS...55..171R}{4\,ff}{Table\,I} of
\citetads{1984A&AS...55..171R}, 
and the flash-spectrum line intensities in
\linkadspage{1968ApJS...15..275D}{51\,ff}{Table\,3} of
\citetads{1968ApJS...15..275D} 
that I got laboriously punched on Hollerith cards in the 1980s.

Following the example of the Utrecht Atlas, the Jungfraujoch intensity
atlas and the Sacramento Peak irradiance atlas were published as
printed tracings, the first on very large sheets in a thick unwieldy
binder, the second as practical letter-size booklet with useful MMH
line identifications added in the margin.\footnote{My
\href{https://robrutten.nl/rrweb/rjr-archive/dircontent.html}{file
archive} supplies
\href{https://robrutten.nl/rrweb/rjr-archive/atlases/beckers-fluxatlas.pdf}{a
scan} of this handy graphical atlas. 
Print it if you want a compact solar spectrum atlas with line
identifications at hand in your next observing run.  
Your line at wavelength {\tt wav} in \AA\ is on pdf page {\tt
fix(wav-3800)/20+7}. 
I lifted an original in 1977 from the stack in the Sacramento Peak
library and used it in all my spectroscopic observing since,
annotating every line I liked, and still cherish it.
Earlier Sacramento Peak Observatory and Utrecht Observatory had a copy
of the Utrecht Atlas with the stronger lines likewise labeled with
their MMH identification -- but all laboriously handwritten per human
``computer'' (job label then, it was E.B.J. van der Zalm)
instead of Calcomp-drawn (\rrref{footnote}{note:calcomp}) per
electronic computer. 
Later I hoarded the Utrecht one; it is now at
\href{https://www.sonnenborgh.nl}{Museum Sterrenwacht Sonnenborgh}.} 
Both atlases became also available on magnetic tape
enabling distilling the clean line lists above, but the
digital market was taken over by J.W.\,Brault's Fourier Transform
Spectrometer (FTS) at the Kitt Peak McMath telescope delivering
precise atlases still available on the \href{https://nso.edu}{NSO
website} (if you master the non-trivial art of navigating that -- but
see the Vitas collection below). 

The FTS atlas that I use habitually, not available at NSO, is the
``Neckel'' disk-center intensity atlas named after
\citetads{1999SoPh..184..421N} 
announcing its availability.\footnote{Earlier I retyped the
announcement in \citetads{1999SoPh..184..421N} 
into ADS but it became Springerwalled. 
This intensity atlas and its ``flux'' companion (actually
disk-averaged intensity ${\cal F}_\lambda/\pi$
\RTSApage{31}{Eq.\,2.7}) measured in
watt\,cm$^{-2}$\,ster\,$^{-1}\,\AA^{-1}$ (not cm$^{-1}$ as Neckel
specified) are still available at \url{ftp.hs.uni-hamburg.de} per {\tt
cd pub/outgoing/FTS-Atlas} when using {\tt lftp} (not by clicking
since Firefox and Chrome don't open ftp links anymore). 
Both are in the \href{https://robrutten.nl/rrweb/rjr-archive/atlases/vitascollection/1984_Neckel_and_Labs.zip}{Vitas IDL save file} in my
\href{https://robrutten.nl/rrweb/rjr-archive/dircontent.html}{solar
file archive}; the intensity atlas is there also as
\href{https://robrutten.nl/rrweb/rjr-archive/atlases/neckel-intatlas.txt}
{ascii file} with
\href{https://robrutten.nl/rrweb/rjr-archive/00-README.txt}{explanation}.}
He supplied older FTS records from Brault \& Testerman in
absolute units following
\citetads{1984SoPh...90..205N}, 
enabling conversion into brightness temperature for comparing
differing spectral regions (\rrref{endnote}{note:Tb}).
\citetads{2016A&A...590A.118D} compared this Fourier spectrometry
atlas and the earlier Jungfraujoch grating spectrometry atlas in
detail. 
When while writing this I could not revive my old magnetic-tape files of
the Jungfraujoch and Sacramento Peak atlases I asked author N.\,Vitas
whether he has them digitally and yes he did -- both these and many
more,
a collection now shared in my
\href{https://robrutten.nl/rrweb/rjr-archive/dircontent.html}{solar
file archive}
(\href{https://robrutten.nl/rrweb/rjr-archive/atlases/vitascollection/atlases_overview.ods}{his inventory}).} 
and the ultraviolet spectrum atlas of 
\citetads{2001A&A...375..591C}. 

\paragraphrr{Matter--radiation interactions}  
Photons live 
forever\footnote{\citetads{2013PhRvL.111b1801H} 
\linkadspage{2013PhRvL.111b1801H}{4}{concludes}
that they live at least 3 years -- in their rest frame,
meaning $10^{18}$~years for us.}
after their creation unless they meet fermions and interact.
Also their creation is fermion business.
For atoms, ions and molecules the photon-involving processes are
bound-bound transitions producing spectral lines and bound-free and
free-free transitions producing continua. 
In addition there are Thomson scattering of existing photons by free
electrons and Rayleigh scattering of existing photons by bound
electrons. 
These processes can provide local emissivity or extinction changing the
intensity in a given beam, in particular the intensity along the
``line-of-sight'' towards our telescope. 
Other photon-producing or photon-affecting processes may be ignored
for the solar atmosphere unless you are a radio astronomer (cyclotron,
synchrotron, plasma radiation) or high-energy astronomer (pair
annihilation). 

Photons being bosons means that they like to sit in the same place and
therefore that any interaction producing or scattering a photon has a
stimulated addition in which the new or scattered photon coherently
accompanies similar passing radiation that enhances the interaction
probability.  
This contribution scales with $\exp\,(-h\nu/kT)$, negligible in the Wien
limit \SSIpage{3}. 
For scattering it has the same probability between extinction and
emissivity (pairs {\bf c} and {\bf g} in \rrref{Fig.}{fig:pairs}).

\paragraphrr{Bound-bound, bound-free, free-free  interactions}
The five bound-bound interactions are spontaneous photo-deexcitation,
photo-excitation, induced photo-deexcitation, collisional excitation
and collisional deexcitation measured with the Einstein coefficients
$A_{ul}$, $B_{lu}$, $B_{ul}$, $C_{lu}$, $C_{ul}$. 
The radiative first three are transition properties that do not depend
on circumstances (by keeping the local photon supply outside the $B$
definitions). 
Einstein therefore derived their ratios assuming strict detailed
balance (valid only in ideal TE) and declared these valid anywhere so
that $A_{ul}$ (or the classical oscillator strength $f$) is the single
radiative probability parameter \SSFpage{70},
\ISSFpage{23}{Sect.\,5.1}, \IARTpage{60}{Chapt.\,5},
\RTSApage{39}{Sect.\,2.3.1}.

The same five hold for bound-free ionization and recombination with
appropriate nomenclature, and the same five hold for free-free
interactions (but collisional three-body up or down transitions have
no radiative interest).

\paragraphrr{Pair combinations}
Combining the above interactions in successive pairs is so important
that I do not refer to \linkrjrpage{2019SoPh..294..165R}{3}{Fig.\,1} of
\citetads{2019SoPh..294..165R} 
but copy it here in \rrref{Fig.}{fig:pairs}.
It came from \SSFpage{105} and is an extension of
\RTSApage{85}{Fig.\,3.3} with multi-level detours.
\CaII\ and \Halpha\ detour examples are shown in
\linkrjrpage{2019SoPh..294..165R}{4}{Fig.\,2} of  
\citetads{2019SoPh..294..165R}. 
With these added this is a complete inventory of how line photons can
be locally created and modified in bound-bound manner.\footnote{In 3D
teaching I draw these cartoons on the whiteboard as in
\rrref{Fig.}{fig:weihai} and point to them all the time. 
In the blackboard era I drew them with chalk and threw pieces at
inattentive students until hitting
\linkrjrpage{2018NTvN-Minnaertportret}{2}{Minnaert}.} 
Below I refer to these pairs by their alphabetic labels.
Similar pair diagrams can be drawn for bound-free transitions 
(some in \rrref{Fig.}{fig:weihai}).

\paragraphrr{Equilibria}
\SSFpage{69} summarizes bound-bound equilibria with the 2-level
simplification at left in \rrref{Fig.}{fig:pairs}.
It assumes that any excitation is followed by de-excitation in the
same transition, ignoring multi-level NLW and so restricting RT to a
single wavelength (CS) or the narrow spectral band covering a single
transition (CRD). 

TE = {\em ``thermodynamic equilibrium''\/} describes a homogeneous
isothermal fully-enclosed gas where nothing ever happens and where
every type of excitation occurs just as frequently as its de-exciting
counterpart (``detailed balance''): as many collisions down as up, as
many radiative transitions down as up, with all rates defined by the
temperature. 
Maybe this boring paradise occurs approximately at the center of a
black hole (bar slow Hawking losses), but not within the Sun since its
non-enclosedness implies net outward energy leak from the hot core to
the cold universe with the accompanying outward temperature decline
through the entire Sun pushing the fusion-produced energy surplus to
eventually escape
\rrendnote{solar neutrinos}{note:neutrinos}{%
Fortunately the fusion-released gamma rays escape by slow outward
diffusion converting them into more agreeable sunlight
\RTSApage{110}{Shu quote} plus enthralling dynamo action causing
magnetic flux escape and harmonious humming enabling internal sounding
(see never-cited grand/t overview in
\citeads{2008ASPC..383..315A}). 
In contrast, the neutrinos that are also released in hydrogen fusion
are virtually unstoppable and escape directly from the solar core. 
For them we do not need RT theory, nor worry about the $10^{11}$
passing every second through our thumbnails.
They posed a
\href{https://math.ucr.edu/home/baez/physics/ParticleAndNuclear/solar_neutrino.html}{big
problem} last century when too few were detected. 
It was not a fault of the measurement nor of stellar structure theory
but the blame was on particle physics: part of the solar neutrinos
``oscillate'' into another flavor on their way here. 
Photons, once they escape, fulfill their task of information carrying
and information-transferring detectability much better.\footnote{Solar
neutrinos may convey information on scotogenic dark matter beyond my
and your solar physics interest (\citeads{2021arXiv210505613D}).}}.

SE = {\em ``statistical equilibrium''\/} foregoes detailed balancing
but requires overall balancing.
All level populations are instantaneously balanced at the current
ambient temperature, density and radiation without memory of earlier
circumstances.
For each level the total rate equation \SSFpage{75}
\RTSApage{52}{Eq.\,2.100} summing changes to its population ends up
zero: every moment as much coming in as going out, with excesses
between coming in or going out some particular way compensated by
excesses going out or coming in another way.
The SE assumption underlies all following equilibria except NSE.

LTE = {\em ``local thermodynamic equilibrium''\/} assumes TE at the
local temperature while permitting temperature and density gradients
and small leaks, clearly a cheap cheat. 
Its definition is that Saha-Boltzmann (SB) partitioning applies to all
populations, as for element Schadeenium\footnote{Naming: which 
\label{note:schoonschip} you can't pronounce. 
My Utrecht colleague
\href{https://robrutten.nl/nieuwenhuijzenshots/1961-1965/Sonnenborgh-Aert-Schadee-2.jpg}{A.\,Schadee}
invented this didactic element.
M.J.G.\,Veltman called his revolutionary algebra solver for the CERN
CDC 6600 ``schoonschip'' (clean sweep, shipshape) to annoy his 
colleagues on pronunciation.}
in my recommended SSA\,2
\linkpdfpage{https://robrutten.nl/rrweb/rjr-edu/exercises/ssa/ssa.pdf}{9}{Cecilia
Payne exercise}. 
It is often thought that LTE is defined as equality of the source
function $S\!_\nu$ to the Planck function $B_\nu$ but this is a
corollary \RTSApage{43}{Sect.\,2.3.2} \RTSApage{45}{Eq.\,2.7.3}. 
Colleagues that think this usually ignore or are not aware of NLTE
departures in opacities (\eg\ for ALMA,
\rrref{endnote}{note:ALMA-temp}).

LTE holds closely within the Sun and to some extent for many lines and
continua in the low photosphere. 
The condition is that collisions both up and down heavily outweigh
photons up and down in \rrref{Fig.}{fig:pairs} and its bound-free
counterparts.
Photon destruction {\bf a} and photon creation {\bf d} + {\bf e} must
dominate over spontaneous and stimulated scattering, also for all
steps in the photon conversion sequences at right. Thus, the photons
must behave as ``honorary gas particles'' (Castor), likewise boxed in
to local circumstances; what we observe must be only a small leak.
In SB partitioning the level population ratios sense only the
temperature \RTSApage{49}{Eq.\,2.86} because the dominating up and
down collisions both require one collider so that their density cancels.
The ionization stage ratios sense temperature more complicatedly by
folding the Boltzmann factor with the Maxwell distribution for the
caught electron into the Saha distribution \RTSApage{50}{Eq.\,2.88};
they scale inversely with the electron density from the need to catch
one.

NLTE = {\em ``no local thermodynamic equilibrium''\/} is a misnomer. 
Usually it means relaxing SB while assuming SE with accounting for
NLS, less often NLW, much less often NLT. 
Often named and written non-LTE.

CE = {\em ``coronal equilibrium''\/} assumes SE but low density and
absence of local irradiation so that only {\bf d} and its bound-free
counterpart remain. 
All ups are collisional and all downs are spontaneously radiative from
lack of impinging particles and photons.\footnote{Likewise, {\bf d}
and its bound-free counterpart govern the beautiful polar light in
aurorae (but the colliding electrons and protons are not thermally but
electrically accelerated). 
At sufficient height in the Earth's atmosphere restoring collisions do
not compete.
Nitrogen-atom recombination shines blue-violet.
Nitrogen and oxygen in molecules shine red and green by
radiative deexcitation from metastable levels.
These lines were first called ``geocoronal'' akin to
solar ``coronium'' lines (\rrref{endnote}{note:forbidden}). 
There are no detections of exoplanet aurorae yet.} 
Every photon is locally created by a collision obeying the local
kinetic temperature. 
They occur infrequently but in their ground state ions can wait
endlessly for suited colliders. 
After eventual excitation no collider or stimulating photon arrives
within the $1/A_{ul}$ decay probability for spontaneous deexcitation.
The resulting photon escapes outward to space and may enter our
telescope (fat chance), or it drowns inward in the Sun, or it gets
bound-free scattered in other gas (more below). 
In this case the level ratios sense both temperature and collider
density since there is only the required up collision without frequent
down collision required for LTE.
The stage ratios sense only temperature because each ionization is
collisional and each recombination is radiative, both requiring a
single electron catch.   

CE and SB are thin versus thick (tenuous versus dense gas) SE
partitioning extremes.  
\SSXpage{174} compares them for photospheric electron density; the SB
peaks shift left by about -0.05 in $\log(T)$ per tenfold $N_\rme$
reduction.

A CE complication and boon is that dielectronic excitation must be
included beyond the single-electron jump pairings in
\rrref{Fig.}{fig:pairs}.
The reason is that at coronal temperatures the mean Maxwellian energy
is far above the desired standstill value for the free electron to be
caught at ionization threshold in one-electron recombination (peak
probability in hydrogenic Kramers
$\alpha_\nu^{\rm bf}\!\sim\!\,\nu^{-3}$ edge decay).
The shift of the Maxwell peak to larger energy for higher temperature
gives $1/T$ dependence to radiative recombination rates. 
With dielectronic excitation this too large thermal energy is cut down
by using a large part for simultaneous bound-bound excitation of
another electron, both in dielectronic ionization up and in dielectronic
recombination down.
They are multi-level additions behaving in CE as pair {\bf d}: up only
collisions, down only spontaneous photons
\rrendnote{dielectronic ionization and recombination}{note:dielectronic}{%
Unstable levels: both processes excite one electron in a normal
bound-bound transition and put a second electron into a level above
the nominal ionization limit of the lower-stage ground state. 
This is possible because the orbitals are boosted by the first
excitation; they may also be non-hydrogenic affected.
Such a state is unstable because {\em autoionization\/} may occur,
freeing the electron from there into escape.
This usually leaves the fresh ion in its ground state and then the
freed electron carries off the energy excess of the unstable state
above that, but if the fresh ion is excited it decays, in CE always
per line photon. 

The other path out of the unstable level is decay of one (or both) of
the excited electrons, in CE also always per line photon, into a
configuration with the second electron in a regular bound level below
the nominal ionization limit ({\em ``radiative stabilization''\/}).

Autoionization often has higher probability than spontaneous
stabilization.

{\em Dielectronic ionization\/}: an energetic collider, usually
hitting the lower ion in its ground state, kicks up two
electrons with the first one taking a good part of its kinetic energy. 
With the remainder the second may ionize bound-free but is more likely
put into a high level. 
If this is still below the ground-state limit both electrons decay
radiatively, the second likely in a cascade, together a two-electron
multi-level version of pair {\bf d} converting kinetic energy into
line radiation without ionization. 
However, if the second electron reaches an unstable level above the
nominal limit then it may be freed by autoionization before its
radiative decay, a sequence called ``excitation-autoionization'' by
\citetads{2007A&A...466..771D}. 
The other still bound excited electron decays to the ground state of
the new ion, also producing one or more line photons. 
In this double-excitation manner the energetic collider yields higher
ionization probability than for one-electron collisional ionization by
losing energy that escapes as line photons.

{\em Dielectronic recombination\/}: in the reverse sequence the
energetic colliding free electron kicks a bound electron up to
a high level and is itself captured into an unstable state which
stabilizes already before re-auto-ionization through spontaneous
radiative decay of one of the two excited electrons, often the lower
one, to a stable level of the lower stage. 
If that is still excited it decays radiatively to the ground state;
the other electron also, usually per multi-level cascade. 
Depending on the branching ratios some decays may temporarily end up
in a metastable level comparable to \CaIR\ in the \CaII\ detour
example in \linkrjrpage{2019SoPh..294..165R}{4}{Fig.\,2} of
\citetads{2019SoPh..294..165R} 
but then not go down from there collisionally as there but CE-wise
radiatively in a forbidden transition such as the optical coronal
emission lines of \rrref{endnote}{note:forbidden}. 

{\em Radiative result\/}: no continuum but line photons galore.}.
The temperature-only sensitivity of CE stage partitioning is therefore
maintained; the rate ratio between direct and dielectronic depends
only on temperature (Maxwell peak shift).   
The complication is that many possible bound-bound transitions to 
high levels must be included and evaluated with CHIANTI 
(\rrref{endnote}{note:CHIANTI}).
The boon is that the dielectronic variants produce photons in many
lines including the diagnostics in the AIA EUV passbands and help keep
the corona cool through substantial radiative losses.
Classical \linkadspage{1964ApJ...139..776B}{4}{Fig.\,1} of
\citetads{1964ApJ...139..776B} 
shows the high-temperature importance for \HeI\ ionization.
The radiative rate drops as $1/T$ but the dielectronic rate first
peaks two-orders-of-magnitude higher before its $1/T$ decay sets in. 
Classical \linkadspage{1969MNRAS.142..501J}{20}{Fig.\,3} of
\citetads{1969MNRAS.142..501J}\footnote{Carole Jordan is the CE
counterpart of SB Cecilia Payne, moving to Oxford from Culham rather
than to Cambridge USA from Cambridge UK.
My SSA exercises should become Annie Cannon -- Cecilia Payne -- Carole
Jordan, likely using \href{https://www.chiantidatabase.org}{CHIANTI},
and I should also add a Walter Grotrian \SSXpage{6} exercise to make you
appreciate that during totality you bask in ordinary photospheric
sunlight -- but your cleanest ever (\rrref{footnote}{note:clean}).}
shows its importance for coronal iron partitioning (the lower panel
shows partitioning without it from
\citeads{1964ApJS....8..307H}). 
The ``Jordan versus Payne'' CE--SB comparisons on \SSXpage{174}
include it also.
A nice summary following
\citetads{1964ApJ...139..776B} 
is given in \linkadspage{1968slf..book.....J}{144}{Sect.\,6.2.5} of
\citetads{1968slf..book.....J} 
\rrendnote{\CaII\ \HK\ core reversals}{note:HKreversals}{%
On \linkadspage{1968slf..book.....J}{146}{page 128} Jefferies cited
the suggestion of
\citetads{1964ApJ...140..384G} 
that the double emission peaks of the \CaII\ \HK\ core reversals may
be due to dielectronic recombination and also the demonstration by
\citetads{1965SAOSR.174..405N} 
that this does not work for CRD in \HK.
Noyes used Avrett's brandnew CRD scattering code producing the
canonical $\sqrt{\varepsilon}$ graphs of \SSFpage{80}
in the belief that CRD is a better approximation than CS which he had
initially assumed in
\citetads{1964AJ.....69R.542G}, 
but actually \HK\ are PRD lines.
Goldberg's mechanism seems to be collisional dielectronic ionization
of \CaI\ leaving \CaII\ ions excited in the upper levels of \HK\ --
but I note that \CaI\ is a minority species and think the idea was a
red herring.

It does illustrate how during decades the \CaII\ \HK\ reversals were
the most enigmatic features in the optical spectrum by being the only
Fraunhofer-line departures from regular bell shape. 
This extended literature (\eg\ early 
\citeads{1913ApJ....38..292E}) 
was outstandingly reviewed by
\citetads{1970PASP...82..169L} 
(but not mentioning the Goldberg--Noyes dielectronic angle).
\citetads{1967AJ.....72..784A} 
had suggested that the \Htwo\ and \Ktwo\ dips map a temperature
minimum as in NLTE cartoon \SSFpage{37}, but then
\citetads{1975ApJ...199..724S} 
showed that PRD upsets this notion as in cartoon \SSFpage{89} with
their \linkadspage{1975ApJ...199..724S}{7}{Fig.\,9} copied in
\SSFpage{92}.
In addition, such plane-parallel modeling cannot explain the
observed asymmetry between the violet and red peaks 
(\rrref{Sect.}{sec:Avrett}).

The \CaII\ \HK\ core reversals also govern the famous cool-star
relation in \linkadspage{1957ApJ...125..661W}{9}{Fig.\,1} of
\citetads{1957ApJ...125..661W} 
which to my knowledge remains unexplained.  
Doing so will make you famous! 
Since the main contribution to the peaks in solar irradiance is from
activity-measuring calcium network and plage on classic
spectroheliograms (plus internetwork acoustics as \HtwoV\ and \KtwoV\
grains, see \rrref{endnote}{note:gillespie}) I suggest an approach
along the lines of
\citetads{1976A&A....53..341S} 
and \citetads{1979ApJ...228..509A} 
but applied to fluxtube atmospheres with attention to the slow network
waves in the center panels of
\linkrjrpage{1993ApJ...414..345L}{4}{Fig.\,3} of
\citetads{1993ApJ...414..345L} 
and fluxtube shocks as in
\citetads{2011ApJ...730L..24K} 
while subtracting or including the basal flux contribution
(\rrref{endnote}{note:basal}).}.

NSE = {\em ``no statistical equilibrium''\/}\footnote{Naming: I used
to write non-E, also in earlier versions of this compendium. 
Others use NEQ. 
I switched to NSE analogously to NLTE.  
If you write non-LTE you might write non-SE.} admits NLT temporal
memory in populations, \ie\ not all level population rate equations
\RTSApage{52}{Eq.\,2.100} sum to zero.
The term ``non-equilibrium ionization'' (often abbreviated to NEQ)
usually concerns hydrogen where the actual NLT agent is slow
collisional settling of \Lyalpha\ in cooling gas whereas H ionization
occurs in an SE Balmer loop (more below).

\paragraphrr{Bound-bound  redistribution}
On \linkadspage{1929MNRAS..89..620E}{2}{page~2}
\citetads{1929MNRAS..89..620E} 
asked ``the crucial question whether light absorbed in one part of a
line is re-emitted in precisely the same part of the line'': to what
extent a resonance-scattered photon remembers the frequency it had in
the preceding photo-excitation (pairs {\bf b, c, f, g}). 
The one extreme is that it does precisely in ``coherent'' scattering
(CS). 
In Eddington's days this was thought to be the rule, but in his
landmark thesis\footnote{Scanned 
by A.V.~Sukhorukov and put on ADS by me.}
\citetads{1942PhDT........12H} 
showed that most solar lines in the visible obey the other extreme,
suffering complete redistribution (CRD) meaning complete loss of
memory and representing a new sample of the line extinction profile
rather than $\delta$-function frequency conservation.
CRD is indeed valid for most lines except the strongest (large
extinction) whose photons escape high in the atmosphere where
collisions governing collisional redistribution (``damping'') are
rare. 
These lines must be described by partial redistribution (PRD) which
combines Doppler redistribution in the core (CS in the frame of the
atom but an redistributing average over the observed
individually-Doppler-shifted photon ensemble) with coherency in the
inner wings and collisional redistribution in deeply-formed hence
collision-rich outer wings \SSFpage{89}.
The principal example is \Lyalpha. 
Other well-known PRD lines are \MgII\ \hk\ and \CaII\ \HK\
\SSFpage{92}, \SSFpage{93}.
My personal example is \BaII\ 4554\,\AA\
\rrendnote{\BaII\,4554\,\AA\/ and \MnI\ lines}{note:4554}{%
\BaII\ 4554\,\AA\ showed PRD signature near the
limb in my eclipse observation in
\linkrjrpage{1978SoPh...56..237R}{17}{Fig.\,9} of
\citetads{1978SoPh...56..237R}. 
It was confirmed in
\linkrjrpage{1979ApJ...231..277R}{4}{Figs.\,1\,--\,3} of
\citetads{1979ApJ...231..277R} 
and \linkadspage{1992A&A...265..268U}{5}{Fig.\,6} and
\linkadspage{1992A&A...265..268U}{7}{Fig.\,11} of
\citetads{1992A&A...265..268U}. 
The line is a valuable spectropolarimetry diagnostic
(\eg\
\linkpdfpage{https://robrutten.nl/uso-dwingeloo-2009/lectures/Keller-Spectropolarimetry.pdf}{4}{C.U.\,Keller course},
\citeads{2007ApJ...666..588B}) 
and also a good Doppler diagnostic of the upper photosphere
(\citeads{2009A&A...506.1393S}, 
\citeads{2009A&A...506.1405K}) 
because it combines large atomic mass, hence small thermal broadening,
with considerable core broadening by isotope splitting and hyperfine
structure making it suited for filter instruments and so yielding the
astounding Dopplergram in
\linkrjrpage{2001AAp...378..251S}{4}{Fig.\,5} of
\citetads{2001A&A...378..251S}. 

Other lines with large hyperfine broadening are the \MnI\ ones
recommended by G.\,Elste to W.C.\,Livingston as candidates for
long-term spectral irradiance monitoring because they are insensitive
to infamous microturbulence.
Livingston found larger cycle-dependent variation in
\MnI\,5394.7\,\AA\ than for other atomic lines
(\linkadspage{2007ApJ...657.1137L}{9}{Fig.\,16} of
\citeads{2007ApJ...657.1137L}). 
This enhanced sensitivity was attributed by
\citetads{2001A&A...369L..13D} 
to pumping of \MnI\,5394.7\,\AA\ by \MgIIk\ but erroneously assuming
CRD for \MgIIk\ in the spectral synthesis of their demonstration. 
\citetads{2009A&A...499..301V} 
showed that the actual reason is that hyperfine-broadened lines indeed
lack the ``microturbulent'' thermal and granular Doppler smearing
through which all narrower photospheric lines lose such sensitivity
(MURaM demonstration in their
\linkrjrpage{2009AAp...499..301V}{8}{Fig.\,6} and
\linkrjrpage{2009AAp...499..301V}{9}{Fig.\,7}).

In a review of this issue in
\citetweb{https://robrutten.nl/rrweb/rjr-pubs/2011bandung.pdf}
{2011bandung} 
\linkrjrpage{2011bandung}{2}{Fig.\,1} summarizes Livingston's results
and \linkrjrpage{2011bandung}{3}{Fig.\,2} shows that in plage the
\MnI\ blend (at -0.7\,\AA\ from \MgIIk\ center) is not pumped at all
because the blend is at the wavelength where the PRD source function
departs most from the CRD one and sits in the dip outside the \MgII\
peak even in strong plage.
The apparent \MnI\ over \FeI\ brightening of plage in
\linkrjrpage{2011bandung}{2}{Fig.\,1} in scans from
\citetads{2004IAUS..223..645M} 
results from mean-field brightness normalization. 
The surrounding granulation is actually darker in the \MnI\ line than
in the turbulence-sensitive \FeI\ line while these lines show the same
actual plage brightness because the fluxtube ``holes'' schematized in
\linkrjrpage{2011bandung}{5}{Fig.\,4} reach as deep because both \FeI\
and \MnI\ ionize away.}.
Other strong lines suffer less or no PRD
\rrendnote{detour and cross redistribution}{note:detourdist}{%
\CaIIHK\ are the strongest lines in the visible and of the dominant
ionization stage so that the two-level pairs in
\rrref{Fig.}{fig:pairs} suffice. 
In the ALC7 star they have $\varepsilon\tapprox\,10^{-4}$
\SSXpage{88}, as do \MgIIhk\ \SSXpage{89} while \Lyalpha\ reaches
$\varepsilon\tapprox\,10^{-6}$ \SSXpage{90}.
The other strong lines in the visible are the Balmer lines, the \CaII\
infrared triplet, the \MgIb\ triplet, and the \NaID\ and \KI\
resonance doublets.
These are all scattering lines with $\varepsilon\le\,10^{-2}$ but from
minority stages\footnote{For
\label{note:Htop}
Balmer lines regard only the hydrogen top with $n\tis\,2$ as ``ground
level'' because its population is fully controlled by \Lyalpha\ doing
its own thing (\rrref{Sect.}{sec:Oslo}), making this top a
low-abundance severe-minority stage since $n_2/n_\rmH <10^{-7}$
\SSXpage{91} while the proton fraction $n_\rmp/n_\rmH >10^{-4}$ where
the \Halpha\ core forms. 
In this fixed-bottom minority-stage top \Halpha\ is a resonance line
comparable to
the alkali resonance lines (\rrref{endnote}{note:suction})
but scattering higher while absent in the upper photosphere
(\rrref{endnote}{note:photoelectric}).}
with sizable contributions by the detour pairs in
\rrref{Fig.}{fig:pairs}, for \Halpha\ ionization-recombination loops
as at right in \linkrjrpage{2019SoPh..294..165R}{4}{Fig.\,2} of
\citetads{2019SoPh..294..165R} 
contributing up to 10\% \SSXpage{96}.
These detour contributions have no coherency.  
In addition, the Balmer lines have wide Doppler redistribution from
the small hydrogen mass and their outer wings are formed deeply
were collisional Stark redistribution is large
(\rrref{endnote}{note:quietHa}).
For \NaIDtwo\
\citetads{1992A&A...265..268U} 
found PRD signatures only near the limb, none for \KI\,7699\,\AA. 
The much stronger \MgI\ 2852\,\AA\ resonance line has significant PRD
effects (\citeads{1977ApJ...216..654C}). 

Cross redistribution occurs for the \CaII\ triplet lines.
In the detour sketch at left in
\linkrjrpage{2019SoPh..294..165R}{4}{Fig.\,2} of
\citetads{2019SoPh..294..165R} 
this concerns the consecutive loop 3933 up, 8542 down and back up,
3933 down.
It was analyzed by
\citetads{1989A&A...213..360U} 
who showed in \linkadspage{1989A&A...213..360U}{7}{Fig.\,2} that PRD
frequency dependence of the \CaIIK\ source function affects the \CaIR\
source function strongly through the Wien-reversing sensitivity
transcription (\rrref{endnote}{note:pumping}).
However, since \HK\ are much stronger these large PRD effects do not
affect emergent \CaIR\ profiles from plane-parallel atmospheres as
modeled there because in these the infrared lines are formed well
below the heights of PRD source function splitting for \HK. 
This is likely not the case when both \HK\ and the triplet lines form
together in chromospheric features without such height separation
(\eg\ spicules-II and return fibrils,
\rrref{endnote}{note:spicules-II}).}.

I must concede that PRD theory remains ``not yet'' in
\RTSApage{92}{Sect.\,3.4.3}.
Classic:
\citetads{1962MNRAS.125...21H} 
\SSFpage{90}.
Clearest explanation so far:
\linkadspage{1968slf..book.....J}{110}{Chapter 5} of
\citetads{1968slf..book.....J} 
summarized in \SSFpage{91}. 
The introduction to
\citetads{2017A&A...597A..46S} 
is a good literature overview. 
Recent developments are listed in \SSFpage{94}.
Next to frequency redistribution there is also angle redistribution to
worry about for moving gas due to its Doppler anisotropy -- what gas
in the solar atmosphere doesn't move? 
See \citetads{2012A&A...543A.109L}. 

\paragraphrr{Bound-free  redistribution}
Bound-free transitions are similar to bound-bound transitions and may
be described by the same rate equations \RTSApage{68}{Sect.\,3.2.3}.  
Resonance scattering (radiative up and down in the same transition,
pairs {\bf b}, {\bf c}, {\bf f}, {\bf g}) is also the same except that
the bound-free spectral extent is much wider and that recombination
catches an electron as a fresh sample of the kinetic energy
distribution without remembering the energy of the electron kicked out
in preceding ionization, hence pairwise only memory of ionization to
the threshold but not of the excess above it.
Therefore there is always CRD over the ionization edge.

A particular case is bound-free scattering of EUV line photons imaged
in narrow bands as by EIT, TRACE, AIA.
They may be extincted in bound-free photo-ionization of \HI, \HeI\ or
\HeII, most likely followed by spontaneous photo-recombination because
the colliding-particle and suited-photon densities are too low for
other bound-free equivalents than the one for scattering pair {\bf b}. 
Through CRD over the edge such a bound-free scattered photon not only
gets a random redirection but also a new wavelength, most likely near
the ionization threshold (probability for \HI\ given by Kramers
$\nu^{-3}$ decay \RTSApage{45}{Eq.\,2.74}), hence far from the line.
Bound-free scattering so redirects radiation out of the line of sight
and also out of the instrument passband; it may locally darken
narrow-band images\footnote{Often misnamed ``absorption'' instead of
extinction from misinterpretation as ``true'' absorption as does
radiation-blocking Mercury in transit.}
\rrendnote{dark EUV features}{note:rain}{%
When you see the same black features on bright backgrounds (on the
disk but outside no-emissivity coronal holes) in multiple EUV
wavelengths then out-of-the-passband scattering is the most probable
mechanism. 
Cartoon in \SSXpage{10} (where ``volume blocking'' is a silly misnomer);
I drew it as \linkrjrpage{1999ASPC..184..181R}{12}{Fig.\,10} of
\citetads{1999ASPC..184..181R} 
to answer a TRACE question by C.J.\,Schrijver. 
It shows that when black features are also present in \HeI\,584\,\AA\
images (not in AIA alas) the agent is photoionization in the \HI\
Lyman continuum. 
For shorter wavelengths the agent can also be \HeI\ or \HeII\
ionization (\cf\ \citeads{2014ApJ...782...87C}). 

This may occur in filaments seen in EUV lines as in the IBIS\,--\,SDO
active-region blinker of \rrref{endnote}{note:IBIS}. 
Other such features are ``condensation'' (meaning recombination)
clouds, \eg\ in coronal rain
(\citeads{2012ApJ...745..152A}, 
\citeads{2015ApJ...806...81A}) 
visible in \Halpha\ via NSE recombination cascade giving large
opacity on the disk and large emissivity off-limb as in the temporal
cycling of \citetads{2020A&A...633A..11F} 
(\linkadspage{2020A&A...633A..11F}{10}{Fig.\,7}).
More in \rrref{endnote}{note:blobs}.}.  

\paragraphrr{Free-free  redistribution}
Every collider is a fresh sample of the kinetic energy distribution
giving redistribution over the probability $\nu^{-3}$ decay
\RTSApage{46}{Eq.\,2.76}, a continuum without threshold.  
In addition there is no memory of a preceding transition so that pair
combinations as in \rrref{Fig.}{fig:pairs} make no sense.
If the kinetic energy distribution is the Maxwell distribution defined
by temperature then that fully defines both emissivity
(Brehmstrahlung) and extinction so that their ratio (source function)
is the Planck function.
Thus, free-free transitions always have $S\!_\nu\tis\,B_\nu$ because
they are intrinsically collisional.

A particular case is ALMA for which at longer wavelengths the main
agent is hydrogen free-free (electrons meeting protons) \SSXpage{32}.
ALMA is therefore often appreciated as a thermometer, but it is as
often not appreciated that the extinction is likely NLT controlled and
may be orders of magnitude above the local SB value with scattering
aureoles, just as for \Halpha.
ALMA is a thermometer but without easy knowledge where you stick it
\rrendnote{ALMA as thermometer}{note:ALMA-temp}{%
The extinction graphs on \SSXpage{32} are a selection of the tableau
in \linkrjrpage{2017AAp...598A..89R}{4}{Fig.\,1} of
\citetads{2017A&A...598A..89R}. 
They specify SB extinction for the hydrogen mm continua in comparison
with \Lyalpha, \Halpha\ and \CaIR. 
The \HI\ bf and ff continua and \Halpha\ share an initial steep
Boltzmann rise, less steep for \CaIR\ and \Hmin\ ff, followed by
slower decay or leveling out where hydrogen is fully ionized (lower
panels).
For non-SB NLTE this steep rise desteepens somewhat, pivoting around
the Balmer continuum brightness temperature of 5300\,K because that
defines radiative overionization from level $n\tis\,2$ at cooler and
underionization at hotter temperatures; these NLTE corrections can
reach 2-3\,dex slope modification of the 10-dex rise
(\rrref{Sect.}{sec:Oslo}).  
The free-free curves increase with the passband wavelength by the
Rayleigh-Jeans factor $\lambda^2$ in
$\alpha_\lambda^{\rm ff}\tsim N_\rme\,N_\rmp\,T^{-3/2}\lambda^2$
\RTSApage{47}{Eq.\,2.79}. 
The upshot is that hot features that reach optical thickness in
\Halpha\ are likely thicker to much thicker in the ALMA passbands.
In the last column of \linkrjrpage{2017AAp...598A..89R}{4}{Fig.\,1}
hot features become even thicker than in \Lyalpha, still optically
thick at only 100-km geometrical thickness at hydrogen density only
$10^{11}$\,cm$^{-3}$.

The assumption of SB LTE in these curves holds for hot conditions
giving fast collisional equilibriating, for the hydrogen diagnostics
because their equilibrium is regulated by \Lyalpha\
(\rrref{Sect.}{sec:Oslo}) which defines the very large Boltzmann
increase with NLTE ionization as minor slope correction. 
However, in post-heating cooling gas the actual extinction hangs per
NSE retardance for minutes at the high values reached previously
because \Lyalpha\ settles slower (\rrref{Sect.}{sec:Oslo}, the 
PSBE of \linkrjrpage{2017AAp...598A..89R}{2}{Sect.\,2.1} of
\citeads{2017A&A...598A..89R}). 
This explains the large opaqueness of cooling \Halpha\ return fibrils
constituting chromospheric \Halpha\ network canopies
(\rrref{Sect.}{sec:chromosphere}) and their telltale differences with
\CaIR\ fibrils (\rrref{endnote}{note:nonEchrom}).  
For fibrils in \Halpha\ and \CaIR\ the intensity senses
$\sqrt{\varepsilon}$ scattering with fibril contrast variations that
are defined primarily by individual Doppler modulation for
non-retarded \CaIR\ with its $\sqrt{m_{\rm Ca}/m_\rmH} \tis\,6.3$ less
thermal broadening and by individual retardance history for \Halpha,
whereas for ALMA fibril contrasts instead represent actual
instantaneous temperatures (\rrref{endnote}{note:ALMA-chrom}).
Hence, the scenes appear different although sampling the same
features, with \Halpha\ core-width the closest to ALMA.

My prediction in \citetads{2017A&A...598A..89R} 
with a tutorial in
\citetads{2017IAUS..327....1R} 
was therefore that ALMA shows chromospheric canopies wherever these
are present in \Halpha\ (\rrref{endnote}{note:ALMA-chrom}).
The good pattern match between ALMA intensity and \Halpha\ core width
in \linkadspage{2019ApJ...881...99M}{6}{Fig.\,4} of
\citetads{2019ApJ...881...99M} 
confirmed it.\footnote{In
\linkrjrpage{2017AAp...598A..89R}{9}{prediction 2} of
\citetads{2017A&A...598A..89R} 
I wrongly claimed dark-dark correspondence between ALMA and \Halpha\
images. 
It should instead have been bright-bright correspondence between ALMA
intensity and \Halpha\ core width as indicator of kinetic temperature,
evident in the bottom rows of
\linkrjrpage{2009AAp...503..577C}{7}{Fig.\,6} of
\citetads{2009A&A...503..577C}. 
The next-best indicator there is \CaIR\ Doppler-shifted minimum
intensity but there is no correlation for \Halpha\ minimum intensity
which is history-modulated rather than thermally sensitive.
I had plotted these correlation diagrams myself and should have
remembered them. 
So it goes.  
Indeed my ALMA to GONG \Halpha\ pattern alignment in
\rrref{endnote}{note:ALMA-GONG} uses dark-bright instead of dark-dark
correspondence in quiet areas, similarly to my GONG \Halpha\ pattern
alignment to AIA 304\,\AA\ there. 
Active regions instead show bright-bright correspondences between
these pairs, as evident for \Halpha\ core intensity and
\HeII\,304\,\AA\ in the IBIS\,--\,SDO blinker of
\rrref{endnote}{note:IBIS}.} 
Earlier predictions based on plane-parallel modeling and various
simulations suggested instead that ALMA samples lower ``layers'' and
shocks but these efforts erred in assuming SE hydrogen ionization
(\rrref{Sect.}{sec:Oslo}).   

Thus, while ALMA is a thermometer thanks to free-free source function
LTE, due to NLT extinction NLTE it more likely samples \Halpha\
canopies than internetwork acoustics
(\rrref{endnote}{note:ALMA-waves}) in the clapotisphere underneath
(\rrref{endnote}{note:clapotis}) -- a blessing because we understand
the latter for decades but do not understand the chromosphere.
ALMA furnishes transition-radiation diagnostics that sample
chromospheric canopies also seen in \Halpha\ and in \HeII\,304\,\AA.
It samples their temperatures around $\tau_{\nu\mu}\tis\,1$ but these
thick-feature sampling depths suffer NLS \Lyalpha\ surround scattering
and vary NLT-wise with the local temperature history
(\rrref{endnote}{note:ALMA-chrom}).}.

\paragraphrr{Thomson and Rayleigh scattering}
In terms of \rrref{Fig.}{fig:pairs} these scatterings are described by
pairs {\bf b, c} and {\bf f, g}, none of the others. 
In the Thomson limit (short of Compton) they still suffer Doppler
redistribution.
Pairs {\bf c} and {\bf g} cancel in occurrence probability, averaged
for Doppler redistribution.  

A particular case is Thomson scattering by free electrons making the
white-light corona so beautifully visible during eclipses and with
LASCO
\rrendnote{white-light corona}{note:wlcorona}{%
The reason why we must travel far to see the corona\footnote{Personal
history: after starting my career with spectroscopy at three eclipses
my wife and I became positively corona-infected eclipse tourists. 
Our traveling in 2020 to my 15th to complete my three Saros 142
periods was covid-canceled.} is evident from the visibility diagram in
\SSXpage{5} after \linkadspage{1953sun..book..207V}{51}{Fig.\,6} in
the classic review 
by \citetads{1953sun..book..207V}. 

The pearly white color and linear polarization of the K corona and its
apparent radiation temperature
(\citeads{1938PASP...50..225M}) 
suggest photospheric light that is bent towards the viewer without
wavelength selectivity (unlike blue-favoring Rayleigh scattering),
hence Thomson scattering, Mie scattering, or reflection without
refraction.
The absence of photospheric Fraunhofer lines in the coronal spectrum,
unlike their presence in the spectrum scattered daily by your nose,
was a large disappointment for the first observers bringing
spectroscopes to totality, played a role in the debate whether the
corona is real or an illusion, and remained enigmatic during many
decades. 
Thomson and Mie scattering are coherent so retain frequencies and
hence should retain Fraunhofer lines.
Then \citetads{1931ZA......3..199G} 
noticed that \CaII\ \HK\ leave a washed-out dip and
\linkadspage{1931ZA......3..199G}{22}{proposed here} that their
100\,\AA\ smearing corresponds to 7500~\kms\ Doppler redistribution
\SSXpage{6}.
Later \citetads{1976SoPh...48....3C} 
refined this \HK\ washout into estimation of coronal temperature, but
at the time Grotrian excluded thermal broadening with a reminder that
the solar-atmosphere temperature is at most 6000\,K while this
deep-photosphere value gives only 550~\kms. 
He did not specify the corresponding broadening temperature
$6000\times(7500/550)^2\tis\,1.1\,10^6$\,K -- too outrageous since
everybody knew that heat flows necessarily from hot to cool so further
out from the deep may only be cooler -- but hinted
(\linkadspage{1931ZA......3..199G}{22}{footnote 3}) that some sort of
nonthermal ionization might play a role. 
  
The after-all second-law-challenging attribution to million-K thermal
electron motion came after Edl\'en's high-ionization term inventory
and identification of the enigmatic bright ``coronium'' lines
\SSXpage{8} also requiring such high temperature for ionization 
(\rrref{endnote}{note:forbidden}).
Detection of synchrotron radio emission was the clincher for the
greatest solar physics discovery so far.
Explaining million-K coronal temperatures beyond incantation
``nanoflare heating'' citing Parker's
(\citeyearads{1972ApJ...174..499P}) 
small-scale reconnection of twined, braided, knotted fields remains a
challenge while dielectronic line emission helps keeping the corona
cool (\rrref{endnote}{note:dielectronic}).

For smooth-disk illumination the scattered brightness is locally
proportional to $N_\rme$, but as pointed out by
\citetads{1930ZA......1..209M} 
limb darkening and possibly spots must be accounted for in the local
scattering source function $S\!_\nu\tis\,J_\nu$, and also the
integration along the entire line of sight with problematic confusion
between different structures, as evident in the marvelous
\href{http://www.zam.fme.vutbr.cz/~druck/eclipse}{Druckm\"uller
eclipse images}. 
Nevertheless, coronal Thomson scattering presents the most
straightforward solar-atmosphere density diagnostic even while its
Doppler erasure portrays the greatest solar enigma.\footnote{Also
\label{note:clean}
appreciate when admiring the eclipse corona that you are illuminated
by your cleanest sunshine ever: pearly white, no dark lines, beautiful
polarization
(\linkadspage{2020ApOpt..59F..71S}{3}{Fig.\,2} of
\citeads{2020ApOpt..59F..71S}) 
with Minnaert and Van de Hulst single-coronal-scattering
and multiple-scattering-sky cancelation points above and below
(\linkadspage{2020ApOpt..59F..71S}{5}{Fig.\,3e}).}

The outer-corona F component in \SSXpage{5} is from Mie scattering by
dust particles that move slower, maintaining Fraunhofer-line
signatures.}.%
 
\paragraphrr{Extinction and emissivity}
The extinction coefficient\footnote{Naming: often called ``absorption
coefficient'' but I follow Zwaan in using ``extinction'' because the
photon is taken out of the particular beam (direction and frequency)
but may still exist in another direction (scattering pairs {\bf b} and
{\bf c}) or at another frequency (conversion pair {\bf h}). 
Absorption is then only pair {\bf a}, sometimes called ``true
absorption''.} $\alpha_\nu$ and the emissivity\footnote{Naming: Zwaan,
Jefferies, Mihalas and others called this ``emission coefficient''. 
I followed them in IART and also while writing RTSA in 1995, but at an
Oslo school that summer P.G.\,Judge used ``emissivity'' and I realized
that that is a better name because it is not a coefficient, not a
fraction of something but measuring new radiation.} $j_\nu$ are the
macroscopic ways of quantifying the above processes locally along a
given beam with given intensity
\SSFpage{23}, \ISSFpage{9}{Sect.\,3.2}, \IARTpage{34}{Chapter 3},
\RTSApage{32}{Sect.\,2.1.2}. 
Extinction is measured as cross-section per cubic cm
(cm$^{-1}$\,=\,cm$^2$/cm$^3$), emissivity as new photon energy per
cubic cm. 
Each sums all processes acting at the wavelength of interest:
(multiple overlapping) lines (bound-bound) and continua (bound-free,
free-free, Thomson, Rayleigh). 
A photospheric overview is in \SSFpage{68}
\rrendnote{extinction diagrams, H-minus}{note:extcont}{%
Diagrams of bound-free and free-free He and He extinction for
different stellar atmospheres assuming SB are shown in the four
Figs.\,8.5\footnote{Erratum: Fig.\,8.5a has (in Gray's words) a small
error of $10^{-26}$.} 
of \citetads{2005oasp.book.....G} 
and in \SSFpage{53}.
Three earlier versions in
\citetads{1992oasp.book.....G} 
are copied into \RTSApage{211}{Fig.\,8.14}.
In both sets the first panel is for the Sun and shows the broad hump
of \Hmin\ bound-free extinction around 1\,$\mu$m, unlike Kramers-law
$\lambda^3$ decay below the 1.6\,$\mu$m threshold because \Hmin\ is
not a one-electron but a two-electron system, and the regular
$\sim\!\lambda^3$ increase for \Hmin\ free-free extinction which
dominates beyond the threshold. 
For hotter stars (other panels) the atomic \HI\ bound-free edges take over.

The telltale \Hmin\ pattern was first shown observationally in later
forgotten \linkadspage{1923ApJ....58..113L}{19}{Fig.\,4} of
\citetads{1923ApJ....58..113L} 
and then in \linkadspage{1936ZA.....11..132M}{14}{Fig.\,4} of
\citetads{1936ZA.....11..132M}, 
see also (if you read French)
\citetads{1946AnAp....9...69C}.\footnote{History: 
while writing this I was pointed by Minnaert in
\linkadspage{1946AnAp....9...69C}{1}{footnote\,1} in
\citetads{1946AnAp....9...69C} 
to \citetads{1923ApJ....58..113L}, 
a formidable publication.
``Scattering or real absorption?'' is the key question here too; I
admire his ``collustrivity function'' (\ie\ $J_\lambda$, his $J$ is our $I$)
and wonder whether
his \linkadspage{1923ApJ....58..113L}{7}{Eq.\,15} is an early
Eddington-Barbier -- but yet after its succinct statement in
\linkadspage{1921MNRAS..81..361M}{9}{Eqs.\,36-37} of
\citetads{1921MNRAS..81..361M}, 
his first and yet more formidable publication (see
\citeads{2018OAst...27...76P}).} 
On \linkadspage{1930MNRAS..91..139P}{28}{p.\,166}
\citetads{1930MNRAS..91..139P} 
already proposed \Hmin\ free-free as main stellar-atmosphere opacity
agent but also this was forgotten until
\citetads{1939ApJ....89..295W} 
resuggested it, eventually proven by the landmark lengthy
computation\footnote{History: at the bottom of
\linkadspage{1946ApJ...104..430C}{2}{p.\,431}
\citetads{1946ApJ...104..430C} 
wrote ``In any case, to improve on the Hartree approximation would
require an amount of numerical work which will be several fold; and
the task is immense even as it is'' with footnote: ``the present work
has required the numerical integration of 63 radial functions and the
evaluation of 523 infinite integrals, not to mention the computation
of numerous auxiliary functions and tables. (All this work was done
with a Marchant.)''.
Their Marchant was a desktop mechanical calculator containing hundreds
of gears which could add, multiply, and even divide by repeatingly
cranking a handle. 
Chandrasekhar spent months of computer time on \Hmin, literally
cranking.
I still used a Marchant for being more precise than my three-digit
Darmstadt slide rule in computing 1966 eclipse parameters from
Besselian elements before learning ALGOL 60 for the 1970 eclipse and
was happily surprised when in 1972 the HP-35 pocket calculator changed
our world (nowadays called a phone because your advanced pocket
computer does that too).}
by \citetads{1946ApJ...104..430C} 
producing their \linkadspage{1946ApJ...104..430C}{10}{Fig.\,3}.
It is treated in practical
\linkpdfpage{https://robrutten.nl/rrweb/rjr-edu/exercises/ssb/ssb.pdf}{13}{SSB\,2}.

The classic comprehensive extinction ``confusograms'' in
\linkadspage{1951ZA.....28...81V}{17}{Figs.\,I--XV} of
\citetads{1951ZA.....28...81V} 
were redrawn in English in
\citetads{1973itsa.book.....N}, 
with the solar one shown in \SSFpage{54} and reprinted in
\RTSApage{199}{Fig.\,8.5} and four stellar ones in
\RTSApage{212}{Figs.\,8.15\,--\,8.18}. 
Their not-so-trivial interpretation\footnote{When E.\,Vitense wrote
her textbook
\citetads{1989isa2.book.....B} 
she didn't copy her thesis graphs but simplified them in Fig.\,7.8 =
\SSIpage{7}.}
is treated in \RTSApage{252}{Problem 10}.

I have emulation programs and comparable resulting diagrams for both
Gray's and Vitense's graphs in my
\href{https://robrutten.nl/rridl/edulib/dircontent.html}{IDL edulib}.
An overview of total extinction including lines in the FALC atmosphere
is shown in \linkrjrpage{2019arXiv190804624R}{13}{Fig.\,6} of
\citetads{2019SoPh..294..165R}. 
The minima are treated in \rrref{endnote}{note:deepest}.}.

The ultraviolet bound-free edges of \SiI, \MgI, \FeI\ and \AlI\
contribute the electrons needed for \Hmin\ by pairing relatively large
abundance with low ionization energy \SSFpage{56} 
\RTSApage{164}{Table + Fig.\,7.1}, with their joint
abundance giving $N_\rme\tapprox\,10^{-4}N_\rmH$ and keeping the
photospheric gas near-neutral.\footnote{Gaseous matter is mostly
plasma throughout the universe but in cool-star photospheres almost
all electrons and in our surroundings all electrons are tied to
nuclei.
We live in an extraordinary neutral environment energized by an
extraordinary near-neutral environment.}

Line broadening and the Voigt function are summarized in \SSFpage{64}
and detailed in \RTSApage{72}{Sect.\,3.3}.

The strongest disk-center optical
lines are \CaII\ \HK\ and the Balmer lines
(largest dips in the top-right panel of
\linkrjrpage{2019arXiv190804624R}{18}{Fig.\,10} of
\citeads{2019SoPh..294..165R}; 
extinction comparisons in \SSFpage{60} and \SSFpage{62}).

Optical thickness $\tau_\nu\!\equiv\!\int \alpha_\nu \dif s$ is summed
extinction along the beam, with $\tau_\nu\tis\,1$ the 1/e decay
separator between optically thin and thick (transparent or intransparent)
\ISSFpage{12}{Eq.\,20}. 
This is a property of the gas parcel (structure, cloud, layer, shell,
object), not of a spectral feature.
Effective thickness likewise measures transparency for monofrequently
scattering radiation \IARTpage{103}{Sect.\,7.3.3}
\RTSApage{58}{Sect.\,2.6.3}.\footnote{Naming:
\label{note:thinthick} ``optical thickness'' is also used for
non-optical EUV or radio wavelengths.
It expresses the degree of blocking reached by extinction of whatever
sort of radiation in counterpart to geometrical thickness. 
``Degree of intransparency'' would have been better, with limits
transparent versus intransparent = opaque, and with translucent for
effectively thin (or the nice ``foggy'' of
\citeads{1905ApJ....21....1S}). 
You should use ``optical depth'' only for integrated extinction along
the line of sight into your feature or radially into your 1D, 2D or 3D
model star -- too often it is used wrongly for optical thickness.
Colleagues that call a spectral line thick or thin I call
``thinnies'' derogatorily.
How can a line be thin -- set a width threshold in \AA\ or Hz or
cm$^{-1}$ or \kms? 
A physical structure can be thin or thick or effectively thin or thick
in a spectral feature (maybe the Sun is thin in neutrino lines) but no
line is ever thin or thick. 
Never say {\em ``this line is thin''\/} or earn my scorn!}

A particular point is that photons contributed by stimulated emission
are not counted positively as contribution to emissivity but
negatively as contribution to extinction.
It facilitates the description because extinction ({\bf a} + {\bf b} +
{\bf c} + {\bf h}) and stimulated emission ({\bf e} + {\bf g} + {\bf
j}) both need beam photons and therefore similar quantification
(Einstein $B$ values \SSFpage{70}). 
Moreover, with this subtraction the cancelation of stimulated
scattering pairs {\bf c} and {\bf g} is effectuated. 
If this isn't done the extinction would be overestimated by including
{\bf c} and optical depth scales in the Rayleigh-Jeans domain would
suggest larger opaqueness and further out escape than actually the
case.

For coronal spectrum formation one usually assumes CE throughout which
implies that all extinction is negligible and emissivities can be
evaluated without RT
\rrendnote{optical coronal lines}{note:forbidden}{%
CE requires pair {\bf d} photon creation exclusively, but in coronal
gas the Maxwell peak is far beyond feeble optical-transition energies
implying low collisional excitation rate.
The optical coronal-line photons come from dielectronic radiative
decay (\rrref{endnote}{note:dielectronic}) and mostly from pair {\bf
f} photo-excitation by the irradiating photospheric continuum
(``resonance fluorescence'').
These lines are usually named ``forbidden emission lines'' but they
appear naturally in emission off the limb as any off-limb line does
from lack of continuum brightness along the line of sight
(\RTSApage{168}{Fig.\,7.2}). 
They are ``forbidden'' because only magnetic dipole and electric
quadrupole lines without parity change (list in
\linkpdfpage{https://robrutten.nl/rrweb/rjr-archive/linelists/aq1976.pdf}
{89}{paragraph 30} of \citeads{1976asqu.book.....A}) 
provide such small jump energies in high-ionization stages.
Where these provide the only radiation-down decay path they make their
upper level ``metastable'' and long-lived in the absence of collisions
that otherwise upset waiting for eventual radiative deexcitation. 
Then the waiting does not matter and they photo-deexcite just as the
high-probability (``permitted'') EUV transitions in dielectronic
coronal-line producing cascades (\rrref{endnote}{note:dielectronic}). 
Visible not because of being forbidden but notwithstanding
this.\footnote{History:
\citetads{1927MNRAS..88..134E} 
formulated this on \linkadspage{1927MNRAS..88..134E}{2}{p.\,135} as
{\em ``A forbidden line starting from state (2) becomes more or less
equivalent to an ordinary line starting from state (3)''} and added to
the collision-avoiding low-density requirement posed by
\citetads{1927Natur.120..473B} 
that there should be no upward photo-excitation from the metastable
level during its long wait for radiative decay.
EUV irradiation from the cool solar disk underneath is indeed
negligible. 
In addition, these lines are NLTE-Wiened into isotropy including
our sideways way (\rrref{footnote}{note:wiening}).}

The bright lines observed during totality since the 1860s were
attributed to invoked element ``coronium'' until
\citetads{1939NW.....27..214G}, 
inspired by the attribution of all nebulium lines of nova RR Pictoris
to \FeVII\ by
\citetads{1939Natur.143..374B}, 
suggested from the high-ionization term inventory of
\citetads{1937ZPhy..104..407E} that the red line at 6374\,\AA\ is due
to \FeX\ and tentatively from
\citetads{1937ZPhy..104..188E} 
that the green line at 5303\,\AA\ is due to \FeXIV.\footnote{Greedy
present: Edl\'en (\citeyearads{1937ZPhy..104..188E},
\citeyearads{1937ZPhy..104..407E}) and
\citetads{1939NW.....27..214G} 
are now paywalled by Springer, for me at 38.10\,\euro\ each (per half
page for the last) without translation of their German, citation
linking or any other service. 
I doubt that Edl\'en and Grotrian published their work to help
overflow Springer's scavenger coffers eighty years later.} 
As in \citetads{1931ZA......3..199G} 
(\rrref{endnote}{note:wlcorona}) he did not dare to claim high
temperature, only hinted again that these identifications might fit
the speculation that the outer atmosphere is far from thermal
equilibrium.
There are dozens more than these two; eleven observed per coronagraph
are listed on \linkadspage{1939MNRAS..99..580L}{11}{p.\,590} of
\citetads{1939MNRAS..99..580L} 
and more were eventually identified by
\citetads{1943ZA.....22...30E}.\footnote{History: already fifteen in
Edl\'en (1941) in {\em Arkiv Math.\ Astron.\ Fysik\ 28B\/} that I did
not find on ADS but clinched high coronal temperature (with a bolder
claim from Alfv\'en, see
\citeads{2018FrASS...5....9R}) 
undoing ``coronium'' -- he got the Royal Society gold medal for it
(\citeads{1945MNRAS.105..138M}, 
a recommended address starting on nebulium and
geocoronal lines).}
Their widths are commensurate with coronal loop temperatures and
dynamics (\eg\
\citeads{2006ApJ...639..475S}). 

Since then infrared coronal lines, in particular \FeXIII\,10747\,\AA\
and 10798\,\AA, got more attention in the quest for
coronal magnetometry because Zeeman splitting widens
$\propto\!\!\lambda^2$ while Doppler broadening widens
$\propto\!\!\lambda$ (\eg\
\citeads{1977ApJ...214..632H}, 
\citeads{1982ApJ...255..764Q}, 
\citeads{1998ApJ...500.1009J}, 
\citeads{2007Sci...317.1192T}, 
\citeads{2008ApJ...680.1496L}, 
\citeads{2016ApJ...819L..36K}) 
with high hopes for DKIST and the proposed
COSMO (\citeads{2016JGRA..121.7470T}) 
reviewed by
\citetads{2018ApJ...852...52D}.}. 
It then suffices to compute and sum all emissivities along the line of
sight per wavelength
\rrendnote{emission measure, DEM}{note:EM}{%
Colleagues studying the coronal EUV spectrum usually do not worry
about non-detected photons.\footnote{Solar physicists used to be
either thin or thick. 
Thin ones don't care about photons that do not fly directly from
collisional creation to their telescope; they excelled in atomic
physics and later in plasma physics. 
Thick ones care about photons after creation and not in the line of
sight; they excelled in 2-level scattering and later in MHD. 
They now grow together.
The Sun doesn't care and magnanimously sends photons to both types
adhering to her kind \linkrjrpage{1990IAUS..138..501R}{9}{``Principle
of Solar Communicativity''} formulated at the only solar IAU symposium
ever in the USSR
(\citeads{1990IAUS..138..501R}).} 
Unhindered travel to the telescope, not counting extinction but only
emissivity in photon production per pair {\bf d}.
This must be evaluated for all lines within the passband\footnote{For
\label{note:UVimspect} 
spectral-passband imagers as SDO/AIA usually a complex multi-temperature
combination from registering multiple lines. 
Spectrometers as Hinode's EIS register spectrally dispersed bands but
do not inform precisely where from -- slitjaw imaging as by IRIS is
highly desirable -- or miss the more interesting happening just off
the slit. 
It is a pity that Fabry-P\'erot imaging spectroscopy is not feasible
at short wavelengths; SST/CHROMIS is the shortest
(\citeads{2019A&A...626A..55S}). 
My paradise dream is slitless 2D imaging spectrometry resolving
wavelength across \Lyalpha\ (continued in
\rrref{endnote}{note:Lyalpha}).}
at every location along the line of sight and integrated, with
confusion an issue.
Locally the emissivity scales with the product $n_l \, N_\rme$ of
lower-level population density and electron density, the latter
representing the required collider for excitation per {\bf d} which is
most likely an electron.  
Regard a resonance line from an ion ground state.
Its population enters as $(n_l/N_\rmE)\,A_\rmE\,N_\rmH$ with element E
abundance $A_\rmE \tis\ N_\rmE/N_\rmH$, hence local emissivity
$\sim gf\,A_\rmE \,G_{ij}(T) \,N_\rme \,N_\rmH$ where $G_{ij}$ is a
conglomerate function of temperature describing the element
partitioning defining $n_l/N_\rmE$ for this line $ij$ and needing all
sophistication of CHIANTI (\rrref{endnote}{note:CHIANTI}) for
evaluation within the CE assumption. 
For a review see
\citetads{2018LRSP...15....5D}. 
If the line has an excited lower level function $G_{ij}$ depends also
on electron density. 
Integration over the volume of an unresolved emitting structure (say a
loop) yields emission
$\sim gf\,A_\rmE \int G_{ij}(T) \, N_\rme \, N_\rmH \dif V$ where
$N_\rmH$ is often replaced by $N_\rme$ taking
$N_\rmH = N_\rmp \approx N_\rme\,(1-2\,\rmB)$ with B the
helium/hydrogen abundance ratio
(\linkadspage{1964SSRv....3..816P}{20}{Sect.\,5} of
\citeads{1964SSRv....3..816P}). 
The density dependencies $N_\rme \, N_\rmH$ may be removed using the
differential emission measure
$\mbox{DEM}(T) \equiv N_\rme N_\rmH \,(\rmd s /\rmd T)$ in
cm$^{-5}$\,K$^{-1}$, a combined local gradient measure that defines
the emission measure integrated over $s$ along the line of sight as
$\mbox{EM} \equiv \int N_\rme \, N_\rmH \dif s = \int \mbox{DEM}(T)
\dif T$
in cm$^{-5}$
(\linkadspage{2018LRSP...15....5D}{113}{Sect.\,7} 
of \citeads{2018LRSP...15....5D}). 
With this density removal the emission in the line from a feature
becomes $ \sim f\,A_\rmE \int \mbox{DEM}(T) \,\, g\,G_{ij}(T) \dif T$
where the Gaunt factor $g$ is under the integral because it depends on
temperature (\linkadspage{1978ApJ...225..641W}{3}{Eq.\,2} of
\citeads{1978ApJ...225..641W}). 
}
\rrendnote{multi-DEM analysis}{note:DEM}{%
The different AIA passbands are often characterized as each sampling a
specific temperature with larger brightness implying more gas of that
temperature. 
Of course this is a simplification because each passband is wide and
contains multiple lines. 
Recognizing this it is tempting to exploit multiple AIA EUV
passbands\footnote{Usually 
excluding 304\,\AA\ for ``being too optically thick'' for CE-based
CHIANTI (\rrref{endnote}{note:CHIANTI}) -- oops
(\rrref{footnote}{note:thinthick}).}
with their different temperature sensitivities for constraining the
temperature of a feature or just a pixel in ``DEM analysis'' as in
\citetads{2015ApJ...807..143C} where the validations against ``ground
truth'' simulation inputs in
\linkadspage{2015ApJ...807..143C}{8}{Figs.\,6\,--\,9} are impressive.
However, on smaller scales then large AR-like features as modeled
there multi-temperature confusion along the line of sight poses
problems.
In addition, for rapidly evolving or short lived features the
assumption of SE within CHIANTI's CE assumption is likely invalid
because equilibriating is slow at low electron densities, a similar
warning as treated extensively for hydrogen diagnostics including ALMA
of denser cooling cooler gas in the remainder of this compendium after
the thin\,--\,thick separation point (p.\,\pageref{sec:SP}).
}.
This is done with the CHIANTI package
\rrendnote{CHIANTI}{note:CHIANTI}{%
\href{https://www.chiantidatabase.org}{CHIANTI website}.
(Name suggestion: {\em Coronal Holism In
Appreciating Non-Thick Ionization\/}.)
Community effort by
\href{https://www.chiantidatabase.org/contacts.html}{these
colleagues}.
It handles coronal spectroscopy with a large transition inventory for
evaluating CE ionization equilibria, radiative losses, DEMs and more. 
The need to include numerous lines contributed by dielectronic
processes (\rrref{endnote}{note:dielectronic}) makes it 
ever-growing.\footnote{Example 
from P.R.\,Young for He-like \OVII: currently CHIANTI
specifies 577 levels, 127 below the ionization threshold and
autoionization rates specified from 128 up.
Since CHIANTI v9 the 25 levels of \OVIII\ are added also in evaluating
dielectronic \OVII\ rates.}}.

\paragraph*{Thin\,--\,thick separation point.}~~\label{sec:SP} 
If your problem is so optically thin in all its aspects that assuming
CE suffices you may stop reading here.  
Only emissivities without extinctions for you, perhaps extinctions without
emissivities as the bound-free out-of-the-passband scattering above, no
source functions.
Endnotes \ref{note:dielectronic}, \ref{note:rain},
\ref{note:wlcorona}, \ref{note:EM}, \ref{note:DEM}, \ref{note:CHIANTI},
\ref{note:suction}, \ref{note:CEcloud} may cater to you (but beware
of my scorn in \rrref{footnote}{note:thinthick}).

\paragraphrr{Source function}
I wonder who gave this key non-thin quantity its well-suited name in
describing weighted local addition of new photons to a given beam by
dividing local emissivity and extinction:
\begin{equation}
  S\!_\nu \equiv j_\nu/\alpha_\nu 
  \mbox{~~~~~~~~~~~~~~}
  S\!_\nu^{\rm tot}= \frac{\sum j_\nu}{\sum \alpha_\nu},
  \label{eq:S}
\end{equation} 
where the first is per spectral feature as the line in
\rrref{Fig.}{fig:quantities}, the second the total for overlapping
lines and continua.  
Their emissivities and extinctions add linearly but
the combination ratio weighs component source functions \SSFpage{23}
\ISSFpage{16}{Fig.\,7} \RTSApage{32}{Sect.\,2.1.2}.\footnote{The
frequency-independence for CRD across a line holds only for $S\!_{\nu_0}^l$, 
not for $S\!_\nu^{\rm tot}$ weighted with the continuum contribution
\RTSApage{33}{Eq.\,2.23}.}
It seems like dividing cows and horses since
their dimensions are just [cm$^{-1}$] versus [erg\,cm$^{-3}$ s$^{-1}$
Hz$^{-1}$ sr$^{-1}$], but the resulting dimension is that of intensity
\SSFpage{23} which 'tis all about.

The source function is a key quantity for optically thick colleagues
(mostly studying the photosphere and chromosphere) because it
separates local physics and surround physics, or atomic perspective
and environmental quality.
Take a strong spectral line, meaning large line extinction
coefficient. 
Its local addition of many bound-bound transitions to continuous
transitions at its wavelength adds a high narrow spike in extinction
versus wavelength. 
This is local physics: the required particles must be present in
sufficient number set by elemental abundance, partitioning over
ionization stages and over atomic levels within stages that is
primarily set by temperature and density (closer to SB than to CE),
and by the transition probability.
In the emissivity there is a similar spike given by lookalike
particles in the upper level. 
In the ratio, however, this spike may vanish completely -- as for LTE
with the featureless Planck function as ratio. 
Then the ratio is completely set by the temperature, an environmental
rather than an atomic physics quantity. 
When LTE is not valid the local environment also senses nonlocalness,
surely in space (NLS), likely also in wavelength (NLW), possibly also
in time (NLT). 
Thus, optically thick colleagues evaluate local atomic properties to
define the extinction which defines where their signal comes from,
then evaluate the source function to see what environmental effects
may make it depart from the local temperature. 
This split between asking ``Where?'' and ``What?'' is often described
as ``accounting for NLTE'' but too often meaning only for the second.

However, such split interpretation gets mixed when environmental
effects also affect the extinction. 
This happens when that is controlled by other transitions sensitive to
source function nonlocalness.
The extinction of optical atomic lines being defined by scattering
bound-free ultraviolet continua is one example, the dominance of
\Lyalpha\ in controlling \Halpha\ extinction and ALMA \HI\ free-free
extinction is another. 
In such cases, opacity NLTE gets more important than source function
NLTE. 
Colleagues that see NLTE only as a source function issue may miss the
crux of their problem.

Also, the source function is a composite because it is a
process property.
Multiple processes operating at a given wavelength each have their own
source function and combine weightedly via the sums in
\rrref{Eq.}{eq:S}.
In the solar photosphere optical lines often combine a scattering line
source function with a near-LTE continuous source function because the
latter is dominated by \Hmin\ bound-free interactions. 
These are not necessarily LTE as is the case for free-free
interactions. 
Ultraviolet bound-free continua get far from LTE but in the optical
and infrared the ambient thermal kinetic energy is similar to or
exceeds the transition energy and at sufficiently deep formation
collisions reign, making LTE a good approximation
\rrendnote{deepest wavelengths}{note:deepest}{%
The FALC extinction spectra in
\linkrjrpage{2019arXiv190804624R}{13}{Fig.\,6} of
\citetads{2019SoPh..294..165R} 
show the canonical minimum at the 1.6\,$\mu$m threshold of \Hmin\
bound-free extinction where one sees deepest into the Sun (\eg\
\citeads{1979A&A....77..257L}). 
A nicer display is Avrett's holistic sketch in \SSXpage{36} from
\citetads{1990IAUS..138....3A} 
plotting brightness temperature (\rrref{endnote}{note:Tb}) through the
entire VALIIIC spectrum.\footnote{I used it as RTSA exam by
white-pasting labels and then ask explanation of every curve (my
teaching is always graphic).  Try yourself?}
Both left and right of the double-hump maximum at 1\,$\mu$m the
extinction and formation height increase, sampling the photospheric
temperature decrease until the VALIIIC minimum above which lines and
edges reverse into emission, with line crowding in the blue and UV
sketched schematically.
The shallow dip between the humps (continuum enlarged in \SSIpage{5}
and \SSFpage{55}) spans the visible.
The second hump is the opacity minimum at 4000\,\AA. 
At shorter wavelengths metal ionization edges steeply increase the
summed continuous extinction \SSIpage{7}.

The 4000\,\AA\ extinction minimum provides nominally less deep
sampling than the 1.6\,$\mu$m dip \SSFpage{55}, but effectively better
because contribution functions steepen and intensity contrasts
increase with the Wien response to temperature variations
(\citeads{1989SoPh..124...15A}). 
In addition, the diffraction limit is also four times sharper (but
needs yet better seeing since the Fried aperture
$r_0\tsim\,\lambda\,^{6/5}$).  
Hence, SST/CHROMIS 4000\,\AA\ imaging is currently the best in
observing deepest granulation
(\citeads{2019A&A...626A..55S}). 
For deep imaging of intergranular MCs the SST 3950\,\AA\ wide-band
filter is yet better because it samples the overlapping outer wings of
\CaIIHK\ between these lines where reduction of collisional damping by
lower density produces wing opacity reduction, hence hole deepening in
MCs brightening them into increased contrast.
See the SST sample in \linkrjrpage{2020smvc.book...29R}{4}{Fig.\,2} of
\citetads{2020smvc.book...29R}.\footnote{The 
sharpness difference between the SST and HMI magnetograms in
\linkrjrpage{2020smvc.book...29R}{4}{Fig.\,2} is dramatic. 
However, also the SST lacks magnetic sensitivity; only the Hinode SP
reaches non-kilogauss internetwork fields thanks to long integration
in seeing-free space (\rrref{endnote}{note:basal}).}}.
The \NaID\ lines are the darkest lines in the optical spectrum
(\rrref{endnote}{note:NaD}) because they are photospheric
$\sqrt{\varepsilon}$ scatterers (\rrref{endnote}{note:NaD}).
In hotter stars where hydrogen is ionized the reverse occurs: weak LTE
lines appear in emission on the dark Thomson-scattering continuum
(\linkadspage{1992A&A...262..209R}{215}{Fig.\,1} of
\citeads{1992A&A...262..209R}). 
The solar \CaII\ \HK\ wings also contain weak emission lines but these
are from NLTE interlocked scattering
(\rrref{endnote}{note:interlocking}) with \FeII~3969.4~\AA\ close to
\Hepsilon\ the weirdest: it responds deeper than its quasi-continuum
near-LTE background (\linkrjrpage{1980ApJ...241..374C}{9}{Fig.\,8} of
\citeads{1980ApJ...241..374C}). 

\paragraphrr{Line source function}
Most classic treatments of NLTE scattering in the line source function
assume a gas of 2-level atoms to keep the description monochromatic
(CS) or limited to a single line (CRD). 
I did the same in RTSA, with a long derivation inventing ``sharp-line
atoms'' enabling monofrequent use of the Einstein coefficients in
\RTSApage{84}{Sect.\,3.4}, an elegant derivation by Zwaan of the
$\sqrt{\varepsilon}$ law in \RTSApage{112}{Sect.\,4.3.1}, and the
classic numerical CRD results for this law in an isothermal atmosphere
of \citetads{1965SAOSR.174..101A} 
in \RTSApage{128}{Fig.\,4.12}
\rrendnote{square root epsilon law}{note:sqrteps}{%
This T-shirt-qualifying law says
$S\!_\nu(0)\tis\,\sqrt{\varepsilon_\nu}\,B_\nu(T)$ at the top of an
isothermal constant-$\varepsilon$ 2-level-atom plane-parallel
non-irradiated atmosphere \RTSApage{117}{Eq.\,4.81}, as shown in
Avrett's \RTSApage{128}{Fig.\,4.21}. 
The emergent intensity \RTSApage{117}{Eq.\,4.84} is similarly low:
scattering lines get dark.
It seems a straightforward result from simple linear-looking
\rrref{Eq.}{eq:S_CS} and \rrref{Eq.}{eq:S_CRD} for such a simple
atmosphere -- but see Huben{\'y}
(\citeyearads{1987A&A...185..332H}, 
\citeyearads{1987A&A...185..336H}) renaming it the Rybicki equation
and analyzing why and how it works.\footnote{History: written near
Vienna without library or computer access, the year may hint why. 
I assigned these publications as student presentation topic in my RTSA
course because I did not understand them and instead added two pages
explanation myself \RTSApage{117}{discussion} which Zwaan said he did
not understand. 
I also taught the two-level definition of $\varepsilon$ wrongly in
IART (now correct in \IARTpage{101}{Eq.\,7.10}) until
\href{https://robrutten.nl/rrweb/rjr-pubs/2003-thomas-epsilon.pdf} {I
finally grasped} that the cancelation of {\bf c} and {\bf g} even
applies to no-memory Thomson scattering because photons are bosons.}}.
Isothermal and 2-level seems far-fetched but the $\sqrt{\varepsilon}$
law is often applicable 
\rrendnote{validity of $\sqrt{\varepsilon}$ law}{note:flatSl}{%
The ALC7 ``chromosphere'' is near-isothermal
\SSXpage{72} with near-constant $\varepsilon$ in lines as \NaIDone\
\SSXpage{80} from increasing hydrogen ionization and hence
near-constant (relatively increasing) electron density defining the
collision frequency handling destruction pair {\bf a}.  
The gas density drops exponentially outward \SSXpage{72} but electrons
dominate the collision frequency by moving much faster than atoms. 
Avrett's canonical isothermal + constant $\sqrt{\varepsilon}$
demonstrations \SSFpage{80} therefore apply well to strong lines from
ALC7.

In addition, strong lines experience an atmosphere as more isothermal
than the continuum because their total $\tau$ scales are compressed in
height with respect to the continuum $\tau$ scale due to their steeper
buildup \SSFpage{38}, flattening their $B(\tau)$ gradient
\SSFpage{85}. 
Where the temperature decays obeying RE as in the
upper photosphere (\rrref{endnote}{note:REphot}), lines see yet shallower
$B_\nu$ gradients.

Hence line core darkening by $\sqrt{\varepsilon}$ scattering applies
to many strong lines.}
and yet more often instructive.
An equation summary is shown in \SSFpage{79} and as
\RTSApage{275}{concluding ``RT rap''}.

Two-level scattering theory was mostly developed for 1D radial
geometry but lateral scattering can cause feature smoothing and aureoles
in brightness and in opacity 
\rrendnote{scatter aureoles}{note:aureoles}{%
Surround scattering causes brightness and/or opacity aureoles around
small bright structures giving them visibility if they are intrinsically
smaller than the telescope resolution.
Examples are hot spicule-II tips (probably Alv\'enic heating,
\rrref{endnote}{note:spicules-II}), photospheric reconnection in EBs
(\rrref{endnote}{note:EBs}), chromospheric reconnection in CBP loop
tops (\rrref{Sect.}{sec:chromosphere}).

Didactically exemplary for brightness aureoles is the \HI\ Lyman-line
penetration into the upper photosphere from the abrupt overlying
``chromosphere'' of the FCHHT-B Fontenla star
(\rrref{Sect.}{sec:Avrett}).
It is compared with the ALC7 Pandora star in \SSXpage{97}, with two
key plots repeated in \SSXpage{171}.
Its steep temperature jump represents a one-sided heated feature
producing downward NLTE \Lyalpha\ and \Lybeta\ scattering radiation
across the FCHHT-B temperature minimum, constituting closely similar
declining-intensity $J_\nu$ aureoles extending over 400\,km.
These lines have no opacity aureoles ($b_1\tis\,1$ at left in
\SSXpage{171}) because \HI\ is very much the majority ionization stage
with virtually all hydrogen atoms in the ground level.
Similar scattering brightness aureoles without opacity aureoles occur
laterally around small hot features for other majority-stage two-level
scattering resonance lines including \MgIIhk\ and \CaIIHK. 

In contrast, \Halpha\ does have both NLTE opacity aureole boosting
from \Lyalpha\ surround irradiation and NLTE brightness aureole
boosting from its own surround irradiation. 
The $b_2$ peak at left in \SSXpage{171} partially cancels the deep
population dip in the Boltzmann fraction (dashed versus dotted curve
at left), but this dip remains so deep that \Halpha\ has constant
$J_\nu$ across the entire FCHHT-B temperature minimum equal to the
``chromosphere''-onset value with $\tau\tis\,10$ at $h\tis\,1000$\,km.
This gap-filling $J_\nu$ represents a plateau in the
$\sqrt{\varepsilon}$ scattering decline beginning at the much deeper
thermalization depth (\cf\ ALC7 formation in \SSXpage{91}) and is
built up by radiation scattering outward from below and scattering
back from the FCHHT-B ``chromosphere''
(\citeads{2012A&A...540A..86R}). 
This FCHHT-B \Halpha\ scattering plateau is exemplary for \Halpha\
aureoles around small hot features, but for Ellerman bombs only in the
far \Halpha\ wings since the core is obscured by overlying \Halpha\
fibril canopies (\rrref{endnote}{note:EBs},
\citeads{2016A&A...590A.124R}). 

Exemplary for opacity aureoles without brightness aureoles are hot
features observed in the ALMA continua that contain large \Lyalpha\
intensity and send out \Lyalpha\ photons scattering laterally,
spreading into cool surroundings and there enhancing hydrogen
ionization and \HI\ free-free opacity for ALMA. 
Photospheric Ellerman bombs are obscured by overlying chromospheric
canopies also for ALMA, but if such opacity aureoles around higher-up
reconnection events reach optical thickness they permit measurement of
surround temperatures.

Exemplary for absence of both aureole brightness boosting and aureole
opacity boosting are photospheric MCs (\rrref{endnote}{note:MCs}).
Their brightness is LTE deep-hole radiation, not from local heating
but from relative evacuation by magnetic pressure and enhanced by
low-density ionization for neutral-metal lines, dissociation for
molecular bands (as the CH G-band around 4310\,\AA\ and the CN band
below 3883\,\AA) and reduction of collisional damping in outer wings
of strong lines.
They do not spread brightness-boosting or opacity-boosting photons
around, which makes the sharp imaging of MCs in these diagnostics
ideal for showcasing optical telescope performance, as with the SST
sample between \CaIIHK\ in
\linkrjrpage{2020smvc.book...29R}{4}{Fig.\,2} of
\citetads{2020smvc.book...29R}.}.

Two-level scattering source function evaluation became a key part of
numerical spectrum synthesis with many programs following the key
concepts of opposite-direction averaging of
\citetads{1964CR....258.3189F} 
\RTSApage{137}{Sect.\,5.2} and of operator splitting of
\citetads{1973ApJ...185..621C}, 
leading to approximate lambda iteration (ALI) and similar methods
\RTSApage{145}{Sect.\,5.3.2}. 
Some of these programs are detailed below in
\rrref{Sects.}{sec:Holweger}\rrref{\,--}{sec:Avrett}.\footnote{An online tool
for hands-on experimentation with various methods is at
\href{http://rttools.irap.omp.eu}{http://rttools.irap.omp.eu} and
described by
\citetads{2016EJPh...37a5603L}.  
It produces plots as \RTSApage{147}{Fig.\,5.2} on your screen.}

The extension with multi-level detours is still lacking in my courses,
so I add that here. 
The best descriptions so far are in
\linkadspage{1968slf..book.....J}{199}{Sect.\,8.1} of
\citetads{1968slf..book.....J}\footnote{Erratum: 
remove minus in the equation below Eq.\,8.8.}
and in \linkadspage{1971A&A....10...54C}{1}{Sect.\,II} of
\citetads{1971A&A....10...54C}, 
both using ratios rather than fractions as probability parameters.
Split the line extinction coefficient into the destruction (a for
absorption), scattering (s), and detour (d) contributions of
\rrref{Fig.}{fig:pairs}:
\begin{equation}
  \alpha_\nu^l \equiv  
    \alpha_\nu^\rma +\alpha_\nu^\rms + \alpha_\nu^\rmd
  \hspace{10mm}
  \varepsilon_\nu \equiv \alpha_\nu^\rma / \alpha_\nu^l
  \hspace{10mm}
   \eta_\nu \equiv \alpha_\nu^\rmd / \alpha_\nu^l 
   \label{eq:epseta} 
\end{equation}
where $\varepsilon$ is the collisional destruction probability of an
extincted photon and $\eta$ is its detour conversion probability.
With these, the general line source 
function becomes 
\begin{equation}
 S\!_\nu^l = (1 - \varepsilon_\nu - \eta_\nu) \, J_\nu 
  ~+~ \varepsilon_\nu B_\nu(T) ~+~ \eta_\nu S\!_\nu^\rmd 
\label{eq:S_CS} 
\end{equation}
for CS and  
\begin{equation}
 S\!_{\nu_0}^l =  (1 - \varepsilon_{\nu_0} 
     - \eta_{\nu_0}) \, \overline{J_{\nu_0}} 
     ~+~ \varepsilon_{\nu_0} B_{\nu_0}(T) 
     ~+~ \eta_{\nu_0} S\!_{\nu_0}^\rmd
\label{eq:S_CRD}
\end{equation}
for CRD with
$\overline{J_{\nu_0}} \equiv (1/4\pi)\int\!\!\int I_\nu
\varphi(\nu\!-\!\nu_0) \dif\nu \dif\Omega$
the ``mean mean'' intensity averaged over all directions and the line
profile, with $\nu_0$ the line-center frequency and also used as line
identifier.

Equations \ref{eq:S_CS} and \ref{eq:S_CRD} describe physical
contributions to local NLTE divergence between $S^l$ and $B$. 
In each the first term represents the reservoir of photons that
contribute new photons to the beam by direct scattering ({\bf f} and
{\bf g}), the second describes collisional beam-photon creation ({\bf
d} and {\bf e}), the third the contribution of new beam photons via
detours ({\bf i} and {\bf j}) which themselves may contain
collisional, scattering and also such bound-free steps. 

A convenient formal way to quantify NLTE effects is to use
NLTE population departure coefficients $b \!\equiv\! n/n_{\rm LTE}$ 
for the lower ($b_l$) and upper ($b_u$) level
\rrendnote{Menzel departure coefficients}{note:bdef}{%
Be aware that Avrett and Fontenla used the Menzel definition of
population departure coefficients that applies the next-ion population
as normalization and is described carefully in VALIII
\linkadspage{1981ApJS...45..635V}{29} {Eqs.\,13\,--\,18}.
I use the more intuitive Zwaan definition \RTSApage{53}{Eq.\,2.104}
\SSFpage{73} using the total element population as
normalization.\footnote{I used to write these as $\beta_l$ and
$\beta_u$ following \citetads{1972SoPh...23..265W} 
but stopped because colleagues use $b$ wrongly anyhow.}
Colleagues may think to use the latter while using the former not
realizing that between the two the values reverse for majority species
as \HI\ -- which led
\citetads{2009ApJ...707..482F} 
to misinterpret their own results
(\citeads{2012A&A...540A..86R}). 
Various other authors (not to be named) reversed the meaning of VALIII
\linkadspage{1981ApJS...45..635V}{30}{Fig.\,30}, ignoring equation
(18) above it, so that its $b_1$ curve implies chromospheric hydrogen
ground-state population ten times the total gas density.}.
In the  Wien approximation (negligible stimulated  emission) these yield
the simple relations \SSFpage{73}:
\begin{equation}
  \alpha^l\approx b_l\,\alpha^l_{\rm LTE} 
  \mbox{~~~~~~~~~~~~}
  S\!_{\nu_0}^l\approx (b_u/b_l)\,B_{\nu_0}(T)
  \label{eq:b}
\end{equation}
which directly show that source function LTE $b_u/b_l \tapprox 1$ does not
imply opacity LTE $b_l \approx 1$ or vice-versa.
The formation graphs for many lines from ALC7 starting at \SSXpage{78}
employ these 
(\rrref{endnote}{note:ALC7plots}).

Equations \ref{eq:S_CS} and \ref{eq:S_CRD} show that \acp{LTE}
$S\tis\,B$ equality holds when $\varepsilon\tis\,1, \eta\tis\,0$
and/or $J\tis\,S^\rmd\tis\,B$, both closely correct below the standard
$h\tis\,0$ surface at $\tau_{5000}^c\tis\,1$. 
Above it $\varepsilon$ becomes small from lower electron density
\rrendnote{epsilon is small}{note:epssmall}{%
Another T-shirt slogan
for this course.
Its value sinks from unity in the deep photosphere to
$\varepsilon\tapprox\,10^{-2}-10^{-6}$ for \NaI\ \SSXpage{80}, \CaIIK\
\SSXpage{88}, \MgIIk\ \SSXpage{89}, \Lyalpha\ \SSXpage{90} making the
line source function mostly $J$ hence NLS.
For estimating emergent intensity with the Eddington-Barbier
approximation you then need to know the local intensity in all
directions defined by source function values elsewhere.
If the $\eta_{\nu_0} S\!_{\nu_0}^\rmd$ contribution is significant you
also need to know the intensity in all directions in other transitions
with other wavelengths.
Worse, the same may hold for evaluating the $\tau_\nu(h)$ scale needed
in your Eddington-Barbier interpretation or your line synthesis
results or in your ``inversion'' code.
RT is easy only when LTE holds everywhere at all participating
wavelengths for all $S$ and $\alpha$ processes or when there is no RT
as in CE.

The smallness of $\varepsilon$ is also numerically challenging.
The source term $\varepsilon\,B$ may seem neglectable but it isn't
because without it there are no photons to scatter.  In the early days
of electronic computing it was tempting to stop \lambdop\ iteration
\SSFpage{101} when the convergence reached tiny steps at the machine
precision -- but they were not negligible because still far from the
answer needing many more \SSFpage{103}.  This was remedied by operator
splitting in approximate \lambdop\ iteration (ALI) \SSFpage{102}
\RTSApage{145}{Sect.\,5.3.2} and convergence acceleration
\RTSApage{150}{} \SSFpage{103}.

Also, at small $\varepsilon$ the up and down radiative rates may be
large but nearly cancel; then using their ``net radiative bracket''
normalized difference can be numerically better and useful
(\citeads{1960ApJ...131..429T}, 
\linkadspage{1968slf..book.....J}{151}{Eq.\,6.75\,ff} of
\citeads{1968slf..book.....J}, 
\linkadspage{2002ApJ...572..626C}{6}{Eq.\,10} of
\citeads{2002ApJ...572..626C}).} 
while $\eta$ is usually smaller.
Most chromospheric lines and
bound--free continua are heavily scattering with $S\tapprox\,J$ (but
free--free continua always have $S\tis\,B$ because each interaction is
collisional). 
For such scatterers the two-level simplification $\eta\tis0$ is a good
approximation but weaker spectral features may suffer detour
complications
\rrendnote{photon pumping, lasering, Wien
transcription}{note:pumping}{%
Photon pumping is multi-level NLW interaction where one transition may
feed another photons and Wien sensitivity. 
Classic nighttime cases are planetary-nebulae and nebulium lines
\SSXpage{142\,ff}. 
The two detour loops in \linkrjrpage{2019SoPh..294..165R}{4}{Fig.\,2}
of \citetads{2019SoPh..294..165R} 
are good solar examples.

At right the Balmer continuum overionizes hydrogen from level
$n\tis\,2$ in the upper photosphere because it escapes much deeper and
at short wavelengths where the large Wien sensitivity produces $J_\nu
\!>\! B_\nu$ as in \SSXpage{76} for the ALC7 star.  Hence the ALC7
proton density is ``pumped up'' with respect to the local SB value
(the $n\tis\,2$ level has $b_2\tapprox\,1$ because \Lyalpha\ is
thermalized (lefthand plot \SSXpage{76}).  This photon excess
contributes together with Brackett, Paschen and Balmer-line photon
losses (suction, \rrref{endnote}{note:suction}) into downward
overpopulation flow through the \HI\ Rydberg levels
(\linkrjrpage{1994IAUS..154..309R}{6}{Fig.\,3} with explanation in
\citeads{1994IAUS..154..309R}). 
Hotter features in the deep photosphere cause larger hydrogen
overionization higher up and also sideways in cooler surroundings.

The $S^l/B\tapprox\,b_u/b_l$ ratio may so be controlled in the Wien
domain and transcribe to the Rayleigh-Jeans domain where given
departure ratio from unity has larger effect as plotted in
\SSFpage{74} from \RTSApage{54}{Fig.\,2.8}. 
Blow-up and then negative lasering is reached at right (line
extinction and source function both become negative for sufficient
$b_u/b_l$ \SSFpage{74} \RTSApage{55}{Eq,\,2.116} but emissivity
remains positive). 
Such blow-up (but not yet lasering) is reached in the \MgI\,12\,micron
lines (\rrref{endnote}{note:MgI12micron}).

For \FeII\ lines long-wavelength $S^l/B$ blow-up from pumping in the
ultraviolet with transcription of the exponential Wien sensitivity
holding there to linear Rayleigh-Jeans sensitivity is illustrated in
\linkrjrpage{1980ApJ...241..374C}{6}{Fig.\,5} of
\citetads{1980ApJ...241..374C} 
and in \linkrjrpage{1988ASSL..138..185R}{10}{Fig.\,3} of
\citetads{1988ASSL..138..185R}. 
In the latter the high peak of the excitation temperature for \FeII\
lines at 7000\,\AA\ arises because it must accommodate the imposed
$b_u/b_l$ ratio, here $b_u$ overpopulation excess because
$b_l \tapprox\,1$ for dominant-stage ground states. 
The $b_u$ overpopulation results from radiative $J_\nu\!>\!B_\nu$
feeding in the ultraviolet
(\linkrjrpage{1988ASSL..138..185R}{5}{Fig.\,1}). 
At smaller temperature sensitivity larger $T_{\rm exc}$ excess is
needed for the imposed $b_u/b_l$ ratio.
 
The lefthand detour loop in
\linkrjrpage{2019SoPh..294..165R}{4}{Fig.\,2} of
\citetads{2019SoPh..294..165R} 
is for \CaII. 
Similar transcription of the Wien sensitivity of the PRD-split source
function of \CaIIK\ to \CaIR\ is evident for the wing source functions
(label g) of \CaIIK\ in panel c and of \CaIR\ in panel g of
\linkadspage{1989A&A...213..360U}{7}{Fig.\,2} of
\citetads{1989A&A...213..360U}.} 
\rrendnote{photon loss, no nanoflame heating, photon
suction}{note:suction}{%
``Photon loss'' means losing the energy content of a photon altogether
from the object (star, atmosphere, feature), as ``neutrino losses''
means energy losses from a stellar core. 
In both the LTE and CE extremes the energy loss is direct because the
emergent photon was locally created collisionally (bound-bound pair
{\bf d} or in LTE {\bf e}, their bound-free equivalents, free-free). 
Where the Maxwell distribution holds this is direct loss of local
thermal energy.

For LTE it should be a negligible leak (otherwise LTE is not valid);
an example is forbidden \MgI\,4571\,\AA\ from ALC7 with
$S^l\tapprox\,B$ out to $\log(\tau)\tapprox\,-4$ in \SSXpage{82} (but
not its opacity which is heavily depleted by ultraviolet
overionization).

For CE summing all emergent photons from an EUV-bright feature
quantifies radiative losses that measure required heating for a stable
structure obeying SE (otherwise CE is not valid). 
Hence CHIANTI-computed radiative loss curves following classic
\linkadspage{1989ApJ...338.1176C}{6}{Fig.\,2} of
\citetads{1989ApJ...338.1176C} 
specify required steady coronal heating for stable emission. 
However, it is a mistake to count summed brightness of short-lived
bright features as contribution to local heating: to the contrary,
this energy heats the telescope detector while the many more photons
missing that either illuminate the universe unremarkably or drown
unnoticed in the photosphere. 
Only the non-radiation-lost remainder of the feature heating may
count. 
What you see is what <you> get, not the gas; what you see does not
tell you what that gets. 
Observing small flaring flames 
spoils their nanoflare heating candidacy unless their emission is
demonstrably a small fraction.

For a two-level scattering line (initial {\bf d} or {\bf e} followed
by many {\bf b} or {\bf c} steps with small chance of collisional
destruction) the actual photon losses occur near the mathematical
$\tau\tis\,0$ surface, in the last scattering around depth
$\tau\tis\,1$, but the losses start being felt already near the
thermalization depth.  There the radiation is nearly isotropic (fully
in the isothermal simplification) and becomes more and more
anisotropic further out, ending up purely outward directed at the
surface.  The scattering decline towards the $\sqrt{\varepsilon}$
value there may be seen as inward-diffusing photon losses.  The same
holds within effectively thick features rather than a plane-parallel
atmosphere.

For two-level atoms the losses deplete the upper level with larger
fractional $b_u$ decrease than the corresponding fractional $b_l$
increase of the lower level due to the Boltzmann ratio between them:
in two-level atoms the lower level is the main population reservoir
with $n_1\!\gg\!n_2$ and therefore close to LTE because there is
nowhere else to go so that nearly the whole abundance sits in
$n\tis\,1$. 
The $b$ curves in the first plot of \SSXpage{89} for \MgIIk\ in ALC7
are the textbook example. 
These ions are not two-level atoms but close to it because Mg is
predominantly once-ionized throughout the ALC7 photosphere and
``chromosphere'', as shown by the fractional lower-level population
curve in the same plot.
Hence \MgII\ \hk\ suffer NLS two-level scattering only, no NLW
interlocking to other wavelengths nor NLT memories in time.

``Photon suction'' is a NLW multi-level detour phenomenon, done by strong
lines when the population reservoir is in the next stage, as for most
neutral metals in the photosphere. 
It is the reverse of photon pumping in which incoming photons
overexcite higher levels from a lower population reservoir
(\rrref{endnote}{note:pumping}); in photon suction outgoing photons
draw population from a higher reservoir down into lower levels. 

The alkali resonance lines are the prime example.  Their formation in the
VALIIIC and HOLMUL stars was detailed 
by \citetads{1992A&A...265..237B}. 
\linkrjrpage{1992AAp...265..237B}{8}{Figure\,6} 
shows ground-level departure curves $b_1$ (still called $\beta$
following \citeads{1972SoPh...23..265W}) 
of \KI\ and \NaI\ increasing slightly until the height where
ultraviolet underionization causes steep rises. 
This slight increase was surprising because neutral metals suffer
ultraviolet overionization in the upper photosphere by sampling the
hot Balmer continuum coming from below, evident in the ALC7 population
panels for \MgI\,4571\,\AA\ in \SSXpage{82} and \FeI\,6301.5\,\AA\ in
\SSXpage{83}. 
Such pumping may happen from excited levels, but the alkali
ground-state ionization edges at 2856\,\AA\ for \KI\ and 2411\,\AA\
for \NaI\ suffer it also.
For them the overionization is fully compensated by population
replenishment from the low-lying next-ion population reservoir, drawn
from there in a net recombination flow along high levels that is
driven by the photon losses in the resonance lines\footnote{The
compensation was earlier noted in
\linkrjrpage{1991AAp...244..501G}{3}{Fig.\,1} of
\citetads{1991A&A...244..501G} 
but there incomplete from lacking high levels in the model atom.}. 
If one starts a NLTE modeling iteration with LTE populations the
ultraviolet $J_\nu>B_\nu$ excesses deplete the ground state but the
NLTE scattering losses diffuse inward and draw recombination
population until SE is reached with full suction compensation for the
alkalis with their low ionization energies. 
A two-level-plus-continuum demonstration is shown in
\linkrjrpage{1994IAUS..154..309R}{4}{Fig.\,2} of
\citetads{1994IAUS..154..309R}, 
with explanation. 
It was made by J.H.M.J.\,Bruls for the cover of
\citetnoweb{1992Bruls..PhDthesis}.}.

\paragraphrr{Transfer equation along the beam}
The differential and integral forms are
\begin{equation}
   \frac{\rmd I_\nu}{\rmd s} = j_\nu \!-\! \alpha_\nu \,I_\nu 
\mbox{~~~~~~~~}
I_\nu (\tau_\nu) = I_\nu (0) \ep{-\tau_\nu} \!+\!\!\!
  \int_0^{\tau_\nu} \!\!\!\!S\!_\nu (t_\nu) \ep{-(\tau_\nu - t_\nu)} \dif t_\nu
\end{equation}
with geometrical thickness $s$ and optical thickness
$\tau_\nu\!\equiv\!\int \alpha_\nu \dif s$ along the beam direction
through the medium \ISSFpage{12}{Sect.\,3.3}.

\ISSFpage{14}{Fig.\,5} treats a spectral line from a homogeneous
isothermal cloud with LTE processes, say a galactic \HI\ cloud showing
the 21-cm line to a radio astronomer.\footnote{The
version in \RTSApage{36}{Fig.\,2.2} is the most-shown
of my didactic {\tt xfig} diagrams (sometimes with sloppy attribution to
\href{https://robrutten.nl/robshots/rene.jpg}{R.G.M.\,Rutten}), nowadays
as Python animation.}
Appreciate how the line grows or sinks with increasing cloud opacity
until it hits the $I_\nu \tis\,S\!_\nu \tis\,B_\nu$ optically thick
limit (more formal graph in \IARTpage{42}{Fig.\,3.3}).
Thick isothermal LTE clouds show no lines (as ``black holes have no
hair'').

Such simple clouds became the basis for ``cloud modeling'' of
\Halpha\ mottles on the disk and are also much used for filaments and
prominences.
Assuming the cloud homogeneous with Gaussian (``microturbulent'') line
broadening yields the classic cloud parameters \SSXpage{2}:
line-center optical thickness, line-of-sight velocity, source
function, Doppler width, incident intensity profile from below/behind
\rrendnote{cloud modeling}{note:cloudmodeling}{%
For a review see \citetads{2007ASPC..368..217T}. 
Cloud modeling was started by J.M.\,Beckers in his 1964
thesis\footnote{Remembrance:
\label{note:beckers}
\href{https:robrutten.nl/rrweb/rjr-pubstuff/lar-2/Jacques-Beckers.jpg}
{Jacques\,M.\,Beckers} (1934\,--\,2021) studied at Utrecht and got his
PhD there with M.G.J.\,Minnaert as formal adviser but
doing the work in Sydney with R.\,Giovanelli as effective adviser. 
At his request I put a scan of his thesis on ADS:
\citetads{1964PhDT........83B}. 
It was printed by the US Air Force because he had already moved to
Sunspot where I befriended him in 1977. 
Then a solar physics paradise
(\citeads{2018arXiv180408709R}), 
now a ghost town.
His summer student G.D.\,Nelson (later the nontethered astronaut
\href{https://en.wikipedia.org/wiki/Manned_Maneuvering_Unit}{grappling
and cartwheeling} the Solar Maximum Mission when his million-dollar
wrench did not fit) told me ``he is a fountain of ideas''. 
He was.
At the time he repudiated the Alfv\'en-wave refrigeration of umbrae
proposed by \citetads{1974SoPh...36..249P} 
by measuring coronal line widths
(\citeads{1977ApJ...215..356B}), 
repudiated the polar vortex proposed by
\citetads{1977GApFD...8...93G} 
by measuring polar \FeI-line Dopplershifts
(\citeads{1978ApJ...224L.143B}), 
and joined T.M.\,Brown in developing the Fourier Tachometer
(\citeads{1981siwn.conf..150B}, 
\citeads{2013ASPC..478...93B}) 
that with solid polarization encoding by
\citetads{1981siwn.conf..155E} 
became the Michelson heart of the GONG stations, SOHO/MDI and SDO/HMI. 
Later he participated in the Hubble-repairing
\href{https://www.stsci.edu/files/live/sites/www/files/home/hst/documentation/_documents/AStrategyforRecovery.pdf}{COSTAR
design}. 
His off-axis parabola design with DOT-like venting to avoid evacuation
for CLEAR (\citeads{1998SPIE.3352..588B}) 
became the DKIST strawman.}. 
Further complexity was added by \eg\
\begin{itemize} \vspace{-1ex} \itemsep=-0.5ex

\item \citetads{1999A&A...345..618M}: 
cloud model look-up tables for ``inversion'';

\item \citetads{1999A&A...346..322H} 
and \citetads{1996A&A...309..275M}: 
source function variation along the line of sight. 
This was already done by
\citetads{1968SoPh....3..367B} 
in a famous spicule review by using NLTE population tables for
hydrogen from
\citetads{1967AuJPh..20...81G} 
-- although warning against assuming SE below
\linkadspage{1968SoPh....3..367B}{32}{Eq.\,10}, an early precursor of
\citetads{2002ApJ...572..626C} 
dynamic H ionization indeed holding for spicules
(\rrref{endnote}{note:spicules-II});

\item \citetads{2014ApJ...780..109C}: 
second cloud model for the incident profile;

\item \citetads{2015A&A...579A..16H}: 
multi-thread filament RT with mutual irradiation.
\end{itemize} \vspace*{-0.5ex}%
For the incident profile often a field average or a nearby subfield
average is used but this is wrong in the case of backscattering as
observed for \Halpha\ filaments as sub-filament brightening towards
the limb by
\citetads{1975SoPh...45..119K}, 
proposed for \Halpha\ fibrils by
\citetads{2004A&A...418.1131A}, 
and occurring in 1D standard models
(\rrref{endnote}{note:photoelectric}). 
The near-equality of the dashed profiles in the lower panels of their
\linkrjrpage{2012AAp...540A..86R}{7}{Fig.\,8} made
\citetads{2012A&A...540A..86R} 
suggest as \linkrjrpage{2012AAp...540A..86R}{9}{recipe} to use the
outward intensity profile in a RE model at optical depth equal to the
cloud’s thickness as impinging background profile \SSXpage{3}. 
This recipe improves the finding of
\citetads{2010MmSAI..81..769B} 
that a RE emergent-intensity estimate works best.}  
\rrendnote{coronal clouds}{note:CEcloud}{The simplest clouds are fully
transparent as CE assumes for coronal structures, \eg\ loops. 
No RT whatsoever, no interest in source functions -- but let me
nevertheless cast these non-clouds into optically-thick RT terms for
comparison.
The homogeneous thin cloud solution \ISSFpage{13}{Eq.\,27} has no
impinging intensity from behind ($I_\nu(0)\tapprox\,0$) in CE hence
$I_\nu(D)\tis\,j_\nu\,D$ with $D$ the geometrical cloud thickness
along the line of sight. 
This holds for Thomson scattering producing the white-light off-limb
corona (with $S\!_\nu\tapprox\,J_\nu$, \rrref{endnote}{note:wlcorona})
and also for thermally made EUV lines except the strongest. 
In terms of \rrref{Eq.}{eq:S_CS} (invalid for no RT but artificially
illustrative by counting escaping photons into $\alpha_\nu^\rmd$ with
$\eta_\nu^\rmd\tapprox\,1$ for CE) both $J_\nu\tapprox\,0$ and
$S\!_\nu^\rmd\tapprox\,0$ so that
$S\!_\nu^l\tapprox\,\varepsilon_\nu\,B_\nu(T)$ or
$j_\nu^l\tapprox\,\alpha^\rma_\nu\,B_\nu(T)$ with exceedingly small
$\varepsilon_\nu$ and $\alpha^\rma_\nu$ illustrating lack of
collisional deexcitation to compete with escape.

Even though all line photons are thermally created the intensity stays
far below saturation to the Planck function for lack of local
enclosure. 
Coronal clouds are far hotter than the photosphere and their radiation
is thermally defined by their temperature just as for the optical
continuum from the photosphere \SSIpage{4} -- but in contrast to the
photosphere coronal clouds radiate far below their thermal capacity
because they let all new photons depart instead of boxing them
in.\footnote{Example: 
\label{note:wiening}
the off-limb optical coronal lines of \rrref{endnote}{note:forbidden}
are Rayleigh-Jeans if one enters their million-Kelvin CHIANTI
ionization-stage temperature in the $\exp(h\nu/kT)\tis\,1$ threshold
test \SSIpage{3} -- but this Planck notion makes no sense for
their source-function-free thin formation. 
In LTE their emissivity would be
$j_\nu^{\rm \,LTE}\tis\,\alpha_\nu^{\rm LTE}\,B_\nu(T_\rme)$, in NLTE
it is
$j_\nu\tapprox\,(b_u/b_l)\,\alpha_\nu\,B_\nu(T_\rme)$
but here with large $b_l$ (the CE/SB curve ratio in \SSXpage{174} at the 
pertinent CE peak) and $b_u/b_l$ of order $10^{-8}$.
These lines are controlled non-collisionally; the actual source for
their emissivity in scattering contributions {\bf f} and {\bf g} is
the irradiating photospheric continuum from the
nearly-$2\pi$-filling Sun underneath.
Fortunately this nonlocal 5000\,K radiation is not Rayleigh-Jeans but
Wien-favors spontaneous pair {\bf f} with $1/4\pi$
isotropic chance to be redirected our way, whereas stimulated {\bf g}
photons would go into the nearly-$2\pi$
opposite outer space missing the tangential line of sight to Earth.
We see these lines thanks to NLTE Wiening.
}}.

\paragraphrr{Transfer equation in optical depth}
For very optically thick objects as the Sun there is no interest in
what impinges the other side since nothing comes through. 
We therefore flip $\tau$ to measure optical depth instead, either
along the line of sight ($\tau_{\nu\mu}$) or more often radially into
the Sun against height ($\tau_\nu$). 
With this flip the radiative transfer equation
upsets its source to negative\footnote{The source function can go negative in lasering (\rrref{endnote}{note:pumping}).}:
\begin{equation}
  \frac{\rmd I_{\nu\mu}}{\rmd \tau_{\nu\mu}} = I_{\nu\mu} - S\!_{\nu}
  \mbox{~~~~~~~~~~~~~~}
  \mu \, \frac{\rmd I_\nu}{\rmd \tau_\nu} = I_\nu - S\!_\nu,
\end{equation}
the first along the line of sight, the second for viewing angle
$\mu\tis\,\cos(\theta)$ into a plane-parallel atmosphere \SSFpage{23}
\RTSApage{97}{Eqs.\,4.10, 4.11}, \RTSApage{105}{Eqs.\,4.28-4.31}.
The corresponding integral ``formal solutions'' for the intensity
$I_\nu^+(0,\mu)$ that emerges from a non-irradiated atmosphere are
\begin{equation}
  \int_0^\infty \!\! S\!_\nu(\tau_{\nu\mu}) \, \ep{-\tau_{\nu\mu}} \dif \tau_{\nu\mu}
  \mbox{~~~~~~~~~~~~~~}
  (1/\mu)\int_0^\infty \!\! S\!_\nu(\tau_\nu) \, \ep{-\tau_\nu/\mu} \dif \tau_\nu,
\end{equation}
which are simple and easy to evaluate for LTE but complex and
nonlinear when the NLTE scattering and detour terms in
\rrref{Eqs.}{eq:epseta}\,--\,\ref{eq:S_CRD} contribute to $\alpha_\nu$
and $S\!_\nu$ involving radiation in other directions, at other
frequencies, in other transitions, at other times.
Since \rrref{Eqs.}{eq:b} express NLTE deviations in populations an equation
system results combining radiative transfer for multiple to many
angles and frequencies and population variations for multiple to many
species and levels \RTSApage{52}{Sect.\,2.6.1}. 
Solution is necessarily iterative \RTSApage{133}{Chapt.\,5}, or
nowadays through machine learning taught from validated iterative
solutions.

The recipe for quick interpretation is to apply the
Eddington-Barbier\footnote{Unjustly named
(\citeads{2018OAst...27...76P}) 
-- as often with Eddington.} approximation
\begin{equation}
  I_\nu^+(\tau_\nu\tis\,0,\mu\tis\,1) \approx S\!_\nu(\tau_\nu\tis\,1)
\end{equation}
with $\tau_\nu$ and $S\!_\nu$ the totals \ISSFpage{13}{Eq.\,35},
\IARTpage{45}{321}, \RTSApage{38}{Eq.\,2.44},
\RTSApage{106}{Eq.\,4.32}, \SSFpage{29}, \SSFpage{51}.
It simplifies the split between asking ``Where?'' and
``What?'' to a single location.\footnote{When you see
$I_\nu\tapprox\,S\!_\nu(\tau_\nu \tis\,2/3)$ you read text from an
unresolved nighttime colleague confusing flux and intensity
\SSFpage{29} \RTSApage{31}{Eq.\,2.7}. 
They also add a telltale $\pi$ in their $B_\nu$ definition.}

The split is valid in LTE making this recipe easiest for LTE in which
emergent line profiles map temperature variation with height through
Planck-function sensitivity conversion and with extinction-coefficient
profile weighting \SSIpage{16}. 
Absorption lines map outward declining source functions, emission
lines increasing source functions. 
Avrett sketched this sign change for VALIIIC lines in \SSXpage{36}
with flips from sampling its temperature minimum near
1600\,\AA\footnote{The ultraviolet flip from absorption to emission
was already described by
\citetads{1963BAN....17..209D}. 
It shows well in \SSXpage{34} from
\citetads{1974psus.book.....S}. 
IART examination questions: explain the sign of the Lyman and \HeI\
bound-free edges and how it fits the general intensity drop to lower
wavelength.} 
and near 160\,$\mu$m. 
The latter is closer to LTE.

Absorption lines go together with limb darkening \SSFpage{51}
\RTSApage{39}{Fig.\,2.4}, emission lines with limb brightening. 
Just beyond the limb the first flash-reverse into the latter
\RTSApage{168}{Fig.\,7.2}.\footnote{\CaII\ \HK\ and the Balmer lines
are also the strongest in the optical flash spectrum, but there out of
the blue (yellow rather) joined by \HeIDthree\ which is virtually
absent on the disk (\rrref{endnote}{note:HeI}).}

The ``four-panel'' diagrams in \SSFpage{31\,ff} are a cartoon course
on Eddington-Barbier spectrum formation with increasing complexity.
You should blink the absorption\,--\,emission pairs by full-page
flipping.  Then do the exam on \SSFpage{40}, first for \NaIDtwo\
(answer $h\tapprox\,500$\,km) and then for the blend.  If you decide
$h \tapprox\ 150$\,~km for the blend by reading off where its center
intensity equals the source function just as for \NaIDtwo\ you are a
million times wrong 
\rrendnote{\NaID, \MgIb, \CaIR\ heights}{note:NaD}{%
[{\em Spoiler:\/} million difference in line identification
\SSFpage{41}.] 
The same \linkadspage{1992A&A...265..268U}{5}{Fig.\,4} of
\citetads{1992A&A...265..268U} 
is also the subject of less simple but recommended
\RTSApage{250}{Problem 7} on \NaIDtwo\ line formation in the FALC star
(\rrref{Sect.}{sec:Avrett}) [{\em Spoiler:\/}
\href{https://robrutten.nl/rrweb/rjr-siucourses/afyc_anwers_2003.ps.gz}
{my answers to most RTSA problems}]. 
The \NaID\ lines are the deepest in the visible
(\linkrjrpage{1984AApS...55..143R}{17}{Fig.\,4} of
\citeads{1984A&AS...55..143R}, 
\linkrjrpage{2019arXiv190804624R} {18}{Fig.\,10} of
\citeads{2019SoPh..294..165R})
\footnote{History: \label{note:calcomp} the first was drawn with a
Calcomp plotter, a mechanical device moving a pen with an ink
cartridge transversely across a rotating drum with paper chart from a
large roll, both motions computer-controlled in 0.01\,inch steps. 
Each strip was a meter long and photographically reduced to
photo-offset print size.
Earlier the entire university computer (one for all,
\rrref{endnote}{note:mainframes}) was required for such plotting. 
For job scheduling per estimated duration I first had to count all
steps in a two-minute job (\rrref{endnote}{note:mainframes}). 
The operators liked plotting these crowded spectral charts for giving
them long breaks between mounting and dismounting the magnetic tape
with the Jungfraujoch Atlas (\rrref{endnote}{note:atlases}).
They also liked the endless runs of M.J.G.\,Veltman's {\em
schoonschip\/} (\rrref{footnote}{note:schoonschip}) by G.\,'t~Hooft
earning their joint electroweak renormalization Nobel by eventually
producing only zeroes.}. 
The \NaID\ lines are often called chromospheric. 
Your Eddington-Barbier estimate $h\tapprox\,500$\,km with \SSFpage{40}
puts the \NaIDtwo\ formation just above the temperature minimum of the
FALC star, but the line source function shows a $\sqrt{\varepsilon}$
scattering decline as if the FALC star has no non-RE heating
whatsoever. 
The photons emerging there were made much deeper, near thermalization
depth in the photosphere, and scattered out from there ignoring the
temperature rise and then escape. 
Intensity variations of the core therefore correspond to photospheric
temperature variations, not at the $\tau\tapprox\,1$ Eddington-Barbier
height.
This scattering is quantified for \NaIDone\ in the similar ALC7 star
in \SSXpage{85} where the $S,B,J$ graph shows optical depth locations,
destruction probability, thermalization length and decoupling height
(\SSXpage{80}, \rrref{endnote}{note:ALC7plots}). 
The \NaID\ lines are the deepest because they are the strongest
scattering lines of which the emergent intensity is fully controlled
within the photosphere.
However, in measuring Doppler or Zeeman modulation the
Eddington-Barbier estimate for the inner wings may be appropriate
because these signals are encoded at the last scattering.

Similar FALC results in
\linkrjrpage{2011AAp...531A..17R}{11}{Fig.\,13} of
\citetads{2011A&A...531A..17R} 
compare \NaIDone\ formation with \MgIbtwo\ and \CaIR\ formation. 
\MgIbtwo\ forms similar to \NaIDone, decoupling slightly higher, but
the source function of \CaIR\ decouples above the FALC temperature
minimum so that this line does convey FALC-``chromosphere'' signature. 
The same is seen for the ALC7 star in \SSXpage{27} and in comparing
\SSXpage{85} \SSXpage {84} \SSXpage{87}.

\citetads{2011A&A...531A..17R} 
used high-quality SST observations to compare the three lines in
intensity in \linkrjrpage{2011AAp...531A..17R}{5}{Fig.\,4} in
comparison with simulation results in
\linkrjrpage{2011AAp...531A..17R}{6}{Fig.\,5} and those smeared to
observation-resolution in
\linkrjrpage{2011AAp...531A..17R}{7}{Fig.\,6}, and then observed and
simulated line-core Dopplergrams in
\linkrjrpage{2011AAp...531A..17R}{8}{Fig.\,7}. 
These figures confirm non-chromospheric sampling by \NaIDone\ and
\MgIbtwo\ whereas \CaIR\ does show chromospheric signatures
(\rrref{endnote}{note:nonEchrom}).
The main difference between the first two is that the \NaID\ lines
have higher Doppler sensitivity than the \MgIb\ lines by having
narrower cores with steeper flanks. 
This explains the remarkable difference between their wings in
sampling upper-photosphere reversed granulation in
\linkrjrpage{2011AAp...531A..17R}{2}{Fig.\,1}.

The steep \NaID\ flanks also make \NaID\ Dopplergrams a sharp proxy
for localizing upper-photosphere shocks in MCs, as summarized in
\SSXpage{54}.\footnote{Repeating the corollary on
\linkrjrpage{2011AAp...531A..17R}{15)}{page 15} of
\citetads{2011A&A...531A..17R} 
that Fourier analysis of full-disk \NaID\ Doppler sampling with the
proposed GOLF-NG space instrument
(\citeads{2012ASPC..462..240T}) 
update will not achieve chromospheric helioseismology as advertised,
whereas hoped-for $g$-mode detection will be hampered by
upper-photosphere reversed granulation, fluxtube shocks and
``shutter'' modulation by overlying features as in
\linkrjrpage{2007ApJ...654.1128D}{6}{Figs\,5 and 6} of
\citetads{2007ApJ...654.1128D}.}}. 
Be aware that use of the Eddington-Barbier approximation is often a
fallacy
\rrendnote{contribution, response, sensitivity} {note:response}{%
When using ``the formation height'' you may earn my scorn.
Only thin clouds at specific height (as in
\rrref{endnote}{note:CEcloud}) share their height between different
formation quantities. 
For thick objects a common characterizing escape height may be
applicable when LTE holds for all contributing processes, but for
scattering lines the intensity is set by temperature much deeper and
more widely within the atmosphere or within an effectively thick
structure than the Doppler and Zeeman response to line-of-sight
velocities and magnetism (as in the \NaID\ example of
\rrref{endnote}{note:NaD}). 
Also, for weak lines the emergent intensity is dominated by background
continuum processes rather than line processes (exemplified by their
factor-million height ratio in \SSFpage{41} with
Schuster-Schwarzschild cloud modeling appropriate for the telluric
blends).

In plane-parallel atmospheres the vertically emerging intensity is
$I_\nu\tis\,\int S\!_\nu\exp(-\tau_\nu)\dif\tau_\nu$ with integrand or
``contribution function'' $S\!_\nu\exp(-\tau_\nu)$ sketched in
\SSFpage{29}, an extended plume. 
The photons do not ``escape from $\tau_\nu\tis\,1$'' but throughout
this plume. 
Such contribution plumes are plotted against height, \ie\
$\rmd I/\rmd h = j_\nu(h)\exp(-\tau_\nu(h))$, for many wavelengths in
the VALIIIC spectrum in fabulous many-page Fig.\,36 
\linkadspage{1981ApJS...45..635V}{33}{Fig.\,36} of 
of \citetads{1981ApJS...45..635V} 
with excerpts in SSX. 
The first one in \SSXpage{61} for the 500\,nm continuum shows the
expected contribution shape with initial $j_\nu(-h)$ rise set
primarily by near-exponential inward density increase until stopped by
the steep $\exp(-\tau)$ decrease and with
$I(1,0)\tapprox\,S(\log\tau\tis\,0)$.
The last one in \SSXpage{67} for $\lambda\tis\,40$ \,nm shows weird
$\rmd I/\rmd h$ shape without peak near $\log \tau\tis\,0$ and two
orders-of-magnitude deviation from the Eddington-Barbier value.

Earlier, sophistication beyond this simple integrand was based on
weighting functions (\eg\
\citeads{1951AnAp...14..115P}) 
and deriving such for line depression rather than emergent intensity
(\citeads{1974SoPh...37...43G}). 
The most informative format became Carlsson's integrand breakdown
rewriting $\rmd I/\rmd h$ into
$[S\!_\nu]\,\,[\tau_\nu\exp(-\tau_\nu)]\,\,[\rmd (\ln(\tau_\nu))/\rmd
z]$
in \linkadspage{1994chdy.conf...47C}{15}{Eq.\,6} of
\citetads{1994chdy.conf...47C} 
and grey-scale charting these three terms and their product as
function of height and wavelength through the line of interest
(\linkadspage{1994chdy.conf...47C}{16}{Fig.\,8\,ff},
\rrref{endnote}{note:4pdiagrams}).

In the meantime ``response functions'' were formulated that estimate
the effect of perturbations of a given quantity on an emergent line
profile (classics \citeads{1971SoPh...20....3M}, 
\citeads{1975SoPh...43..289B}, 
\citeads{1986A&A...163..135M}) 
but nowadays this is done by numerical sensitivity analysis, for
example the ``multi-Multi'' analysis in
\linkrjrpage{1992AAp...253..567C}{14}{Fig.\,12} of
\citetads{1992A&A...253..567C} 
varying many transitions in their \MgI\ model atom one by one
(\rrref{endnote}{note:MgI12micron}), the successive step-function
temperature perturbation to quantify UV intensity response of
\citetads{2005ApJ...625..556F} 
and the NLTE-nonE versus LTE simulation-run comparison in
\linkrjrpage{2007AAp...473..625L}{7}{Fig.\,4} of
\citetads{2007A&A...473..625L} 
(\rrref{Sect.}{sec:Oslo}).

Scorn-avoiding moral: even if your superduper simulation reproduces
actual observations you still have to delve per sensitivity analysis
into your simulation to diagnose how and why. 
Yet deeper (non-blindly) when you use a simulation to teach machine
learning.
As difficult as diagnosing observations but with the advantages that
you can check below $\tau\tis\,1$ and know the inputs. 
\citetads{1994chdy.conf...47C} 
(more formally
\citeads{1997ApJ...481..500C}) 
remains the canonical example: intricate observed patterns, convincing
reproduction with data-driven simulation
(\rrref{endnote}{note:CSshocks}), four-panel breakdowns
(\rrref{endnote}{note:4pdiagrams}).}.  

Be also aware that solar-atmosphere structures may be optically thick
without being effectively thick so that radiation scatters through.
Filaments or fibrils may appear intransparent in \Halpha\ while
illumination from behind is still important. 
This is even the case for \Halpha\ in the ``chromosphere'' plateau of
1D standard models, as in \SSXpage{91} which not only shows
$S \approx J > B$ across the ALC7 temperature minimum from
backscattering 
but also has
photospheric photons scattering through the model ``chromosphere''. 
This caveat was beautifully demonstrated in
\linkadspage{2012ApJ...749..136L}{7}{Fig.\,7} of
\citetads{2012ApJ...749..136L} 
by comparing 1D \Halpha\ synthesis for a Bifrost
(\rrref{endnote}{note:Bifrost}) simulation with MULTI
\rrendnote{MULTI program}{note:MULTI}{%
This NLTE spectral synthesis code of
\citetads{1986UppOR..33.....C} 
is obviously named for going beyond the 2-level approximation. 
It initially used linearization following
\citetads{1985JCoPh..59...56S} 
with the brilliant Scharmer operator \RTSApage{148}{Eq.\,5.52} (yes,
the same as the brilliant SST builder,
\rrref{endnote}{note:grancontrast}) to obtain $I_\nu$ rather than
$J_\nu$ from $S\!_\nu$ with a local two-way along-the-beam
Eddington-Barbier approximation. 
It was later extended with the faster combination of the diagonal OAB
operator of
\citetads{1986JQSRT..35..431O} 
and convergence acceleration of \citetads{1974JChPh..61.2680N}
\RTSApage{150}{}. 
Convergence comparison in \SSFpage{103}.}
and 3D \Halpha\ synthesis with MULTI3D 
\rrendnote{MULTI3D program}{note:MULTI3D}{%
Extension of MULTI to 3D parallel computation by
\citetads{2009ASPC..415...87L}. 
It uses short characteristics with further-away contributions
corrected by iteration.}.
In the columnar 1D line synthesis the deep-photosphere granulation is
better visible than lower-contrast chromospheric fibrils but in 3D
synthesis the granular contrast is smoothed away by lateral
scattering
\rrendnote{no Thomas photoelectric control}{note:photoelectric}{%
\Halpha\ is a special line from $n\tis\,2$ at 10.2\,eV sitting so high
up in the \linkadspage{1968pgda.book.....M}{9}{\HI\ Grotrian
diagram}\footnote{Naming: Grotrian
\label{note:Grotrian}
diagrams are term diagrams including transitions (named after
\citeweb{https://openlibrary.org/books/OL27041734M}
{Grotrian1928diagrams}). 
The {\em Partial Grotrian diagrams of astrophysical
interest\/} of
\citetads{1968pgda.book.....M} 
is a reprint of Appendix A of \citetads{1956lcea.book.....M} 
prepared by C.E.\,Moore and is a venerable classic of astronomical
spectroscopy. 
The Utrecht Sonnenborgh library had a dozen; I stole one and fifty
years later still have it on my desk even though it is now
\href{https://ned.ipac.caltech.edu/level5/Ewald/Grotrian/Moore.pdf}
{ADS-available}. 
In the scan the \linkadspage{1968pgda.book.....M}{9}{\HI\ diagram}
misses the horizontal $n\tis\,2$ line (and the bottom of Pfund)
present in the original; is there a conspiracy theory why this
happened to the most important level of observational solar physics?
More complete term and Grotrian diagrams are in the fat volumes of
Bashkin and Stoner (\citeyearads{1975aelg.book.....B},
\citeyearads{1978aelg.book.....B}, \citeyearads{1981aelg.book.....B},
\citeyearads{1982aelg.book.....B}).
I also stole Moore's famous RMT multiplet table
(\citeads{1959mtai.book.....M}) 
but also got that per magnetic tape (and now supply it in my
\href{https://robrutten.nl/rrweb/rjr-archive/dircontent.html}{archive})
although she forbade digitization because machine access would seduce
lazy scientists to non-understanding automation -- she wouldn't have
applauded machine learning.
See her on this 1963
\href{https://robrutten.nl/nieuwenhuijzenshots/sym1963/1963-sym-18-33-Jan-Oort_Donald-Menzel_Charlotte-Moore_Marcel-Minnaert_Albrecht-Unsoeld.jpg}
{Minnaert farewell photo} amid famous colleagues.
\href{https://robrutten.nl/nieuwenhuijzenshots/sym1963/1963-sym-17-11_3rd-row_Donald-Menzel_Richard-Thomas_John-Jefferies.jpg}{This
one} shows R.N.\,Thomas between D.H.\,Menzel and J.T.\,Jefferies (with
a box of slides, \rrref{endnote}{note:projection}) in front of
E.H.\,Avrett.
More photos by H.~Nieuwenhuijzen in his
\href{https://robrutten.nl/nieuwenhuijzenshots/sym1963/album.html}
{Minnaert farewell album} and 
\href{https://robrutten.nl/nieuwenhuijzenshots/2012-SIU-afscheid/album.html}
{Utrecht farewell album}.}, close to the 13.6\,eV \HI\ ionization
limit. 
It has a sizable $\eta S^\rmd$ contribution from bound-free detours as
the loop shown in \linkrjrpage{2019SoPh..294..165R}{4}{Fig.\,2} of
\citetads{2019SoPh..294..165R}. 
This was famously called {\em ``photoelectric control''\/} by
\citetads{1957ApJ...125..260T} 
with a schematic source function diagram in
\linkadspage{1959ApJ...129..401J}{5}{Fig.\,3} of
\citetads{1959ApJ...129..401J} 
that was reprinted in Fig.\,12-9 of
\citetads{1970stat.book.....M} 
and Fig.\,11-11 of
\citetads{1978stat.book.....M}. 
It looks similar to the \Halpha\ source function in ALC7 with
$S\!_\nu^l$ leveling out in the photosphere higher than $B_\nu$ in the
temperature minimum, indeed similarly to the Balmer continuum as if it
follows that \SSXpage{76}. 

However, this ``photoelectric control'' designation was wrong because
even \Halpha\ is mostly scattering as shown in
\linkrjrpage{2012AAp...540A..86R}{9}{Fig.\,12} of
\citetads{2012A&A...540A..86R}: 
while the detour photon contribution in the FCHHT-B ``chromosphere''
of \citetads{2009ApJ...707..482F} 
is much larger than the collisional-creation contribution it is still
much smaller than the scattering contribution. 
Even \Halpha\ is foremost a two-level scattering line; the
upper-photosphere $S\tapprox\,J>B$ excess is due to backscattering
from the ``chromosphere'' (\rrref{endnote}{note:aureoles}). 
Such excess does not occur in the RE atmosphere at left in
\linkrjrpage{2012AAp...540A..86R}{6}{Fig.\,7} of
\citetads{2012A&A...540A..86R} 
whereas with ``photoelectric control'' by $\eta S^\rmd$ it should
since the ultraviolet continua in
\linkrjrpage{2012AAp...540A..86R}{5}{Fig.\,5} there have similar
$S\tapprox\,J > B$ excess with or without a chromosphere.
 
Thomas' ``photoelectric control'' was a red herring.
There are some lines as the \MgII\ triplet around \MgII\ \hk\ for
which source function enhancement in cooling recombining gas (possibly
NSE) producing emission is likely, but the many Thomas-citing
typecastings of \Halpha\ as ``photoelectric'' were misled.} 
\rrendnote{\Halpha\ photosphere and chromosphere visibilities}
{note:Halpha-high}{%
The 10\,eV \Lyalpha\ excitation energy of the $n\tis\,2$ lower level
of \Halpha\ produces both extraordinary invisibility of photospheric
fine structure and extraordinary visibility of chromospheric fine
structure in this line and in its Balmer companions. 
No wonder since \Lyalpha\ is the most important transition in the
entire solar spectrum.
I summarize these visibilities here; more in \rrref{Sect.}{sec:Oslo}
and \rrref{Sect.}{sec:chromosphere}.

Invisibility: the exceedingly deep and wide small-Boltzmann-ratio
opacity gap across the cool upper photosphere
(\rrref{endnote}{note:quietHa}) not only permits deep back-scattering
penetration (\rrref{endnote}{note:aureoles} and
\rrref{endnote}{note:photoelectric}) but also implies that most
outward-directed \Halpha\ photons created in the deep photosphere
travel up in arbitrary direction until hitting opaque chromospheric
fibrils at least 1000\,km higher, and then mostly scatter back or
through the fibrils with $\sqrt{\varepsilon}$ darkening (the ALC7
``chromosphere'' only reaches thickness $\log(\tau)\tapprox 1.3$ in
\Halpha\ with $\log(\varepsilon)\!<\!-3$ and thermalization in the
deep photosphere \SSXpage{91}).
The granulation pattern has larger contrast than fibrilar intensity
modulation but is erased by this erratic-direction gap crossing over
larger-than-granule distance. 
Columnar 1D spectral synthesis wrongly maintains it, hence the
striking 1D\,--\,3D differences in
\linkadspage{2012ApJ...749..136L}{7}{Fig.\,7} of
\citetads{2012ApJ...749..136L}. 
You see chromospheric \Halpha\ fibrils in photons that were created in
the granulation far underneath but smoothed its signature on their way
up.

Visibility: being \linkadspage{1968pgda.book.....M}{9}{fed by \Lyalpha}
gives extraordinary opacity in its outer wings to spicule-II
chromospheric heating events (\rrref{endnote}{note:spicules-II}) and
in its core to cooling chromospheric return fibrils
(\rrref{endnote}{note:nonEchrom}), the first from heating and the
second from cooling with NSE opacity retardance (\rrref{Sect.}{sec:Oslo}) that
controls the location of its $\sqrt{\varepsilon}$ source function
decay and $\sqrt{\varepsilon}$ darkness (\rrref{endnote}{note:Ha-RE}).
The ALMA \HI\ free-free continua share in this boosting but sample local
temperature rather than scattering decline
(\rrref{endnote}{note:ALMA-temp}).
Such propagating heating and cooling events are likely smaller than
telescope resolutions intrinsically, but they gain wider \Halpha\ and
ALMA opacity from \Lyalpha\ NLS surround scattering and lingering
track opacity from NLT \Lyalpha\ retardation
(\rrref{endnote}{note:aureoles}, 
\citeads{2016A&A...590A.124R}, 
\citeyearads{2017A&A...598A..89R}). 

Optical \HeI\ lines sit at twice higher excitation energy
(\linkadspage{1968pgda.book.....M}{11}{\HeI\ Grotrian diagram}) and
even add coronal visibility (\rrref{endnote}{note:HeI}).}.

The most important equation of classical 1D 2-level scattering is the
Schwarzschild equation 
\RTSApage{98}{Eq.\,4.14} defining the \lambdop\ operator
\RTSApage{101}{Eq.\,4.20}; a synopsis is given after
\linkrjrpage{2019arXiv190804624R}{9}{Eq.\,10} of
\citetads{2019SoPh..294..165R}. 
Its cutoff at the surface produces outward $J_\nu < S\!_\nu$
divergence for small inward increase of $S\!_\nu(\tau_\nu)$ but
outward $J_\nu > S\!_\nu$ divergence for steep increase
\RTSApage{100}{Fig.\,4.2}, \RTSApage{103}{Fig.\,4.4},
\RTSApage{121}{Fig.\,4.9}). 
Be aware that steep horizontal gradients are similarly important in 3D
radiative transfer. 
Steep horizontal gradients occur for example already in the
photosphere in and above granulation and in and around MCs
constituting network and plage
\rrendnote{MCs, network, plage, faculae}{note:MCs}{%
For MCs (magnetic concentrations, small upright kilogauss
``fluxtubes'') and their visibility as ``bright point'' I refer to the
extensive caption of \linkrjrpage{2020LingAstRep...1R}{91}{Fig.\,73}
of
\href{https://robrutten.nl/rrweb/rjr-pubs/2020LingAstRep...1R.pdf}{LAR-1}.
They are bright from being holes in the surface, brighter when deeper,
not signifying heating but cooling of fluxtube walls by radiation
leaks that increase solar irradiance at Earth when there are more
(\citeads{1977SoPh...55....3S}). 

They appear brighter from hole deepening in the G band around
4308\,\AA\ due to CH lines and in the CN bandhead at 3883\,\AA\
(\rrref{endnote}{note:gillespie}) through molecule dissociation, in
outer wings of strong lines (Leenaarts
\citeyearads{2006A&A...449.1209L}, 
\citeyearads{2006A&A...452L..15L}) 
through less collisional damping.
Sharp images from optical telescopes reaching the Scharmer threshold
(\rrref{footnote}{note:Scharmer}) show them on subarcsecond scales in
intergranular lanes with irregular and fast-varying morphology
(``flowers'' in
\citeads{2004A&A...428..613B}). 
Any HMI magnetogram shows larger ones sprinkled as small black and
white grains in irregular patterns all over the disk, denser in active
regions but fewer towards the limb where their upright Zeeman
signature is not aligned with the line of sight. 
The corresponding bright grains in AIA 1700 and 1600\,\AA\ chart them
better but unsigned and shifting limbward
(\rrref{endnote}{note:1600-1700}). 
Away from active regions the MC patterns constitute the somewhat
cellular magnetic network\footnote{Only partially outlining the
downflow borders of supergranulation cells. 
The latter appear more regular towards the limb in HMI Dopplergrams
since the cell borders are populated only incompletely with MCs. 
At Utrecht C.J.\,Schrijver once handed a large magnetogram to solar
colleagues requesting to draw the magnetic network. 
The results varied so much (from sloppy by me to meticulous by Zwaan)
that he concluded there is no objectively definable magnetic network. 
Nor is there a clear distinction between active network and small
plage (``plagette'').}.

Near the limb the small MCs are known as faculae, more evident because
they brighten limbward from slanted viewing into hot granules and
bunch up in foreshortening.
Faculae were already visible in white-light images that did not reach
the high resolution of modern telescopes.
This fortunate visibility enabled N.R.\,Sheeley to use their numbers
on historical Mount Wilson full-disk photographs as polar field proxy
(Sheeley \citeyearads{2008ApJ...680.1553S} 
and earlier; \citeads{2012ApJ...753..146M}). 
Larger patches of MC concentrations were traditionally recognized on
\CaIIHK\ spectroheliograms and called flocculi (Hale and Ellerman) and
plage (Deslandres). 
Their chromospheric appearance is coarser, hence more evident, than
underlying photospheric MCs but corresponds closely to the MC surface
grouping into network and plage. 
The ultraviolet MC grains are photospheric, as in Gillespie's classic
CN spectroheliogram of \rrref{endnote}{note:gillespie}.}.  
Obviously they occur in and around the multitude of small dynamic
structures constituting the higher atmosphere.

\SSFpage{108} gives a summary of the key line formation equations.

\section{LTE:~ Holweger models} \label{sec:Holweger}
The classic Holweger models
(\citeads{1967ZA.....65..365H} 
and HOLMUL of \citeads{1974SoPh...39...19H}) 
were a pinnacle of solar LTE interpretation
\rrendnote{Bilderberg study week}{note:Bilderberg}{%
History: the ``International study week on the quiet
photosphere'' was organized by Kees de Jager
(\rrref{footnote}{note:CdeJ}) in 1967 at the Bilderberg hotel near
Arnhem (famous from more famous Bilderberg conferences) in order to
jointly derive the Bilderberg Continuum Atmosphere of
\citetads{1968SoPh....3....5G} 
which later got improved to the HSRA of
\citetads{1971SoPh...18..347G}. 
The proceedings became the first issue of {\em Solar Physics\/} volume 3,
reprinted separately as
\citetads{1967sqpl.conf.....D}. 
Kees had organized the well-attended Minnaert farewell symposium at
Utrecht in 1963
(\href{https://robrutten.nl/nieuwenhuijzenshots/sym1963/album.html}
{photos}) and edited its proceedings starting Reidel's Astrophysics
and Space Science Library
(\citeads{1965ASSL....1.....D}). 
He turned the Bilderberg successor into a true workshop, sharing out
homework beforehand, setting up working groups on various topics, and
inviting promising youngsters
(\linkadspage{1996SoPh..169..233B}{2}{Sect.\,2} of
\citeads{1996SoPh..169..233B}) 
and also his yet younger Utrecht students. 
The participants (fierce but friendly contestants) sat half-circling
the speaker at the blackboard with me in the center to operate the
slide projector (\rrref{endnote}{note:projection}). 
I also was the notulist, daily spreading progress notes per stencil
mimeograph\footnote{History: a stencil was a sheet of thin material
that fitted a normal typewriter for typing without ink ribbon, making
letter-shaped holes. 
Mistakes were filled in with lacquer for retyping. 
The mimeograph duplicated the stencil by forcing ink through the holes
onto paper sheets, with the stencil mounted on a hand-cranked rotating
inking drum rolling over fresh paper sheets pushed from an input bin
to an output bin.
All offices handled duplication this way (if not photographically)
until Xerox plain-paper copiers became widespread. 
During the meeting O.\,Gingerich taught me how to add graphs showing
the converging BCA model stratification by plotting results from his
pre-computed stacks of hydrostatic equilibrium tables on graph paper
and overlay the stencil for scratching the curves with a needle pencil
into it, held together against a window as backlight.}. 
I later described this extraordinary expert meeting, the best I
attended, in \citetads{2002JAD.....8....8R} 
with a \linkrjrpage{2002JAD.....8....8R}{5}{group photo} including
Holweger (more Bilderberg photographs at the bottom of my
\href{https://robrutten.nl/Rob_s_astronomer_shots.html}{astronomershots}).
\href{https://robrutten.nl/astronomershots/album1967/edith-muller.jpg}
{E.A.\,M\"uller} declared at its conclusion that she wouldn't use the
new model with its chromosphere requiring complex NLTE to handle that
but would instead stick to Holweger's model without chromosphere for
its reliably well-working LTE. 
She then did so with HOLMUL
(\citeads{1974SoPh...39...19H}).}. 
They replaced curve of growth\footnote{Treated in practical
\linkpdfpage{https://robrutten.nl/rrweb/rjr-edu/exercises/ssa/ssa.pdf}{28}{SSA\,3}.
\linkrjrpage{1983AAp...117...21R}{6}{Fig.\,4} of
\citetads{1983A&A...117...21R} 
may have been the last solar \FeI\ one.} 
abundance determination by individual line-by-line fitting which
became an abundance industry
\rrendnote{classic abundance determination}{note:abundance}{%
Holweger used the ALGOL ``Kieler program'' of
\citetads{1966AAHam...8...26B} 
for LTE spectral synthesis in a classical approach: impose a $T(h)$
``model'' relation, evaluate density stratifications for the given
element mix by assuming hydrostatic equilibrium
\RTSApage{166}{Sect.\,7.2.3}, compute continuum extinction from \Hmin\
and extinction of a given line per SB ($b_u\tis\,b_l\tis\,1$ in
\RTSApage{224}{Eq.\,9.6}), use the integral solution of the transport
equation \RTSApage{38}{Eq.\,2.43} to obtain the emergent profile. 
He used line areas (``equivalent width'' \RTSApage{224}{Sect.\,9.1.2})
as quality gauge, repeated this for many optical lines, and derived a
best-fit temperature and corresponding density stratification by
trial-and-error manual $T(h)$ adjustment.

His many followers in classic abundance determination then took these
stratifications (``model atmosphere'') for granted and did similar
line fitting for different lines of a given element to establish its
abundance, with much debate about the reliability of oscillator
strength (transition probability) $g\!\,f$ values.}.  
It employed critical ad-hoc adjustment parameters: fake line
broadening for duplicating observed line broadening with
``microturbulence'' \RTSApage{82}{Eq.\,3.80} (adding height-dependence
as in \linkadspage{1981ApJS...45..635V}{19}{Fig.\,11} of
\citeads{1981ApJS...45..635V}) 
and ``macroturbulence'' \RTSApage{83}{Eq.3.81} (with
``radial-tangential'' anisotropy added by
\citeads{1977ApJ...218..530G}), 
and fake wing extension with a collisional damping ``enhancement
factor''. 
The Holweger models ignored the existence of the solar chromosphere to
avoid self-reversals in strong optical lines
\rrendnote{\MgI\ 12-micron emission lines}{note:MgI12micron}{%
Assuming LTE implies predicting self-reversing chromospheric cores in
\eg\ the \NaID\ lines treated in practical
\linkpdfpage{https://robrutten.nl/rrweb/rjr-edu/exercises/ssb/ssb.pdf}{22}{SSB\,3}
but also in the strong \FeI\ lines that Holweger employed. 
Their absence was attributed correctly to NLTE scattering by others
but Holweger insisted tenaciously that NLTE computer programs fatally
overestimate NLTE departures by not including sufficient hydrogen-atom
collisions \SSXpage{39} and that the actual chromosphere starts
higher. 
When he refereed 
\citetads{1992A&A...253..567C} 
explaining the enigmatic \MgI\ emission lines at
12~micron this was also his main complaint, but the Rydberg
recombination modeled there was already collision-dominated.

The formation of these striking features is interesting.  
Naturally they were first attributed\footnote{History:
\label{note:whitepaste} they show high peaks in emission at the center
of absorption troughs.
Being unusual they were blamed on some instrumental defect and
white-pasted by
\citetads{1980STIN...8031298G} 
in their infrared spectrum atlas. 
Brault and Testerman had already spotted these intriguing features
(\linkadspage{1983ApJ...269L..61B}{2}{Fig.\,1} of
\citeads{1983ApJ...269L..61B}) 
at Kitt Peak; Brault therefore phoned them to ask why their atlas
showed gaps precisely at these locations. 
Answer: ``everybody knows that Fraunhofer lines are in absorption''. 
\citetads{1981ApJ...247L..97M} 
(including two more Murcrays) then sensed their gaffe, dewhitened them
and claimed discovery (without mentioning Brault) with fancy
attribution to Si\,\specchar{viii} and Mg\,\specchar{viii} but not
daring to exclude telluric origin. 
So it goes.
They were identified as solar \MgI\ $n\tis\,7$--6 transitions by
\citetads{1983ApJ...275L..11C}.} 
to chromospheric formation
(\citeads{1983ApJ...269L..61B}) 
but they are photospheric, combining NLW photon pumping in
photo-ionization (\rrref{endnote}{note:pumping}), NLW photon suction
(\rrref{endnote}{note:suction}) and a diffusive radiative-collisional
Rydberg population departure flow. 
In the Rydberg top this is dominated by collisional decays, the
opposite of coronal dielectronic radiative-down
(\rrref{endnote}{note:dielectronic}). 
See the cartoon explanations in
\citetads{1994IAUS..154..309R} 
including Mats' ``Rydberg flows for kayakers'' in
\linkrjrpage{1994IAUS..154..309R}{7}{Fig.\,4}
(\href{https://robrutten.nl/rrweb/rjr-edu/manuals/student-report/example.pdf}
{better quality}). 
\citetads{1992A&A...253..567C} 
obtained excellent fits in
\linkrjrpage{1992AAp...253..567C}{10}{Fig.\,6} with a 1D RE model
atmosphere and likewise in
\linkrjrpage{1992AAp...253..567C}{15}{Fig.\,14} with the non-RE MACKKL
model. 
These fits are uncommonly good because these lines form in the upper
photosphere where RE reigns and 1D models apply best
(\rrref{endnote}{note:REphot}).
 
Actually,
\citetads{1987A&A...173..375L} 
had come close to model these emission lines properly but their \MgI\
model atom did not reach high enough up the collision-dominated
Rydberg ladder. 
Their study was wildly and wrongly attacked by
\citetads{1989ApJ...340..571Z}, 
in turn severely rebutted by me in
\citetads{1992A&A...253..567C}. 
This impressed Zirin so much that he saw me as potential successor. 
I saw him as thin (\rrref{footnote}{note:thinthick}) because in
the abstract he stated that these \MgI\ lines are thin. 
Wrong language (\cf\ \rrref{footnote}{note:Zirin}) and wrong altogether
because these lines are the most beautiful $S(\tau)$ mappers in the
solar spectrum (\linkrjrpage{1994IAUS..154..309R}{9}{Fig.\,5}).},
but nevertheless HOLMUL became highly popular\footnote{ADS shows
\href{https://ui.adsabs.harvard.edu/abs/1974SoPh...39...19H/metrics}{900$^+$
citations} but declining rate this century.} 
with abundance determiners because it worked so well, giving them
smaller spread for multiple lines than other standard models of the
time
\rrendnote{plane-parallel atmosphere files}{note:ppmodels}{%
I copied files with plane-parallel model atmospheres from the 1980s
that I found I still have into my
\href{https://robrutten.nl/rrweb/rjr-archive/dircontent.html} {solar
file archive} started while writing \rrref{endnote}{note:atlases}.
These include HOLMUL, the HSRA model of
\citetads{1971SoPh...18..347G}, 
the thesis model of \citetads{1972PhDT.........7L}, 
the VALIIM model of \citetads{1976ApJS...30....1V}, 
the radiative-equilibrium model of
\citetads{1976A&AS...23...37B}, 
the models VALIIIA\,--\,VALIIIF of \citetads{1981ApJS...45..635V} 
and the MACKKL model of \citetads{1986ApJ...306..284M}. 
The FALC model of \citetads{1993ApJ...406..319F} 
is supplied and analyzed in
\href{https://robrutten.nl/rrweb/rjr-edu/exercises/ssb/dircontent.html}
{practical SSB1}. 
All four FAL models of
\citetads{1993ApJ...406..319F} 
are available in RH (\rrref{endnote}{note:RH}) input format in the
\href{https://github.com/han-uitenbroek/RH/tree/master/Atmos}{RH
github Atmos} directory, but my new
\href{https://robrutten.nl/rrweb/rjr-archive/dircontent.html} {solar
file archive} adds some more including VALIIIC, MACKKL, models of
\citetads{2009ApJ...707..482F} 
and ALC7 of \citetads{2008ApJS..175..229A}.}. 
No wonder because the model was made as best fit to photospheric
lines, in particular of \FeI, so that fitting similar lines with the
model was a self-fulfilling prophecy\footnote{Not ``garbage-in =
garbage-out'' but model-out $\approx$ model-in propelling its
assumptions and limitations.
The modern equivalent is machine learning: teaching it with specific
sets of models or simulations delivers answers propelling (if not
advocating) their good ingredients but also their errors.}, 
in particular for the much debated Fe abundance itself (\eg\
\citeads{1996A&A...305..325K}). 
Earlier \citetads{1982A&A...115..104R} 
showed that such model fitting is also self-correcting
\rrendnote{NLTE masking}{note:masking}{%
For a steep outward temperature decline the outward increasing
ultraviolet $J\!>\!B$ excess from scattering and \lambdop\ results in
NLW-type NLTE overionization of minority-species metals \SSFpage{111}. 
Instead adopting Saha ionization then puts the height assigned to
$\tau\tis\,1$ too high so that the fitted temperature gradient gets
too shallow and the same as for locations with less steep temperature
decline. 
Different surface structures with different gradients (as granules and
lanes) can so be fitted erroneously by a single shallow gradient.

Additionally, the neglect of source function NLTE for higher-up
scattering in stronger lines results in undoing an actual
chromospheric temperature rise, masking that also.

\linkrjrpage{1982AAp...115..104R}{8}{Fig.\,8} of
\citetads{1982A&A...115..104R} 
shows how a HOLMUL-like model so results from the steeper-gradient
HSRA model of \citetads{1971SoPh...18..347G}, 
summary in \SSXpage{40}. 
A cartoon is shown in \linkrjrpage{1988ASSL..138..185R}{14}{Fig.\,4}
of \citetads{1988ASSL..138..185R}.}. 

So-called ``inversion'' codes (I find ``best-fit'' better suited) are
automated versions of Holweger's trial-and-error model derivation
procedure 
\rrendnote{Milne-Eddington approximation}{note:ME}{%
Inverters often apply the further simplification of assuming
Milne-Eddington line formation, \ie\ constancy of the
line-to-continuum extinction ratio with height (Holweger instead
evaluated continuum and line extinction in detail).
This assumption was tested with a MURaM simulation and LTE line
synthesis for comparable weak \FeI\ lines at different excitation
energy in \linkrjrpage{2009AAp...499..301V}{10}{Fig.\,8} of
\citetads{2009A&A...499..301V}; 
summary in \SSXpage{41}.
It is bad at low excitation, better at higher excitation as already
evident in \linkrjrpage{1984AApS...55..143R}{16}{Fig.\,3} of
\citetads{1984A&AS...55..143R}.}. 
Copying Holweger's fads also copies the fallacies including NLTE
masking, an important failure because NLTE scattering in the
ultraviolet bound-free edges controls the opacity of minority-species
atomic lines throughout the spectrum. 
In the optical these often have source functions close to LTE through
interlocking
\rrendnote{interlocking \FeI, \CeII, \FeII\
lines}{note:interlocking}{%
Spectral species as \eg\ \FeI\ and \FeII\ are rich in bound-bound
transitions. 
Optical multiplets are mostly high up but connected by ultraviolet
transitions to levels lower down (ground state or low-lying metastable
levels), comparable to the \CaII\ triangle in
\linkrjrpage{2019SoPh..294..165R}{4}{Fig.\,2} of
\citetads{2019SoPh..294..165R} 
but in large number. 
With their small lower-level excitation energy these ultraviolet
transitions tend to be strong, maintaining $S^l\tapprox\,B$ further
out as in the righthand cartoon of \SSFpage{111} and forcing
$b_u\tapprox\,b_l$ on their upper levels at the deeper heights where
their weaker optical siblings reach $\tau\tapprox\,1$.
The low \FeI\ levels are all strongly collisionally coupled as evident
in VALIII \linkadspage{1981ApJS...45..635V}{32}{Fig.\,34}.
When the weaker higher-up lines so share strong-line $b_l$ in their
upper and lower levels they have LTE source functions -- but their
opacities scale with $b_l$ and sense ultraviolet overionization in the
upper photosphere as in the lefthand cartoon of \SSFpage{111}),
\SSXpage{42} and also evident in VALIII
\linkadspage{1981ApJS...45..635V}{32}{Fig.\,34}. 
Summary: Fe lines suffer NLS and NLW. 
Their NLTE formation cannot be modeled by accounting only for NLS
scattering in the line itself. 

Other interlocking lines occur in lanthanide rare earths. 
They have multiplets with many members that overlap closely in energy. 
\CeII\ has most lines, peaking at 10 lines/\AA\ around \CaIIHK. 
These are weak lines but scatter strongly with much crosstalk. 
They show up in emission inside the limb both outside and within the
extended wings of \CaII\ \HK\ which was explained in the classic
analysis of \citetads{1971A&A....10...64C} 
following on \citetads{1971A&A....10...54C} 
and his earlier thesis studies. 
They appear bland from wide multi-wavelength scattering producing
their $J$, whereas similar \FeII\ blends in the wings of \HK\
(\citeads{1980A&AS...39..415R}) 
show much spatial variation due to pumping by particular ultraviolet
transitions (\citeads{1980ApJ...241..374C}) 
and so gaining ultraviolet Wien sensitivity
(\rrref{endnote}{note:pumping}).}  
but their interlocked opacities have NLTE deficits up to an order of
magnitude from ultraviolet over\-ionization set by steep temperature
gradients in deeper layers \SSXpage{42}. 
This fallacy has been long ignored but was at last addressed by
\citetads{2020A&A...633A.157S}. 
Worse: in the actual 3D solar photosphere matter and radiation
gradients also occur and act horizontally, requiring 3D line synthesis
(see \citeads{2021A&A...647A..46S}). 
Proper quantification of such ultraviolet extinction control is
severely hampered by the dense ultraviolet line haze with its own NLTE
properties
\rrendnote{line haze modeling}{note:haze}{%
Photospheric species as \FeI\ and \FeII\,\footnote{The classic
Grotrian diagrams of
\citetads{1968pgda.book.....M} 
(\rrref{footnote}{note:Grotrian}) are only partial, with their
\linkadspage{1968pgda.book.....M}{66}{\FeI} and
\linkadspage{1968pgda.book.....M}{65}{\FeII} charts showing only major
systems. 
Complete ones would be black all over.} 
are immensely rich in bound-bound transitions with dense line crowding
into the ``veiling line haze'' of
\citetads{1972SoPh...22...64L} 
and \citetads{1980A&A....90..239G}, 
densest in the violet and ultraviolet (first extinction curve in
\linkrjrpage{2019arXiv190804624R}{13}{Fig.\,3} of
\citeads{2019SoPh..294..165R}). 
It is discussed as major obstacle in
\linkrjrpage{2019arXiv190804624R}{17}{Sect.\,6.1} of
\citetads{2019SoPh..294..165R} 
together with various recipes to treat it in spectral synthesis,
varying from the simple Zwaan-inspired
\linkrjrpage{1992AAp...265..237B}{3}{Fig.\,2 fudge} of
\citetads{1992A&A...265..237B} 
to brute-force solving all NLTE population equations for all
transitions in
\citetads{2015ApJ...809..157F}. 
Uitenbroek's RH code (\rrref{endnote}{note:RH}) offers both these and
also LTE and equivalent two-level scattering synthesis for all blend
lines not treated in NLTE detail. 
The latter two are compared with observations in
\linkrjrpage{2019arXiv190804624R}{18}{Fig.\,10} of
\citetads{2019SoPh..294..165R}, 
showing that both fail. 
The Avrett-Loeser Pandora code uses an imposed gradual transition from
$S\!_\nu^l\tis\,B_\nu$ to $S\!_\nu^l\tis\,J_\nu$ with SB opacities
(\rrref{endnote}{note:Pandora}). 
At the end of the section
(\linkrjrpage{2019arXiv190804624R}{19}{p.\,19}) I suggested a recipe
solving for NLTE populations in a representative simple atom and
applying the resulting departures to all blends.}.
Stronger lines need NLTE modeling not only for their extinction but
also for their source function set by their own scattering or by
non-2-level detour transitions. 
When ``inversion'' codes are proudly advertised as NLTE usually only
the modeled-line resonance scattering is meant, not the
extinction-affecting scattering in the bound-free ionizing continua nor
resonance scattering in the line haze affecting those and neglecting
LTE forcing by stronger interlocked ultraviolet lines
\rrendnote{\NiI\,6768, \MgI\,4571, \FeI\,6302, \FeI\,6173\,\AA}
{note:6302-4571}{%
Exemplary neutral-metal lines from the photosphere.
No NLT but all four suffer NLS and NLW in their opacities while they
vary in NLS and NLW in their source functions.

\NiI\,6767.8\,\AA\ is an intercombination line with small
$A_{ul}\tis\,3.3\,10^5$\,s$^{-1}$
(\linkpdfpage{https://robrutten.nl/rrweb/rjr-archive/linelists/1988JPCRD..17S....F.pdf}{382}{No.\,102}
in \citeads{1988JPCRD..17S....F})\footnote{Transition 
probabilities: NIST provides a
\href{https://physics.nist.gov/cgi-bin/ASBib1/TransProbBib.cgi}
{bibliography engine}. 
I supply pdf scans of some classical NBS compilations including this
one in my
\href{https://robrutten.nl/rrweb/rjr-archive/dircontent.html}{file
archive}.}
and small magnetic and temperature sensitivity. 
It lies in a clean spectral stretch where no other lines (Fraunhofer
or telluric) affect modulation encoding of Dopplershift in a Michelson
tachometer (\rrref{footnote}{note:beckers}) and was therefore selected
for helioseismology with the tachometers in the GONG stations and in
SOHO/MDI.
\citetads{1993A&A...269..509B} 
analyzed its formation in the VALIIIC star and a Nordlund-Stein
granulation simulation. 
His \linkadspage{1993A&A...269..509B}{2}{Grotrian diagram} shows that
the line connects singlet and triplet term systems rich in connected
and permitted optical and ultraviolet lines.
\linkadspage{1993A&A...269..509B}{4}{Figure\,4} shows its VALIIIC
formation in the three-panel format of
\linkrjrpage{1992AAp...265..237B}{8}{Figure\,6} of
\citetads{1992A&A...265..237B} 
(inspiring my \SSXpage{78\,ff} line-formation triptychs
described in \rrref{endnote}{note:ALC7plots}).
The top panel of his \linkadspage{1993A&A...269..509B}{4}{Figure\,4}
shows shared population departure dips from ultraviolet
overionization.
The line has strong resonance scattering (compare $\varepsilon_2$ for
its
\href{https:robrutten.nl/rrweb/rjr-pubstuff/lar-2/ssx-alc7-ni6768.pdf}
{ALC7 formation} with that of \NaIDone\ in \SSXpage{85}), but in the
center panel of Bruls' \linkadspage{1993A&A...269..509B}{4}{Figure\,4}
the chromospheric $S^l$ does not lie near $J$ nor between $B$ and $J$
evidencing dominance of the detour term in \rrref{eq.}{eq:S_CRD}.
Interlocking to stronger scattering lines results in yet steeper
outward $S^l$ decay that starts in the photosphere and higher up does
not care about $B$ or $J$ at the line itself. 
Interlocking of the lower level to close companions causes curve
bunch-up in the top panel and affects the line opacity. 
The line has low temperature sensitivity because $S^l$ is collectively
set by deeply escaping 3D-averaged radiation in the interlocking
lines.
$S^l$ lies just above $B$ in the upper photosphere but this apparent
near-LTE behavior is coincidental for this particular $B(h)$ decline. 
The later Avrett models have hotter upper photospheres with
$B\!>\!S^l$
(\href{https:robrutten.nl/rrweb/rjr-pubstuff/lar-2/ssx-alc7-falp-ni6768.pdf}
{comparison ALC7--FALP}).\footnote{Moral: source function LTE means
dominance of collisional processes delivering Boltzmann upper-lower
population ratio with corollary $S^l\tis\,B$ \RTSApage{45}{Eq.\,2.73},
but actual $S\tapprox\,B$ does not guarantee collisional control and
may be pseudo-LTE (nowadays fake LTE).
The $S\tapprox\,B$ around
\linkadspage{1976ApJS...30....1V}{43}{2070\,\AA} in
\citetads{1976ApJS...30....1V} 
is wrong LTE.}

\MgI~4571.1\,\AA\ is also an intersystem line
(\linkadspage{1968pgda.book.....M}{20}{Grotrian diagram}) with very small
$A_{ul}\tis\,4.3\,10^2$\,s$^{-1}$
(\linkpdfpage{https://robrutten.nl/rrweb/rjr-archive/linelists/1969atp..book.....W.pdf}{68}{No.\,1}
in \citeads{1969atp..book.....W}), 
but similar strength by being from the \MgI\ ground level.
It suffers similar ultraviolet overionization but less ultraviolet
interlocking and no resonance scattering.
Its ALC7 formation \SSXpage{82} shows $\varepsilon\tapprox\,1$,
$b_u\approx\,b_l$ and $S^l\tapprox\,B$ up to extraordinary height.
The same holds for FALC in \SSXpage{103} and hotter FALP in
\SSXpage{129}.
These robust equalities imply near-strict source function LTE.
This line so represents an extraordinary upper-photosphere
thermometer
(\citeads{1972SoPh...23...18W}, 
\citeads{2009ApJ...696.1892L}),\footnote{I 
used \MgI\,4571\,\AA\ in
\citetads{1973SoPh...28..347R} 
to measure pole-equator temperature difference in the then fierce
general relativity debate on the solar oblateness observed by
\citetads{1967PhRvL..18..313D} but I was not aware of the better
measurements in the \CaIIK\ wing (both $\alpha_\nu$ and $S\!_\nu$ in
LTE) by
\citetads{1972ApJ...171L..71A}. 
This debate likely killed Dicke's sharing in the 1978 half-Nobel for
the cosmic microwave radiation -- he explained to Penzias and Wilson
that the persistent noise they registered with their Dicke radiometer,
scooping his search for it, was Big Bang leftover and co-authored the
discovery.}
but it is less trivial to find where it samples temperature (same for
ALMA, \rrref{endnote}{note:ALMA-temp}).
``Inversion'' fitting delivers temperatures but for proper whereabouts
estimation it requires a pertinent Mg model atom (at the least
``quintessential'' one of
\citeads{2009ApJ...696.1892L}) 
plus the other electron donors and a line-haze recipe because its
opacity is controlled in the ultraviolet and much out of LTE as evident
from the population departure dips at left in \SSXpage{82}. 
In plane-parallel modeling of such $S^l\tapprox\,B$ lines the
chromospheric temperature rise maps into core reversal, more
evident near the limb, similar to the \MgI\,12\,$\mu$m lines of
\rrref{endnote}{note:MgI12micron}, and in the onset towards the flash
spectrum such lines should show the core in emission between
absorption dips.  
This pattern was not evident in my heavily smeared eclipse spectra in
\linkrjrpage{1977SoPh...51....3R}{21}{Fig.\,14} of
\citetads{1977SoPh...51....3R}, 
but later (visiting Tucson) I found it present on the
\href{https://robrutten.nl/rrweb/rjr-archive/observations/KPNO/limbspectrum_4571.png}{original
limb spectrum plate} of
\citetads{1968ApJS...17....1P}, 
now added to my
\href{https://robrutten.nl/rrweb/rjr-archive/dircontent.html}{file
archive}.
This beautiful telltale plate does show this quite unusual 
core-emission-plus-wing-dip pattern, with much spatial variation along
the tangential slit.
The nearby strong \FeII\ and \TiII\ lines show emission wings and the
weaker \FeII\ lines full emission, also with spatial variation,
whereas the \CeII, \LaII\ and \SmII\ rare earth lines show
Canfield-style blander emission (\rrref{endnote}{note:interlocking}).  
Similar limb emission differences occur for the blends in \CaII\ \HK\
(\citeads{1980A&AS...39..415R}). 

\FeI\ 6301.5\,\AA\ and its 
twin \FeI\ 6302.5\,\AA\,\footnote{Multiplet
816, $\chi_l\tis\,3.65$\,eV, for 
6301.5\,\AA\ $A_{ul}\tis\,3.2\,10^6$\,s$^{-1}$,
$g_{\rm eff}\tis\,1.667$, for 6302.5\,\AA\ 
$A_{ul}\tis\,1.8\,10^6$\,s$^{-1}$, $g_{\rm eff}\tis\,2.5$.}
constitute a famous polarimetry doublet (\eg\ Hinode/SOT-SP, Lites
\etal\
\citeyearads{2007PASJ...59S.571L}, 
\citeyearads{2008ApJ...672.1237L}, 
\citeads{2007ApJ...659.1726K}, 
\citeads{2010A&A...517A..37S}, 
\citeads{2013A&A...556A.113S}). 
In the ALC7 star \SSXpage{83} \FeI\ 6301.5\,\AA\ suffers opacity NLTE
from ultraviolet overionization and $S^l\!<\!B$ from combined
scattering and interlocking.
It forms similarly in the hotter FALP model \SSXpage{130}.

\FeI\,6173.3\,\AA\,\footnote{Multiplet 62, lower-level excitation energy
$\chi_l\tis\,2.2$\,eV, small $A_{ul}\tis\,2.3\,10^5$\,s$^{-1}$ 
(\linkpdfpage{https://robrutten.nl/rrweb/rjr-archive/linelists/1988JPCRD..17S....F.pdf}{33}{No.\,71}
in \citeads{1988JPCRD..17S....F}), 
Land\'e factor $g_{\rm eff}\tis\,2.5$
(\linkadspage{1973SoPh...28....9H}{4}{Table\,III} of
\citeads{1973SoPh...28....9H}), 
\linkadspage{1968pgda.book.....M}{66}{Grotrian diagram}.}
is used by SDO/HMI as both Doppler and Zeeman diagnostic.
Its
\href{https:robrutten.nl/rrweb/rjr-pubstuff/lar-2/ssx-alc7-fe6173.pdf}
{ALC7 formation} shows the usual NLTE extinction depletion in the ALC7
photosphere. 
Ultraviolet NLTE source-function interlocking is evident in the ALC7
chromosphere by $S^l$ lying well below $B$ and $J$.   
In the ALC7 photosphere this line scatters as much as \FeI\,6302\,\AA\
(similar $\varepsilon$ decay), but interlocking to stronger yet
thermalized companions enforces $b_u\tapprox\,b_l$ and hence
$S^l\tapprox\,B$. 
This is inherent LTE, not coincidental as for \NiI\,6768\,\AA, because
these near-equalities occur also in the hotter
\href{https:robrutten.nl/rrweb/rjr-pubstuff/lar-2/ssx-alc7-falp-fe6173.pdf}
{FALP photosphere}.

For these four example lines and most other neutral-metal lines the
large opacity deficits from ultraviolet over\-ionization imply that
NLTE ``inversion'' fitting should always use RH-like NLTE computation
(\rrref{endnote}{note:RH}) with a model atom for the element of
interest that is appropriate to quantify both scattering and
interlocking and also reaches high enough for ionization-recombination
sequences. 
In addition, the computation must have sizable electron-donor atoms
and a suited line-haze recipe to quantify $J_\nu$ across the element's
ionization edges.

When instead LTE inversion is tried the predicted line from a given
temperature stratification gets too deep from incorrect ionization
estimation whereas actual $S^l\!<\!B$ from scattering and/or
interlocking makes it too shallow -- double-error mutual compensation
in Holweger-like ``inversion'' fitting as Holweger's NLTE masking
(\rrref{endnote}{note:masking}).

Be aware that these didactic demonstrations are all static 1D
plane-parallel.  
The real photosphere is not -- it is dynamically 3D fine-structured
with steep small-scale sideways gradients causing similar irradiation
$J\!\neq\!B$ opacity and source function departures from LTE values.}. 

Holweger's assumptions of 1D modeling, LTE and no chromosphere were
also revived in SATIRE irradiance modeling for network and plage
(\citeads{1999A&A...345..635U}) 
which therefore suffers similarly from mistreating the
opacity-affecting NLTE ultraviolet continua and their NLTE line haze. 
More explanation in \linkrjrpage{2019arXiv190804624R}{11}{Sect.\,4} of
\citetads{2019SoPh..294..165R}. 

The Asplund revolution 
(\citeads{2009ARA&A..47..481A})\footnote{At
\href{https://ui.adsabs.harvard.edu/abs/2009ARA\&A..47..481A/metrics}
{over 5000 citations} one of the most cited publications of solar physics
but still less than the 8000$^+$ for its pre-revolution predecessor 
\citetads{1989GeCoA..53..197A}. 
If you desire citations you must cater to non-solar colleagues. 
My most-cited article (\citeads{1994A&A...288..860C}) 
was non-solar: stellar lithium abundance determiners outnumber solar
spectrum analysts. 
It was also the most boring, tabling minor NLTE corrections eagerly
cited as negligible.} ended the HOLMUL popularity, primarily by
including simulated granulation
(\citeads{2009LRSP....6....2N} 
with the foundation laid in
\citeads{1984ssdp.conf..181N}) 
\rrendnote{illustrious quartet}{note:grancontrast}{%
\AA.\,Nordlund initially had a habit of publishing his revolutionary
granulation simulations in non-page-limited proceedings (\eg\
non-scanned Nordlund
\citeyearads{1985MPARp.212..101N}, 
\citeyearads{1985MPARp.212....1N}). 
In scanned ADS-available \citetads{1984ssdp.conf..174N} 
he argued that the large difference in granulation contrast between
his numerical simulation and observations was to be blamed on the
latter, becoming the prototype of the numerical simulator habit to
blame observations if these do not match the wonderful and necessarily
correct computation\footnote{Generally it is better that
simulations do {\em not} reproduce observations so that there is evidently
something to learn without temptation to claim correctness from
happenstance data reproduction (\rrref{endnote}{note:Ha-RE}).} 
-- but actually Nordlund was right as eventually demonstrated by
\citetads{2019A&A...626A..55S}. 

Nordlund stems from B.\,Gustafsson at Uppsala who also suggested RT
theory as venue to G.B.\,Scharmer at Stockholm who then became thesis
adviser to Gustafsson's student M.\,Carlsson
(\rrref{endnote}{note:MULTI}). 
Nordlund and Carlsson each began fruitful collaborations with
R.F.\,Stein at East Lansing who has Scandinavian roots. 
This illustrious Nordic quartet eventually co-authored
\citetads{2004ApJ...610L.137C}. 
M.\,Asplund is also a Gustafsson pupil.\footnote{Gustafsson's
astronomer production reminds me of C.\,Zwaan at Utrecht (PhD adviser
for me, H.C.\,Spruit, A.\,Greve, A.A.\,van Ballegooijen,
C.J.\,Schrijver, R.G.M.\,Rutten, K.L.\,Harvey-Angle, L.H.\,Strous,
K.F.\,Tapping and others who left astronomy; PhD co-adviser for
H.\,Uitenbroek, J.H.M.J.\,Bruls, N.M.H.\,Hoekzema). 
Utrecht University has a
\href{https://robrutten.nl/Utrecht_solar.html}{great past} but
\href{https://robrutten.nl/Closure_Utrecht.html}{no present} in solar
physics. 
So it goes.}}.  
A major aspect is that its temperature inhomogeneity upsets the
notion that spatially-averaged intensity may be modeled as
spatially-averaged temperature because at optical and shorter
wavelengths this notion is undone by the Wien nonlinearity of the
Planck function
(\citeads{2011ApJ...736...69U}). 
The simulated granulation undid the need for turbulence fudging and
better atomic physics
(\citeads{2000yCat..41420467B}) 
undid the need for damping fudging.

Another Asplund ingredient was NLTE spectrum synthesis but yet without
detailed spectral synthesis of ultraviolet overionization weakening
atomic minority-species lines as reviewed for \FeI\ by
\citetads{1988ASSL..138..185R} 
and quantified from granular/intergranular 1D models by
\citetads{2001ApJ...550..970S}. 
Such synthesis should be 3D to include also the steep horizontal
temperature gradients (\linkadspage{2001ApJ...550..970S}{6}{Fig.\,4})
by which hotter granules irradiate cooler intergranular lanes in the
ultraviolet where the Balmer continuum escapes. 
This irradiation also causes line weakening and therefore abundance
underestimation when neglected; correction may give a shift back
towards the classic ``high'' values
(\citeads{1998SSRv...85..161G}, 
non-Springerwalled
\citeads{1996ASPC...99..117G}) 
desired by helioseismologists (review in
\linkadspage{2021LRSP...18....2C}{118}{Sect.\,6} of arXived
\citeads{2021LRSP...18....2C}). 

\paragraph*{Photosphere\,--\,chromosphere separation point.}~~\label{sec:SP2} 
If your interest is purely photospheric you may stop reading here --
missing the fun.
Endnotes \ref{note:unsold}, \ref{note:REphot}, \ref{note:p-modes},
\ref{note:1600-1700}, \ref{note:CO}, \ref{note:basal} may cater to
you.

\section{NLS:~ Auer \& Mihalas modeling} \label{sec:AuerMihalas}
NLS is the realm of 2-level modeling treated extensively in
\RTSApage{95}{Chapt.\,4}, numerical developments around the \lambdop\
operator summarized in \RTSApage{133}{Chapt.\,5}, and numerical PRD
formalisms ``not yet'' in RTSA. 
\SSFpage{111} shows a cartoon summary also shown as
\linkrjrpage{2019arXiv190804624R}{10}{Fig.\,4} of
\citetads{2019SoPh..294..165R} 
and discussed in \linkrjrpage{2019arXiv190804624R}{11}{Sect.\,4}
there.
Non-local scattering makes $J$ depart from $B$ in setting $S$
according to \rrref{Eq.}{eq:S_CS} and \rrref{Eq.}{eq:S_CRD}, for
bound-free continua and for lines of increasing strength (or different
parts of strong PRD lines). 
The ultraviolet continua gain $S \tapprox\ J$ excess over the
photospheric $B$ decline per \lambdop\ \SSFpage{86}. 
Strong lines obtain low $S$ per \lambdop\ from $\sqrt{\varepsilon}$
scattering by seeing actual temperature gradients as near-isothermal
from $\tau$ scale compression (\rrref{endnote}{note:flatSl}).

The classic warm-star \HI\ analyses of Auer and Mihalas
(\citeyearads{1969ApJ...156..157A}, 
\citeyearads{1969ApJ...156..681A}) 
advertised here were an early pinnacle of 2-level modeling 
\rrendnote{LTE diehard}{note:unsold}{%
I wonder what A.\,Uns\"old, who at Kiel had told Holweger to produce a
thesis proving that LTE suits the solar spectrum, made of the
Auer\,--\,Mihalas results, representing extremes of the then rampant
LTE-versus-NLTE debates. 
In 1967 he had written to Minnaert that Holweger had definitely proven
the LTE veracity of the solar spectrum (in a recommendation for his
attending the Bilderberg meeting of \rrref{endnote}{note:Bilderberg})
but the Auer--Mihalas results did upset his LTE bible {\em ``Physik
der Sternatmosph\"aren''\/}
(\citeads{1955QB461.U55......}, 
\href{https://ui.adsabs.harvard.edu/abs/1955psmb.book.....U/metrics}
{900$^+$ citations}) of which the Utrecht Sonnenborgh library had many
copies for students (some stolen). 
The translation by H.\,Panofsky and A.K.\,Pierce was never printed; I
believe that Uns\"old insisted that any astrophysicist instead could
and should learn German to read his book. 
I have it (honestly from eBay) and can read it but haven't opened it
in years. 
So it goes. 
I know that ``diehard'' is Anglo-Dutch, not German, but the word fits
him well. 
He is at left in
\href{https://robrutten.nl/nieuwenhuijzenshots/sym1963/1963-sym-18-16_Albrecht-Unsoeld_Andrei-Severny.jpg}{this
photo}.}.
Not purely 2-level since adding the \HI\ continuum level and so
including the Lyman and Balmer continua to \Lyalpha\ in the first
publication and to \Halpha\ in the second. 
The key results were atmospheric temperature stratifications copied in
\citetads{1970stat.book.....M} 
and in \RTSApage{187}{Fig.\,7.10}. 
A large didactic boon is that the effects of these transitions are
well separated in height and therefore distinguishable. 
``Big whopper'' \RTSApage{254}{exercise12} poses five pages of hard
questions on their graphs that I challenge you to
answer.\footnote{{\em [Spoiler\/}:
\href{https://robrutten.nl/rrweb/rjr-siucourses/afyc_anwers_2003.ps.gz}{my
answers} to most RTSA problems including this one.]} 
The hardest is to understand: {\em ``Amazingly, the photon losses in
the subordinate Balmer-$\alpha$ line, located in the low-energy red
part of the spectrum, cause heating of the whole outer atmosphere of
this hot star.''\/} \RTSApage{189}{}. 
Because their graphs did not include $J$\footnote{Any NLTE study
should always include $J$ in figures showing $S$ and $B$, \cf\ 
\SSXpage{80}.}
 we did a multi-level re-do
including Avrett-style \linkrjrpage{2003ASPC..288..130W}{5}{$S, B, J$
graphs} and \linkrjrpage{2003ASPC..288..130W}{6}{radiative
heating/cooling graphs} in
\citetads{2003ASPC..288..130W}, 
giving as answer: {\em ``The photon losses in \Baalpha\ suck
population from the proton reservoir through a collisionally-dominated
Rydberg recombination flow and so boost the outward temperature
rise''\/} \linkrjrpage{2003ASPC..288..130W}{7}{}. 
All yours to understand.

\section{NLS+NLW:~ Avrett models} \label{sec:Avrett}
The various solar-atmosphere models masterminded by E.H.\,Avrett
(called ``Pandora stars'' below 
\rrendnote{Pandora program}{note:Pandora}{%
NLTE spectral synthesis code of
\citetads{1992ASPC...26..489A}; 
more
\href{https://www.cfa.harvard.edu/~avrett/pandora/modeling-with-pandora.pdf}
{here} and \href{https://lweb.cfa.harvard.edu/~avrett/pandora}{here}
and
\href{https://lweb.cfa.harvard.edu/~avrett/pandora/pandora-f-list.txt}{here}.
Similarly to Holweger's method it applies hydrostatic plane-parallel
equilibrium to a trial manually-fitted $T(h)$ temperature
stratification but uses SE NLTE for radiation evaluation in a giant
``equivalent two-level'' iteration loop treating each bound-bound or
bound-free transition that is explicitly taken into account (not added
as background opacity) with two-level rate equations, letting
iteration take care of multi-level crosstalk. 
The equation system is large enough to earn the
\href{https://lweb.cfa.harvard.edu/~avrett/pandora/pandora-f-list.txt}
{ominous name}.

Whereas Holweger concentrated on fitting observed optical lines,
mostly from \FeI, Avrett concentrated on fitting observed disk-center
continua, especially in the ultraviolet. 
For the line haze (\rrref{endnote}{note:haze}) Avrett added increasing
numbers of lines from the ever-growing list of
\citetads{2009AIPC.1171...43K} 
and treated them with an imposed ad-hoc gradual source function
transition from $S\tis\,B$ in the model photosphere to the continuum
$S\tis\,J$ in the model ``chromosphere'' to avoid non-observed core
reversals that otherwise show up above the temperature minimum in LTE
sampling (\linkadspage{2008ApJS..175..229A}{15}{p\,243\,ff} of
\citeads{2008ApJS..175..229A}). 
Seems reasonable, but using LTE opacities is wrong for over-ionized
minority species. 
The brute force solution is to include all lines with full NLTE
synthesis as attempted by
\citetads{2015ApJ...809..157F} 
in irradiance modeling.

I use the ALC7 Pandora star of
\citetads{2008ApJS..175..229A} 
for my spectrum-formation demonstrations \SSXpage{72\,ff}. 
It isn't the Sun but it is well-suited to showcase its RTSA physics
and it is self-consistent within this physics.\footnote{J.\,Fontenla
diverged from Avrett after
\citetads{1993ApJ...406..319F}, 
writing a Pandora alternative that is in principle equivalent but was
used less self-consistently by enhancing the output similarity to the
non-1D non-static non-SE spectra of the actual Sun with ad-hoc tricks.
The first was to add best-fit pressure in the upper ``chromosphere''
beyond the turbulent pressure defined by
\linkadspage{1981ApJS...45..635V}{19}{observed microturbulence} in the
Pandora stars by invoking the Farley-Buneman instability
(\citeads{2005A&A...442.1099F}, 
\citeads{2008A&A...480..839F}, 
\citeads{2014ApJ...783..128M}, 
\citeads{2018ApJ...857..129F}) 
-- but see \citetads{2009ApJ...706L..12G}. 
The second (\citeads{2007ApJ...667.1243F}) 
was to add best-fit non-gravitational acceleration extending the model
photosphere to reproduce dark mean-spectrum CO lines that are in
reality not formed in plane-parallel SE (\rrref{endnote}{note:CO}).}

Pandora is
\href{https://www.cfa.harvard.edu/~avrett/pandora}{web-available} but
documented only in many pages of coding requests and specifications %
(including Fortran variable names) to R.\,Loeser that reside in
Avrett's CfA office. 
P.\,Heinzel holds copies in Ondrejov.})  
are all classic pinnacles of plane-parallel (1D) NLTE modeling: VALIIM
of \citetads{1976ApJS...30....1V}, 
the VALIII models, in particular quiet VALIIIC, of
\citetads{1981ApJS...45..635V}, 
MACKKL of \citetads{1986ApJ...306..284M}, 
the FAL models, in particular quiet FALC, 
of \citetads{1993ApJ...406..319F} 
and quiet ALC7 of \citetads{2008ApJS..175..229A} 
with an update in
\citetads{2015ApJ...811...87A}.\footnote{VALIIIC 
now has
\href{https://ui.adsabs.harvard.edu/abs/1981ApJS...45..635V/metrics}
{2100$^+$ citations} at increasing rate. 
Practical
\linkpdfpage{https://robrutten.nl/rrweb/rjr-edu/exercises/ssb/ssb.pdf}{5}{SSB\,1}
dissects the FALC stratifications with comparison to our own
atmosphere which is transparent to sunlight by missing \Hmin\
treated in the continuum part of
\href{https://robrutten.nl/rrweb/rjr-edu/lectures/rutten_ssi_lec.pdf}{SSI}
and in practical
\linkpdfpage{https://robrutten.nl/rrweb/rjr-edu/exercises/ssb/ssb.pdf}{13}{SSB\,2}.
Practical
\linkpdfpage{https://robrutten.nl/rrweb/rjr-edu/exercises/ssb/ssb.pdf}{22}{SSB\,3}
makes you synthesize the \NaID\ lines from FALC assuming LTE and hence
misrepresenting their cores.} 
Their strength is comprehensive inclusion of many ultraviolet transitions. 
A weakness is ignoring Wien nonlinearity in assigning observed
intensities to averaged temperatures of inhomogeneities, the same
mistake as in HOLMUL abundance determination rectified in the
hydrodynamic Asplund revolution.

NLW was still lacking in VALIIM which only accounted for scattering in
the \SiI\ continuum so that its
\linkadspage{1976ApJS...30....1V}{38}{Fig.\,23} multi-page analogon to
multi-page VALIII Fig.\,36 
shows erroneous $S\!_\nu\tapprox\,B_\nu$ in the ultraviolet above the
\SiI\ threshold at 1682\,\AA\ where other electron-donor bound-free
contributions dominate \SSXpage{68}. 
This was remedied in VALIIIC but overestimating their effect because
not enough line-haze blending lines were included, resulting in a too
steep upper-photosphere temperature decline. 
Including more and more lines from the growing Kurucz tabulations
(\eg\ \citeads{2009AIPC.1171...43K}) 
then brought the upper-photosphere back up, announced in
\citetads{1984BAAS...16..450A}, 
shown as model C$^\prime$ in
\linkadspage{1985cdm..proc...67A}{25}{Fig.\,17} of
\citetads{1985cdm..proc...67A}, 
and tabulated in MACKKL; since then it has remained the same. 
It is remarkably similar to HOLMUL and even to Kurucz or Uppsala
LTE--RE models \RTSApage{169}{Fig.\,7.3}
\rrendnote{RE upper photosphere}{note:REphot}{%
The reason for the near-equality of the upper photosphere in all
quiet-Sun 1D models since MACKKL to the older HOLMUL and RE models is
that this is the most homogeneous domain of all. 
The granular convection has stopped. 
Acoustic waves do not yet shock. 
Magnetic fields are still mostly confined to slender fluxtubes at
small density in the modeled ``quiet'' areas.
Gravity waves do not yet couple into them
(\rrref{endnote}{note:waves}).

In addition, the bulk of the escaping solar radiation is in the
optical and forces radiative equilibrium (RE). 
In ``grey'' RE \RTSApage{173}{Sect.\,7.3.2} the continuum gets
gradient $S(\tau)\tsim\,(1+1.5\,\tau)$ \RTSApage{176}{Eq.\,7.43}
because for that gradient $\lambdop(S)\tapprox\,S$ \SSFpage{45}. 
The continuum in the optical is dominated by \Hmin\ which has near-LTE
$S\!_\nu\tapprox\,B_\nu$ in the bound-free range and strict LTE
$S\!_\nu\tis\,B_\nu$ in the free-free range. 
Most lines in the optical also have $S\!_\nu\tapprox\,B_\nu$
(\rrref{endnote}{note:interlocking}). 
Hence the temperature gradient is set LTE-wise by the RE condition
\RTSApage{174}{Eq.\,7.33}.

Indeed, more modern Bifrost (\rrref{endnote}{note:Bifrost})
simulations obtain about the same average temperature decay with
relatively small spread at these heights \SSXpage{59}.  Profile
fitting with 1D models is least suspect for lines formed there, as the
\MgI\ 12\,micron lines of \rrref{endnote}{note:MgI12micron}.}.  

I love the Avrett models, in particular VALIIIC because of the many
diagrams in \citetads{1981ApJS...45..635V} and especially its
many-page 
\linkadspage{1981ApJS...45..635V}{33\,ff}{Fig.\,36} 
continuum formation diagrams (some copied in
\RTSApage{204}{}\,--\,\RTSApage{206}{} and in
\SSXpage{61}\,--\,\SSXpage{67}). 
Since VALIIIC obeys all the standard RT equations, meaning all those
in RTSA, these graphs describing spectral continua arising from the
VALIIIC density and temperature stratifications provide valuable
insights on spectrum formation and its intricacies. 
They are of enormous didactic value because {\em everything\/} in
Avrett's output graphs is fully understandable from the input physics
$\approx$ RTSA physics
\rrendnote{Fe ionization in VALIIIC}{note:VAL-FeII}{%
RTSA teacher J.\,T.\,Wright (Penn State) has pointed out that in VALII
\linkadspage{1976ApJS...30....1V}{37}{Fig.\,22} the electron donor
ranking is Mg--Si--Fe but VALIII
\linkadspage{1981ApJS...45..635V}{67}{Fig.\,47} has Fe on top -- and
reported that his students could not reproduce the latter.
I checked with the RH code (\rrref{endnote}{note:RH}) and also got Mg
on top, just as for MACKKL and newer Pandora stars.

What boosted \FeII\ in the VALIII publication? 
The steep VALIIIC upper-photosphere temperature decline desteepened in
subsequent Pandora stars does not define donor ionizations directly
because these follow scattering source functions given by
\RTSApage{93}{Eq.\,3.109}.
As in \SSXpage{42} they are primarily controlled by the temperature at
their deeper effective escape where all Pandora stars are about the
same \SSXpage{58}. 
The less steep upper-photosphere decline of the newer Pandora stars
boosts these continua only indirectly with larger (but there minor)
$\varepsilon\overline{B}$ contribution to $\overline{S}$ and via
\lambdop\ into slightly higher $\overline{J}$, giving somewhat larger
ionization to all donors.
The corresponding NLTE corrections are largest for VALIIIC
(\linkadspage{1981ApJS...45..635V}{32}{Fig.\,34}), smaller for the
newer Pandora stars because their $B_\nu$ stays closer to $J_\nu$
(compare \FeI\,6301.5\,\AA\ formation between FALC and the warmer FALP
in \SSXpage{117}) -- but all rank Mg on top in my RH modeling.
I then suspected that this rogue non-RH-reproduced \FeII\ boosting in
VALIIIC came from details of line haze inclusion but trials putting
the entire Kurucz line haze to absent, LTE or two-level scattering in
RH (for comparisons see
\linkrjrpage{2019arXiv190804624R}{18}{Fig.\,10} and its discussion in
\citeads{2019SoPh..294..165R}) 
all gave Mg on top.

The actual Sun has Mg on top, showcased by the observed dominance of
the \MgI\ lines around 12\,$\mu$m (\rrref{endnote}{note:MgI12micron}).
The companion \FeI\ Rydberg lines did not steal the similar \MgI\
7.4\,$\mu$m trough-and-peak show in
\linkadspage{1995A&A...301..593S}{5}{Fig.\,1} of
\citetads{1995A&A...301..593S} 
modeled so very well in \linkrjrpage{1994IAUS..154..309R}{9}{Fig.\,6}
of \citetads{1994IAUS..154..309R} 
with a MACKKL-like RE photosphere
(\linkrjrpage{1992AAp...253..567C}{6}{Fig.\,2} of
\citeads{1992A&A...253..567C}). 
Because these fits of upper-photospheric lines represent the best of
all solar spectral-line syntheses with 1D standard models
(\rrref{endnote}{note:REphot}) I also tried RH
on this chromosphere-free 1D star and again got Mg on top as champion
electron donor reaching 35\% from 200\,km up to the atmosphere top.}.
The same holds for the newer Avrett models and his latest ALC7. 
Therefore I made many ALC7 demonstration figures: \SSXpage{72} showing
ALC7 stratifications, \SSXpage{75} ultraviolet continua, \SSXpage{76}
hydrogen line formation, \SSXpage{77} overview of strong-line
formation, then \SSXpage{78\,ff} explaining triptych formats
\rrendnote{triptychs ALC7 line formation}{note:ALC7plots}{%
The displays in \SSXpage{78\,ff}
show a hundred triptychs with populations, source functions, emergent
profiles for representative lines of the ALC7 star and comparable
plane-parallel stars using intercomparable graph formats and scales.

The first graph \SSXpage{79} shows lower and upper level population
departure coefficients $b_l$ and $b_u$ \SSFpage{73}. 
The other curves are NLTE and LTE fractional populations. 
Their divergence follows the $\log(b_l)$ departure from zero
(\rrref{Eq.}{eq:b}).

The second graph \SSXpage{80} shows $B$, $J$ and $S^l$ (these must
always be shown all three together). 
They are on equivalent temperature scales, deWiened to enable
comparisons between different lines across the spectrum
(\rrref{endnote}{note:Tb}). 
On logarithmic intensity scales the $S^l$ separations from $B$ would
equal the logarithmic separations of $b_u$ from $b_l$ in the first
graph. 
The thick solid curve usually lies between the other two, then
identifying it as
$S^l\tapprox\, (1-\varepsilon_2)\,J+\varepsilon_2\,B$.
For PRD lines it is split between monofrequent samplings. 
The $\log(\tau)$ marks are defined by the lower-level fractional
population, abundance and density drop. 
The other two curves measure scattering governing $S^l-B$ splits.

The third graph \SSXpage{81} is the computed emergent profile on the
same equivalent-temperature scale as the second graph to enable
checking the Eddington-Barbier approximation.  I did not add observed
disk-center atlas profiles in fear of the Ha--Ha Erlebnis in
\rrref{endnote}{note:Ha-RE}: you might be tempted to overclaim a good
fit as observation-proven model truth.}  
followed by such plots for thirteen ALC7 lines in \SSXpage{82\,ff}. 
They were all made
\rrendnote{my IDL programs}{note:IDL}{%
Under my \href{https://robrutten.nl/Recipes_IDL.html}{Recipes for IDL}
with this
\href{https://robrutten.nl/rridl/dircontent.html}{inventory}. 
RH plot programs in
\href{https://robrutten.nl/rridl/rhlib/dircontent.html}{rhlib}, LTE
line formation programs in
\href{https://robrutten.nl/rridl/ltelib/dircontent.html}{ltelib}, SDO
stuff in
\href{https://robrutten.nl/rridl/sdolib/dircontent.html}{sdolib}, SST
data-handling programs in
\href{https://robrutten.nl/rridl/sstlib/dircontent.html}{sstlib},
image sequence browser
\href{https://robrutten.nl/rridl/imagelib/showex.pro}{\tt showex} in
\href{https://robrutten.nl/rridl/imagelib/dircontent.html}{imagelib},
etc. 
Installation is described in
\href{https://robrutten.nl/rridl/00-README/sdo-manual.html}{this SDO
manual}. 
I even offer a
\href{https://robrutten.nl/rrweb/rjr-edu/manuals/idl-simple-manual.html}{beginner
IDL manual}. 
For my SDO pipeline (\rrref{endnote}{note:SDO-STX}) do my
\href{https://robrutten.nl/rrweb/sdo-demo/instruction.html}{alignment
practical}.}
with the 1D version of github-public
\href{https://github.com/han-uitenbroek/RH}{RH} 
\rrendnote{RH program}{note:RH}{%
The RH spectrum synthesis code of
\citetads{2001ApJ...557..389U} 
is named after \citetads{1992A&A...262..209R} 
and follows their scheme of multi-level approximate lambda iteration
\RTSApage{145}{Sect.\,5.3.2} not iterating \lambdop\ for $J_\nu$ but
the $\Psi$ operator for $j_\nu$. 
It permits overlapping lines, includes PRD and full-Stokes options,
exists in 1D, 2D, 3D, spherical, and Cartesian versions, and also in
parallel multi-column ``1.5D''
(\citeads{2015A&A...574A...3P}). 
It has various options to treat the worrisome ultraviolet line haze
(\rrref{endnote}{note:haze}).

I supply some plane-parallel 1D ``standard'' model atmospheres in RH
format beyond the
\href{https://github.com/han-uitenbroek/RH/tree/master/Atmos}
{github-supplied} FAL models of
\citetads{1993ApJ...406..319F} 
in my \href{https://robrutten.nl/rrweb/rjr-archive/dircontent.html}
{solar file archive} (\rrref{endnote}{note:ppmodels}).
My RH-output plotting programs are in my 
\href{https://robrutten.nl/rridl/rhlib/dircontent.html}{RH IDL lib}.

For the ALC7 spectrum-formation graphs starting at \SSXpage{78} I used
RH 1D version-2 with H, He, Si, Al, Mg, Fe, Ca, Na, and Ba active, C,
N, O, S, and Ni passive, and with 20\,m\AA\ sampling of 343\,000 lines
between 1000 and 8000\,\AA\ in the atomic and molecular line list of
\citetads{2009AIPC.1171...43K}. 
\linkrjrpage{2019arXiv190804624R}{13}{Fig.\,6} of
\citetads{2019SoPh..294..165R} 
gives an overview of their extinction in the FALC star. 
The line haze is prominent between $\log \lambda \tis\,3$ and 4.}.  
They represent a detailed view of how all these lines form in the ALC7
atmosphere. 
In addition \SSXpage{98\,ff} show ALC7 formation comparisons between
comparable pairs of lines and \SSXpage{102\,ff} compare the formation
of twelve lines between three Avrett models.\footnote{ALC7\,--\,FALC in
\SSXpage{102\,ff}, FALC\,--\,FALP in \SSXpage{115\,ff}, ALC7\,--\,FALP
in \SSXpage{128\,ff}, each with twelve line buttons at bottom. 
The display titles may return to the previously shown display. 
The {\em start\/} button at bottom-left should return to the SSX
contents overview.} 
Any student of thick solar spectrum formation should be able to
understand and appreciate all these curves in all their gory detail. 
I invite you to study and explain:
\begin{itemize} \vspace{-0.8ex} \itemsep=0.3ex
\item $N_\rme$ stratification;
\item Eddington-Barbier validity in all lines;
\item $\varepsilon$ decline in scattering lines;
\item PRD-split source functions of \BaII~4554, \CaIIK, \MgIIk,
\Lyalpha;
\item thermalization depth for all lines, in particular \Halpha;
\item NLTE extinction dip of \MgI~4571, \FeI~6301.5, \MgIbtwo\ but not
for \NaDone, \HeI~10830; 
;RR GONG and HMIlines not in SSX-2020
\item LTE extinction of \MgIIk, \Halpha, \Hbeta, \HeI~584;
\item slight rise in NLTE over-extinction for \Lyalpha;
\item line formation differences between different Pandora stars.
\end{itemize} However -- the stern caveat is that I love the Avrett
models only for representing educational solar-analog ``Pandora stars''
\SSXpage{58}. 
These computationally existing stars extend infinitely plane-parallel
(1D) without any inhomogeneities or any magnetism or any sort of waves
or any type of reconnection -- no granules, spicules, floccules, let
be filaments, flares, CMEs, and whatever else that makes our kind Sun
non-plane-parallel interesting. 
The Pandora stars are a marvelous boon to a RT teacher like me but they
also had and have detrimental effects on far too many colleagues
misled into accepting their stratifications as a valid average over
small fluctuations around realistic means. 
The first figure of \citetads{1981ApJS...45..635V} 
in \SSXpage{60} 
is a double champion of solar physics: the most-shown figure but also
the most misinterpreted. 
It misled many solar physicists into thinking it describes the Sun
instead of an unrealistic plane-parallel-layer star. 
{\em The solar atmosphere is not stably layered but dynamically
structured.}

Take \Halpha\ as example. 
In his VALIIIC diagram \SSXpage{60} Avrett, admirably careful, drew
its core coming from the VALIIIC ``chromosphere'', its wing from the low
VALIIIC photosphere with an extended gap in between that is not
present in his formation spans of \CaIIK\ and \MgIIk. 
He did so because at low temperature the Boltzmann excitation and
therefore the extinction of \Halpha\ are negligible.
Correct -- but the actual \Halpha\ core in actual quiet Sun mostly
samples opaque fibril canopies\footnote{Naming: ``canopies''
spreading around magnetic features over field-free surroundings were
introduced as magnetic ones from multi-line magnetometry including
\CaIR\ and \Halpha\ by
\citetads{1982SoPh...79..267G}. 
They warned that \Halpha\ fibril contrasts likely mark opacity
or source function variations rather than lateral field variations.  
A good rendering is in the simulation-assisted inversion map in
\linkrjrpage{2017ApJS..229...11J}{9}{Fig.\,7} of
\citetads{2017ApJS..229...11J}.} 
reaching much higher
\rrendnote{quiet-Sun \Halpha\ scenes}{note:quietHa}{%
Note the deep population chasm in the dashed curve in the first panel
of \SSXpage{91}, first described by
\citetads{1972SoPh...22..344S} 
and compared with \CaIIH\ in \SSFpage{61} from
\citetads{2006A&A...449.1209L}. 
Observationally the gap is evident when you shift the SST/CRISP
\Halpha\ passband from line center to a wing for a quiet-Sun target.
Up to about $\Delta \lambda \tis\,\pm0.5$\,\AA\ the scene usually
shows fibrils extending from network and then you suddenly drop into
deep-photosphere granulation, skipping the middle and upper
photosphere. 
This opacity skip explains why the equivalent widths of the Balmer
lines are so much smaller than for \CaII\ \HK\ with their
wide-stretching outer wings and clapotispheric cell grain shock
signatures
(\rrref{endnotes}{note:gillespie}\rrref{\,--}{note:clapotis}) in their
inner wings.
\Halpha\ has $\sqrt{m_{\rm Ca}/m_\rmH} \tis\,6.3$ wider thermal core
broadening and wider Holtsmark-distribution damping-wing broadening
but it lacks opacity underneath its fibrilar NLT canopies.
These are more opaque in \Halpha\ than in \CaIR\
(\rrref{endnote}{note:nonEchrom}) but likely less than in the centers
of \CaII\ \HK\ with the latter showing richer fine structure thanks to
larger Dopplershift sensitivity from PRD source function splitting and
less smoothing from opacity aureoles (\rrref{endnote}{note:aureoles}).

Only in rare utterly quiet areas you may see somewhat higher-up
reversed granulation in \Halpha\ near
$\Delta \lambda \tis\,\pm0.5$\,\AA.
I offer two SST examples. 
The first is a very quiet target corner in the data of
\citetads{2007ApJ...660L.169R} 
in this
\href{https://robrutten.nl/rrweb/rjr-movies/2006-06-18-quiet-ca-hawr.avi}{\CaIIH\,+\,\Halpha\
blue-wing movie} where reversed \Halpha\ at
$\Delta \lambda \tis\,-0.45$\,\AA\ shows reversed granulation as grey
pancakes underlying narrow fast-changing filamentary structures that
represent fine-structured ``mushroom'' aftermaths from the
internetwork shocks showing as \CaII\ \HtwoV\ grains in the parallel
movie at left.

The second is a similarly quiet SST target (in a disk-center coronal
hole) that you may inspect yourself by doing my
\href{https://robrutten.nl/rrweb/sdo-demo/instruction.html}{alignment
practical} (\rrref{endnote}{note:SDO-STX}). 
Its {\tt showex} commands (\rrref{footnote}{note:showex}) for its SST
data enable \Halpha\ profile scanning to inspect scene changes,
\Halpha\ wing blinking against reversed granulation in the \CaIR\
wings to identify it in \Halpha, and core comparisons to show the
larger opacity and extent of the fibrilar canopies in \Halpha\ than in
\CaIR\ (\rrref{endnote}{note:nonEchrom}).}.

Worse, Pandora-star ``chromospheres'' do not represent average
canopy-fibril temperatures but primarily ultraviolet radiation
temperatures reached in acoustic shocks in internetwork regions
underneath chromospheric canopies. 
The shocks result from upward propagating waves excited by the 3-min
components of the photospheric $p$-mode interference pattern
\rrendnote{seismology success story}{note:p-modes}{%
I summarize this history here because the acoustic $p$-mode
oscillations affect solar lines throughout the visible and ultraviolet
everywhere while their identification and usage were the most exciting
development in my half-century in solar physics (more in my
decades-old
\linkpdfpage{https://robrutten.nl/rrweb/rjr-edu/lectures/rutten\_5min\_obs\_lec.pdf}{1}
{5-minute lecture}).
In the daily HMI Dopplergram movies the 5-minute oscillation
discovered by
\citetads{1962ApJ...135..474L}\footnote{Best-ever 
\label{note:Leightoncs}
publication in observational solar physics (\cf\
\citeads{1999ApJ...525C.962G}): 
discovery of supergranulation, reversed granulation, 5-minute
photospheric oscillation, \Halpha\ return fibrils around network, and
more.
Not at all ``preliminary'' as the title claimed for this Part~I, and
all based on ingenious photographic spectroheliogram processing as
analog precursor to IDL permitting Dopplergram subtraction, temporal
differencing, space-time correlation and more with detector size well
beyond the 4K$\times$4K sensors of SDO. 
Fortuitously the resolution was just right for the $p$-mode
interference blobs and the heliograph scan duration over the image of
4 min was just right for two-wing Doppler mode scanning
(\linkadspage{1962ApJ...135..474L}{2}{Fig.\,1}) with quick
scan-direction reversal to produce time-delay correlation plates
permitting the discovery of pattern reproduction after 5 minutes
(\linkadspage{1962ApJ...135..474L}{22}{Fig.\,14}) and also the
everywhere dominance of this periodic motion field
(\linkadspage{1962ApJ...135..474L}{25}{Fig.\,16}).
Part II followed with the discovery of the 3-minute chromospheric
oscillation (\citeads{1963ApJ...138..631N}, 
see \linkadspage{1967IAUS...28..293N}{7}{Fig.\,1} of
\citeads{1967IAUS...28..293N}). 
Part III identified the chromospheric network and the magnetic network
as outlining supergranulation cell borders
(\citeads{1964ApJ...140.1120S}).} 
appears everywhere as twinkling patches (clearest at disk center and
smaller than supergranulation cells seen limbward) of apparently
coherent oscillation somewhat larger than granules\footnote{A DKIST
show-off granulation movie will need \komega\ filtering as in
P.\,S{\"u}tterlin's
\href{https://robrutten.nl/rridl/dotlib/conefilt.pro}{\tt
conefilt.pro} for the DOT.}. 
The patches are the surface interference pattern of many solar
$p$-modes and contribute ``radial macroturbulence'' to the wash-out of
narrower photospheric lines than \MnI\ ones
(\rrref{endnote}{note:4554}).
AIA\,1700\,\AA\ movies show the 3-min components which are not
evanescent but above the local cutoff frequency, propagate up and
appear as fast-moving erratically-distributed coming and going
brightness wisps in internetwork regions. 
They piston the higher-up Carlsson-Stein shocks
(\rrref{endnote}{note:CSshocks}),

Both the photospheric and clapotispheric
(\rrref{endnote}{note:clapotis}) oscillation patterns appear random
and chaotic. 
\citetads{1973SoPh...31...23W} 
studied apparent phase patterns of the 5-minute oscillation and
concluded that they are totally random: {\em ``any description of the
observed motion in terms of simple deterministic functions will be
inadequate''\/}. 
Pure chaos -- so much for future helioseismology!
I was the referee, found it convincing, and gave a Utrecht colloquium
reviewing 5-minute oscillations but only the discovery part (showing
35-mm slides (\rrref{endnote}{note:projection}) of the plates in
\citeads{1962ApJ...135..474L}) 
and the confusion part (speculating about gravity waves turning into
Alfv\'enic waves after
\citeads{1967IAUS...28..429L}) 
itemized in
\linkpdfpage{https://robrutten.nl/rrweb/rjr-edu/lectures/rutten\_5min\_obs\_lec.pdf}{1}{this
later display} but yet without the enlightenment part because I was
not aware of the marvelous prediction of
\citetads{1970ApJ...162..993U}\footnote{The 
\label{note:ulrichack} 
\linkadspage{1970ApJ...162..993U}{10}{acknowledgment} thanks
R.F.\,Christy (earlier his thesis adviser) for suggesting that the
self-excited oscillations might be more than numerical instability of
Christy's stellar-structure code (the Nobel for starting
helioseismology should go to that).
E.N.\,Frazier is acknowledged for suggesting that the observed
5-minute oscillation might be the surface manifestation.}.

At the time the same held for F.-L.\,Deubner who was then contesting
Frazier's finding of substructure in the \komega\ diagram
(\citeads{1972SoPh...22..263D})\footnote{The 
telltale markings are by R.W.\,Noyes whose {\em Solar Physics\/}
volumes were scanned by ADS.}.
However, Deubner was asked to referee
\citetads{1973SoPh...32...31W} 
and so became aware of
\citetads{1970ApJ...162..993U} 
and also \citetads{1972ApJ...177L..87W} 
stating that the Sun is spherical and that Ulrich's modes must be
spherical harmonics.  
Deubner then recognized the parabolic mode-location curves in Ulrich's
predictive \komega\ diagram in
\linkadspage{1970ApJ...162..993U}{6}{Fig.\,2} as the fish to catch.
He knew from \linkadspage{1966AnAp...29..153M}{17}{Fig.\,6} of the
authoritative (in beautiful French) \komega\ description of
\citetads{1966AnAp...29..153M}\footnote{History: Deubner and Mein both
completed conservatory schooling and hesitated between being musician
or astronomer. 
They performed the Beethoven Op.\,105.2 and Debussy cello-piano
sonatas at the Zwaan retirement workshop
(\citeads{1994ASIC..433.....R}) 
after rehearsing at {\em Lingezicht\/}.}
that long duration and large extent were the way to \komega\
resolution and fast cadence and fine sampling to \komega\ extent.
He found the ridges, best in his hand-drawn
\linkadspage{1975A&A....44..371D}{4}{note-in-press diagram} in
\citetads{1975A&A....44..371D} 
adding the detailed spherical-harmonic predictions of
\citetads{1975PASJ...27..581A}. 
He showed it first at a conference in Nice, nicely apologizing to
Frazier
(\citeads{1976pmas.conf..259D})\footnote{McIntyre 
was not involved but wrote an ADS-misleading superfluous comment on
terrestrial seismology. 
\citetads{1999ApJ...525C1199B} 
did better but missed out on Christy and Frazier
(\rrref{footnote}{note:ulrichack}).}, where it became the proceedings
cover (\citeads{1976pmas.conf.....C}) 
and made him famous\footnote{Scooping Ulrich and Rhodes who also had
detected the ridges, while
\citetads{1970ApJ...162..993U} 
may have scooped
\citetads{1971ApL.....7..191L}.} 
in establishing that the photospheric $p$-mode pattern pistoning
chromospheric shocks appears random but actually is a
\linkpdfpage{https://robrutten.nl/rrweb/rjr-edu/lectures/rutten\_5min\_obs\_lec.pdf}{50}{most
regular and beautiful harmonic chord}.\footnote{Since my 1989
popular-astronomy hagiography
\href{https://robrutten.nl/rrweb/rjr-pubs/1989vakidioot-zonzingt.pdf}
{``De zon zingt''} I bring
\href{https://robrutten.nl/robshots/rob_flute.jpg}{my flute} to public
lectures ``The Singing Sun'' to demonstrate overtone harmonics.}
Since then helioseismology became mature.\footnote{It vindicated
stellar structure and evolution theory without upsetting these,
spawned asteroseismology and enabled far-side solar activity mapping
but it did not identify the solar activity dynamo. 
Presently it seems mired in the solar oxygen abundance, a spectrum
formation issue.}}
\rrendnote{no gravity-modes success story}{note:g-modes}{%
The alkali resonance lines and some others are suited to build
gas-filled resonance cell detectors in which atoms resonance-scatter
optical sunlight, either irradiance for full-disk profile sampling
following the initial application for \SrI\,4607.3\,\AA\ of
\citetads{1961PhRvL...7..437B} 
or with imaging using magneto-optical encoding following \"Ohman
(1960, not on ADS) and
\citetads{1968SoPh....3..618C} 
and mostly developed by A.\,Cacciani
(\href{https://www.robrutten.nl/bibfiles/ads/abstracts/solabs_cacciani.html}
{bibliography}). 
Using full-disk resonance-cell \NaI\ and \KI\ Dopplershift detectors
in around-the-world networks and in space (SOHO/GOLF) became a large
industry in $g$-mode-searching low-$l$ helioseismology but a romantic
overpromise/overclaim saga rather than a success story as in
\rrref{endnote}{note:p-modes}. 
For the latest claim/reject debate read
\linkadspage{2021LRSP...18....2C}{85}{p.\,85--86} of
\citetads{2021LRSP...18....2C} 
concluding ``evidence for solar g modes remains uncertain''.}. 

The warning bell for shocks upsetting plane-parallel modeling were the
double reversals in the cores of \CaII\ \HK\
(\rrref{endnote}{note:HKreversals}). 
The Pandora stars give fairly good renderings of the mean disk-center
\HK\ profiles but any plane-parallel model can do
this (\cf\ Ha-Ha line formation in \rrref{endnote}{note:Ha-RE} for
\Halpha), even the mean reversals when including PRD 
(\linkadspage{1985cdm..proc...67A}{26}{Fig.\,18} of
\citeads{1985cdm..proc...67A}). 
However, they cannot reproduce spectral-atlas mean-profile reversal
asymmetries: the violet-side peaks in \HK\ are higher than the
red-side peaks (the red \CaII\ \HtwoR\ peak is barely visible in the
atlas profile in \linkrjrpage{2011AAp...531A..17R}{10}{Fig.\,11} of
\citeads{2011A&A...531A..17R}). 
For decades spectroheliograms had suggested \CaII\ \HtwoV\ and \KtwoV\
``cell grains'' as cause
\rrendnote{\CaII\ \KtwoV\ and CN spectroheliograms} {note:gillespie}{%
\CaII\ \KtwoV\ ``cell grains'' are ubiquitous in high-quality
narrow-passband \KtwoV\ imaging as in the ``Selected
spectroheliograms'' of \citet{Title1966a} 
(which I have on my desk but cannot find in ADS) and in this
remarkable
\href{https://robrutten.nl/rrweb/rjr-archive/observations/KPNO/10468-2-pos.jpg}
{\KtwoV\ spectroheliogram} taken in 1975 by B.\,Gillespie with the
East Auxiliary of the McMath telescope at Kitt Peak. 
It adorned the cover of
\citetads{1985cdm..proc.....L} 
(negative) and was partly shown (positive) in Fig.\,3.9 (wrongly
labeled K center) of
\citetads{1988assu.book.....Z} 
and in \linkrjrpage{1991mcch.conf...48R}{2}{Fig.\,1} of
\citetads{1991mcch.conf...48R}, 
\linkrjrpage{2007ASPC..368...27R}{4}{Fig.\,2} of
\citetads{2007ASPC..368...27R} 
and elsewhere.\footnote{Whenever W.C.\,Livingston was asked for an
illustrative chromosphere image he sent a piece of this one.} 
Later J.W.\,Harvey sent me a CD-ROM duplicate; while writing this I
added this
\href{https://robrutten.nl/rrweb/rjr-archive/observations/KPNO/10468-2.tif}{original
negative} in my
\href{https://robrutten.nl/rrweb/rjr-archive/dircontent.html}{solar
file archive}.

The spectroheliograph had a dual-beam arrangement as at Mount Wilson
(\rrref{footnote}{note:Leightoncs}); Gillespie took a synchronous
\href{https://robrutten.nl/rrweb/rjr-archive/observations/KPNO/10468-4-pos.jpg}{companion
spectroheliogram} in the CN\,3883\,\AA\ bandhead, also now in my file
archive courtesy J.W.\,Harvey.
I co-aligned the pair with
\href{https://robrutten.nl/rrweb/rjr-archive/observations/KPNO/align_gillespie_ca_cn.idl}{this
IDL job} into
\href{https://robrutten.nl/rrweb/rjr-archive/observations/KPNO/gillespie_ca.fits}{gillespie\_ca.fits}
and
\href{https://robrutten.nl/rrweb/rjr-archive/observations/KPNO/gillespie_cn.fits}{gillespie\_cn.fits}
and advise zoom-in blinking.\footnote{The
\label{note:showex}
\href{https://robrutten.nl/rrweb/rjr-archive/observations/KPNO/align_gillespie_ca_cn.idl}{align
job} is a good example of iterative alignment of images with
corresponding patterns but no close similarity in detail. 
My SDO\,--\,STX cross- and co-alignment pipeline
(\rrref{endnote}{note:SDO-STX}) is built on this technique using
SolarSoft {\tt auto\_align\_images.pro} from T.R.\,Metcalf
(I call it metcalving).
Movie browser {\tt showex.pro} in my IDL library
(\rrref{endnote}{note:IDL}) can load many concurrent movies and offers
auto-blinking between selectable movie pairs at selectable speed and
zoom-in magnification with scatter-diagram correlation and more.
My SDO\,--\,SST
\href{https://robrutten.nl/rrweb/sdo-demo/instruction.html}{alignment
practical} demonstrates its use.
It can also load equal-size images and it can be run from the terminal
command line (script in
\href{https://robrutten.nl/rridl/00-README/sdo-manual.html}{my SDO
manual}): {\tt showex gillespie\_ca.fits gillespie\_cn.fits}.}
The \KtwoV\ image shows bright cell grains
(\citeads{1991SoPh..134...15R}) 
so ubiquitous and numerous that it is hard to make out the calcium
network except for diffuse fibril spreading around it (much more,
covering most internetwork, if the spectroheliogram had selected
\Kthree\ center).
Some of these bright grains mark migrating internetwork MCs as the
``persistent flasher'' of
\citetads{1994ASIC..433..251B} 
(\rrref{endnote}{note:basal}) but the majority are Carlsson-Stein
acoustic shock markers (\rrref{endnote}{note:CSshocks}). 
The CN image shows yet more smaller bright grains but these are
regular MC holes deepened by CN dissociation
(\rrref{endnote}{note:MCs}).
Towards disk center (lower-right part of the image) they mostly
coincide with calcium network grains appearing brighter and coarser;
the few internetwork ones with bright \KtwoV\ counterparts are likely
flashers.
These are embedded in grey small-scale internetwork patterns partly
marking reversed granulation and gravity waves
(\rrref{endnote}{note:waves}) but mostly interference patterns of
three-minute waves on their way up to shocking height
(\rrref{endnote}{note:clapotis}).}.
These small bright grains occur intermittently but ubiquitously in
internetwork cell interiors and are not magnetic (although some
authors tenaciously claimed this). 
The wide confusing literature on this phenomenon upsetting
plane-parallel modeling, not a minor deviation to be glossed over but
a key phenomenon requiring understanding, was reviewed by
\citetads{1991SoPh..134...15R} 
concluding \linkrjrpage{1991SoPh..134...15R}{51}{here} that these
grains are an acoustic interference phenomenon along the lines of
\citetads{1974SoPh...38..109L}. 
This was then brilliantly proven with observation-driven RADYN
simulations by
\citetads{1997ApJ...481..500C} 
\rrendnote{Carlsson-Stein shocks}{note:CSshocks}{%
In \citetads{1994chdy.conf...47C} 
they first described their marvelous numerical reproduction of
observed internetwork shocks in
\linkrjrpage{1993ApJ...414..345L}{4}{Fig.\,2} of
\citetads{1993ApJ...414..345L} 
using a data-driven RADYN simulation. 
Summed over space and time these shocks contribute (as ``\HtwoV\ and
\KtwoV\ cell grains'') to the asymmetry between the violet and red
\CaII\ \HK\ core reversals (\rrref{endnote}{note:HKreversals}).
The characteristic intricate shock development pattern had been
established in \linkadspage{1983ApJ...272..355C}{9}{Fig.\,4} of
\citetads{1983ApJ...272..355C} 
and was sketched in \linkrjrpage{1991SoPh..134...15R}{7}{Fig.\,2} of
\citetads{1991SoPh..134...15R}. 
The resulting excellent match of the observed and simulated \CaII\
\HtwoV\ grain sequences in 
\linkadspage{1994chdy.conf...47C}{23} {Fig.\,14} 
adorned the proceedings cover. 
These well-known results were then formally published in
\citetads{1995ApJ...440L..29C} 
and
\citetads{1997ApJ...481..500C}. 
More in \rrref{endnote}{note:INshocksHa}.}.
They showed that upward-propagating three-minute waves pistoned by
the $p$-mode pattern maxima convert non-linearly
into shocks producing cell grains
\rrendnote{Carlsson four-panel breakdown}{note:4pdiagrams}{%
\linkadspage{1994chdy.conf...47C}{16}{Figs\,8\,--\,12} of
\citetads{1994chdy.conf...47C} 
also introduced the informative Carlsson four-panel spectral-feature
breakdown diagrams. 
For \CaIIH\ one is shown in \SSXpage{156}, four in
\linkadspage{1997ApJ...481..500C}{5}
{Figs.\,4}\,--\,\linkadspage{1997ApJ...481..500C}{8}{7} of
\citetads{1997ApJ...481..500C}, 
three for \NaIDone\ and one for \CaIR\ in
\linkrjrpage{2010ApJ...709.1362L}{8}{Fig.\,9} of
\citetads{2010ApJ...709.1362L}. 
I recommend studying these -- by understanding them you
join a select club!

In my courses they became examination material. 
An astute student noted that the second panel of \SSXpage{156}
exhibits the wrong assumption of CRD. 
Carlsson and Stein argued that their neglect of \CaIIH\ PRD was
compensated by not including \MgII\ \hk.

These diagrams clearly demonstrate the importance of NLT effects even
though the spectral synthesis assumed SE. 
The observed bright \CaII\ \HtwoV\ grains
(\rrref{endnote}{note:HKreversals}) which the simulation reproduced so
admirably exist only because previous shocks define the opacity
distribution along the line of sight. 
The bright \HtwoV\ grain represents an upward traveling acoustic wave
at too low height to show shock heating at its wavelength if the
higher atmosphere were stationary; it gains its bright visibility from
large higher-up downdraft after the earlier passage of a wave that
shocked higher up. 
The downdraft pulls the line core redward and uncovers the grain.}
and that the ultraviolet radiation temperatures are Wien-weighted to
the shock maxima and not linear temperature averages. 
In \linkadspage{1994chdy.conf...47C}{8}{Fig.\,4} of
\citetads{1994chdy.conf...47C} 
the upgoing waves are still wave-like in the \AlI\ edge and fully
shocking in the Lyman continuum.
Their model-building experiment in
\linkadspage{1994chdy.conf...47C}{11}{Fig.\,5} (summary in
\SSXpage{157}) demonstrated that the nonlinear Wien weighting skews
the apparent average over internetwork shocks into an Avrett-style
apparent ``chromosphere'' which is not the chromosphere but mimics the
underlying internetwork-shock domain
\rrendnote{clapotisphere}{note:clapotis}{%
I called the above-the-photosphere under-the-chromosphere internetwork
domain filled with Carlsson-Stein shocks ``clapotisphere'' in
\linkrjrpage{1995ESASP.376a.151R}{7}{Fig.\,12} 
of \citetads{1995ESASP.376a.151R}. 
``Clapotis'' on sea charts marks locations with ``wild'' waves.
In regular wind-driven ocean waves material motion is mostly vertical;
the same holds for the 5-minute oscillation of the solar surface. 
Sea clapotis results from pattern interference by wave reflection off
capes and harbor quays or in meeting of different swell systems
growing into ``wild'' nonlinearity.
The chromospheric 3-min shocks are vertically NLT interfering in
complex fashion, wild enough to need Carlsson breakdown diagrams
(\rrref{endnote}{note:4pdiagrams}).
The photospheric $p$-mode oscillations that drive the shocks are also
complex interference patterns but outstandingly regular, linear, and
precisely predictable (\rrref{endnote}{note:p-modes})}.

In AIA 1700 and 1600\,\AA\ movies these acoustic internetwork waves
appear as fast-moving fast-evolving wisps of emissivity amid more stable
network consisting of slower-evolving collections of bright grains
marking MCs (\rrref{endnote}{note:MCs}).
The 1600\,\AA\ images sample the acoustic wisps slightly higher
because the summed photo-ionization extinction of electron donors
\MgI, \FeI, \SiI, and \AlI\ increases for shorter wavelength (compare
emergence heights and main contributors on the successive pages of
VALIII Fig.\,36 from \linkadspage{1981ApJS...45..635V}{37}{page 670}).
This height difference produces the Fourier phase differences in the
upper panel of \linkrjrpage{2001AAp...379.1052K}{18}{Fig.\,18} of
\citetads{2001A&A...379.1052K} 
reproduced per RADYN simulation by
\citetads{2005ApJ...625..556F} 
\rrendnote{acoustic and gravity wave heating}{note:waves}{%
\citetads{2005Natur.435..919F} 
also laid the long-advocated attribution of coronal heating to acoustic
waves (\eg\
\citeads{2003dysu.book..181U}) 
to rest (but see
\citeads{2021A&A...652A..43Y} 
for field-guided wave heating of the chromosphere).
An earlier candidate for coronal heating were gravity waves proposed
by
\citetads{1963ApJ...137..914W}\footnote{History: E.N.\,Parker's 
first PhD student, plugging heating by gravity waves with Parker's
blessing (before nanoflares).} who introduced the ``diagnostic''
\linkpdfpage{https://robrutten.nl/rrweb/rjr-edu/lectures/rutten_5min_obs_lec.pdf}{22}{$k\!-\!\omega$
diagram} in solar physics following geophysics
\citetads{1960CaJPh..38.1441H} 
(\rrref{endnote}{note:p-modes}).
While I doubt that they heat the corona (but see
\citeads{2008ApJ...681L.125S}) 
they transform relatively easy into Alfv\'enic torsion waves
(\linkadspage{1967IAUS...28..429L}{19}{p19\,ff} of
\citeads{1967IAUS...28..429L})\footnote{A 
worthwhile read, followed by an outstanding discussion edited by
R.N.\,Thomas, the best discussion record ever. 
See in particular \linkadspage{1967IAUS...28..429L}{40}{p\,40\,ff}
on spicule vorticity.}
and so may play a role in spicule-II heating of the quiet chromosphere
(\rrref{endnote}{note:spicules-II}).
More on gravity waves in \rrref{endnote}{note:CO}.}. 

ALMA samples these internetwork wave patterns also and may measure
their actual temperatures linearly, but only in the very quietest
areas
\rrendnote{internetwork acoustics with ALMA}{note:ALMA-waves}{%
Various SE-assuming simulations have predicted that ALMA samples
acoustic internetwork waves on their way up to become shocks in the
under-the-canopy clapotisphere (\rrref{endnote}{note:clapotis})
whereas I have predicted that these are obscured by fibrilar canopies
wherever these are opaque in \Halpha\
(\rrref{endnote}{note:ALMA-temp}).
Recently
\citetads{2021RSPTA.37900185E} 
modeled ALMA response using the Bifrost simulation of
\citetads{2016A&A...585A...4C} 
but since this lacks opaque internetwork \Halpha\ canopies
(\rrref{endnote}{note:Bifroststar}) this modeling applies only to very
quiet non-shielded areas as the ones offered in
\rrref{endnote}{note:quietHa} and described in
\rrref{endnote}{note:INshocksHa} and likely sampled by datasets D6 and
D9 of \citetads{2021RSPTA.37900174J} 
showing enhanced 4--8\,Mhz power in
\linkadspage{2021RSPTA.37900174J}{21}{Fig.\,11}. 

AIA 304\,\AA\ images can assist in identifying such rare canopy-free
locations.
Even there traditional pixel-by pixel Fourier power, phase difference
and coherency comparisons (as \eg\ for internetwork acoustics in TRACE
UV image sequences by
\citeads{2001A&A...379.1052K}) 
are misleading because the rapidly expanding thin shells seen in the
two offered \Halpha\ sequences imply the need for follow-the-motion
along-the-feature tracking, as is the case for canopy fibril dynamics
(\rrref{endnote}{note:ALMA-chrom}).}.

The MCs outlining network also appear with brighter contrast at
1600\,\AA\ than at 1700\,\AA. 
Plane-parallel colleagues attribute this enhancement similarly to increased
donor-ionization opacity; standard ``network'' and ``plage'' models
therefore have less steep upper-photosphere temperature decay than
their ``quiet'' companions for internetwork. 
However, the actual fluxtube brightness is photospheric
hole-in-the-surface radiation, with deeper holes at shorter
wavelength from lower Balmer continuum extinction,
the major opacity source at neutral-metal ionization from reduced density.
This is also known for decades (\rrref{endnote}{note:MCs} and 
the caption of
\linkrjrpage{2020LingAstRep...1R}{91}{Fig.\,73} in
\href{https://robrutten.nl/rrweb/rjr-pubs/2020LingAstRep...1R.pdf}{LAR-1})
\rrendnote{MC shifts 1600--1700\,\AA}{note:1600-1700}{%
Away from disk-center magnetic bright points in AIA 1600\,\AA\ images
shift limbward from their 1700\,\AA\ counterparts.  
These shifts do not result from higher formation but from increasing
hole transparency giving larger hole depth in top-down viewing
and further penetration into granules behind in slanted viewing.
They are shown, quantified, undone and explained in
\linkrjrpage{2020LingAstRep...1R} {88}{Figs.\,70--73} of
\href{https://robrutten.nl/rrweb/rjr-pubs/2020LingAstRep...1R.pdf}{LAR-1}.
My SDO pipeline (\rrref{endnote}{note:SDO-STX}) has an
option to unshift and subtract them to bring out flaring active-region
fibrils (\citeads{2015ApJ...812...11V}, 
\citeads{2016A&A...590A.124R}, 
\rrref{endnote}{note:EBs})
that are brightened independently in 1600\,\AA\ images by emission in
\CIV\ lines.}.
Actual MCs are close to RE throughout the photosphere without evidence
of kinetic heating, as expected for structures that inhibit convection
(\linkrjrpage{2005AAp...437.1069S}{7}{Fig.\,7} of
\citeads{2005A&A...437.1069S}). 

Thus, Pandora-star ``chromospheres'' that are primarily based on
ultraviolet continua actually represent, non-linearly, complex
interference patterns of clapotispheric shocks plus misinterpreted
network surface holes. 
Nothing to do with the solar on-disk chromosphere in \Halpha\ that
carries the name originally given by
\citetads{1868RSPS...17..131L} 
for the colorful off-limb appearance of the Balmer lines and
\HeIDthree\
\rrendnote{Lockyer chromosphere, \HeIDthree, flash
color}{note:Lockyer}{%
I typed \citetads{1868RSPS...17..131L} 
into ADS, remain its main citer so far, and look forward to write
``less refrangible'' in publications \SSXpage{12\,ff}.
Lockyer's low total of
\href{https://ui.adsabs.harvard.edu/search/fl=identifier\%2C\%5Bcitations\%5D\%2Cabstract\%2Caff\%2Cauthor\%2Cbibcode\%2Ccitation_count\%2Ccomment\%2Cdoi\%2Cid\%2Ckeyword\%2Cpage\%2Cproperty\%2Cpub\%2Cpub_raw\%2Cpubdate\%2Cpubnote\%2Cread_count\%2Ctitle\%2Cvolume\%2Clinks_data\%2Cesources\%2Cdata\%2Cemail\%2Cdoctype&p_=0&q=database\%3A\%20astronomy\%20author\%3A\%22Lockyer\%2C\%20J.\%20Norman\%22\%20&rows=25&sort=date\%20desc\%2C\%20bibcode\%20desc&start=0/metrics}{ADS
citations} (but steepening reads) shows how citation ranking fails
for the first professor in astrophysics and the founding editor
(during 50 years!) of {\em Nature\/}. 
I recommend his delightful
\href{https://books.google.nl/books/about/The_Spectroscope_and_Its_Applications.html?id=EbANQlPbaS0C&redir_esc=y}{``The
Spectroscope and Its Applications''}.

\citetads{1868RSPS...17..131L} 
discovered \HeIDthree\ (as did ``Indian observer'' Janssen shortly
before him). 
It contributed its splendid yellow color (familiar to us from sodium
street lights) to the chromatic beauty of the emission lines he knew
as prominence lines but saw everywhere just outside the limb with his
new (but ``incomplete'' and delayed beyond Janssen) spectroscope
\SSXpage{12}, concluding that {\em ``prominences are merely local
aggregations of a gaseous medium which entirely envelopes the sun''}.

This was well before he saw himself the beautiful purple-pink of the
flash chromosphere at eclipses, but he knew the visually-integrated
color from prominences and chose the name to distinguish it from {\em
``the white light-giving photosphere''\/} and the {\em ``cool
absorbing atmosphere''\/} -- a reversing-layer Fraunhofer-line
interpretation prior to Schuster-Schwarzschild modeling treated in the
\linkpdfpage{https://robrutten.nl/rrweb/rjr-edu/exercises/ssa/ssa.pdf}{31}{SSA\,3.3}
practical, the opposite to Milne-Eddington modeling
(\rrref{endnote}{note:ME}).

The chromospheric flash color at eclipses during a few seconds after
second and before third contact (longer near the edge of the totality
strip where Houtgast went on purpose) sums Lockyer's lines dominated
by \Halpha, plus white electron scattering (Sect.\,7 on walled
pdf page 8 of \citeads{2009SoPh..254...89J}). 
The beautiful color represents \HI, \HeI\ and freed H and He 
electrons: enjoy it yet more by appreciating it as the stuff the
universe is made of.}

Yet worse, actual \Halpha\ canopies in quiet areas are mostly made by
spicules-II (\rrref{endnote}{note:spicules-II}) that reach heights
around 7000\,km
(\citeads{2014ApJ...792L..15P}) 
far beyond any Pandora-star atmosphere.
On the disk they are observed as outer \Halpha-wing RBEs and RREs
(\rrref{endnote}{note:spicules-II}) -- so much for ascribing
outer-wing intensities to deep-photosphere sampling and so much for
assigning brightness to heating since they heat to EUV visibility
(\citeads{2016ApJ...820..124H}) 
but are dark.
Their subsequent cooling recombining return flows produce most of the
dark \Halpha\ line-core fibrils (``coarse mottles'') around
network
(\citeads{2019A&A...632A..96R}). 
Both types of ubiquitous dynamic \Halpha\ structure are transparent in
the ultraviolet continua used for constructing Pandora stars
which therefore have no chromosphere, misusing that name
for the modeled clapotisphere underneath
\rrendnote{Ha--Ha scattering}{note:Ha-RE}{%
Pandora stars yield \Halpha\ disk-center profiles that correspond
reasonably well to the observed disk-center atlas profile
(VALIII \linkadspage{1981ApJS...45..635V}{26}{Fig\,25\,--\,26})
Naively this may be seen as vindication of the model
``chromosphere''. 
However, about the same profile is also produced in a NLTE RH
computation of \Halpha\ from an 1D RE atmosphere having no
chromosphere whatsoever, only an RE photospheric temperature decline
(\SSXpage{3},
\linkrjrpage{2012AAp...540A..86R}{6}{Figs\,7--8} of
\citeads{2012A&A...540A..86R}). 

The reason is that \Halpha\ makes its own $\sqrt{\varepsilon}$
scattering decline, more or less the same in \Halpha\ optical depth
scale irrespective of geometrical height scale and presence or absence
of an upper-photosphere opacity gap.
I call this ``Ha--Ha'' line formation.
It similarly implies that actual \Halpha\ from higher-located
optically and effectively thick fibrils samples similar scattering
declines. 
Where these appear darker they are not necessarily cooler, just more
opaque.

Furthermore, in the ALC7 atmosphere \CaIR\ has nearly the same
$\sqrt{\varepsilon}$ decline with closely the same $\tau$ sampling
(comparison in \SSXpage{29}). 
The solar-atlas profiles of \Halpha\ and \CaIR\ in
\linkrjrpage{2011AAp...531A..17R}{10}{Fig.\,10} of
\citetads{2011A&A...531A..17R} 
differ substantially but mostly from different thermal broadening and
\Halpha's abnormal opacity gap, both also known to ALC7 so that it
reproduces both profiles reasonably well -- but so does a Kurucz RE
atmosphere without chromosphere and so do actual higher-up
chromospheric fibrils (ha--ha).

In quiet-Sun resolving observations the actual nonthermal motions
yield different scenes in the cores of these lines, except for their
time-averaged core widths which sense mean temperature.
In the wings both lines show spicules-II but \Halpha\ more and longer.
More in \rrref{endnote}{note:nonEchrom}.}.
They also do not contain ``cool clouds'' observed in CO lines
\rrendnote{cool COmosphere}{note:CO}{%
Named in \citetads{1994ApJ...423..806W} 
by T.R.\,Ayres who since his participation in $^{13}C$-seeking
\citetads{1972ApJ...171..615H} 
eloquently promoted CO coolness (``thermal bifurcation'' in
\citeads{1981ApJ...244.1064A}, 
``dichotomy'' in \citeads{1986HiA.....7..425A}, 
``chilling truth'' in
\citeads{1990ASPC....9..106A}, 
``heart of darkness'' in
\citeads{1994Sci...263...64S}, 
``through the lens of 3D spectrum synthesis'' in
\citeads{2013ApJ...765...46A}). 

Cool indeed since
\citetads{1972BAAS....4..389N} 
found brightness temperature 3500\,K near the limb in the CO
4.7\,$\mu$m fundamental band, clashing with the 4170\,K minimum of the
then-leading HSRA model of \citetads{1971SoPh...18..347G} 
and the same value in later VALIIIC, the lowest of all Avrett models
\SSXpage{58}. 
\citetads{1981ApJ...245.1124A} 
confirmed these; 
\citetads{1986ApJ...304..542A} 
used spectroscopy of the 2.3\,$\mu$m first-overtone and 4.5\,$\mu$m
fundamental CO bands together with \CaIIK\ to define competing models
COOLC and FLUXT in their
\linkadspage{1986ApJ...304..542A}{12}{Fig.\,12} for quiet Sun and
plage, the first undoing Avrett's VALC$^\prime$ quiet ``chromosphere''
shown in \linkadspage{1985cdm..proc...67A}{25}{Fig.\,17} of
\citetads{1985cdm..proc...67A} 
into monotonous temperature decay down to 3500\,K as in an RE model,
the second similar to but everywhere hotter than the VALIIIP plage
model -- indeed bifurcating into multi-component requiring fill-factor
weighting.
It was redone in 
\linkadspage{1990ApJ...363..705A}{9}{Fig.\,4} of
\citetads{1990ApJ...363..705A} 
and again in \linkadspage{1996ApJ...460.1042A}{12}{Fig.\,8}
of \citetads{1996ApJ...460.1042A} 
with a COOLC model bending the lower MACKKL chromosphere
down into their COmosphere with minimum 3000\,K at height 850\,km.
\citetads{2007ApJ...667.1243F} 
invoked and imposed non-gravitational acceleration to mimic that in
their \linkadspage{2007ApJ...667.1243F}{5}{Fig.\,1}, also in the
FCHHT-B star of \citetads{2009ApJ...707..482F} 
that I use didactically per \SSXpage{171} in
\rrref{endnote}{note:aureoles}.

Of course, wishful 1D plane-parallel SE model adjustment must be
replaced by diagnosing 3D(t) CO line formation in actual
solar-atmosphere structures. 
There is a zoo of candidate cooling downdrafts: in intergranular lanes
in the low photosphere, from granular overshoot in the middle
photosphere, in the five-minute oscillation there, after acoustic
three-minute shocks in higher-up internetwork clapotisphere, from
gravity waves also under chromospheric canopies, in quiet-Sun fibrils
constituting internetwork canopies, cooling rain along
longer chromospheric fibrils connecting active regions.

\citetads{1972BAAS....4..389N} 
initially attributed their cool CO cores to intergranular lanes but
quickly detected five-minute modulation in
\citetads{1972ApJ...176L..89N}. 
This was also reported by \citetads{1996ApJ...460.1042A} 
and studied by Uitenbroek
(\citeyearads{2000ApJ...531..571U}, 
\citeyearads{2000ApJ...536..481U}, 
\citeyearads{2004IAUS..219..103U}). 
Diagnostic-diagram $p$-mode ridges are clearly present in
\linkadspage{2011ApJ...734...47P}{6}{Fig.\,5} of
\citetads{2011ApJ...734...47P}. 
However, this modulation is only over a few 100\,K and may just represent
the evanescent undulations that all upper-photosphere gas undergoes.
It shows up as a significant but relatively minor acoustic modulation
peak in the wide power distribution with smaller contribution at
larger height in \linkadspage{2011ApJ...734...47P}{5}{Fig.\,4} of
\citetads{2011ApJ...734...47P}. 
It does not exclude cooling downdrafts higher up that deposit
cool gas in deep pools (``clouds'') sloshed up-down in the surface
interference pattern of the global oscillations.

The topic seems still wide open. 
The scarce CO spectroheliograms 
so far (\linkadspage{1996ApJ...460.1042A}{6}{Fig.\,3} of
\citeads{1996ApJ...460.1042A} but missing color I think;
\linkadspage{2000ApJ...531..571U}{4}{Fig.\,2} and
\linkadspage{2000ApJ...531..571U}{5}{Fig.\,3} of
\citeads{2000ApJ...531..571U}; 
\linkpdfpage{https://arxiv.org/pdf/1008.5375.pdf}{3}{Fig.\,3} of
\citeads{2010AN....331..589S}) 
have insufficient spatial and temporal resolution and extent, but at
that they suggest ubiquitous presence of grainy cool clouds measuring
multiple arcsec and lasting a few minutes in under-the-canopy
internetwork (longer at low-pass filtering as in
\linkadspage{2000ApJ...531..571U}{7}{Fig.\,7} of
\citeads{2000ApJ...531..571U}) 
and with off-limb extent to about 1000\,km height. 

Chromosphere simulations as
\citetads{2007A&A...473..625L} 
(\rrref{Sect.}{sec:Oslo}) and
\citetads{2011A&A...530A.124L} 
(\rrref{endnote}{note:runaway}) suggest cooling return flows from
under-the-canopy clapotispheric shocks (endnotes
\rrref{}{note:CSshocks},\rrref{}{note:clapotis}) with wide-blob
``cloud'' morphology. 
However, the scarce observations of internetwork shock aftermaths in
\Halpha\ (visible through the NLT overextinction of
\rrref{Sect.}{sec:Oslo} but requiring targeting rare canopy-free
internetwork areas as in the SST movie and the SST data linked in
\rrref{endnote}{note:quietHa}) indicate narrow rapidly changing
filamentary mushroom strands (suggesting magnetic structuring to me)
requiring high spatial and temporal resolution. 
High resolution is likely also required for higher-up
magnetism-channeled downdrafts feeding collection into deeper clouds,
possibly longer-lived from steady rain-down.
The detection of molecular lines in IRIS plage spectra by
\citetads{2014A&A...569L...7S} 
is a warning telltale.

In addition to these candidates I use my author prerogative to plug
gravity waves as another contribution candidate, through comparison to
mid-ultraviolet diagnostics. 
Gravity waves are copiously excited in granular overshoot at
frequencies below the Brunt-V\"ais\"a\l\"a limit
(\linkpdfpage{https://robrutten.nl/rrweb/rjr-edu/lectures/rutten_5min_the_lec.pdf}{8}{p.\,8\,ff}
of my
\href{https://robrutten.nl/rrweb/rjr-edu/lectures/rutten_5min_the_lec.pdf}
{oscillation theory lectures}) which reaches about 5\,mHz
(\linkadspage{2017ApJ...835..148V}{8}{Fig.\,7} of
\citeads{2017ApJ...835..148V}) 
but unlike stay-local overshoot they propagate slantedly up and away
and interfere laterally up to breaking height (Mihalas and Toomre
\citeyearads{1981ApJ...249..349M}, 
\citeyearads{1982ApJ...263..386M}). 
The TRACE 1700\,\AA\ diagnostic diagram (power as function of
horizontal wavenumber and frequency or reversely wavelength and
period) in the upper panel of
\linkrjrpage{2003AAp...407..735R}{3}{Fig.\,3} of
\citetads{2003A&A...407..735R} 
shows prominent $p$-mode ridges with an extended three-minute tail
beyond the five-minute peak in the mean power spectrum along the
righthand side, but much larger low-frequency and mesoscale (around
5\,arcsec wavelength) power contribution located below the Lamb line. 
Its location, the corresponding negative $1700\!-\!1600$\,\AA\ phase
difference in the lower panel
and the tilting and partially randomizing scatter clouds in adjacent
\linkrjrpage{2003AAp...407..735R}{3}{Fig.\,2} suggest atmospheric
gravity waves.  
In the striking one-hour-averaged 1700\,\AA\ inverse-intensity image
in \linkadspage{2010arXiv1012.1196R}{12}{Fig.\,5} of
\citetads{2010arXiv1012.1196R} 
the grainy long-lived brightest (coolest) ``clouds'' attributed to
gravity waves measure a few arcsec and occur ubiquitously in the
network hearts where canopies reach highest, fewer towards surrounding
network -- similarly to reported CO clouds and with similar low-pass
filtered timeslices (last panel of
\linkrjrpage{2003AAp...407..735R}{2}{Fig.\,1} of
\citeads{2003A&A...407..735R}). 

These cool clouds are unlikely to be visible in \Halpha\ anywhere for
lack of prior heating and ionization supplying NLT opacity. 
The best optical candidates are neutral-stage resonance lines.
In my archived
\href{https://robrutten.nl/rrweb/rjr-archive/linelists/dunnetal-bin10AA.png}
{summary plot} of the eclipse spectra of
\citetads{1968ApJS...15..275D} 
the \NaID\ lines (just right from far-reaching \HeIDthree) extend
beyond the 1000\,km of the CO lines.  
There are dark internetwork clouds in the IBIS \NaIDone\ observations
of \citetads{2010ApJ...709.1362L} 
and the SST \NaDone\ observations of
\citetads{2011A&A...531A..17R} 
but for these the corresponding simulations suffered lack of opacity
from electron-density underestimation by not implementing NLTE
ultraviolet overionization of the metal donors (see
\linkrjrpage{2010ApJ...709.1362L}{7}{p\,1368} and
\linkrjrpage{2011AAp...531A..17R}{14}{p.\,14}, respectively).
The \MgI\ 4571.1\,\AA\ thermometer line
(\rrref{endnote}{note:6302-4571}) probably does not reach high enough
since it has only 700\,km off-limb extent in my
\href{https://robrutten.nl/rrweb/rjr-archive/linelists/dunnetal-reordered.dat}
{overview table} from
\citetads{1968ApJS...15..275D}. 

What's next? 
Probably simulations and DKIST. 
Simulations should properly handle NLTE and NLT
ionization-recombination and emulate internetwork canopies beyond the
Bifrost star of \rrref{endnote}{note:Bifroststar} to gain canopy
blocking and magnetically guided rain.
DKIST will gain from infrared Fabry-P\'erot imaging spectroscopy
sampling beyond heliographic slit scanning.
ALMA may evaluate some cool-cloud temperatures but will generally not
see these underneath obscuring canopies, only at the very quietest
locations (those without \Halpha\ canopies and showing three-minute
oscillations to ALMA, see endnotes \rrref{}{note:ALMA-temp} and
\rrref{}{note:quietHa}).}.

Yet worse, there even exist publications (not to be named) that used
the VALIIIC diagram in \SSXpage{60} 
to define ``height differences'' as the 400\,km there between \CaII\
\Kthree\ and the core of \Halpha\ to identify and measure wave
propagation and explain coronal heating. 
Maybe there exist plane-parallel stars in a parallel plain universe
where this is a viable tactic (but do those have coronae?).

\section{NLS+NLW+NLT:~ Oslo simulations} \label{sec:Oslo}
NLT is the current frontier in solar spectrum interpretation, not just
in underlying MHD phenomena as the Carlsson-Stein shocks
(\rrref{endnote}{note:4pdiagrams}) but also by not assuming SE in
spectrum synthesis.
Anything you see may not be so much how it appears now but rather what
happened before.\footnote{Outdoors example: airliner contrails signify
that aircraft engines passed before. 
If you don't know that aircraft exist you may have to invoke 
\href{https://robrutten.nl/robshots/lgm-roswell.jpg}{Roswell flying
saucers} to explain long white fibrils on our sky.}

The key publications I summarize here are the 1D(t) HD RADYN analysis
of acoustic shocks by
\citetads{2002ApJ...572..626C} 
and the 2D(t) MHD Stagger simulation of network and internetwork
shocks by \citetads{2007A&A...473..625L}. 
The first showcases the physics, the second demonstrates the effects.
Neither includes spectrum synthesis but the hydrogen rate analyses of
the first and the hydrogen population results of the second make these
classics of solar spectrum formation.

\paragraphrr{1D HD RADYN simulation}
\citetads{2002ApJ...572..626C} 
used their RADYN code to study hydrogen partitioning in acoustic
shocks as those of
\citetads{1997ApJ...481..500C} 
in an exemplary analysis.
Key results are summarized in \SSXpage{165}. 

The main agent is \Lyalpha. 
They found that detailed radiative balance is a good approximation for
\Lyalpha\ at shock heights (the discussion on VALIII
\linkadspage{1981ApJS...45..635V}{29}{p.\,662} concerns larger
heights). 
It became a tractability assumption in multi-D MHD simulations
\rrendnote{chromospheric radiation shortcuts}{note:shortcuts}{%
Multi-dimensional MHD simulation codes need tractability shortcuts to
speed up beyond slow 1D RADYN evaluations.
Various shortcuts developed at Oslo have been implemented in the
CO5BOLD code by
\citetads{2006A&A...460..301L}, 
the STAGGER code of
\citetads{2004IAUS..223..385H} 
by \citetads{2007A&A...473..625L}, 
the Bifrost code by \citetads{2011A&A...531A.154G} 
and lately the MURaM code by
\citetads{2022arXiv220403126P}. 
A key one is the multi-group approximation for scattering in the PhD
thesis of \citetads{1998PhDT........34S} 
(not posted but this part is published in
\citeads{2000ApJ...536..465S}). 
It employs tabulated group-mean opacities, scattering probabilities
and Planck functions all based on SB populations.
Further shortcuts to estimate $J_\nu$ locally instead of nonlocal
evaluation were formulated in the MSc thesis of
\citetads{1999MsT..........1S}.\footnote{A 
solar RT course in itself, posted at
\href{http://urn.nb.no/URN:NBN:no-96791}{Oslo} and
\href{https://ui.adsabs.harvard.edu/link_gateway/1999MsT..........1S/AUTHOR_HTML}{ADS}
for this endnote.}
He checked the assumption of detailed radiative balancing for
\Lyalpha\ and other Lyman lines on VALIIIC and with RADYN and added
similarly tested recipes to approximate $J_\nu$ in the scattering
bound-free hydrogen continua by parametrizing their smooth decays
(seen \eg\ in \SSXpage{42}) as exponential decay between their deep
LTE set as $S\tis\,B$ at $\tau_{\nu_0}\tis\,1$ at edge threshold
$\nu_0$ and their scattering emergence limit simply set as
$J_\nu(\tau_\nu\tis\,0) = J_\nu(\tau_\nu\tis\,1)/2$.}.
It likely applies where hydrogen is largely neutral so that \HI\
$n\tis\,1$ has almost all hydrogen and is by far the most populated of
all atomic levels.
Most \Lyalpha\ photons are then created per pair {\bf d}.
Because the probabilities of destruction {\bf a} and Balmer detour
{\bf h} are small they resonance scatter many times without coming far
because their mean free path is small (only centimeters in the
medium-density center panel of \SSXpage{32}) and the thermalization
length also (only kilometers in the center panel of \SSXpage{90}).  
Hence they just scatter endlessly in up-down sequences in close
confinement.
This approximation becomes invalid in hotter features where hydrogen
ionizes largely and also near these by missing hot surround and back
radiation (\rrref{endnote}{note:aureoles}).

\linkadspage{2002ApJ...572..626C}{4}{Fig.\,3} of
\citetads{2002ApJ...572..626C}, 
reproduced in the \SSXpage{165} summary, illustrates \Lyalpha\
photon balancing as negligible net 1-2 rate.   
The fat transitions there (large net photon rate) are photo-ionization
in the Balmer continuum and recombination with $\Delta n\tis\,1$
steps
\rrendnote{Rydberg \HI\ lines}{note:Rydberg}{%
The \HI\ Rydberg ladder produces emission lines in the infrared by
radiative-collisional recombination departure diffusion
(\citeads{1992A&A...259L..53C}) 
similar to the \MgI\ 12-micron lines in
\rrref{endnote}{note:MgI12micron}.
At that wavelength the \HI\ and \MgI\ lines are still apart but they
creep together for higher levels and longer wavelengths.
The Rydberg levels producing lines near 12\,micron have larger
population for \MgI\ than for \HI\
(\linkrjrpage{1994IAUS..154..309R}{2}{Fig.\,1}) of
\citeads{1994IAUS..154..309R}) 
but this likely reverses at ALMA wavelengths through NLT \HI-top
population boosting (\rrref{endnote}{note:ALMA-line}).}
ending with fat \Halpha.  
For given $n_2$ population (highly NSE in post-shock cooling) this
loop obeys SE but it is far out of LTE because the impinging Balmer
continuum originates in the deep photosphere, with radiation
temperature about 5300\,K \SSXpage{76}. 
Where this value exceeds the local temperature photon pumping enhances
ionization.
There are also contributing photon losses in heavily scattering
\Halpha\ \SSXpage{91} which operates within the low-energy 3.6~eV
hydrogen top (\rrref{footnote}{note:Htop}) representing a
minority-species photon-suction loop as in the neutral alkalis
(\rrref{endnote}{note:suction}).\footnote{In the VALIIIC
``chromosphere'' Balmer continuum pumping and \Halpha\ photon losses
represent large but balancing terms in the radiative energy budget
\SSXpage{69}, so that losses in \CaIIHK, the \CaII\ infrared lines and
\MgIIhk\ dominate \SSXpage{70}. 
The total curve in \SSXpage{70} defines the unspecified local heating
needed to maintain the VALIIIC atmosphere
(\rrref{endnote}{note:suction}), much more for its ``chromosphere'' than
for its 10$\times$ thinner transition region.}

The key hydrogen balance for NLT is the \Lyalpha\ collision balance. 
The collisional up and down rates differ much in temperature
sensitivity due to the Boltzmann ratio in their Einstein relation
\SSXpage{166} \RTSApage{43}{p.\,23} \RTSApage{70}{Sect.\,3.2.5}. 
The exponential $C_{lu}$ sensitivity determines the settling time
scale at which \Lyalpha\ reaches Boltzmann equilibrium. 
A few cases are plotted in
\linkadspage{2002ApJ...572..626C}{7}{Fig.\,6} of
\citetads{2002ApJ...572..626C}, 
reproduced in \SSXpage{166}.
In hot shocks the frequent \Lyalpha\ up- and down collisions make the
balance reach Boltzmann equilibrium in seconds, but in subsequent
post-shock cooling gas the settling time scale increases to multiple
or many minutes. 
During this time the large $n_2$ overpopulation also governs large
NSE overionization since the whole top of the hydrogen atom
including the proton population is defined by this lower boundary
condition.
The Balmer-top SE loop just adds yet more overionization with respect
to SB if the temperature drops below 5300\,K. 

\paragraphrr{2D MHD Stagger simulation}
The next step in NSE simulation was
\citetads{2007A&A...473..625L}, 
going from 1D to 2D and from HD to MHD and using the radiation
shortcuts of \rrref{endnote}{note:shortcuts}.
\SSXpage{167} is a result summary. 
The three panels at its top are cutouts showing three of the nine
quantities in \linkrjrpage{2007AAp...473..625L}{4}{Fig.\,1} at a
sample time step. 
They demonstrate that the simulation (evolved after starting from an
LTE initializing simulation) portrays a quiet solar scene with two
opposite-polarity MCs 8\,Mm apart that resemble network fluxtubes,
with internetwork in between showing a canopy-like dome with cool to
very cool gas and less steep upward density decay underneath.  
For me the 9-panel thumbnail in \SSXpage{167} is a button that opens a
movie running the whole simulation. 
Please
\href{https://robrutten.nl/rrweb/rjr-movies/hion2_fig1_movie.mov}{download
it here} and play it. 
It vividly shows how shocks (thin blue filaments in temperature)
travel up along the MCs and more erratic between them, the latter
kicking up the apparent canopy. 
The first represent field-guided dynamic fibrils
\rrendnote{dynamic fibrils}{note:DFs}{%
Field-guided shocked waves in slanted active-region MCs that (in my opinion)
became the first success story in the century-old struggle to
understand the rich zoo of solar \Halpha\ features
(\eg\ \citeads{2006ApJ...647L..73H}, 
\citeads{2007ApJ...655..624D}, 
\citeads{2007ApJ...666.1277H} 
\SSXpage{22}).
Their slant contributes cut-off frequency lowering
(\linkadspage{1973SoPh...30...47M}{6}{Eq.\,4} of
\citeads{1973SoPh...30...47M}; 
\linkadspage{1977A&A....55..239B}{2}{Fig.\,1} of
\citeads{1977A&A....55..239B}). 
Their signature is repetitive parabolic extension and (non-ballistic)
retraction along their length in $x-t$ timeslices
(\linkadspage{2007ApJ...655..624D}{7}{Fig.\,7} of
\citeads{2007ApJ...655..624D}). 
Many are visible as dark up-and-down dancing stalks in phased rows in
this magnificent 5-s cadence
\href{https:robrutten.nl/rrweb/rjr-pubstuff/lar-2/halpha-soup-half.mpg}{SST/SOUP
\Halpha\ movie} (courtesy L.H.M.~Rouppe van der Voort)\footnote{For
Acrobat the 19\,MB size may exceed its limit, needing manual download
from
\url{https:robrutten.nl/rrweb/rjr-pubstuff/lar-2/halpha-soup-half.mpg}.}.
The rows map MC arrangements in active-region network and plage; the
phasing along rows is set by the interference-pattern undulations of
the combined $p$-modes (\rrref{endnote}{note:p-modes}).
Their tops may heat to high temperature and show up as bright grains
in the \SiIV\ lines in IRIS 1400\,\AA\ slitjaw images
(\citeads{2016ApJ...817..124S}). 
Short ones are observed in sunspots
(\citeads{2013ApJ...776...56R}).} 
as evident from their characteristic time-slice parabolas in the
lefthand column of \linkrjrpage{2007AAp...473..625L}{6}{Fig.\,3}.
The internetwork shocks are acoustic Carlsson-Stein shocks
(\rrref{endnote}{note:CSshocks}).

The last panel of the three in \SSXpage{167} and of the nine in
\linkrjrpage{2007AAp...473..625L}{4}{Fig.\,1} and its movie version
show the NLTE departure coefficient $b_2$ of \HI\ level $n\tis\,2$,
upper level of \Lyalpha\ and lower level of \Halpha.
Its variations are enormous, reaching values up to $b_2\tis\,10^{12}$
in cooling post-shock blobs\footnote{This panel also shows high-up
green arches. 
I think that these are artifacts stemming from the shortcut assumption 
(\rrref{endnote}{note:shortcuts}) that \Lyalpha\ obeys detailed radiative
balance (valid lower down).}.
This is not NLTE at the 0.1-1~dex level traditionally regarded as
significant in solar and stellar abundance studies -- this is 12\,dex (!)
of retarded overpopulation 
\rrendnote{runaway cooling}{note:runaway}{%
The \HI\ $b_2$ values get so exceedingly large because the
clapotispheric post-shock gas cools to very low temperatures, down to
the 2400\,K limit imposed by
\citetads{2007A&A...473..625L} 
to avoid having to implement extensive molecule formation -- which
contributes further cooling as by the many CO lines
(\citeads{1981ApJ...244.1064A}). 
Simulation cooling was detailed by
\citetads{2011A&A...530A.124L} 
with the conclusion that it cannot be avoided unless there is yet
unidentified Joule heating in upper-photosphere internetwork. 
An issue obviously related to Ayres' COmosphere of
\rrref{endnote}{note:CO}. 

However, even if current simulations overestimate cooling and
therefore the corresponding \HI\ $b_2$ values, these then still
express that the $n\tis\,2$ population maintains the high values
obtained in hot precursor phases. 
Whether the actual cooling reaches $b_2\tis\,10^{12}$ or say ``only''
$10^6$ does not matter. 
It is anyhow the champion solar NLTE departure for denser-than-CE gas
and essential for understanding hydrogen ionization, the \Halpha\
chromosphere and ALMA diagnostics (\rrref{Sect.}{sec:chromosphere}).}.
The next to last panel of
\linkrjrpage{2007AAp...473..625L}{4}{Fig.\,1} and its movie version
shows the departure for the \HI\ continuum level ($n\tis\,6$ in the
atom model) and illustrates that Balmer continuum pumping adds up to
another 3\,dex for the coolest clouds (instantaneously).

\linkrjrpage{2007AAp...473..625L}{5}{Fig.\,2} of
\citetads{2007A&A...473..625L} 
is also best studied in its
\href{https://robrutten.nl/rrweb/rjr-movies/hion2_fig2_movie.mov}{movie
version}.  
Its panels show behavior of the $y$-axis quantity along the dashed
vertical lines in the fourth panel of
\linkrjrpage{2007AAp...473..625L}{4}{Fig.\,1}, at left sampling the
lefthand MC and at center sampling internetwork.
In the lower three rows of the
\href{https://robrutten.nl/rrweb/rjr-movies/hion2_fig2_movie.mov}{Fig.\,2
movie} the thick curves are the simulation results, the thin curves
the values obtained by assuming LTE instead so that the curve
separations correspond to the departure coefficients in Fig.\,1 and
its movie version.
Playing this movie shows shocks running up with time in the
temperature curves in the lefthand top panel.
The actual shocks in the Fig.\,1 movie tend to go slanted in the
internetwork, so that the
righthand Fig.\,2 movie column samplings are not
along-the-shock path histories
\rrendnote{internetwork shocks in \CaIIH\ and \Halpha}{note:INshocksHa}{%
When Mats Carlsson hunted in the data of
\citetads{1993ApJ...414..345L} 
he found only few pixels with multi-cycle spectrum-versus-time
behavior matching his RADYN output for the same pixel and selected the
best for
\linkadspage{1997ApJ...481..500C}{19}{Fig.\,16} of
\citetads{1997ApJ...481..500C}, 
showing two less-well matching samplings in
\linkadspage{1997ApJ...481..500C}{20}{Fig.\,17}.
The first two are columns labeled 110 and 132 in the observed time
slices in \linkrjrpage{2008SoPh..251..533R}{4}{Fig.\,2} of
\citetads{2008SoPh..251..533R}, 
the latter two columns 149 and 30. 
The latter shocks may have been too slanted to produce vertical-column
NLT signature as in the 1D simulation.

The quiet-target SST movie linked in
\rrref{endnote}{note:quietHa} and
\href{https://robrutten.nl/rrweb/rjr-movies/2006-06-18-quiet-ca-hawr.avi}{here
again} shows fast-varying threads constituting \Halpha\ mushroom
shells following on and around \CaII\ \HtwoV\ grains, indeed not
radial along the line of sight: internetwork shocks clearly are 3D(t)
rather than 1D or 2D phenomena even if the $p$-mode wave pistoning is
primarily radial.
Per-pixel Fourier analysis to establish vertical phase relations is
invalid for the resulting shocks and their mushroom aftermaths. 
The cadence of the movie is 7 seconds but faster is clearly desirable,
confirming the need for 1\,s spectral \Halpha\ imaging with DKIST
(\rrref{endnote}{note:telescopes}).

The {\tt showex} SST displays in my
\href{https://robrutten.nl/rrweb/sdo-demo/instruction.html}{alignment
practical} show \Halpha\ mushroom threads well in the upper-left
quiet corner of the SST field but at slower cadence.}
but sample different shocks at different times while the lefthand
movie column samples dynamic fibril shocks better along their
near-vertical paths.
The movie strikingly demonstrates that the ionization fraction, proton
density and $n_2$ population density in the lower rows reach LTE
values in shocks but thereafter do not drop steeply along with the
temperature. 
Instead they hang roughly at the value they got in the
shock.\footnote{Yet \label{note:PSBE} another acronym: in
\linkrjrpage{2017AAp...598A..89R}{2}{Sect.\,2.1} of
\citetads{2017A&A...598A..89R} 
I distilled a PSBE = ``post-SB-extinction'' recipe from this result.
Since \Halpha\ gains SB extinction in heating features and maintains
these high values in subsequent cooling this recipe is to evaluate the
hottest SB extinction for a parcel of gas and maintain that
subsequently.} 
There should be eventual settling to the LTE value, but long before
that the next shock arrives: shocks repeat faster than the settling
time scale. 
The persistently-high $n_2$ population results in the dark appearance
of dynamic fibrils (seen well in the
\href{https:robrutten.nl/rrweb/rjr-pubstuff/lar-2/halpha-soup-half.mpg}{SST/SOUP
\Halpha\ movie} of \rrref{endnote}{note:DFs}) throughout the
downstroke phase of their timeslice parabolas whereas they would fade
fast with the lowering temperature if SE LTE were valid.

The second row shows that hydrogen generally does not ionize fully in
the shocks but only a few percent. 
However, this already means two orders of magnitude increase of the
electron density which in neutral-hydrogen gas sits at the $10^{-4}$
abundance fraction of the electron donor metals (VALIII
\linkadspage{1981ApJS...45..635V}{67}{Fig.\,47} \SSXpage{72}).
This increase in collision frequency appears sufficient to obtain LTE
Boltzmann equilibrium in \Lyalpha\ within the shocks, giving LTE
Balmer extinction (bottom panels of the
\href{https://robrutten.nl/rrweb/rjr-movies/hion2_fig2_movie.mov}{Fig.\,2
movie}).
Another condition for this equilibriating is that a feature should be
thick enough to contain \Lyalpha, optically thicker than the
thermalization length from its surface.
At partial ionization hydrogen retains sufficient $n_1$ opacity
to keep these shocks effectively thick in \Lyalpha.

The upshot for the \HI\ top up from $n\tis\,2$, in particular \Halpha,
is that it combines all three types of nonlocalness: NLS in \Halpha\
from its Ha--Ha resonance scattering (\rrref{endnote}{note:Ha-RE}),
NLW from \Lyalpha-controlled SB extinction in heated gas, NLT from the
slow supra-SB \Lyalpha\ settling in cooling gas. 
The combination makes the dynamically refurbished quiet-Sun
chromosphere observed in \Halpha\ and likewise with ALMA a NLS+NLW+NLT
phenomenon (\rrref{Sect.}{sec:chromosphere}).

\paragraphrr{3D MHD Bifrost simulations}
Since these classic simulations the Oslo simulators went to 3D(t) MHD
Bifrost
\rrendnote{Bifrost program}{note:Bifrost}{%
This MHD simulation code of
\citetads{2011A&A...531A.154G} 
is named after old-Norse Bifr\"ost in the Icelandic Edda: {\em burning
rainbow bridge between Earth (Midgard) and the realm of gods
(Asgard\/} -- not too solid since built from fire, water and air but
slightly more inviting than Pandora's box).
Reference lists (until 2018 at the time of writing) are given in
\SSXpage{160\,ff}.
ADS abstract search {\em "Bifrost" and "solar"\/} gives publication
count and Hirsch $N\tis\,85$, $h\tis\,19$, solid enough for a solid
permanent position. 
{\em "MURaM" and "solar"\/} $N\tis\,96$, $h\tis\,20$ idem for the MHD
simulation program of
\citetads{2005A&A...429..335V}. 
Maybe you should simulate.} 
of \citetads{2011A&A...531A.154G} 
\SSXpage{159}.
The cubic Bifrost stars are far more solar-like than the
plane-parallel Pandora stars by being 3D(t) harboring dynamism and
magnetism.
Unlike the Pandora stars they contain realistic granulation,
overshooting granulation and gravity waves, acoustic waves (with box
modes as $p$-mode surrogate) and shocks, MCs arranged in network
patterns, dynamic fibrils, bipolar-network-connecting chromospheric
fibrils, and reconnective Ellerman bombs (EB). 
However, they still lack other solar features and their cube sizes are
yet too small for supergranulation, active regions, and larger-scale
coronal fine structure
\rrendnote{public Bifrost star}{note:Bifroststar}{%
Many Bifrost studies employ the 10-min sequence of snapshots of the
active-network simulation made public by
\citetads{2016A&A...585A...4C}. 
At currently 115 ADS citations it has become a ``standard model''
successor to the Pandora stars, after 35 years a worthy 3D(t) MHD
successor to the 1D static VALIIIC presenting a more realistic
rendering of active network than any plane-parallel model.
It shares the didactic virtue of the Pandora stars in permitting
experiments, tests, explanations.
However, it remains a solar analog, not the Sun, and just as the
Pandora stars which it presently replaces as canonical solar surrogate
it may mislead those who think that the Sun adheres to the model and
may similarly mislead use for teaching machine learning.
 
This particular Bifrost star has chromospheric fibrils connecting two
opposite-polarity plage patches, analyzed by Leenaarts \etal\
(\citeyearads{2012ApJ...749..136L}, 
\citeyearads{2015ApJ...802..136L}) 
as \Halpha\ fibril emulations.
However, their short-loop bipolar-rooted configuration resembles CBPs
but without CBP-like heating, perhaps from lack of further emergence
and/or small-scale reconnection as do occur in the more active Bifrost
star of \citetads{2019A&A...626A..33H} 
used by \citetads{2020A&A...643A..27F} 
which also sports Ellerman bombs (\rrref{endnote}{note:EBs}) and UV
bursts (\citeads{2018SSRv..214..120Y}). 

The public Bifrost star also lacks spicules-II and return fibrils
constituting opaque internetwork canopies in \Halpha, so missing an
important quiet-chromosphere ingredient unlike what the actual Sun
would show around such network.
This shortcoming may be due to lack of \eg\ resolution, small-scale
vorticity, small-scale charged-neutral separation (\cf\
Mart{\'{\i}}nez-Sykora \etal\
\citeyearads{2017ApJ...847...36M}, 
\citeyearads{2020ApJ...889...95M}; 
N{\'o}brega-Siverio \etal\
\citeyearads{2020A&A...633A..66N}, 
\citeyearads{2020A&A...638A..79N}; 
\citeads{2021A&A...654A..51B}; 
\citeads{2021RSPTA.37900176K}). 
}.

Bifrost spectral synthesis has mostly concentrated on the IRIS
diagnostics. 
Obviously it beats Pandora-star spectral synthesis in NLS, although
most analyses rely for tractability on columnar modeling with the
RH\,1.5D program of
\citetads{2015A&A...574A...3P}. 
The Pandora stars handle NLW better with respect to ultraviolet
ionization and the problematic ultraviolet line haze
(\rrref{endnote}{note:haze}).
Various Bifrost runs included NLT for hydrogen (as in the classics
above), but not in subsequent spectral synthesis except the NLT
analysis of EUV helium lines by
\citetads{2017A&A...597A.102G}. 
Recognizing that the chromosphere is a domain of intermittent
ionization-recombination cycling between neutral and charged the
emphasis is now on combining NSE hydrogen with multi-fluid ambipolar
diffusion and Hall effect (Mart\'{\i}nez-Sykora \etal\
\citeyearads{2017ApJ...847...36M}, 
\citeyearads{2020ApJ...889...95M}, 
N\'obrega-Siverio \etal\
\citeyearads{2020A&A...633A..66N}, 
\citeyearads{2020A&A...638A..79N}). 

\section{NLS+NLW+NLT chromosphere spectrum} 
\label{sec:chromosphere}
SE is usually a sound assumption for thick spectral modeling of
photospheric structures (SE NLTE) and for thin spectral modeling of
coronal structures (SE CE). 
Chromospheric structures in between are the hardest to model.
In these hydrogen varies between neutral and ionized
and NLT reigns next to NLS and NLW in cycling them through thick and
thin
\rrendnote{photosphere, chromosphere, corona: hydrogen ionization}
{note:ionization}{%
The Sun is 90\% hydrogen. 
Neutral in the photosphere, ionized in the corona. 
The chromosphere in between is the domain of partial hydrogen
ionization, complexly small-scale and time-dependent.
A radiative-equilibrium atmosphere has outward temperature decline
without hydrogen ionization and without further interest: much solar
atmosphere physics is about ionization, essentially of hydrogen.  
The large threshold per particle provides a substantial sink of energy
in heating and gain of energy in cooling.
Caveats:
\begin{itemize}   \vspace*{-1.2ex} \itemsep=-0.8ex
\item the sink and gain can be non-local radiation, even for
optically thick features when effectively thin;
\item the key threshold is not 13.6\,eV as commonly coded but the 10.2\,eV
\Lyalpha\ jump because the rest concerns the Balmer continuum from/to
elsewhere (\rrref{Sect.}{sec:Oslo}); 
\item not plane-parallel time-independent as in Pandora stars but
small-scale and rapid in explosive events,
repetitively cycling in ubiquitous heating events as spicules-II;
\item not obeying SE as in most modeling but significantly retarded in
cooling phases (\rrref{Sect.}{sec:Oslo}).
\end{itemize} \vspace*{-1.2ex} %
At photospheric densities and temperatures where hydrogen is neutral a
simple recipe (following Mihalas in \RTSApage{166}{Sect.\,7.2.2}) is
to set $N_\rme\tis\,10^{-4}\,N_\rmH$, the contribution from the
electron donors. 
At chromospheric densities NSE (non-E) evaluation is mandatory for
hydrogen ionization due to the large \Lyalpha\ jump
(\rrref{Sect.}{sec:Oslo}). 
Simulations hence must treat \HI\ ionization with multilevel RT as in
\citetads{2007A&A...473..625L}. 
Reasonable shortcuts are to assume detailed radiative balancing in
\Lyalpha\ and $T_{\rm rad}\tis\,5300$\,K for the impinging Balmer
continuum. 
A snapshot from such NSE simulations may yet utilize RH
(\rrref{endnote}{note:RH}) to portray spectral diagnostics by using
the time-lagged electron density from the simulation.  
For hydrogen diagnostics of cooling gas the retarded hydrogen
$n\tis\,2$ population must also be imposed, instead of the much
smaller instantaneous value found by RH assuming SE.
SE is then still valid for the top of the hydrogen atom from
$n\tis\,2$ including Balmer photoionization, Rydberg recombination
cascade, \Halpha\ photon suction. 
Similar retarded He-top treatment is likely required for He
diagnostics including AIA \HeII\,304\,\AA.}
\rrendnote{no transition region}{note:noTR}{%
I skip the so-called ``transition region'' because none such exists as
a separable domain in the solar atmosphere. 
Plane-parallel stars need such a layer to furnish substantial lines
with formation temperatures around $10^5$\,K as observed in the solar
spectrum. 
The Sun doesn't have a spherical shell like that. 
But it is okay to write ``TR line'' if you mean transition radiation
(\rrref{endnote}{note:sheaths}).

The shocks in the
\href{https://robrutten.nl/rrweb/rjr-movies/hion2_fig1_movie.mov}{Fig\,1
movie} of \citetads{2007A&A...473..625L} 
lift the corona (let's define that as fully ionized hydrogen) above
them tremendously up and down, doing so effortlessly with a steep
space-time-varying temperature jump as instantaneous highly-warped
envelope.
The simulated corona there reaches deeper in the MCs.
This simulation describes quiet chromosphere; in active regions the
outbursts are yet further from 1D shell description.
The dynamics in this simulation are mostly acoustic shocks with more
field guiding in the network, less in the internetwork. 
The simulation does not have spicule-II jets emanating from network
which would lift and warp the internetwork transition to the corona
yet more.
This transition is not a shell but a highly dynamic thin envelope of
the chromosphere (\rrref{endnote}{note:nonEchrom}) locally producing
transition radiation (\rrref{endnote}{note:sheaths}).}.
For a review see
\citetads{2019ARA&A..57..189C}. 

\paragraphrr{Quiet chromosphere}
Away from active regions the chromosphere extends above and from
clusters of photospheric MCs making up plage and network, the latter
partially outlining clapotispheric internetwork ``cell interiors''
with occasional MCs on their way to cell boundaries
\rrendnote{internetwork fields and basal flux}{note:basal}{%
Internetwork\footnote{Naming: H.\,Zirin
\label{note:Zirin}
and (ex-)associates used
``intranetwork'' arguing that one drives interstates on, not off, the
highway while not appreciating that intravenous means within veins
rather than between them. 
Zirin's wife was a linguist and corrected his books -- but only the
text, not the equations so that he blamed her for the Planck functions
in Eqs.\,4.29\,--\,4.32 of
\citetads{1988assu.book.....Z} 
inspiring \IARTpage{51}{Question\,4.8}.
All four are wrong; his Rayleigh-Jeans limit goes infinite with
wavelength whereas Planck put the bosonic -1 in his best-fit
distribution to avoid Wien infinity ({\rrref{endnote}{note:bosons}}).}
MCs riding supergranular outflows to cell boundaries as the
``persistent flasher'' of
\citetads{1994ASIC..433..251B} 
are traditionally identified as disjoint members of newly emerged tiny
bipolar ephemeral regions
(\citeads{1973SoPh...32..389H}) 
on their way to sustain quiet network
(\citeads{1997ApJ...487..424S}, 
Sect.\,5.1.2 of Schrijver and Zwaan
(\citeyearads{2000ssma.book.....S}, 
\citeyearads{2008ssma.book.....S}). 
Such quiet network tends to be small-scale bipolar whereas plage in
active regions and plage emerging in smaller failed active regions
without sunspots tends to be unipolar obeying the large-scale dipole
nature of active region emergence. 
Subsequent shredding into more open plage and active network then adds
bipolarity.

More recently another type of internetwork field became attributed to
local field production by a granular-scale near-surface turbulent
dynamo.
Its action was observationally suggested from Hanle depolarization
(\citeads{2004Natur.430..326T}) 
and by the detection of abundant primarily horizontal weak fields at
granular scales in sensitive full-Stokes spectropolarimetry with
Hinode (Lites et al.\ \citeyearads{2008ApJ...672.1237L}, 
\citeyearads{2017ApJ...835...14L}) 
and simulated by \citetads{2010ApJ...714.1606P}. 
The primarily horizontal signature implies that opposite-polarity
closing loops occur on granular scales within the photosphere, a lower
and denser ``magnetic carpet'' than the traditional network-scale one
made by ephemeral regions and invoked to supply coronal heating
(\citeads{2001ApJ...561..427S}, 
\citeads{2002ApJ...576..533P}). 
No direct chromospheric interest therefore, but subsequent
supergranular-flow migration giving frequent encounters causing
cancelation or enhancement up to fluxtube collapse led
\citetads{2014ApJ...797...49G} 
to suggest that this superficial dynamo action competes with ephemeral
active region emergence in quiet-network replenishment

Since this contribution should operate similarly in any star with
convective granulation and supergranulation, even stars without any
internal dynamo, it may contribute cycle-independently to building
network and hence to cool-star basal flux in addition to acoustic
brightening.\footnote{Naming: defining a minimal level of
chromospheric emission in solar-type stars was a major activity at
Utrecht initiated by C.J.\,Schrijver by plotting X-ray fluxes from the
Einstein satellite and UV fluxes from the IUE satellite against
emission in \CaII\ \HK\ measured at Mount Wilson (manually on log-log
graph paper) and noticing the need to include a ubiquitous
chromospheric pedestal (\eg\
\citeads{1981SSRv...30..191M}, 
\citeads{1982AdSpR...2i.243S}, 
\citeads{1986LNP...254...19Z}, 
\citeads{1987A&A...172..111S}, 
\citeads{1991A&A...252..203R}) 
with an overview in \linkadspage{1991A&A...252..203R}{16}{Fig.\,5} of
the not-me last. 
Schrijver named it ``basal flux''. 
\citetads{1991ApJS...76..383D} 
summarized the Mount Wilson monitoring; reviews are by Schrijver
(\citeyearads{1987LNP...291..135S}, 
\citeyearads{1995A&ARv...6..181S}) 
and in Sect.\,2.7 of Schrijver and Zwaan
(\citeyearads{2000ssma.book.....S}, 
\citeyearads{2008ssma.book.....S}).} 
In solar quiet-Sun ultraviolet images (TRACE 1550, 1600, 1700\,\AA;
SDO 1600, 1700\,\AA; IRIS 1400\,\AA\ slitjaws) network (also quiet
network) usually shows larger brightness contribution (from MC hole
deepening) than internetwork acoustics. 
When network produced by the surface dynamo outweighs the acoustic
contribution by internetwork shocks it may dominate the basal flux
observed at these wavelengths (Table\,2.7 on p.\,68 of
\citeads{2008ssma.book.....S}). 
\citetads{1989ApJ...341.1035S} 
found that basal flux in \CaII\ \HK\ from solar-type stars equals the
solar flux in supergranular cell centers dominated by internetwork
shocks, but \citetads{1992A&A...258..507S} 
found additional basal contribution in weak network and plage.
With a cool-star survey
\citetads{2021ApJ...910...71A} 
also found evidence that basal flux measures the presence of network
not made by an activity dynamo.}.

In the optical the quiet chromosphere is seen only in Balmer lines,
\CaII\ lines and \HeI\ lines. 
The Lockyer definition implies it is made up by the fibrils seen
around network in the \Halpha\ core and with lesser extent in \CaIR\
(\rrref{endnote}{note:nonEchrom}).\footnote{\CaII\ \HK\ are richer in
sampling chromospheric canopies than \Halpha, as in
\linkrjrpage{2017ApJS..229...11J}{5}{Fig.\,2} of
\citetads{2017ApJS..229...11J} 
and in \linkadspage{2019A&A...631L...5B}{4}{Fig.\,3} of
\citetads{2019A&A...631L...5B}, 
partly thanks to PRD source function splitting but also 
because \HK\ do not skip the clapotisphere in their
inner wings and have six times less wide thermal broadening.}
\Halpha\ fibrils around network are largely made by
spicules-II
\rrendnote{spicules-II}{note:spicules-II}{%
In my (not so humble) opinion these are the main agents causing the
fibrilar \Halpha\ and sheath-radiation \HeII\,304\,\AA\ quiet-Sun
chromospheres, being the only candidate agent sufficiently ubiquitous
even in quiet network.
I use the name generically for off-limb and on-disk manifestations.

I noted them as ``straws'' near the limb in
\href{https://robrutten.nl/dot/dotweb/dot-movies/2003-06-18-QS-mu034-ca-core.mpg}{DOT
\CaIIH\ movies} (\linkrjrpage{2006ASPC..354..276R}{2}{Fig.\,1}) of
\citeads{2006ASPC..354..276R}).
They were then detected and measured using Hinode off-limb \CaIIH\
sequences, attributed to Alv\'enic waves and named ``spicules
type-II'' by \citetads{2007PASJ...59S.655D} 
and \citetads{2007Sci...318.1574D}
\footnote{Naming: no bibtex 2007a,\,b coding here because the
first spells Dutch ``de'' so that bibtex classifies a de/De split
personality (but wrongly: Dutch would be B.~de Pontieu but De Pontieu
and De Pontieu, B. as for De Jager, C. for Kees de Jager -- misspelled
as de Jager, C. by ADS in
\href{https://robrutten.nl/bibfiles/ads/bib/dejager.bib} {600$^+$
bibitems}, as many other Dutch). 
The proper spellings B. De Pontieu and L.H.M.~Rouppe van der Voort,
respectively, indicate where they originated. 
Presently ADS offers 4 solar publications by ``van der Voort, L.'',
102 for Kostyk, R.I. and 61 for Kostik, R.I. while half the responses
for ``Rutten, R.'' are not mine. 
What's in a name? 
Using 
my
\href{https://robrutten.nl/bibfiles/ads/namestrings/dircontent.html}
{ADS author strings} may help.
With these I maintain my
\href{https://robrutten.nl/bibfiles/ads/abstracts/dircontent.html}
{solar abstract collection}.}.

Spicules-I reach less high above the limb, crowd into the ``spicule
forest'' in \Halpha\ limb images, and were attributed to active-region
dynamic fibrils (\rrref{endnote}{note:DFs})  
by \citetads{2012ApJ...759...18P}. 

Spicules-II are ubiquitously ejected sideways from quiet-Sun network,
including monopolar network and also in coronal holes. 
On the disk these ejections were likely described earlier as ``fine
mottles'' (\cf\
\citeads{1957BAN....13..133D}), 
as upflows by \citetads{1966PhDT.........1T} 
in his thesis (now on ADS at my request) 
and as outer-wing \Halpha\ ``jets'' by
\citetads{1995ApJ...450..411S} 
and in later Big Bear reports.
More recently they became known as ``rapid blue-wing excursion'' (RBE)
(\citeads{2009ApJ...705..272R}) 
and ``rapid red-wing excursion'' (RRE)
(\citeads{2013ApJ...769...44S}), 
first in the wings of \CaIR\ and then in the outer wings of \Halpha,
with heating up to EUV visibility
(\citeads{2016ApJ...820..124H}). 

Spicule-II aftermaths of cooling recombining return flows cause dark
\Halpha-core fibrils around network (known as ``coarse mottles'' in
spectroheliogram days and listed as discovery 6 in the
\linkadspage{1962ApJ...135..474L}{1}{abstract} of
\citeads{1962ApJ...135..474L}) 
minutes later with their \Halpha\ opacities
probably much enhanced by NLT cool-down retardation as in the
post-shock clouds of
\citetads{2007A&A...473..625L}. 
Evidence are the darkest-darkest time-delay correlations in
\citetads{2019A&A...632A..96R} 
and the ubiquity of these downflowing \Halpha\ red excursions in
\citetads{2021A&A...647A.147B}. 
More in \rrref{endnote}{note:nonEchrom}.

The spicule-II driver mechanism is not known
(\citeads{2012ApJ...759...18P}) 
but because spicules-II exhibit torsion modes near and at the limb
(\citeads{2012ApJ...752L..12D}, 
\citeads{2013ASPC..470...49R}) 
vorticity in granular convection affecting embedded network fluxtubes
seems a candidate for kicking up Alv\'en waves.
The small vorticity drivers feeding upward Alv\'enic swirls along MCs
in the CO5BOLD simulations of
\citetads{2021A&A...649A.121B} 
are promising but should occur slanted because quiet-Sun chromospheric
heating spreads around network (\rrref{endnote}{note:nonEchrom}).
Other or additional agents may be fly-by reconnection as in
\citetads{2012SoPh..278..149M} 
and coupling to gravity waves (\rrref{endnote}{note:waves}).  

Comparing spicules-II with simpler dynamic fibrils in active-region
plage seems a promising topic for simulations. 
The latter (\rrref{endnote}{note:DFs}) are $p$-mode-excited
magneto-acoustic shocks along more crowded and less slanted fluxtubes
as waveguide.
They appear less important in active-region chromosphere heating than
spicules-II are in quiet-Sun chromosphere heating and return-flow
canopy formation.
Since the $p$-mode pattern undulations occur similarly over the whole
solar surface, sloshing waves up into all evacuated fluxtubes, other
agents must add dominance in spicules-II.   
I speculate that the suppression of granular convection in denser
plage (showing as ``abnormal granulation'') 
gives larger relative importance to undulation feeding of dynamic
fibrils whereas spicules-II suffer more small-scale torsion in
fiercer and vortex-richer normal granular convection and may also gain
from gravity-wave coupling by spreading over uninhibited overshooting
granulation in adjacent internetwork.
These agents also operate in unipolar network without flux emergence
or cancelation which in multiple quiet-Sun SST data sets also show
spicules-II, for example the data of my
\href{https://robrutten.nl/rrweb/sdo-demo/instruction.html}{alignment
practical}.\footnote{In conflict with the claim of
\citetads{2019Sci...366..890S} 
that all spicules represent strong-field reconnection, based on a
single data set in which juxtaposition of bipolar MC pairs and RBE
feet looks unconvincing and non-causal to me and which does not show
QSEBs as telltale of quiet-network reconnection.  
The notion also clashes with the attribution of polar plumes rather
than spicules to mixed-polarity network interactions by
\citetads{1995ApJ...452..457W} 
and \citetads{1997ApJ...484L..75W}. 
An obvious test is whether network that is more bipolar and
emergence-rich launches more RBEs.}.}
\rrendnote{RBE--RRE--fibril
ionization-recombination}{note:nonEchrom}{%
RBEs and RREs reach higher degrees of hydrogen ionization than the
shocks in the
\href{https://robrutten.nl/rrweb/rjr-movies/hion2_fig2_movie.mov}{Fig\,2
movie} of \citetads{2007A&A...473..625L}. 
There are more and they extend further in \Halpha\ than in \CaIR\
(\linkadspage{2009ApJ...705..272R}{4}{Fig.\,2} of
\citeads{2009ApJ...705..272R}, 
\linkadspage{2012ApJ...752..108S}{3}{lower-left panel of Fig.\,1} of
\citeads{2012ApJ...752..108S}) 
because \CaII\ ionizes before \HI\footnote{The \CaII\ ionization limit
from the ground state is at 11.9\,eV
(\linkpdfpage{https://robrutten.nl/rrweb/rjr-archive/linelists/aq1976.pdf}
{47}{paragraph 16} of venerable
\citeads{1976asqu.book.....A}); 
the edge from the metastable $3d$ levels
(\linkadspage{1968pgda.book.....M}{16}{Grotrian diagram} in venerable
\citeads{1968pgda.book.....M}) 
covers \Lyalpha\ so that large $J_\nu$ in \Lyalpha\ within heated
features kills their \CaII\ lines.}
and is much less abundant. 
Where their tips reach AIA EUV visibility
(\citeads{2016ApJ...820..124H}) 
hydrogen is nearly or fully ionized.

The resulting recombining return fibrils
(\citeads{2019A&A...632A..96R}) 
are more numerous and also much longer in \Halpha\ than in \CaIR\
(\linkrjrpage{2009AAp...503..577C}{4}{Fig.\,3} of
\citeads{2009A&A...503..577C}); 
larger \Halpha\ opacity is also seen in the aftermath of the exemplary
return contrail in \linkrjrpage{2017AAp...597A.138R}{6}{Fig.\,5} of
\citetads{2017A&A...597A.138R}. 

The bottom row in \linkrjrpage{2009AAp...503..577C}{7}{Fig.\,6} of
\citetads{2009A&A...503..577C} 
shows correlations between \Halpha\ core width and \CaIR\ core width
and profile-minimum intensity, instantaneously and time-averaged per
pixel. 
These diagrams quantify the scene similarities in the first, third and
fourth-panel scenes in their
\linkrjrpage{2009AAp...503..577C}{4}{Fig.\,3} and
\linkrjrpage{2009AAp...503..577C}{5}{Fig.\,4}.
They suggest that both core widths reflect temperature, with
correlation calibration in
\linkrjrpage{2009AAp...503..577C}{9}{Fig.\,9}.
The \CaIR\ Doppler-following minimum intensity also responds to
feature temperature but the \Halpha\ minimum intensity does not
(scatter diagrams in the next higher row in
\linkrjrpage{2009AAp...503..577C}{7}{Fig.\,6}). 
Both are scattering lines with similar outward $\sqrt{\varepsilon}$
$S(\tau)$ decay to the feature surface (\rrref{endnote}{note:Ha-RE})
but they differ strongly in opacity-defined return-fibril mapping with
the \Halpha\ opacity much enhanced by NLT retardation
(\rrref{Sect.}{sec:Oslo}).

You can check these findings yourself with the higher-resolution SST
quiet-Sun data in my
\href{https://robrutten.nl/rrweb/sdo-demo/instruction.html}{alignment
practical}. 
Its last {\tt showex} command browses SST \Halpha\ and \CaIR\ image
sequences sampling many wavelengths and also adding the three core
measures of \citetads{2009A&A...503..577C} 
for both lines. 
Blink these while activating scatter diagrams and also hit the button
for sequence time-averaging. 
The well-correlated width-width blink shows how much further the
\Halpha\ spicules-II reach.
Time-delay blinking \Halpha-0.6\,\AA\ against \Halpha+0.2\,\AA\
confirms
\citetads{2019A&A...632A..96R} 
with yet better data showing striking darkest-darkest correlation
between RBEs and subsequent redshifted core fibrils at about 5-min
delay, and also that these reach close to the spicule-II firing
network.
This suggests that in the cooling returns \Halpha\ gains largest NLT
overopacity keeping it $\sqrt{\varepsilon}$ dark much beyond the
actual lowering temperature.
Blinking the core minima and shifts for the two lines shows that in
\CaIR\ these return fibrils get visible only close to network where
finally \CaII\ recombines from \CaIII. 
Thus, large spicule-II H ionization is followed by cooling return
recombination producing core fibrils that are much more prominent in
the \Halpha\ core.
In contrast, in the ALC7 star the ``chromosphere'' is ha-ha-sampled
the same in the two lines (\rrref{endnote}{note:Ha-RE}).

The few-percent ionization in the shocks in the
\href{https://robrutten.nl/rrweb/rjr-movies/hion2_fig2_movie.mov}{Fig\,2
movie} of \citetads{2007A&A...473..625L} 
with its two-orders of magnitude electron density increase already
suffices to reach near-SB proton densities (3rd movie row).
Hence large electron densities and SB partitioning must also occur in
spicules-II, warranting straightforward SB extinction comparisons as
in \linkrjrpage{2017AAp...597A.138R}{7}{Fig.\,7} of
\citetads{2017A&A...597A.138R} 
for the hot spicule-II phases while recognizing that these high
extinction values persist NLT-wise afterwards in the recombining
return fibrils constituting opaque canopies while instantaneous SB
values drop Boltzmann-steep with their temperature.
Similar SB curves imply the same for the mm wavelengths of ALMA 
(\rrref{endnote}{note:ALMA-temp}).

Summary: in spicule-II heating the hydrogen-top populations and
extinction slide fast up along the SB curves to high values at large
hydrogen ionization, but in subsequent cooling they hang minutes near
these high values. 
They would eventually drop back to the SB value for the actual
temperature but get refreshed before that at the 1.4~min RBE
recurrence rate
(\citeads{2013ApJ...764..164S}). 
The resulting temporary hydrogen-top overpopulations can be large, as
the enormous $b_2\tis\,10^{12}$ values reached in post-shock cooling
in the
\href{https://robrutten.nl/rrweb/rjr-movies/hion2_fig1_movie.mov}{Fig\,1
movie} of \citetads{2007A&A...473..625L} 
(\rrref{Sect.}{sec:Oslo}), and so furnish fibrilar
$\sqrt{ \varepsilon}$ \Halpha\ visibility far beyond the SB
expectation. 
The NLT-opaque fibrilar appearance of the quiet \Halpha\ chromosphere
is due to continuous refurbishment by frequent small-scale ionization
and retarded recombination: {\em the chromosphere ain't stacked in
layers but is dynamically structured and unstuck in
time\/}.\footnote{%
\href{https://en.wikipedia.org/wiki/Slaughterhouse-Five}{Billy
Pilgrim} syndrome. 
Multi-wavelength chromosphere observation is as Tralfamadorian
multi-time viewing; \Halpha\ fibril canopies show what happened
before
(\citeads{2019A&A...632A..96R}). 
So it goes.}}. 

In AIA UV images the quiet chromosphere is transparent. 
They sample clapotisphere underneath and yet deeper in MC holes in the
photospheric surface. 
These are cooler than their surroundings but appear brighter
(\linkrjrpage{2020LingAstRep...1R}{91}{Fig.\,75} of
\href{https://robrutten.nl/rrweb/rjr-pubs/2020LingAstRep...1R.pdf}{LAR-1}).

In AIA 304\,\AA\ images quiet areas show two major chromosphere
constituents: ubiquitous extended heating patches 
(``flocculi'' would be a good descriptor) and sparser brighter
feet of coronal bright points (CBP).
For a quick overview inspect the 10 ``triple'' image sets in
\linkrjrpage{2020LingAstRep...1R}{35}{Figs.\,17}\,--\,\linkrjrpage{2020LingAstRep...1R}{64}{46}
of
\href{https://robrutten.nl/rrweb/rjr-pubs/2020LingAstRep...1R.pdf}{LAR-1}
\rrendnote{quiet-Sun SDO images}{note:triples}{%
I recommend page-blinking the three members of each set in
\linkrjrpage{2020LingAstRep...1R}{35}{Fig.\,17\,--\,46} of
\href{https://robrutten.nl/rrweb/rjr-pubs/2020LingAstRep...1R.pdf}{LAR-1}.
I find it uncanny how well the detector images separate localized
quiet-chromosphere heating (ubiquitous grey patches around any network
but scarcer in coronal holes) from sparser localized coronal heating
(cyan CBP feet in denser bipolar network also in coronal holes). 
This separation implies that quiet-Sun atmospheric heating is not
traditionally ``via the chromosphere to the corona'' (with larger
energy need for the heavier chromosphere) but that it is done by
unrelated mechanisms for quiet chromosphere (Alfv\'enic-wave
spicules-II in my opinion) and quiet corona (reconnective CBPs in my
opinion).}.
The first of each triple is a large-field 193\,\AA\ image showing
coronal activity, coronal holes, and CBPs.
The second is a clipped ``fire detector'' multiplication of precisely
co-aligned 304\,\AA\ and 131\,\AA\ images in which grey patches show
heated quiet chromosphere and cyan-colored pixels mark CBP feet.
The third is the corresponding HMI magnetogram showing what underlies
and probably causes the CBPs.  
The 10 triple sets sample low and high activity and low and high
latitude and permit inspecting multiple coronal holes.

The heated chromosphere patches in the detector images appear
everywhere around network but fewer in coronal holes.
Their surface patterns are cospatial with the fibrilar \Halpha\
canopies around network
(\linkrjrpage{2020LingAstRep...1R}{11}{Appendix~B} of
\href{https://robrutten.nl/rrweb/rjr-pubs/2020LingAstRep...1R.pdf}{LAR-1})
and serve in my SDO pipeline
\rrendnote{SDO\,--\,STX alignment}{note:SDO-STX}{%
My SDO pipeline
(\href{https://robrutten.nl/rridl/00-README/sdo-manual.html}{manual})
consists of
\href{https://robrutten.nl/rridl/sdolib/dircontent.html}{IDL programs}
to request and download SDO sequences for any field, location, time,
duration, for precise co-alignment with image sequences from another
smaller-field higher-resolution telescope (STX = Solar Telescope X).
I started it for the DOT in its final year of operation (2010, the
first year for SDO) and have refined it with many datasets from the SST.

The requesting and downloading gets SDO data for your specified target field
and also for a large disk-center field for internal SDO
cross-alignment.
This is desirable because the different SDO diagnostics are usually
offset from each other over multiple pixels, with
\href{https://robrutten.nl/rridl/00-README/AIA-HMI-2013-12.pdf}{daily
modulation} from earthshine irradiation peaking at local noon around
19\,UT. 
Much larger offsets occur during thermal resettling after SDO eclipses
taking place near equinox. 

The pipeline finds the internal SDO shifts from the large disk-center
fields by cross-correlation in a sequence of best-fitting pairs after
image manipulation to make them more alike and tiling them into many
subfields for zonal averaging with projection-shift corrections using
apparent limb heights determined from full-disk tiling.
Examples for 1600\,--\,1700\,\AA\ cross-alignment are shown in
\linkrjrpage{2020LingAstRep...1R}{88}{Fig.\,70} -
\linkrjrpage{2020LingAstRep...1R}{90}{Fig.\,72} of
\href{https://robrutten.nl/rrweb/rjr-pubs/2020LingAstRep...1R.pdf}{LAR-1}.
The hardest step is to cross-align the AIA EUVs to HMI for which I use
304\,\AA\ images\footnote{Earlier I matched 304\,\AA\ to 1600\,\AA,
later to 1700\,\AA, after matching these to HMI magnetograms, but
direct use of large quiet disk-center fields of 304\,\AA\ images and
magnetograms as anchor pair is better because the magnetograms suffer
smaller limbward projection shifts.
The tiled matching applies apparent 304\,\AA\ height 3600\,km derived
from full-disk azimuthal tile averaging as in
\linkrjrpage{2020LingAstRep...1R}{89}{Fig.\,71} of
\href{https://robrutten.nl/rrweb/rjr-pubs/2020LingAstRep...1R.pdf}{LAR-1}.} 
and magnetograms exploiting their network pattern equality in quiet
areas (\rrref{endnote}{note:sheaths}).
This elaborate cross-correlation procedure usually gives ten times
better SDO cross-alignment than the {\tt aia\_prep.pro}
startoff.\footnote{SDO image offsets are corrected by SolarSoft {\tt
aia\_prep.pro} using ``Master Pointing'' lookup tables determined from
limb fits.
These fits were first done once a week but since early 2019 at 3-hour
cadence and also retro-actively throughout the SDO database.
The 3-hr values are used by {\tt aia\_prep.pro} also for the earlier
dates unless {\tt no\_mpo\_update} is set.
The remaining 1-2\,px shifts vary on time scales of minutes and are
largely removed by my cross-correlation pipeline.
The scales and rotation angles of all AIA diagnostics were determined
by R.A.\,Shine from the 2012 Venus transit using the precise planetary
ephemeris. 
My
\href{https://robrutten.nl/rrweb/sdo-demo/instruction.html}{alignment
practical} for the 2019 Mercury transit suggests that they remain
valid.}
Sample results are shown in
\linkrjrpage{2020LingAstRep...1R}{93}{Fig.\,75} of
\href{https://robrutten.nl/rrweb/rjr-pubs/2020LingAstRep...1R.pdf}{LAR-1}.

The SDO sequences for the target field are first cross-aligned with
the center-field results and then co-aligned to the other telescope
(or vice versa) including correction of differential guiding
drifts.\footnote{The SDO target sequences are collected per JSOC {\tt
im\_patch} cutout either tracking or not tracking the nominal solar
differential rotation at the field center evaluated with SolarSoft
{\tt rot\_xy.pro}. 
The pipeline removes sub-pixel rotation modulation left in these
cutouts.
The SST and other groundbased telescopes correlation-track the
granulation pattern or some magnetic feature near field center which
usually drifts its own way with respect to nominal differential
rotation.} This is easy for groundbased telescopes when the imaging
sequences include granulation (as for SST wide-band MOMFBD
registration, \rrref{endnote}{note:restoration}) and then usually
reaches 0.1\,arcsec precision. 
For IRIS the longest-wavelength slitjaw sequences may correspond to
co-pointed outer-wing \CaIR\ imaging.
ALMA is difficult unless there is co-pointed \Halpha\ imaging
spectroscopy (\rrref{endnote}{note:ALMA-Ha}); otherwise GONG \Halpha\
may help (\rrref{endnote}{note:ALMA-GONG}), in the future SOLIS/FDP
\Halpha\ (\rrref{endnote}{note:ALMA-SOLIS}).

For small fields this pipeline is fast by not handling full-disk
images but only JSOC cutouts. 
All programs use IDL {\tt assoc} to avoid memory loads. 
Inspection is facilitated with browser {\tt showex.pro}
(\rrref{footnote}{note:showex}).

My
\href{https://robrutten.nl/rrweb/sdo-demo/instruction.html}{alignment
practical} instructing use of this pipeline first cross-aligns all SDO
diagnostics during the last Mercury transit with Mercury serving as
check. 
It then co-aligns the quiet-Sun SST data of
\rrref{endnote}{note:quietHa} with SDO. 
The disk-center coronal hole sampled there also sports a few ``SolO
campfires'' (\rrref{endnote}{note:campfires}).}
to cross-align AIA\,304\,\AA\ images directly to HMI magnetograms.
I attribute this component to spicules-II ejected from network.
These also occur in coronal holes.
The lack of corresponding fire-detector network there may be
traditionally ascribed to lack of EUV irradiation from above but may
also be due to lack of coronal contact cooling due to lower coronal
surround density
\rrendnote{transition radiation, sheath ionization, coronal contact
cooling} {note:sheaths}{%
The transition region in Pandora stars is sampled by \HeII\,304\,\AA\
(the \Lyalpha\ of \linkadspage{1968pgda.book.....M}{10}{hydrogen-like
He$^+$}) which is therefore called a transition-region line. 
However, the Sun is not a Pandora star and does not possess a
transition region, neither plane-parallel nor spherical
(\rrref{endnote}{note:noTR}). 
The line emits transition radiation.
Please use TR only for that. 

AIA\,304\,\AA\ images show chromospheric canopies everywhere in quiet
areas outside coronal holes, sampled full-disk over the past decade in
my \href{https://robrutten.nl/rrweb/rjr-pubs/2021LingAstRep...3R.pdf}
{LAR-3 SDO album}.
These transition-temperature canopies extend above and around magnetic
network and so their patterns over the disk are defined by the
underlying network patterns. 
I use this pattern correspondence to co-align AIA\,304\,\AA\ directly
to HMI magnetograms in my alignment pipeline
(\rrref{endnote}{note:SDO-STX}).  
The 304\,\AA\ canopies also resemble the fibrilar \Halpha\ canopies,
not in fibrilar detail but in location pattern, so that one may
co-align reversed-intensity quiet-Sun \Halpha\ (even from GONG,
\rrref{endnote}{note:ALMA-GONG}) and 304\,\AA\ images by pattern
correlation, as shown in blinkable
\linkrjrpage{2020LingAstRep...1R}{29}{Figs.\,11\,--\,12} discussed in
\linkrjrpage{2020LingAstRep...1R}{11}{Appendix~B} of
\href{https://robrutten.nl/rrweb/rjr-pubs/2020LingAstRep...1R.pdf}{LAR-1}.
This pattern similarity extends even to AIA\,131\,\AA\ images, making
the ``fire detector'' multiplication\footnote{Multiplication
emphasizes what they share in common: ``warm'' ($10^4$\,--\,$10^5$\,K)
chromospheric gas and brighter CBP feet.  
The detector algorithm uses iterative quiet-area intensity scaling to
set the greyscale clip and CBP threshold as described in the caption
of \linkrjrpage{2020LingAstRep...1R}{87}{Fig.\,69} in
\href{https://robrutten.nl/rrweb/rjr-pubs/2020LingAstRep...1R.pdf}{LAR-1}.}
of 304 and 131\,\AA\ images in the triple blinkers in
\linkrjrpage{2020LingAstRep...1R}{35}{Fig.\,17\,ff} a good
transition-temperature quiet-chromosphere mapper: the non-cyan grey
clouds everywhere in quiet areas except coronal holes.

The 304\,\AA\ and 131\,\AA\ emission around quiet network firstly
represents hot ionization phases, \ie\ hot tips of spicules-II and
their still hot initial aftermaths. 
This emission marks local heating to ``transition'' temperatures.
The \Halpha\,--\,\CaIR\ core width correlation fit in
\linkrjrpage{2009AAp...503..577C}{9}{Fig.\,9} of
\citetads{2009A&A...503..577C} 
extends as high as 60\,000\,K with many samples in the input
next-to-last panel of \linkrjrpage{2009AAp...503..577C}{7}{Fig.\,6}
reaching yet larger values.\footnote{ Such temperatures may seem
excessively high in terms of \Halpha\ formation in quiet Pandora stars
but it can go worse: \linkrjrpage{2007ASPC..368...27R}{18}{Fig.\,13} of
\citetads{2007ASPC..368...27R} 
shows matching \Halpha\ and 171\,\AA\ brightness in an active region.}
However, the fire detector images in the triple blinkers do not show
wreaths as pearl-studded crowns around magnetic network but fairly
smooth patches extending from it and also covering it. 
They extend further and appear smoother than the corresponding
\Halpha\ patches in
\linkrjrpage{2020LingAstRep...1R}{29}{Figs.\,11\,--\,12}.
Hence tip heating is only part of this ultraviolet chromosphere
mapping.  

Such TR canopies are largely absent in coronal holes although these do
contain spicules-II as exemplified by \Halpha\ RBEs and RREs in the
high-resolution SST {\tt showex} display of a disk-center coronal hole
in my
\href{https://robrutten.nl/rrweb/sdo-demo/instruction.html}{alignment
practical}.  
The \linkrjrpage{2021LingAstRep...3R}{4}{2014\,--\,2015} active-Sun
triptychs in LAR-3 show good correspondence between the many coronal
holes in AIA 304\,\AA\ and 171\,\AA, with the latter somewhat wider.
Also the very quiet \linkrjrpage{2021LingAstRep...3R}{6}{2019}
sampling shows a fair match for the less prominent non-polar holes.

A traditional explanation for quiet-Sun emission in \HeII\,304\,\AA\
is \HeI\ ionization by EUV irradiation from above, as also invoked for
coronal-hole visibility in \HeI\,10830\,\AA\ equivalent width
(\rrref{endnote}{note:HeI}), but generally the 193\,\AA\ images in the
\linkrjrpage{2020LingAstRep...1R}{35}{Fig.\,17\,ff}
LAR-1 triple blinkers show wider-spread diffuse brightness providing blanket
irradiation without network patterning. 

Another candidate is 304\,\AA\ emission from local helium ionization
by kinetic sheath heating in thin skin-like boundary layers around
cooler and denser structures by turbulent mixing with surrounding hot
coronal gas per Kelvin-Helmholtz instability, as proposed for
prominence threads by
\citetads{2019ApJ...885..101H} 
in the form of contact coronal cooling. 
Their article was an eye-opener to me because it suggests another
mechanism to explain that the quiet-Sun chromospheric canopies
observed in \HeII\,304\,\AA\ match those in \Halpha\ so closely in
their patterning across the Sun. 
With this mechanism the first result from turbulent contact cooling of
the corona surrounding the latter.

A RT aspect beyond thin understanding is that if such turbulent sheath
heating by the adjacent corona reaches \HeI\ ionization and
\HeII\,304\,\AA\ emission then it spreads radiatively into the cool
substrate since 304\,\AA\ lies in the bound-free edge beyond the \HeI\
ionization threshold at 504\,\AA\ \RTSApage{196}{Table\,8.1}.
Therefore 304\,\AA\ photons penetrating into the cool structure cause
more ionization, enabling more 304\,\AA\ photons to scatter yet
further, eating their way in as scattering aureole
(\rrref{endnote}{note:aureoles}).
Irradiative ionization likewise affects denser neutral-helium
structures sheath-wise over the penetration depth.
  
Both sheath ionization mechanisms may affect cooling recombining
return fibrils from spicule-II ejections and explain that the
304\,\AA\ canopies extend across network and also that 304\,\AA\ shows
fewer canopies in coronal holes where both lower EUV irradiation and
lower surrounding-corona density lessen sheath ionization.  

I then wondered whether the active-network (``plagette'') \Halpha\
contrail fibril in
\citetads{2017A&A...597A.138R} 
may be exemplary for post-spicule-II \Halpha\ fibrils in quieter
network (\rrref{endnote}{note:spicules-II}) since it inspired their
identification in
\citetads{2019A&A...632A..96R}. 
The images in \linkrjrpage{2017AAp...597A.138R}{2}{Fig.\,1}\,--\,3
suggest that both ionization mechanisms operate.
The first shows fair overall pattern matching of the active patch in
304\,\AA\ with 193\,\AA\ (similar in 211\,\AA, 335\,\AA, 94\,\AA\ but
different in 171\,\AA\ and 131\,\AA) suggesting irradiative
ionization.
The pattern is finer-scale in 304\,\AA\ but higher-density structures
can pick up more radiative ionization and likewise suffer more
turbulent interface heating.
In the cutout sequences in
\linkrjrpage{2017AAp...597A.138R}{3}{Fig.\,2} the fat RBE-like jet in
the first column reaches maximal extent in row 4 where it shows
heated-tip brightening in 304\,\AA\ and 171\,\AA.
In rows 5\,--\,8 the RBE jet retracts and the dark \Halpha\
line-center cooling contrail fibril develops. 
The bright streaks in 193\,\AA\ and in 171\,\AA\
(\linkrjrpage{2017AAp...597A.138R}{4}{Fig.\,3}) break up but the
304\,\AA\ streak lingers longer, suggesting turbulent sheath
ionization.
Both dark and bright \Halpha-center features appear mapped by
304\,\AA\ brightness which may be expected if they have similar gas
density. 

In contrast to this plagette quiet-Sun 304\,\AA\ canopies show no
matching 193\,\AA\ patterns in the
\linkrjrpage{2020LingAstRep...1R}{35}{Fig.\,17\,ff} LAR-1 triple
blinkers so that I suspect that turbulent sheath heating dominates
there. 
This suggests that the ubiquitous quiet-Sun canopies in the AIA
304$\times$131 fire detector images represent chromospheric heating
through kinetic coronal contact cooling as in
\citetads{2019ApJ...885..101H} 
with radiative cooling as in \rrref{endnote}{note:suction}.

In \Halpha\ line center they are finely structured by opacity
dependence on history, \ie\ duration after the slender spicule-II
causing them in their subsequent retarded \Lyalpha\ settling to the
cooling temperature and defining their individual opacity and
$\sqrt\varepsilon$ darkness per fibril.   
Sheath-radiating 304\,\AA\ canopies sense only fibril gas density and
so can appear smoother as they indeed do.

Altogether this suggests that \Halpha\ (and ALMA,
\rrref{endnote}{note:ALMA-chrom}) as well as ``transition'' EUV lines
outline canopies with similar surface patterns by sampling the quiet
chromosphere around network in successive manifestations of
continually repeated small-scale brief heating perturbations followed
by cooling return flows (``dynamic refurbishment'',
\rrref{endnote}{note:nonEchrom}). 

There is no difference in ``height of formation'' between \Halpha,
ALMA and 304\,\AA\ canopies if these are effectively the same, unlike
layered-star stratification.
Observations confirm this equality.
The spicules-II producing \Halpha\ canopies show Balmer-line extent of
Lockyer's chromosphere to 4000\,km above the limb in
\linkadspage{1968ApJS...15..275D}{94}{Table\,3A} of
\citetads{1968ApJS...15..275D} 
and my corresponding
\href{https://robrutten.nl/rrweb/rjr-archive/linelists/dunnetal-decays.png}
{decay plot} while my estimate for the mean height of the 304\,\AA\
chromosphere from my SDO pipeline matching
(\rrref{endnote}{note:SDO-STX}) is 3600\,km (double the Pandora-star
value), also evident for the ALMA chromosphere in
\linkadspage{2018ApJ...863...96Y}{2}{Fig.\,1} of
\citetads{2018ApJ...863...96Y}.}. 

CBPs\footnote{Naming: solar physicists have a habit of calling small
unresolved features ``points'' needing a better name when they get
resolved.
\CaII\ \HtwoV\ and \KtwoV\ cell points became grains, facular points
(or filigree or magnetic bright points or G-band bright points) became
flowers and then MCs.
We need a better term for CBPs (also named X-ray bright points, XRBPs,
XBPs) -- heated small bipolar network-connecting loop groups?
Quiet-Sun EUV patches?}
are sparser but not rare in quiet areas. 
They are best seen in 193\,\AA\ images as small ensembles of
EUV-bright loops located at and connecting bipolar network
(silly-walled LRSP by
\citeads{2019LRSP...16....2M}). 
They appear as mini active regions, are likely due to ongoing flux
emergence and flux assembly into network, and are probably heated by
reconnection near their loop tops with subsequent heating of their
chromospheric feet
\rrendnote{CBP foot visibility}{note:CBPfeet}{%
CBP feet are bright in AIA\,304\,\AA\ and its
\linkrjrpage{2020LingAstRep...1R}{87}{``fire detector''} product with
131\,\AA\ but sharper in IRIS 1400\,\AA\ slitjaw images because \SiIV\
has similar temperature-defined presence as \HeII\ but 10$^3$ lower
abundance. 
This is evident for the complex CBP in the 1400\,\AA\ and 304\,\AA\
columns of \linkrjrpage{2017AAp...597A.138R}{3}{Fig.\,2} of
\citetads{2017A&A...597A.138R}, 
discussed at the end of
\linkrjrpage{2020LingAstRep...1R}{14}{Appendix~B} of
\href{https://robrutten.nl/rrweb/rjr-pubs/2020LingAstRep...1R.pdf}{LAR-1}.\footnote{The
ALMA-detected ``microflare'' of
\citetads{2021ApJ...922..113S} 
had similar large-CBP morphology and similarly occurred between
feet-heated bipolar patches of active network, not as an active-region
outburst.}
The SDO cutouts for a dozen CBPs in
\linkrjrpage{2020LingAstRep...1R}{66}{Fig.\,48\,ff} suggest additional
304\,\AA\ brightness around bright CBP feet due to EUV irradiation
(\rrref{endnote}{note:sheaths}).}
(\citeads{2021A&A...646A.107M}), 
likely by particle beams
(\citeads{2020A&A...643A..27F}, 
\cf\ 
\href{https://robrutten.nl/rrweb/rjr-pubs/2020LingAstRep...1R.pdf}{LAR-1}).
These features can live fairly long, the smallest
\rrendnote{SolO campfires}{note:campfires}{%
Small bright EUV features in the first Solar Orbiter/EUI image that
were press-released named and claimed as new coronal heating agent,
triggering
\citetads{2020arXiv200900376R}. 
I found them already reported as smallest SOHO/EIT CBPs by
\citetads{1998ApJ...501..386F} 
and less sharp but less hullabalooed present throughout the AIA
database without getting attention.
They are the smallest cyan features in the fire detector images in the
ten triple blinkers starting at
\linkrjrpage{2020LingAstRep...1R}{35}{Fig\,17}. 
Counting the many cyan features in these snapshots suggests that at
any moment smaller CBPs and campfires are more numerous than larger
CBPs but the latter live much longer (see fifteen cutout sequences
starting at \linkrjrpage{2020LingAstRep...1R}{66}{Fig.\,48}).

The cyan detector CBP feet are at chromospheric heights and are heated
from above as in
\linkadspage{2020A&A...643A..27F}{13}{Fig.\,12} of
\citetads{2020A&A...643A..27F} 
where features 1 and 2 resemble CBP-loop feet and feature 3 resembles
a small isolated low-lying campfire, loop-less conform the absence of
loop features at campfires in the AIA blinkers, but likewise heated by
a particle beam from above and shining harmlessly as an electric
St.~Elmo's fire.     
It marks, not causes, higher-up reconnective heating as a small
telltale.  
While writing \href{https://robrutten.nl/rrweb/rjr-pubs/2020LingAstRep...1R.pdf}{LAR-1} the work of 
\citetads{2020A&A...643A..27F} 
was an eye-opener to me because it explained how campfires
and larger brushfire feet represent features coming from above, 
as I had tentatively concluded from comparing a dozen
SoLO campfires with their SDO counterparts in
\linkrjrpage{2020LingAstRep...1R}{66}{Fig.\,48\,ff} there.

It is a mistake to sum observed CBP or campfire brightnesses as
contribution to coronal heating since this energy is gone: bright
momentary flames contain no information on non-lost heating
(\rrref{endnote}{note:suction}). 
The actual more steady and diffuse coronal heating patterns in the
AIA\,193\,\AA\ triple images suggest negligible heating from the
smallest campfires and most from the largest CBP reconnection events.}
the shortest, but AIA 193\,\AA\ movies show frequent change suggesting
continuous renewal.
The representative AIA 193\,\AA\ images in the triple blinkers
suggest that CBPs are the main agent in quiet-Sun coronal heating and
cause the diffuse AIA 193\,\AA\ brightness spread widely around them,
together and with contributions from prior CBPs.
Such diffuse brightness is absent in coronal holes where heating does
not spread to surrounding surface areas along closed loops but instead
follows open-field up and out, likely as coronal plumes
(\citeads{1997ApJ...484L..75W}). 

To reproduce such dynamic nature numerically, even for just ``quiet''
chromosphere, poses enormous challenges because the MHD simulations
must produce spicules-II as copiously as actual solar network does and
must then be combined with 3D(t) NLS--NLW--NLT (``NSE'') spectral
synthesis including PRD at least for \Lyalpha\ to recover the observed
\Halpha\ chromosphere
and in addition must produce small-scale bipolar
activity causing and governing CBPs as observed.

ALMA has the potentiality to sample and measure the quiet chromosphere
sharper and faster than its imaging by DKIST 
\rrendnote{quiet chromosphere with ALMA}{note:ALMA-chrom}{%
Outside rare truly quiet internetwork
(\rrref{endnote}{note:ALMA-waves}) ALMA samples chromospheric canopies
that are also but differently sampled by \Halpha\ and \Lyalpha\
(\citeads{2017A&A...598A..89R}) 
and by \HeII\,304\,\AA\ (\rrref{endnote}{note:sheaths}).
The
\linkpdfpage{https://robrutten.nl/rrweb/rjr-pubstuff/lar-2/ibismosaicblinker.pdf}{8}{IBIS
\Halpha\ core-width} scene discussed in \rrref{endnote}{note:IBIS}
represents an analogon of an ALMA active region scene (but mapping
rather than measuring temperature).  
The SST \Halpha\ core-width scene in the final {\tt showex} display of
my
\href{https://robrutten.nl/rrweb/sdo-demo/instruction.html}{alignment
practical} samples a quiet disk-center coronal hole and represents a
quietest-Sun ALMA analogon.

Claiming that longer ALMA wavelengths sample higher layers from the
$\lambda^2$ increase in
$\alpha_\lambda^{\rm ff}\tsim N_\rme\,N_\rmp\,T^{-3/2}\lambda^2$
\RTSApage{47}{Eq.\,2.79} is as plane-parallel wrong as interpreting
\Halpha\ images with the famous VALIIIC diagram \SSXpage{60}
(\rrref{Sect.}{sec:Avrett}). 
The quiet chromosphere is not layered but dynamically structured with
NSE hydrogen-top opacities (\rrref{endnote}{note:nonEchrom}).
ALMA brightness does measure temperature for optically thick features
but the feature proton densities and hence the $\propto\!\lambda^2$
feature opacities and hence the $\tau\tis 1$ feature sampling
locations were set NLT-wise in the recent past, with apparent feature
widening from \Lyalpha\ surround scattering. 
Some features may be thin at 1\,mm and thick at 3\,mm, others so thick
that the measured temperatures sample adjacent outer feature surfaces,
further out at longer wavelength.
The \Halpha\ opacities are also set NLT-wise but \Halpha\ and
\Lyalpha\ obtain relative feature contrasts from variations in
$\sqrt{\varepsilon}$ scattering with deep $\tau_{\rm eff}\tis 1$
signature and from Dopplershifts.

The large NLT opacity enhancements that the hydrogen top ($n\tis\,2$
and higher levels including the ion population) suffers in cooling
return fibrils after spicule-II heating jets
(\rrref{endnote}{note:nonEchrom}) implies that ALMA in quiet areas
likely samples chromospheric canopies as predicted in
\citetads{2017A&A...598A..89R} 
and pattern-matching those in \HeII\,304\,\AA\
(\rrref{endnote}{note:sheaths}). 

Studying the nature and formation of canopy features requires
follow-the-motion along-the-feature tracking with temporal delay
comparisons. 
Frequency-resolving Fourier power, phase difference and coherency
analysis should also be along-the-feature. 

ALMA already excels in obtaining fast cadence and potentially excels
in obtaining angular resolution\footnote{Addition of many longer
baselines may improve ALMA's sharpness beyond SST/CRISP \Halpha\
scenes, also beyond yet sharper SST/CHROMIS \Hbeta, even beyond DKIST
since $D/\lambda\tis\,10\,{\rm km}/1$\,mm equals
$4\,{\rm m}/0.4$\,{\rm $\mu$m} -- but DKIST will likely reach its
diffraction limit only beyond 1\,$\mu$m where seeing with
$r_0\tsim\,\lambda\,^{6/5}$ spoils less but nothing shows the
chromosphere except perhaps CO lines (\rrref{endnote}{note:CO})
. 
DKIST's sharpest chromosphere diagnostics will likely be the \CaII\
lines because these do not suffer apparent feature widening from
NLS surround scattering and NLT opaque-track lingering as the
Balmer lines and ALMA's continua do. 
The \CaII\ lines nicely combine near-SB opacity, without atmospheric
gaps as the Balmer lines have, with PRD-enriched Doppler sensitivity. 

For ALMA asymmetric $(U,V)$-plane filling yielding anisotropic
resolution is an option since the solar scenes consist of long thin
fibrils spread in all directions so that high resolution only
transverse to or along suited ones is already worthwhile, as with the
11-m GISOT design of
\citetads{2004SPIE.5489..491H} 
with an elliptic aperture
(\linkpdfpage{https://robrutten.nl/dot/dotweb/dot-pubs/gisot2004.pdf}{2}{Fig.\,1a})
to minimize prime-focus heat load.} 
while spectral profile resolution in chromospheric lines (\HI\ Balmer
and Lyman, \CaII\ and \MgII) offers larger opacity diversity and
Dopplershift encoding; combination of these diagnostics and/or less
saturated companions (\rrref{footnote}{note:Hbeta}) is desirable.}
\rrendnote{ALMA\,--\,\Halpha\,--\,SDO alignment}{note:ALMA-Ha}{%
Another confirmation that ALMA sees chromospheric canopies in
quiet-Sun areas is that there ALMA images can be co-aligned with
\Halpha\ images through cross-correlation.  
This is easy
on the disk with co-pointed \Halpha\ imaging spectroscopy delivering
temperature-mapping \Halpha\ core width (also shown in the IBIS\,--\,SDO
blinker of \rrref{endnote}{note:IBIS}).
The correspondence is not 1:1 because hot RBE and RRE tips, emanating
furthest from the spicule-II launching network, are not seen in
\Halpha\ core-width imaging which mostly charts cooling return
fibrils. 
Blurring may improve pattern correlation.  

Near the limb ALMA emphasizes hot tips of spicules-II
that appear as upright hedges of
straws, bright in \CaII\ \HK\ 
(\href{https://robrutten.nl/dot/dotweb/dot-movies/2003-06-18-QS-mu034-ca-core.mpg}{DOT
\CaIIH\ movie}, \linkrjrpage{2006ASPC..354..276R}{2}{Fig.\,1}) of
\citeads{2006ASPC..354..276R}),
dark in the \Halpha\ red and blue wings
(\linkrjrpage{2012ApJ...752L..12D}{5}{Fig.\,5} of
\citeads{2012ApJ...752L..12D}, 
\linkrjrpage{2013ASPC..470...49R}{4}{Fig.\,1} of
\citeads{2013ASPC..470...49R}) 
and with bright feet in IRIS 1400\,\AA\ slitjaw images
(\linkadspage{2018ApJ...863...96Y}{2}{Fig.\,1} of
\citeads{2018ApJ...863...96Y}). 
Hence, near-limb ALMA images may be cross-correlated to underneath
summed and reversed \Halpha\ blue-wing and red wing intensity.
Off the limb non-reversed \Halpha\ core intensity of sufficient
resolution may serve for tall spicules-II jutting out from the
spicule-I forest which appears opaque up to 3500\,km height in the
ALMA image and furnishes bound-free scattering blackness
(\rrref{endnote}{note:rain}) in the 171\,\,\AA\ image in
\linkadspage{2018ApJ...863...96Y}{2}{Fig.\,1} of
\citetads{2018ApJ...863...96Y}. 

The first co-pointed ALMA\,--\,\Halpha\ data were with DST/IBIS
(\citeads{2019ApJ...881...99M}) 
but IBIS has left the DST.
The DST is located at 2.5~hours later longitude than ALMA.  
The SST on La Palma is at 3.3~hours earlier, the GST at Big Bear
3.3~hours later, the DKIST on Maui 5.9~hours later. 
Afternoon seeing can be good at La Palma after the Sun has passed the
Caldera while the USA sites usually have best seeing soon after
sunrise.

An attractive alternative is to move the mothballed and relocatable 
\href{https://robrutten.nl/dot/DOT_home.html}{DOT} from La
Palma to ALMA.
It anyhow has to make room for the planned EST and it packs into a few
containers. 
On the disk two-channel speckle bursts in 3 or 5 \Halpha\ wavelengths
with its tunable Lyot filter would deliver qualitative core width
mapping, simultaneous granulation images giving co-registration with
SDO/HMI and other SDOs with my alignment pipeline
(\rrref{endnote}{note:SDO-STX}). 
Inside the limb summed and reversed \Halpha\ wings would do, off-limb
non-reversed \Halpha\ core summation, both with \CaIIH\ to
AIA\,1700\,\AA\ outer-disk co-alignment.  
Continuous co-pointing may then often enable precise co-alignment of
solar ALMA data with all SDO diagnostics. 
DOT mosaicing around the small ALMA field would supply
similar-diagnostic context imaging. 
Pointing mosaicing is straightforward with the DOT because it has a
parallactic mount (no image rotation) and does not use adaptive optics
(speckle restoration suffices at 45\,cm aperture).
Examples in \linkrjrpage{2007ASPC..368...27R}{9}{Fig.\,6} of
\citetads{2007ASPC..368...27R}
and \linkrjrpage{2017AAp...598A..89R}{6}{Fig.\,2} of
\citetads{2017A&A...598A..89R}; 
more in the 
\href{https://robrutten.nl/dot/dotweb/dot-albums/images/album.html}
{DOT image album}.}
\rrendnote{ALMA\,--\,GONG\,--\,SDO alignment}{note:ALMA-GONG}{%
In the absence of co-pointed \Halpha\ spectrometry 
(\rrref{endnote}{note:ALMA-Ha}) co-alignment
may be tried for quiet on-disk scenes with wider-band \Halpha\
core images exhibiting cooling return fibrils that remain dark through
non-SE overopacity. 
With greyscale reversal rough pattern agreement with the fibril
temperatures mapped by ALMA enables co-alignment, easiest for large
quiet fields.
This correspondence is not 1:1 also temporally because the cooling
return fibrils follow a few minutes after the hottest spicule-II tips
(\citeads{2019A&A...632A..96R}). 
Space-time blurring and delayed \Halpha\ sampling may improve pattern
matching.

Such images are almost always available from GONG\footnote{With large
quality variations with time and between stations.
Fortunately ALMA's longitude suits the usually better ones (Big Bear,
Cerro Tololo, Mauna Loa).}. 
Since these are full-disk they may contain a suited large quiet
closed-field area somewhere for precise co-alignment (with \Halpha\
intensity reversed) to simultaneous AIA 304\,\AA\ images as in the
blink for the first SolO/EUI target in
\linkrjrpage{2020LingAstRep...1R}{29}{Figs.\,11\,--\,12} of
\href{https://robrutten.nl/rrweb/rjr-pubs/2020LingAstRep...1R.pdf}{LAR-1}
and discussed in \linkrjrpage{2020LingAstRep...1R}{11}{Appendix\,B}
there. 
The GONG \Halpha\ part (also greyscale-reversed) covering the small
ALMA target field may then serve as ALMA-to-SDO intermediary for
wherever ALMA pointed.

I succeeded in this for the quiet-Sun ALMA data 
of \citetads{2021A&A...652A..92N} 
with suited GONG images sampled in their
\linkadspage{2021A&A...652A..92N}{4}{Fig.\,1}.
I found the second-step ALMA-to-GONG matching easier than the first-step
GONG-to-304\,\AA\ matching, and indeed found good pattern correlation
between the ALMA, GONG and AIA chromospheres. 
This suggests that many ALMA quiet-Sun on-disk sequences can be precisely
co-aligned with all SDO diagnostics via GONG.

For active regions the partial match between the
\linkpdfpage{https://robrutten.nl/rrweb/rjr-pubstuff/lar-2/ibismosaicblinker.pdf}{8}{IBIS \Halpha\ core-width scene} and
\linkpdfpage{https://robrutten.nl/rrweb/rjr-pubstuff/lar-2/ibismosaicblinker.pdf}{10}{AIA 304\,\AA} in the IBIS\,--\,SDO active region blinker of
\rrref{endnote}{note:IBIS}
suggests that direct ALMA to 304\,\AA\ alignment maybe the best to try
but it will be severely hampered by small ALMA field size and the AIA
304\,\AA\ deterioration (\rrref{footnote}{note:SDO-deterioration}).}
\rrendnote{ALMA\,--\,SOLIS\,--\,SDO alignment}{note:ALMA-SOLIS}{%
\linkadspage{2021A&A...651A...6B}{6}{Fig.\,2} of
\citetads{2021A&A...651A...6B} 
taught me unexpectedly that NSO not only furnishes full-disk \Halpha\
monitoring with GONG (\rrref{endnote}{note:ALMA-GONG}) but has done
this also with the Full-Disk Patrol extension of the SOLIS telescope.
Not around-the-world around-the-clock but sporadic during
2012\,--\,2017 with SOLIS first at Kitt Peak and then at Tucson.
Ever since then SOLIS is being moved to a site on the shore of Big
Bear Lake close to the GST -- I hope on a tall
\href{https://robrutten.nl/dot/dotshots/2009/codeso-2009.jpg}{Hammerschlag
tower} as in the
\href{https://nso.edu/wp-content/uploads/2020/09/ngGONG-one-pager-09112019.pdf\#page=2}{mockup
(pdf 2)} for future ng-GONG.

SOLIS/FDP takes \Halpha\ line-core images but also summed-wing images
with 0.5\,\AA\ bandpass at $\pm 0.3\,\AA$ from nominal line center. 
These inner-wing images emphasize the onset of dark outgoing
blueshifted spicules-II and the arrival of incoming redshifted dark
return fibrils and so show the agents producing and constituting
quiet-Sun internetwork canopies more directly than the GONG
\Halpha-core images of \rrref{endnote}{note:ALMA-GONG}.
I therefore expected that they suit better for two-step ALMA-SDO
co-alignment than through GONG. 
This proved correct.
 
I first ran tests with co-alignment program
\href{https://robrutten.nl/rridl/nisplib/nisp_sdo.pro}{\tt
nisp\_sdo.pro} trying both GONG and SOLIS/FDP image sequences from the
\href{https://nso.edu/data/nisp-data}{NSO NISP archive} on 2012-08-15
when the Sun was very quiet around disk center and AIA 304\,\AA\ still
had high S/N. 
I found that indeed the FDP summed-wing \Halpha\ scenes resemble
AIA\,304\,\AA\ more closely than simultaneous GONG images: for the
quiet area at disk center the Pearson cross-correlation coefficient
versus 304\,\AA\ increases from 0.3 for GONG \Halpha\
(first panel \linkrjrpage{2020LingAstRep...1R}{92}{Fig.\,74} of
\href{https://robrutten.nl/rrweb/rjr-pubs/2020LingAstRep...1R.pdf}{LAR-1})
to 0.7 for SOLIS/FDP summed-wing \Halpha. 
Hence co-aligning SOLIS \Halpha\ to AIA\,304\,\AA\ is much easier and
more reliable than for GONG \Halpha.

I then tested the two approaches with the data of
\citetads{2021A&A...651A...6B}. 
Their \linkadspage{2021A&A...651A...6B}{6}{Fig.\,2} shows that the
ALMA target was an active region but with some quieter area at
bottom-right.
As expected, the ALMA appearance there is much closer to reversed
SOLIS summed-wing intensity than to reversed GONG intensity.
I found that precise co-alignment using only a small subfield was easy
with the SOLIS images but impossible with the GONG images, even though
the former show larger image distortions and more image motion 
requiring substantial self-alignment and time averaging.

However, a zonal shift-per-tile analysis as in
\linkrjrpage{2020LingAstRep...1R}{88}{Fig.\,70} of
\href{https://robrutten.nl/rrweb/rjr-pubs/2020LingAstRep...1R.pdf}{LAR-1}
between the simultaneous GONG and SOLIS images shows radially
increasing offset between the two up to 2400\,km difference at the
limb.  
Most of this is projective height difference because the SOLIS/FDP
wing images sample canopy bases while the GONG and 304\,\AA\ images
emphasize outward extensions reaching much higher.
This can be corrected with a {\tt heightdiff} limb offset parameter in
\href{https://robrutten.nl/rridl/nisplib/nisp_sdo.pro}{\tt
nisp\_sdo.pro} but requires precise pixel scale calibration.
So far the archived SOLIS/FDP images are only level 1 without
flatfielding or calibration. 
When SOLIS finally goes into renewed operation at Big Bear which
promises much better seeing and when the image processing improves to
level 2 including precise scale calibration then SOLIS/FDP may become
the best monitoring facility to assist as intermediate step in
co-aligning ALMA observations precisely to SDO.}.

\paragraphrr{Active chromosphere}
For chromospheric activity phenomena I refer to Jongchul Chae's
contribution in this school because I have not worked on these except
EBs and FAFs
\rrendnote{Ellerman bombs and flaring active-region
fibrils}{note:EBs}{%
EBs are active-region MC reconnection events in the low photosphere.
After growing interest in the past decade\footnote{History: I 
\label{note:burst}
participated in \citetads{2011ApJ...736...71W}, 
\citetads{2013JPhCS.440a2007R}, 
\citetads{2013ApJ...774...32V}, 
\citetads{2015ApJ...808..133R}, 
\citetads{2015ApJ...812...11V}, 
\citetads{2016A&A...590A.124R}, 
\citetads{2018SSRv..214..120Y}, 
\citetads{2019A&A...626A...4V} 
and think that I helped boost interest in
\citetads{1917ApJ....46..298E}: 
after half a century of disregard it shot up in
\href{https://ui.adsabs.harvard.edu/abs/1917ApJ....46..298E/metrics}
{ADS citations} passing 100 in its centennial year, near 200 now. 
Moral: if your nice discovery goes unnoticed then relax waiting for
ADS to revive it.} 
and successful simulation by
\citetads{2017ApJ...839...22H} 
and \citetads{2020A&A...633A..58O} 
EBs\footnote{Naming: \label{note:bomber} my laptop got confiscated by
border police spotting the b* word in an EB-modeling email on its
screen. 
It then took them months to reclassify me from terrorist suspect to
\href{https://robrutten.nl/rrweb/rjr-pubs/2022Zenit...2....5R.pdf}
{harmless Ellerman b*er} -- better use b=burst as in
\citetads{2018SSRv..214..120Y}. 
Note that a solar EB releasing 20\,Gt
(\citeads{2015A&A...582A.104R}) 
is a bigger blast than the champion man-made ``Tsar Bomba'' hydrogen
b* of only 50\,Mt.}
now represent an \Halpha\ feature that is largely understood and the
best-established example of reconnection in the photosphere.

Here they are of chromospheric interest because the spectral
signature of these events is complex and instructive.  
EBs appear bright only in the \Halpha\ wings because their
\Halpha\ line cores remain shielded by overlying dark active-region
fibrils (\linkrjrpage{2013JPhCS.440a2007R}{4}{Fig.\,3} of
\citeads{2013JPhCS.440a2007R}). 
They are therefore similarly shielded and invisible for ALMA
(\linkrjrpage{2017AAp...598A..89R}{9}{prediction 9} of
\citetads{2017A&A...598A..89R} 
confirmed by \citeads{2020A&A...643A..41D}). 

EBs are not visible in the \NaID\ and \MgIb\ lines
(\citeads{1917ApJ....46..298E}, 
\citeads{2015ApJ...808..133R}) 
because they are hot enough to fully ionize neutral metals.
Similarly EBs appear bright in UV continua from Balmer continuum
brightening with neutral-atom bound-free opacity ionized away (as in
flare feet showing optical flares).  
\citetads{2019A&A...626A...4V} 
have defined and calibrated an EB detector using AIA\,1700\,\AA\
images which opens the entire SDO database for studying EB occurrence.

In AIA\,1600\,\AA\ images EBs gain additional brightness from the
\CIV\ lines in the passband, similarly to the large \SiIV\
enhancements in \linkrjrpage{2015ApJ...812...11V}{7}{Fig.\,4} and
\linkrjrpage{2015ApJ...812...11V}{9}{Fig.\,8} of
\citetads{2015ApJ...812...11V}. 
The latter figure also shows \MnI, \FeII\ and \NiII\ absorption blends
on high \SiIV, \CII\ and \MgII\ emission peaks.
\linkrjrpage{2016AAp...590A.124R}{5}{Figure\,2} of
\citetads{2016A&A...590A.124R} 
added a schematic EB into the last panel of Fig.\,1 of the simulation
of \citetads{2007A&A...473..625L} 
to show how these can arise from undisturbed EB surroundings, not part
of the phenomenon as suggested for IRIS bombs by
\citetads{2014Sci...346C.315P}. 

Because EBs are hotter and denser than internetwork and dynamic-fibril
shocks in which \Lyalpha\ already reaches collisional equilibrium
their \Halpha\ extinction obeys SB partitioning, permitting the
straightforward SB opacity comparisons in
\linkrjrpage{2016AAp...590A.124R}{9}{Fig.\,5} of
\citetads{2016A&A...590A.124R}. 
In this figure \Halpha\ reaches similar extinction maxima as the
\MgII\ and \SiIV\ IRIS lines, so that the actual EB \Halpha\ profile
under the canopy is probably similar to \MgIIk\ with a high peak
and a small central dip as in
\linkrjrpage{2015ApJ...812...11V}{7}{Fig.\,4} of
\citetads{2015ApJ...812...11V}, 
hence with much larger profile-summed brightness than observed.
It is a mistake to count the observed \Halpha\ EB wing enhancements as
contribution to coronal heating since this energy is radiated away
(\rrref{endnote}{note:suction}) while the \Halpha\ core photons are
boxed in by overlying NLT-opaque canopies but appear insufficient to
brighten these.

Hot EBs contain intense \Lyalpha\ giving \Halpha\ aureole opacity
boosting (\rrref{endnote}{note:aureoles}) and making the feature
visible even if the actual reconnection site is too small.
NSE recombination is slow so that such boosts linger longer, in a wake
if the reconnection progresses upward.  
The Zanstra mechanism for detour photon conversion in planetary
nebulae \SSXpage{144} \IARTpage{137}{Sect.\,8.3.3.1} likely also
operates in their cool surroundings.

Presently EB interest has moved to non-active quiet-Sun QSEBs
(``quiet-Sun Ellerman-like brightening'') marking ubiquitous but
hard-to-detect small-scale photospheric reconnection
(\citeads{2016A&A...592A.100R}, 
\citeads{2017A&A...601A.122D}) 
and best seen in SST/CHROMIS \Hbeta\
(\citeads{2020A&A...641L...5J}). 

FAFs were called ``transient loops'' by
\citetads{2009ApJ...701.1911P}. 
\citetads{2015ApJ...812...11V} 
used ``flaring arch filaments'' but I renamed them to ``flaring
active-region fibril'' in
\citetads{2016A&A...590A.124R} 
to avoid confusion with arch filament systems.
In active regions with much flux emergence they appear in AIA
1600\,\AA\ image sequences with shorter duration and more abrupt
changes than comparatively stable and roundish EBs, have elongated
morphology, and show fast apparent brightness motion along filamentary
strands.
Further description including EB and FAF distinction is in 
\linkrjrpage{2020LingAstRep...1R}{9}{Appendix~A} of
\href{https://robrutten.nl/rrweb/rjr-pubs/2020LingAstRep...1R.pdf}{LAR-1}.}.
Hence no discussion of spot chromospheres, flares, filaments, etc.\
here. 
I restrict this section to two exemplary active region scenes in
different diagnostics.

The first scene is the DST/IBIS\,--\,SDO multi-wavelength image mosaic
of \citetads{2012IAUSS...6E.511C} and Reardon and Cauzzi
(\citeyearads{2012decs.confE..20R}, \citeyearads{2012AAS...22020111R})
showing AR\,1092 near disk center on 2010-08-03 in the early days of
SDO\footnote{The 
\label{note:SDO-deterioration}
deterioration of AIA sensitivities, worst for chromosphere and
prominence telltale 304\,\AA, during the immensely productive first
SDO decade, bread and butter for very many colleagues thanks to its
wonderful open-and-easy access policy, makes me long for AIA-v2 --
with yet more pixels, yet more wavelengths, yet faster?}.  
I received these concurrent co-aligned images in 2011 from
K.P.~Reardon and combined them into
\href{https://robrutten.nl/rrweb/rjr-pubstuff/lar-2/ibismosaicblinker.pdf}
{this pdf blinker}\footnote{In
\label{note:pageblink}
a sequence suited to pairwise page blinking. 
Most pdf readers furnish blinking by flipping pages in-place with the
cursor arrow or page up-down keys. 
Full-page, fit-to-page, presentation-mode viewing may help. 
Multiple-page jumping may be possible with a back key.
Myself I instead use my {\tt showex} browser
(\rrref{footnote}{note:showex}).}. 
The IBIS selection samples \CaIR, \Halpha\ (also \Halpha\ core width
and core-minimum Dopplershift as defined in
\citeads{2009A&A...503..577C}), 
and the (small) equivalent width of \HeIDthree.
The SDO selection adds comparable HMI and AIA diagnostics. 

This IBIS--SDO blinker represents an active-region counterpart to the final
{\tt showex} inspection of the quiet-Sun SST and SDO sequences in my
\href{https://robrutten.nl/rrweb/sdo-demo/instruction.html}{alignment
practical}. 
These two solar displays are utterly different.
The SST set is a time sequence of a small quiet coronal-hole
field with only some network. 
The IBIS set is only a single snapshot, not permitting time-delay
cause-and-effect sequence inspection, but shows a larger active-region
field containing dense unipolar plage, mossy plage
\rrendnote{moss}{note:moss}{%
``Spongy'' active-region plage showing many short
loops and bright footpoints interspersed with darker patches at small
scales in 171\,\AA\ and other EUV lines (\eg\
Berger et al.\ 
\citeyearads{1999ApJ...519L..97B}, 
\citeyearads{1999SoPh..190..409B}, 
\citeads{1999SoPh..190..419D}, 
\citeads{2000ApJ...537..471M}, 
\citeads{2003ApJ...590..502D}). 
\citetads{2020ApJ...903...68P} 
differentiate moss and outflow regions in active-region plage where
the latter likely contribute to the slow wind and non-CME FIP 
generation
(\linkadspage{2020arXiv201008517R}{170}{Fig.\,8.8} of 
\citeads{2021LNP...978.....R}; 
\rrref{endnote}{note:FIP}).}, 
a large regular spot, active-region filaments, and even a small flare
originating in the spot which behaves as a CBP with loop brightening
and feet heating.
Blinking these images represents a vivid course in active-region
phenomenology and also exhibits aspects discussed above
\rrendnote{IBIS--SDO active-region blinker}{note:IBIS}{%
\href{https://robrutten.nl/rrweb/rjr-pubstuff/lar-2/ibismosaicblinker.pdf}
{Download here}. 
I made it in 2011 courtesy K.P.\,Reardon while teaching this course at
LMSAL (Palo Alto) and displayed it ever since in my teaching
escapades.
Most of these images have enhanced grey scales\footnote{Real
scientists do not fake-color intensity scenes: our eyes are good at
appreciating many shades of grey.
When color-coding graphs consider that colleagues may have difficulty
distinguishing red and/or green.}
by bytescaling the square root or the square of the measured intensity
plus clipping.
All have a small area near center magnified four times in the
upper-right corner for more detailed blinking. 

Here are my inventory and tentative interpretations (better blink
yourself! -- how in \rrref{footnote}{note:pageblink}):

{\em HMI Stokes $I$\,--\,$V$ (p1\,--\,p2)\/}: Stokes\,$I$ shows only
granules, pores and the spot, not the network and plage seen in
Stokes\,$V$ (higher resolution would show abnormal granulation in
plage).
Plage brightens in such continuum sampling only near the limb, from MC
transparency in slanted viewing (righthand cartoon in
\linkrjrpage{2020LingAstRep...1R}{91}{Fig.\,73} of
\href{https://robrutten.nl/rrweb/rjr-pubs/2020LingAstRep...1R.pdf}{LAR-1}).
The flare is along connections from the spot to opposite polarity in
the sunspot moat (blink p2\,--\,p6).

{\em HMI Stokes $V$\,--\, AIA\,1700\,\AA\/ (p2\,--\,p3)}: ultraviolet
MC brightening is not from higher-up heating (as commonly thought) but
from magnetic hole deepening (lefthand cartoon and caption of
\linkrjrpage{2020LingAstRep...1R}{91}{Fig.\,73} of
\href{https://robrutten.nl/rrweb/rjr-pubs/2020LingAstRep...1R.pdf}{LAR-1}).
Blinking these disk-center images shows co-spatiality but towards the
limb 1700\,\AA\ bright points shift relatively limbward, 1600\,\AA\
yet more, also not due to offset in formation height but to 
increasing MC transparency  (\rrref{endnote}{note:1600-1700}).
A few bright 1700\,\AA\ points in internetwork mark isolated MCs 
(\rrref{footnote}{note:basal}).
Elsewhere the grey fine structure in 1700\,\AA\ internetwork areas
represents clapotispheric shocks (\rrref{endnote}{note:CSshocks}).

{\em AIA\,1700\,\AA\,--\,\CaIR\ wing $I$ (p3\,--\,p4)\/}: in the
\CaIR\ wing MC brightening in plage and network is hole brightening
from less collisional damping and additional downdrafts, but the inset
suggests much shielding within plage.
Enhanced bright points in the sunspot moat are likely not EBs since
they lack corresponding excess 1700\,\AA\ brightness
(\citeads{2019A&A...626A...4V}) 
and may mark MC downdrafts common in moats
(\citeads{2007A&A...472..607B}). 
Internetwork areas show predominant reversed granulation (inset) and
also sporadic shock brightening in grains
(\linkrjrpage{2011AAp...531A..17R}{13}{Sect.\,5.5} of
\citeads{2011A&A...531A..17R}). 
Slender dark fibrils extending from plage and network represent the
plage counterpart of network RBEs combining blueshift and line
broadening in this blue-wing sampling as in
\linkadspage{2009ApJ...705..272R}{3}{Fig.\,1} of
\citetads{2009ApJ...705..272R}. 

{\em \CaIR\ wing $I$\,--\,core $I$ (p4\,--\,p5)\/}: dramatic
difference. 
The core shows the chromosphere with bright-grained network and plage
hearts signifying onset of heating. 
Magnified blinking shows that many of these dense bright grains are
roughly cospatial with wing MCs, suggesting upward heating, mostly in
unipolar plage (p2).
Blinking to 171\,\AA\ (p11) shows a few areas appearing as EUV moss
(\rrref{endnote}{note:moss}) over bipolar plage (one patch is above the
cutout square but HMI does not show all opposite-polarity field). 
Dark fibrils extend far over internetwork but do not cover all of
that, some internetwork still shows up in the quietest areas with
shock brightening in grains of which some are also visible in the wing
image.

{\em \CaIR\ core $I$\,--\,\Halpha\ core $I$ (p5\,--\,p6)\/}: similar
scenes but all fibrils are much fatter and longer in \Halpha\ which I
attribute to large NSE opacity (\rrref{endnote}{note:nonEchrom}). 
All internetwork is covered by fibrils in \Halpha.
The bright plage grains often correspond well between the two (inset). 
Some may be heated tips of dynamic fibrils observed down-the-throat.
There are long slender dark active-region filaments above polarity
inversion lines (blink either against p2).
The spot flare shows its concurrent loop and feet brightening
similarly in both line cores.

{\em \Halpha\ core $I$\,--\,minimum shift (p6\,--\,p7)\/}: the second
is the shift of the \Halpha\ profile minimum per pixel following
\citetads{2009A&A...503..577C}, 
with blueshift dark.  
The hearts of the plage areas have small-scale grainy appearance with
sizable patches of large blueshift. 
The fibrils extending away show lower-amplitude fibril-to-fibril
modulation, a signature of their intermittent repetitivity in
extending out and shrinking back (called ``breathing'' by a
colleague). 
Blinking the inset shows short dark blueshifted heating onsets; the
longer grey fibrils are return flows.  
The latter show similar but finer structure in \CaIR\ core $I$ (blink
p5\,--\,p7); {\tt showex} zoom-in shows double-branch scatter diagrams
there because Doppler-shifting the narrow \CaIR\ core brightens its
intensity for both blue- and redshift.
In \Halpha\ the umbra shows flashes on their way to running penumbral
waves. 
Only the brightest are visible in \CaIR\ but they are well-known from
\CaIIHK\ as in \linkrjrpage{2003AAp...403..277R}{4}{Figs.\,2\,--\,4} of
\citetads{2003A&A...403..277R}. 
The spot flare shows a bright core-shift feature in the loop
and also a dark shift companion feature further out, at
image edge.  
Diametrically opposite on the other side of the spot there is a
conspicuous bright-dark feature pair that also seems connected.

{\em \Halpha\ core width (p8)\/}: width of the \Halpha\ core per pixel
regardless of Dopplershift following
\citetads{2009A&A...503..577C} 
(\rrref{endnote}{note:nonEchrom}).
This is the most telltale scene in this blinker with respect to
heating.
The plage hearts are grainy and mostly hot. 
The bushes extending away from there are shorter than the fibrils in
core intensity (p6) because these are RBE-type heating onsets.   
Network RBEs reach much further out while heating and accelerating,
but their profile signature moves out from the core to the far wing
(\linkadspage{2009ApJ...705..272R}{6}{Fig.\,5} of
\citeads{2009ApJ...705..272R}). 
The same likely happens for these active-plage fibrils, as for the
active-network fibrils in the data of
\citetads{2017A&A...597A.138R} 
analyzed in
\citetads{2019A&A...632A..96R}. 
Blinking the inset against minimum shift (p7) confirms that the short
dark blueshifted features there are part of these.
The active-region filaments appear about as wide as in \HeII\
304\,\AA, wider than in \Halpha\ core $I$ and \CaIR\ core $I$ which
are affected by Dopplershifts.
The Dopplershift features of the spot flare show heating here, also on
the other side of the spot.

{\em \HeIDthree\,--\,\HeII\,304\,\AA\ (p9\,--\,p10)\/}: the
\HeIDthree\ equivalent-width image is naturally noisy but nevertheless
intriguing.
The two scenes show fair bright-bright correspondence, better further
away from the spot but not 1:1 as one might expect from irradiative
EUV ionization invoked to explain coronal-hole visibility in optical
\HeI\ lines or from turbulent sheath ionization
(\rrref{endnote}{note:sheaths}).
Furthermore, this blink shows significant feature shifts, also in the
inset, for the bright beginnings of dark fibrils; they occur further
out from network in 304\,\AA\ whereas the \HeIDthree-opaquest (bright)
parts remain co-located with corresponding \Halpha\ features. 
These offsets suggest outwardly increasing heating as occurs for
network RBEs in the bottom panels of
\linkadspage{2012ApJ...752..108S}{8}{fig.\,6} of
\citetads{2012ApJ...752..108S}, 
here for fatter fibrils originating from plage.
This blink also indicates that the 304\,\AA\ dark filaments are also
present in \HeIDthree, barely outlined by showing slightly less noise.

{\em \HeII\,304\,\AA\ (p10)\/}: blinking against \Halpha\ core $I$ (p6),
\Halpha\ core Doppler (p7),  
\Halpha\ core width (p8) and \HeIDthree\ equivalent width (p9) are all
interesting.
The first (p10\,--\,p6) shows coarse similarity between \Halpha\ and
304\,\AA: active chromosphere roughly corresponds between these
diagnostics, opposite in sign to the reversed-\Halpha\,--\,304\,\AA\
correlation for quiet closed-field chromosphere
(\rrref{endnote}{note:sheaths}). 
The correlation improves with spatial and temporal averaging (options
in {\tt showex}).
I have used this likeness for direct co-alignment of \Halpha\ to
AIA\,304\,\AA\ for SST targets with much activity and many
active-region filaments and it became an alternative in my SDO
pipeline (\rrref{endnote}{note:SDO-STX}), but usually alignment of
quiet-area AIA\,304\,\AA\ to HMI magnetograms plus SST alignment to
HMI per granulation works better.
The second (p10\,--\,p7) blink shows that the brightest 304\,\AA\ patches
and correspondingly the brightest  \HeIDthree\  opacity patches
match best with the plage hearts where \Halpha\ Doppler is 
fine-scale black-and-white speckled indicating upward heating
(\rrref{endnote}{note:around-upward}).

{\em 171\,\AA\ (p11)\/}: obviously mostly corona, no chromosphere.
Blinking to HMI Stokes $I$ (p1) shows that all long loops connect spot
areas. 
The mossy patches near image center correspond to bright areas in both
He diagnostics (p9 and p10); in \Halpha\ (p6) these may harbor dynamic
fibrils but their identification needs a time sequence.
Otherwise, the grainy areas of chromospheric plage heating in \Halpha\
width (p8) show no coronal counterpart.
The separation between chromospheric and coronal heating in quiet
closed-field areas is already striking (\rrref{endnote}{note:triples});
here it extends to an active region.
Some thick 304\,\AA\ filament parts are also dark in 171\,\AA,
suggesting bound-free scattering as in \rrref{endnote}{note:rain}.}. 

One striking difference between quiet and active scenes is that quiet
closed-field areas permit \Halpha\ and \HeII\,304\,\AA\ co-alignment
by reversing \Halpha\ intensity (\rrref{endnote}{note:sheaths},
\linkrjrpage{2020LingAstRep...1R}{11}{Appendix~B} of
\href{https://robrutten.nl/rrweb/rjr-pubs/2020LingAstRep...1R.pdf}{LAR-1})
whereas active areas permit co-alignment using normal \Halpha\
intensity
(\href{https://robrutten.nl/rrweb/rjr-pubstuff/lar-2/ibismosaicblinker.pdf}
{blinker p6\,--\,p10}).
I attribute this difference to different ratios between roundabout
(``around'') chromosphere heating that dominates in quiet closed-field
areas and upward chromosphere heating that dominates in the hearts of
active network and plage from more numerous and denser MC
clustering
\rrendnote{around and upward heating} {note:around-upward}{%
In quiet areas dark fibrilar \Halpha\ canopies and similar-pattern but
bright \HeII\,304\,\AA\ canopies are made by spicules-II extending
from network (\rrref{endnote}{note:spicules-II}).
The active region in the
\href{https://robrutten.nl/rrweb/rjr-pubstuff/lar-2/ibismosaicblinker.pdf}
{IBIS blinker} shows similar but denser, fatter, longer
\linkpdfpage{https://robrutten.nl/rrweb/rjr-pubstuff/lar-2/ibismosaicblinker.pdf}{6}{\Halpha\
fibril canopies} that are relatively dark in
\linkpdfpage{https://robrutten.nl/rrweb/rjr-pubstuff/lar-2/ibismosaicblinker.pdf}{10}{304\,\AA}. 
It seems likely that such active canopies stem from similar mechanisms
as quiet canopies; I call these ``around heating''. 
However, the hearts of plage and network show dense grainy heating in
\linkpdfpage{https://robrutten.nl/rrweb/rjr-pubstuff/lar-2/ibismosaicblinker.pdf}{8}{\Halpha\
core width}, especially where mostly unipolar, that I call ``upward
heating'' -- meaning at chromospheric heights, not high-reaching as
coronal plumes in coronal holes as in
\citetads{1997ApJ...484L..75W}. 
This heating follows fields that do eventually bend back to the
surface but connect along long coronal loops to faraway activity.

\linkpdfpage{https://robrutten.nl/rrweb/rjr-pubstuff/lar-2/ibismosaicblinker.pdf}{8}{\Halpha\
core-width} shows remarkable difference: grainy bright upward heating
in the plage hearts versus around heating as extended bushes pointing
away from network and plage that seem comparable to spicule-II onsets
(\rrref{endnote}{note:IBIS}).
I attribute this difference to MC clustering and upward field forcing.
Active network and plage contain discrete MCs in larger number and at
larger density than quiet network.
In classic magnetostatic fluxtube cartoons \SSXpage{44} the field
flares out with height; in classic ``Z\"urich wineglass'' models
(\linkadspage{1993A&A...268..736B}{6}{Fig.\,4} of
\citeads{1993A&A...268..736B}) 
neighboring fluxtubes inhibit flaring wide. 
Denser near-unipolar collections so have more upward field, as in the
inversion map in
\linkrjrpage{2017ApJS..229...11J}{9}{Fig.\,7} of
\citetads{2017ApJS..229...11J}. 

I think that the resulting dominance of upward heating over around
heating causes larger brightness in 304\,\AA, outshining the
surrounding fibril canopies that then appear dark in byte-scaled
comparison.   
Where plage and network are more bipolar they have small-scale
short-loop connectivity leading to moss appearance
(\rrref{endnote}{note:moss}).

For both around and upward heating the driving mechanisms are not
known, as is the case for spicules-II producing canopies around quiet
network in closed-field areas (fewer in coronal holes, \eg\
\linkrjrpage{2020LingAstRep...1R}{60}{Fig.\,42} of
\href{https://robrutten.nl/rrweb/rjr-pubs/2020LingAstRep...1R.pdf}{LAR-1}).
However, also for quiet canopies there is observational inclination
dichotomy between more horizontal and more vertical.
On-disk RBEs and RREs are easiest detected in the \Halpha\ wings when
they reach far out from their launching network, therefore
preferentially showing wide-spread around heating, but at the limb
spicules-II are easiest noted and isolated amid their projective
confusion when they are more upward, as shown in the inclination
histograms in \linkadspage{2014ApJ...792L..15P}{5}{Fig.\,4} of
\citetads{2012ApJ...759...18P} 
and seen near the limb \eg\ in this
\href{https://robrutten.nl/dot/dotweb/dot-movies/2003-06-18-QS-mu034-ca-core.mpg}{DOT
straw movie} and in \linkrjrpage{2012ApJ...752L..12D}{5}{Fig.\,5} of
\citetads{2012ApJ...752L..12D}. 
Hence distinction between around versus upward morphology and heating
seems gradual, suggesting similar driving but difference in appearance
and detectability set by variation in dominating field topography
defined by MC clustering density and polarity. 

The CLASP2\,--\,IRIS co-observation of
\citetads{2021SciA....7.8406I} 
indicates that fields keep longitudinal strength up to 300\,Gauss high
in the chromospheric above dense plage, suggesting preponderance of
straight-up fields.
The precise pattern correlation between field strength and \MgIIk\
core intensity and the good correspondence with AIA \HeII\,304\,\AA\
intensity in their \linkadspage{2021arXiv210301583I}{44}{Fig.\,S1}
confirm straight-up heating.

For \HeIDthree\ the sparse and minor on-disk presence in the
\linkpdfpage{https://robrutten.nl/rrweb/rjr-pubstuff/lar-2/ibismosaicblinker.pdf}{9}{IBIS blinker}
versus its bright Balmer-like prominence in the flash spectrum, also
away from active regions, suggests sensitive response to upward
network and plage heating (\rrref{endnote}{note:HeI}).}.
I also wonder whether or how upward chromosphere heating in the hearts
of active network and plage contributes to the slow-wind FIP
composition bias
\rrendnote{FIP effect}{note:FIP}{%
About 3$\times$ relative overabundances compared to photospheric
values in the slow solar wind (SSW) and in solar energetic particles
(SEP) for elements with low neutral-atom ionization ``potential''
(FIP) (\eg\ Meyer \citeyearads{1985ApJS...57..173M}, 
\citeyearads{1991AdSpR..11a.269M}, 
\linkadspage{1989A&A...225..222V}{11}{Fig.\,7} of
\citeads{1989A&A...225..222V}, 
\linkadspage{2020arXiv201008517R}{164}{Fig.\,8.5} of
\citeads{2021LNP...978.....R}). 
These include the most abundant electron-donor elements Mg, Fe, Si, Al
\SSFpage{56} which are predominantly ionized everywhere but amount to
only $10^{-4}\,N_\rmH$ in the photosphere where H and hence the gas
are near-neutral.
This ``FIP effect'' is generally attributed to ion-neutral segregation
involving the degree of line-tying to open field and hence attributed
to low-atmosphere origins (\eg\
\citeads{1989A&A...225..222V}, 
\cf\ \citeads{2019ApJ...879..124L}, 
\citeads{2021LNP...978.....R}). 
Modeling thus must include neutral-ion separation already at low heights.
 
The upper panel of Reames'
\linkadspage{2020arXiv201008517R}{164}{Fig.\,8.5} indicates pivots
near 10\,eV for the SSW values, 14\,eV for the SEP values.
This split suggests that for the SSW the 10.2\,eV \Lyalpha\ jump is
the key in hydrogen ionization by controlling the feeding level of the
Balmer ionization-recombination loop and so defining NSE retardance of
the hydrogen-top (high levels and ion) populations in post-hot cooling
gas, as described for shocks and spicules-II in
\rrref{Sects.}{sec:Oslo} and \rrref{}{sec:chromosphere}.
Presence or absence of frequent hot-cool cycling may be important.
The SEP pivot instead suggests hydrogen ionization directly from the
ground level, hence in more abrupt and fiercer heating events.

The IRIS study of \citetads{2020ApJ...903...68P} 
points to non-mossy upward active-region heating
(\rrref{endnote}{note:around-upward}) in slow-wind generation.
The match in the IBIS\,--\,SDO blinker of ``brightest'' plage-heart patches in
\HeIDthree\ opacity with grainiest \Halpha\ core Dopplershift patches
in plage (p7 versus p9 of the
\href{https://robrutten.nl/rrweb/rjr-pubstuff/lar-2/ibismosaicblinker.pdf}
{blinker}) marking upward heating suggests to test the \HeI\ lines
(\rrref{endnote}{note:HeI}) as potential proxies for locating
upward-heating FIP-offset source areas, possibly thanks to similar
retardance boosting.}.

The enigmatic appearance of the chromosphere in optical \HeI\ lines
may also have mostly to do with upward heating
\rrendnote{optical \HeI\ lines}{note:HeI}{%
The formation of optical \HeI\ lines remains an outstanding enigma.
The main ones are \HeI\,10830\,\AA, weaker \HeIDthree\ at 5876\,\AA,
and \HeI\,6678\,\AA\ ``occasionally making its appearance''
(\citeads{1868RSPS...17..131L}). 
\HeI\,10830\,\AA\ reaches largest extinction because its lower level
is metastable and collects population. 
Its observed depth is claimed as coronal activity mapper in the
optical (\eg\ \linkadspage{1998ASPC..140..325P}{3}{Fig.\,1} of
\citeads{1998ASPC..140..325P}) 
-- but blinking \HeIDthree\ against AIA EUV in the
\href{https://robrutten.nl/rrweb/rjr-pubstuff/lar-2/ibismosaicblinker.pdf}
{IBIS blinker} shows better \HeI\ matching with chromospheric
304\,\AA\ than with coronal 171\,\AA.
The line serves as coronal hole mapper in the McIntosh Archive (\eg\
\citeads{2017SpWea..15.1442W}), 
perhaps better than coronal activity mapping since the AIA fire
detectors (\rrref{endnote}{note:triples}) show less heated
quiet chromosphere in holes, hence less \HeII\ and \HeI-top population.

Lockyer proposed element ``helios'' for \HeIDthree\ which became
helium and remains the only extra-terrestrially discovered element
since coronium \SSXpage{8} and nebulium \SSXpage{142\,ff} went
away. 
I am not aware of a sound explanation why his \HeIDthree\ is so
extraordinary strong off-limb, equal to \Halpha\ in
\linkpdfpage{https://books.google.nl/books/download/The_Spectroscope_and_Its_Applications.pdf?id=EbANQlPbaS0C&output=pdf&sig=ACfU3U3THi5lvx3bWxOCBD0cO2ScPbhV8w}{95}{his
engraved Fig\,42} and ranking between \Hgamma\ and \Hbeta\ in
brightness and off-limb extent in
\linkadspage{1968ApJS...15..275D}{94}{p\,94--95 of Table\,3A} of
\citetads{1968ApJS...15..275D} 
and in my unpublished corresponding 
\href{https://robrutten.nl/rrweb/rjr-archive/linelists/dunnetal-bin10AA.png}
{summary plot} and
\href{https://robrutten.nl/rrweb/rjr-archive/linelists/dunnetal-decays.png}
{decay plot}, much brighter and higher than its \NaID\ neighbors whereas it
is nearly invisible on the disk and hence not even mentioned on
\linkadspage{1966sst..book.....M}{292}{page\,258} of the line table of
\citetads{1966sst..book.....M} 
from the Utrecht Atlas of
\citetads{1940pass.book.....M} 
in which the \NaID\ lines are the darkest in the optical and the
Balmer lines are prominent wide dips
(\linkrjrpage{2019arXiv190804624R}{18}{Fig.\,10} of
\citeads{2019SoPh..294..165R}). 
Similarly, \HeI\,10830\,\AA\ is absent in the ALC7 spectrum
\SSXpage{94}.
How can the Balmer lines in the top of the hydrogen atom and the \HeI\
lines in the top of the helium atom differ so enormously between
transverse and top-down viewing?\,\footnote{History: the
\label{note:Thomas+Athay}
400$^+$-page book by
\citetads{1961psc..book.....A} 
(wrong order at ADS) addressed this question without clear answer. 
It was grandly titled ``Physics of the solar chromosphere'' earning a
scathing review from De Jager \SSXpage{15\,ff} for not mentioning
magnetism nor dynamism. 
De Jager added that Thomas' ``photoelectric control'' producing $S>B$
was the main item of interest but even that was misleading
(\rrref{endnote}{note:photoelectric}).}

Off-limb (no continuum) large emissivity can go with small extinction
through large $\varepsilon B$ and/or large $\eta S^\rmd$ contributions
in \rrref{Eq.}{eq:S_CRD}.
Because \HeIDthree\ shows Balmer-like extent above the limb while the
Balmer extent is dominated by quiet-Sun spicules-II and
\HeII\,304\,\AA\ shows quiet-chromosphere canopies pattern-matching
the spicule-II-made reversed-\Halpha\ ones outside coronal holes
(\rrref{endnote}{note:sheaths}) I speculate that in quiet-Sun around
heating (\rrref{endnote}{note:around-upward}) the \HeI\ top shares in
the spicule-II emissivity phase giving large $\varepsilon B$ off-limb
brightness, but less in the subsequent recombination phase giving the
\HI\ top sufficient NSE extinction to show opaque canopies on
the disk with $\sqrt{\varepsilon}$ darkness in \Halpha\ and with
temperature-defined brightness for ALMA.
Small top-down extinction in resolved on-disk features can go with
large off-limb emissivity from summing many features along the
tangential line of sight.  
The ubiquity of quiet-Sun \HeII\,304\,\AA\ canopies over the surface
attributed to turbulent sheath ionization in \rrref{endnote}{note:sheaths}
suggests indeed that very many features are sampled along the line of
sight to the limb. 
The \HeI\ top containing the optical lines follows suit in
overpopulation.

At more activity the 304\,\AA\,--\,193\,\AA\ match for the active
plagette in \linkrjrpage{2017AAp...597A.138R}{2}{Fig.\,1} of
\citetads{2017A&A...597A.138R} 
is attributed to irradiative ionization in
\rrref{endnote}{note:sheaths}.
The same may hold for the correspondence between \HeIDthree\
absorption and \HeII\,304\,\AA\ brightness for active plage in the
IBIS\,--\,SDO blinker (\rrref{endnote}{note:IBIS}).}.

The second display I link to is an appetizer for things hopefully to
come: comparable active regions in \Halpha\ and \Lyalpha\
in \linkrjrpage{2017AAp...598A..89R}{6}{Fig.\,2} of
\citetads{2017A&A...598A..89R} 
\rrendnote{\Lyalpha\ dream}{note:Lyalpha}{%
Let me dream further re imaging spectroscopy in \Lyalpha\
(\rrref{footnote}{note:UVimspect}; paradise = $5\times5$\,arcmin
field, 0.1\,arcsec resolution, 20 wavelengths all at 1\,s
cadence)\footnote{Yet rosier dreaming: add the same in
\linkadspage{1968pgda.book.....M}{11}{\HeI\,584\,\AA}.}
and compare that with what we know (awake) from SST/CRISP in
\Halpha\footnote{SST/CHROMIS \label{note:Hbeta} \Hbeta\ at
\linkadspage{1968pgda.book.....M}{9}{4861\,\AA}
may be yet better because it images features sharper
($\sim\!\!D/\lambda$) and with less saturation (smaller transition
probability,
\linkpdfpage{https://robrutten.nl/rrweb/rjr-archive/linelists/aq1976.pdf}
{80}{paragraph 29} of \citeads{1976asqu.book.....A}). 
Adding less saturated higher Lyman lines enabling intercomparison is
also desirable.}.
Co-observing with the Chinese ASO-S/LST (\Lyalpha) and CHASE (\Halpha)
space missions may come closest in the near future.

\Lyalpha\ and \Halpha\ are both \HI\,$\alpha$ and even share the
$n\tis\,2$ level but differ tremendously. 
Any feature visible in \Halpha\ has much higher opacity in \Lyalpha\
that will only show its outer surface. 
Within hot and dense features as internetwork shocks, dynamic fibrils,
spicules-II, EBs and EB-like lower-atmosphere reconnection sites
\Lyalpha\ has fast collision-up rates \SSXpage{166} while remaining
boxed-in even though being champion resonance scatterer (many random
steps but of small length \SSXpage{90}). 
The high radiative 1--2 up and down rates then nearly cancel to net
radiative rate zero \SSXpage{165}, the $n\tis2$ level gains
near-Boltzmann population, \Lyalpha\ has $S\!_\nu^l\tapprox\,B_\nu(T)$
and \Halpha\ has SB line extinction. 

\Halpha\ does not show such features in its core when they are
underneath the opaque \Halpha\ fibril canopies spreading from network
and then samples them only in its outer wings, bright for EBs but
Doppler-shifted thermally-broadened dark for spicule-II RBEs and RREs. 
\Halpha\ also skips internetwork shocks by its opacity chasm in the
upper photosphere (\rrref{endnote}{note:quietHa}, \SSXpage{91}).

\Lyalpha\ does not have such gap so that shifting my narrow dream
passband from core to wing means sampling the whole non-coronal
atmosphere down to the low (dark) photosphere (first plot of
\SSXpage{90}). 

In addition, \Lyalpha\ is the worst PRD line in the spectrum (second
plot of \SSXpage{90}) but this is actually a boon because it gives
each part of the line its own source function and response, as in the
\SSFpage{111} cartoon, and hence specific surface-sampling signature
in the dream-resolved emergent profile. 
At each wavelength the emergent intensity does not sample the LTE
source function within hot and dense features but the outer
monofrequent $\sqrt{\varepsilon_\nu}$ scattering decline towards the
feature surface.
As in \rrref{endnote}{note:NaD} the emergent intensity is defined
around $\tau_{\rm eff}\tis\,1$, polarization and Doppler signature
around $\tau\tis\,1$. 
Outer-wing passbands that form in the upper photosphere should be well
modelable with MURaM and Bifrost (\rrref{endnote}{note:REphot}).
What a nice dream!}.
These scenes both sample hydrogen atoms but differ strikingly in
appearance. 
The lines have discordant formation in the ALC7 atmosphere
\SSXpage{28} but nevertheless should sample common ionization features,
I think with bright \Lyalpha\ grains showing dynamic heating and dark
\Halpha\ fibrils showing NLT cooling
\rrendnote{\Lyalpha\ features}{note:Lya-features}{%
The \Lyalpha\ image in \linkrjrpage{2017AAp...598A..89R}{6}{Fig.\,2}
of \citetads{2017A&A...598A..89R} is from the VAULT-II flight
(\citeads{2001SoPh..200...63K}, 
\citeads{2010SoPh..261...53V}, 
\citeads{2007ApJ...664.1214P}) 
and is one of 17 images near disk center. 
A second sequence of four limb images was assembled by
\citetads{2009A&A...499..917K} 
into a
\href{https://robrutten.nl/rrweb/rjr-pubstuff/lar-2/vault2-limb-aligned.mpg}
{limb movie} and a
\href{https://robrutten.nl/rrweb/rjr-pubstuff/lar-2/vault2-disk-aligned.mpg}
{centerward movie}.
They show dense bright grains in plage and active region of which a
few change already during the 51-s movies.
\citetads{2009A&A...499..917K} 
suggested that the latter
 are \Lyalpha\ counterparts to \Halpha\ dynamic
fibrils (\rrref{endnote}{note:DFs}).

The dense more stable grains seem tips of longer-lived upward heating
jets that may also produce the numerous bright grains in non-mossy
hearts of network and plage in the
\linkpdfpage{https://robrutten.nl/rrweb/rjr-pubstuff/lar-2/ibismosaicblinker.pdf}{8}{IBIS
\Halpha\ core-width scene} and contribute the upward chromosphere
heating of \rrref{endnote}{note:around-upward}.

Away from the bright-grain plage areas in the \Lyalpha\ panel of
\linkrjrpage{2017AAp...598A..89R}{6}{Fig.\,2} the emission blobs are
fuzzier and appear remarkably as the short bright bushes extending
towards internetwork in the
\linkpdfpage{https://robrutten.nl/rrweb/rjr-pubstuff/lar-2/ibismosaicblinker.pdf}{8}{IBIS display},
suggesting that they sample similar spicule-II-like
roundabout heating onsets.

The dense NSE \Halpha\ fibril canopies in the lefthand panel of
\linkrjrpage{2017AAp...598A..89R}{6}{Fig.\,2} of
\citetads{2017A&A...598A..89R} 
and in the
\linkpdfpage{https://robrutten.nl/rrweb/rjr-pubstuff/lar-2/ibismosaicblinker.pdf}{6}{IBIS
\Halpha\ image} appear dark not from sampling temperature but from
\Halpha's self-made $\sqrt{\varepsilon}$ scattering decline
(\SSXpage{91}, \rrref{endnote}{note:Ha-RE}) with large NSE
overopacity for cooling recombining hydrogen.
These canopies must be yet more opaque in \Lyalpha\ with
$S^l\tapprox\,b_2B(T)$ superthermal internal source function boosting
but similarly sampling the monochromatic $\sqrt{\varepsilon}$
\Lyalpha\ declines towards the canopy surface, as suggested by dark
quieter areas in \linkrjrpage{2017AAp...598A..89R}{6}{Fig.\,2} of
\citetads{2017A&A...598A..89R} 
where they are not out-radiated by brighter features in profile-summed
brightness. 
The same applies to filaments.}. 
ALMA samples hydrogen ions and may be able to measure both types of
feature at fast enough cadence to establish and correlate
cause--effect delays.
ALMA may also sample hydrogen lines permitting super-sensitive
chromospheric magnetometry 
\rrendnote{Rydberg \HI\ candidate for ALMA}{note:ALMA-line}{%
There may exist chromospheric features with sufficient Rydberg-ladder
(\rrref{endnote}{note:Rydberg}) NLT recombination emissivity for
visibility in the \HI\ 30-$\alpha$ line at 231.901~GHz in ALMA band 6
(\linkrjrpage{2017IAUS..327....1R}{13}{Sect.\,6} of
\citeads{2017IAUS..327....1R}). 
Detection would furnish sensitive chromosphere magnetometry.

\citetads{1992A&A...259L..53C} 
predicted Rydberg lines up to lower level $n\tis\,18$ for the MACKKL
star assuming SE. 
The formation spans between $\tau\tis\,1$ at line center and in the
continuum at the bottom of the third panel of
\linkrjrpage{1992AAp...259L..53C}{2}{Fig.\,1} shift upward with $n$
but remain below the MACKKL temperature minimum. 
Reasonable reproduction of observed profiles is shown in
\linkrjrpage{1992AAp...259L..53C}{3}{Fig.\,3} but for higher
$n\mbox{-}\alpha$ lines the predicted line strengths exceed
corresponding observations in
\linkadspage{2000A&A...357..757C}{5}{Fig.\,6} of
\citetads{2000A&A...357..757C}. 
Reasonable reproduction with Pandora-star SE line synthesis may be
expected for the lower $n$ lines, as for the \MgI\ 12-micron lines of
\rrref{endnote}{note:MgI12micron}, because the upper photosphere is
the domain where Pandora stars come closest to represent a
proper mean of actual reality (\rrref{endnote}{note:REphot}).

Both the modeling and the higher-$n$ observations (including the limb
detection of 21-$\alpha$ by
\citeads{2000A&A...361L..60C}) 
suggested absence of Rydberg lines beyond $n\tapprox\,25$.
However, at mm wavelengths the extinction increase (with \HI\
free-free taking over from \Hmin\ free-free following
\linkrjrpage{2017AAp...598A..89R}{4}{Fig.\,1} of
\citeads{2017A&A...598A..89R}) 
causes formation at chromospheric heights where dynamic NSE opacity
boosting of the hydrogen top applies, not only for \Halpha\ and the
ALMA continua but also for the high-$n$ Rydberg lines in the ALMA
range.
Hence, dynamic long-lasting NLT overextinction of the \HI\ top may
make 30-$\alpha$ yet higher-formed, at sufficiently low density to
bring it out of the reach of collisional ionization lowering, and also
stronger. 
It may be visible.
It may be strongest off-limb since also the Balmer lines extend there
as far and bright as non-NLT \CaII\ \HK\ while weaker on the disk.
Strength prediction may be done with NSE Rydberg synthesis from
numerical NSE simulations that produce spicules-II as copiously as
observed.

Detection of this line enables exciting measurement of
profile-resolving Zeeman splitting because it scales $\sim\!\lambda^2$
while competing Dopplerwidth scales $\sim\!\lambda$ \RTSApage{79}
{Eq.\,3.66}\footnote{Erratum under \RTSApage{79}{Eq.\,3.66}:
$\mbox{FWHM}\tis\,2\sqrt{\ln(2)}\Delta\lambda_\rmD\tapprox\,1.665\Delta\lambda_\rmD$}.
Their ratio produces significant intensity-profile signature already
at 12\,$\mu$m for 300\,--\,500\,Gauss in
\linkrjrpage{1994IAUS..154..309R}{12}{Fig.\,9} of
\citetads{1994IAUS..154..309R}; 
at hundredfold wavelength one Gauss chromospheric sensitivity is in
sight.}.

Finally, there are dark filaments in these active-region scenes
suggesting ``chromospheric'' neutral hydrogen gas in the corona.
Filaments generally appear long-lived but may nevertheless be made
dynamically
\rrendnote{filament blobs}{note:blobs}{%
Filaments and prominences may be seen as {\em ``merely local
aggregations of a gaseous medium which entirely envelopes the sun''\/}
(\citeads{1868RSPS...17..131L}) 
but yet they present the weirdest and toughest features to understand
in the rich solar \Halpha\ zoo. 
How can cool chromospheric gas show up endlessly in the hot corona in
these exceedingly long, slender, high-reaching features exuberantly
rich in fine structure and dynamics? 
Why well visible only in H and He diagnostics?
A nightmare for plane-parallel colleagues. 
Who ordered these?

The
\href{https:robrutten.nl/rrweb/rjr-pubstuff/lar-2/halpha-soup-half.mpg}{SST/SOUP
\Halpha\ movie} of \rrref{endnote}{note:DFs} shows an active-region
filament where small bright bullet-like blobs run intermittently along
its length (first panel of
\linkadspage{2012ApJ...747..129L}{2}{Fig.\,1} in the analysis by
\citeads{2012ApJ...747..129L}). 
Since then I have noted many such fast disturbances running through
filaments in AIA\,304\,\AA\ movies, suggesting frequent
ionization-recombination cycling.
I suspect that they are not disturbances of a mean state but that they
actually make and remake filaments continuously with subsequent
cooling gas in down-raining condensations causing the extraordinary
\HeIDthree-like off-limb visibility of prominences in \Halpha\ and
\HeII\,304\,\AA\ and that this cannot be modeled assuming static SE.
 
For \Halpha\ NLT retardation may cause large cooling-gas
overemissivity off-limb and large overextinction on the disk; let me
repeat (\rrref{Sect.}{sec:Oslo}) that modest hydrogen ionization in
simple internetwork shocks already produces post-shock hydrogen-top
retardation reaching 10$^{12}$ \Halpha\ overopacity (last panel of
\href{https://robrutten.nl/rrweb/rjr-movies/hion2_fig1_movie.mov}{Fig\,1
movie} of \citeads{2007A&A...473..625L}), 
truly in the ``extraordinary'' ballpark.

For \HeII\,304\,\AA\ off-limb brightness a likely candidate is sheath
heating around rain-down blobs in prominence threads by turbulent
Kelvin-Helmholtz coronal cooling as modeled by
\citetads{2019ApJ...885..101H} 
(\rrref{endnote}{note:sheaths}).  
For \HeII\,304\,\AA\ darkness on disk, as for the active-region
filaments in the
\linkpdfpage{https://robrutten.nl/rrweb/rjr-pubstuff/lar-2/ibismosaicblinker.pdf}{10}{blinker
304\,\AA\ image}, bound-free \HI\ scattering is likely
(\rrref{endnote}{note:rain}).

\label{sec:endendnotes}
That filaments may reform in place 
as happened after the exemplary failed Rosetta eruption of
\citetads{2021ApJ...914L...8M} 
also suggests dynamic refurbishment.}.
At high resolution and fast cadence ALMA may also diagnose their
formation including frequent refurbishment and accompanying
retarded-opacity visibility.

\paragraphrr{Chromospheric and coronal heating} 
An opinionated\footnote{Or worse, witness {\em ``\ldots one of the
chief `troublemakers' and `revisionists' in modern solar physics
R.~Rutten\ldots''} on
\linkpdfpage{https://robrutten.nl/rrweb/rjr-archive/books/rbtsa2013.pdf}
{82}{p.\,81} of
\citetweb{https://robrutten.nl/rrweb/rjr-archive/books/rbtsa2013.pdf}
{Teplitskaya2013} found and scanned by A.V.\,Sukhorukov.}
text as this should address {\em ``one of the longest
standing unsolved mysteries in all of
astrophysics''}\,\footnote{J.T. Schmelz, ISSI talk July 2021 for a
covid-zoom audience.}, certainly a text harping on the chromosphere
since generally the blame for coronal heating is sought there -- while
heating the chromosphere to observed temperatures is less glamorous
but requires more energy for its larger mass.

The literature on outer-atmosphere heating is bewilderingly varied.
The traditional emphasis for the coronal part is on loop modeling
following \citetads{1978ApJ...220..643R} 
debating AC (Alfv\'enic wave) versus DC (reconnection) mechanisms,
both dictated by dynamical fluxtube topography in the photosphere.  
Chromospheric heating is attributed to a yet wider variety of agents.
However, regarding the chromosphere as a layer (not dynamically
structured and pervaded) and coronal loops as homogeneous structures
(not dynamically multistrand and multithermal) both seem as misleading
as regarding the photosphere a plane-parallel layer without
convection, waves, magnetism. 

My theme here is spectral evidence.
Blinking the SDO triples (\rrref{endnote}{note:triples}) and the
IBIS\,--\,SDO images (\rrref{endnote}{note:IBIS}) suggests that a more
important split than ``corona versus chromosphere'' is ``quiet versus
active''. 
I summarize my impressions from these representative blinkers.

The SDO triples suggest that quiet chromosphere\footnote{Once again
{\em ad nauseam\/}: not the ``chromosphere'' of Pandora stars
mimicking the Wien-skewed ultraviolet of under-canopy acoustics
(\rrref{endnote}{note:clapotis}) and MC deep-hole shine
(\rrref{endnote}{note:MCs}), but quiet-Sun internetwork canopies of
the \citetads{1868RSPS...17..131L} 
chromosphere observed in \Halpha, \HI\ \Lyalpha, \HeII\ \Lyalpha\ and
also in \MgIIhk, \CaIIHK, \CaIR\ and with ALMA.}
heating is AC-type with spicules-II fired from vorticity-rich network
and producing return-flow fibrilar \Halpha\ and \HeII\,304\,\AA\
canopies, the latter mapped well in the AIA fire detector images.
The major telltale is the good pattern matching of the bright canopies
in AIA\,304\,\AA\ and the dark fibril canopies in \Halpha\ around the
magnetic network in HMI magnetograms, and their ubiquity.
This is ``around'' heating (\rrref{endnote}{note:around-upward}) and
does not heat the corona but may cool it
(\rrref{endnote}{note:sheaths}). 
ALMA may become its best diagnostic
(\rrref{endnote}{note:ALMA-chrom}).

The SDO triples also suggest that quiet corona heating arises chiefly
from CBPs, especially the larger ones, with DC loop-top reconnection
resulting in diffuser coronal connectivity to nearby CBP areas.  
Hence, the tentative suggestion is that the observed CBPs are Parker's
postulated nanoflares.
Their beam-heated feet are also well-mapped in the AIA fire detector
product.

These two quiet-Sun heating agents appear remarkably unrelated, even
though both stem from MC dynamics in network and plage outside active
regions. 
Both are ubiquitous because network is ubiquitous.
Spicule-II quiet-chromosphere heating does not need network bipolarity, CBP
quiet-corona heating does. 
Both act also but less effective in coronal holes where coronal plumes
seem an open-field CBP alternative.
The SDO database offers tremendous material for further
study.\footnote{If
I still had graduate students I would suggest the first five
\linkrjrpage{2020LingAstRep...1R}{9}{research projects} listed in the
\href{https://robrutten.nl/rrweb/rjr-pubs/2020LingAstRep...1R.pdf}{LAR-1}
conclusion.}

The IBIS\,--\,SDO images suggest that active regions add unidentified
``upward'' heating (\rrref{endnote}{note:around-upward}) feeding
long-loop connectivity to distant other active regions, likely with
FIP abundance offsets (\rrref{endnote}{note:FIP}).

\section{Conclusion} \label{sec:conclusion}
Understanding the solar spectrum has come a long way but isn't there
yet. 
The basic physics was well understood by the 1970s -- no dark matter
nor dark energy in the solar atmosphere.\footnote{For a conflicting
speculation search ADS with {\em abs:"axion quark" and "solar
orbiter"} (not to be cited).}
Numerical modeling started about then with the advent of electronic
computing which was welcome because analytic theory had reached its
limit in \citetads{1952bmtp.book.....K}. 
Numerical study provided great insights even though initially mostly
within the misleading straitjacket of plane-parallel atmospheres.
Nowadays the major challenge is to obtain realistic spectrum synthesis
from realistic solar-atmosphere simulations to provide guidance to
interpret high-quality observations conveying solar truth -- with the
proviso that {\em ``simulations are great tools but still toys''\/}
(J.\,Mart{\'{\i}}nez-Sykora).
Full 3D(t) synthesis of all spectral features for every voxel at all
times of a RADYN-style 3D(t) MHD-etc simulation is a bridge too far,
so that the simulation synthesis frontier lies in clever shortcuts
(reviews by \citeads{2019AdSpR..63.1434P} and 
\citeads{2020LRSP...17....3L}; 
NSE simplification on \linkrjrpage{2019SoPh..294..165R}{20}{p.\,20} of
\citeads{2019SoPh..294..165R}). 

Shortcuts as adopting 1.5D columnar synthesis ignoring non-radial
matter and radiation gradients, Holweger-style ``inversion'' suffering
NLTE masking, neglect of ultraviolet NLS scattering and NLW
interlocking, ignoring NLT memory by assuming SE may be as
misleading as believing plane-parallel models -- but just as those
they may be educational steps along the way. 

The tantalizing trophy is spectral understanding of the chromosphere
which I now define as the hectic domain separating dense near-LTE
near-neutral photospheric gas and tenuous near-CE ionized coronal gas,
where hydrogen intermittently ionizes and recombines cycling between
neutral and plasma state with full NLS+NLW+NLT and
HD\,$\leftrightarrow$\,MHD\,$\leftrightarrow$\,plasma complexity:
{\em the chromosphere ain't
stacked in layers but is dynamically structured and unstuck in
time\/} (\rrref{endnote}{note:nonEchrom}). 

Key property: frequent dynamic refurbishment. 

Prime diagnostics: \HI\ Balmer lines, Lyman lines, mm continua, helium
lines -- all enriched by retarded opacities in cooling gas.

Low-hanging fruit: CO clouds, spicule-II heating of quiet
chromosphere, CBP heating of quiet corona. 
In activity: moss, the FIP effect a ripening plum, the Wilson-Bappu
relation overdue.

High-hanging fruit: 150$^+$ years since
\citetads{1868RSPS...17..131L} 
\SSXpage{12\,ff} coined ``chromosphere'' in the dawn of astrophysics
\SSFpage{2\,ff} we Khayyam-wise have no explanation for his bright
yellow ``helios'' line.  
Cracking this nut likely helps to understand prominence visibilities.
 
Every aspect treated here must be considered: the Sun is a most
beautiful complex being.  
We are fortunate that our kind star offers such great inspiring
complexity to keep us on our toes.\\


\begin{footnotesize}
\mbox{}\vspace*{1ex} \mbox{}\\
\noindent
\phantomsection\label{sec:acknowledgements} 
{\em \normalsize Acknowledgements}\\
\addcontentsline{toc}{section}{Acknowledgments} 
\noindent
\mbox{}\!I thank Ram Ajor Maurya and M.K.~Ravi Varma for inviting me
to teach this subject at their
\href{http://www.saihpa2021.nitc.ac.in}{Calicut school}. 
I enjoyed my teaching sessions but suspect that the (unseen) students
came out by the same door as in they went \SSFpage{120}.
I also thank the organizers for requesting this writeup and endorsing
hyperlinks.

Earlier invitations to teach, in Bandung, Mitaka, Palo Alto, Seoul, La
Laguna, Freiburg, Newcastle and Weihai (usually
\href{https://robrutten.nl/rrweb/rjr-elsewherecourses/dircontent.html}
{a full week} for this material) led to displays SSF and SSX; I hope
more will come in a post-covid future.

F.\,Paletou, J.\,Leenaarts, T.M.D.\,Pereira, P.R.\,Young,
C.J.\,Schrijver, P.G.\,Judge and H.N.\,Smitha contributed improvements. 

As always I relied much on ADS. 
ADS also taught me
\href{https://robrutten.nl/Turning_citations.html}{pdf page opening }
and keeps its robust classic linkers active since its 2019 revamp.
ArXiv provides a splendid open-access ADS-linked long-lasting
publication platform.

Stackexchangers solved latex problems.
M.J.\,Rutten started hosting \href{https://robrutten.nl}{robrutten.nl}
for link speed and persistence.

\end{footnotesize}

\begingroup
\def\enotesize{\normalsize} 
\newpage      
\renewcommand*{\enoteheading}{}   
\fancyhead[RE,RO]{\nouppercase{Detailing notes}}%
\section*{Detailing notes}
\label{sec:notes}
\addcontentsline{toc}{section}{Detailing notes} {\em Welcome to the
yellow pages! 
These endnotes are take-or-leave elaborations adding literature,
comments, opinions, history.

Topic selection: personal, favoring topics that I have been involved
in and references and figures which I knew best and came first to mind
in this writeup for the
{Calicut school}.  
With preference for my own displays when these open faster than
ADS.\footnote{Example:
\label{note:speeds}
compare the opening speeds of \SSXpage{60} and
\linkadspage{1981ApJS...45..635V}{3}{Fig.\,1} 
for the most-shown figure of solar physics. 
Both load relatively slow, in 1--3s and 4--10s with my Lingezicht
600\,Mbps. 
Because these speeds and their ratio depend on connection and location
I welcome \href{mailto:robenrietjerutten@gmail.com}{email giving your
values} for these two with your download speed; as thank-you I will
notify you at future arXiv updates. 
When they are slow you may meanwhile be comforted by
\rrref{footnote}{note:slow}.} 

I also sprinkle work-habit reminiscences through these notes because
during my half$\,^+$ century in solar physics the way we work changed
drastically from advances which may appear old-hat granted to you:
photomultipliers, image intensifiers, electronic computing (1960s),
solid-state image detection, writing per computer, internet
communication (1980s), WWW and ADS (1990s), presenting per computer,
Wikipedia (2000s), meeting per computer (2020s).

Navigation: each note ends with a link [Main call] returning to its
calling location in the main text and a link [Back] that also returns there
when you came from there or back to another link that you used --
working in Firefox, Ubuntu acroread and xpdf, macOS Preview but not in
Ubuntu evince, chrome, macOS Safari (try the reader back button
instead; for Acrobat check the
\href{https://helpx.adobe.com/acrobat/using/keyboard-shortcuts.html}{shortcut
table}).
Footnote numbers jump back to their call.}
~~\theendnotes 
\endgroup
\fancyhead[RE,RO]{\nouppercase{Detailing notes}}

\vspace{6ex} 

\begin{footnotesize}

\noindent
\phantomsection\label{sec:version} 
{\em \normalsize Version history}\\
\addcontentsline{toc}{section}{Version history, status, how-to}
\noindent 
\mbox{~~~~}{\em February 5, 2021\/} -- start for NIT Calicut school 
March 8--14.\\
\mbox{~~~~}{\em March 3, 2021\/} --
\href{https://arxiv.org/pdf/2103.02369v1.pdf}{arXiv v1},
pre-school post for attendants, 16 pages.\\
\mbox{~~~~}{\em April 29, 2021\/} --
\parbox[t]{6cm}{\href{https://arxiv.org/pdf/2103.02369v2.pdf}{arXiv
v2}, honoring
\href{https://robrutten.nl/rrweb/rjr-pubs/2021LingAstRep...3R.pdf}
{C. de Jager's 100th birthday}, 27 pages,
moved from UU to \href{https://robrutten.nl}{robrutten.nl}.}\\
\mbox{~~~~}{\em May 7, 2021\/} -- notes $\rightarrow$ endnotes.\\
\mbox{~~~~}{\em May 31, 2021\/} --
\href{https://arxiv.org/pdf/2103.02369v3.pdf}{arXiv v3},
endnotes on yellow pages, 34 pages.\\
\mbox{~~~~}{\em July 20, 2021\/} -- more informative reference format.\\
\mbox{~~~~}{\em July 29, 2021\/} -- reached deep-thought page limit.\\
\mbox{~~~~}{\em August 2, 2021\/} --
\href{https://arxiv.org/pdf/2103.02369v4.pdf}{arXiv v4}, 42 pages.\\
\mbox{~~~~}{\em August 25, 2021\/} --
\href{https://arxiv.org/pdf/2103.02369v5.pdf}{arXiv v5}, 42 pages.\\
\mbox{~~~~}{\em December 13, 2021\/} --
\href{https://arxiv.org/pdf/2103.02369v6.pdf}{arXiv v6}, 50 pages.\\
\mbox{~~~~}{\em April 7, 2022\/} -- 100 endnotes, 150 footnotes:
time to stop.\\
\mbox{~~~~}{\em April 9, 2022\/} --
\href{https://arxiv.org/pdf/2103.02369v7.pdf}{arXiv v7}, 56 pages.\\
\mbox{~~~~}{\em August 14, 2022\/} --
\href{https://arxiv.org/pdf/2103.02369v8.pdf}{arXiv v8}, 62 pages.\\[0.5ex]
ArXiv serves \href{https://arxiv.org/abs/2103.02369}{all posted
versions}.
The evolution can be highlighted with {\tt diffpdf} or source {\tt
latexdiff}.

\mbox{}\\
\noindent
\phantomsection\label{sec:status} 
{\em \normalsize Status}\\
\noindent
I may update this text further (see \rrref{footnote}{note:speeds}
for notifications)
and \href{mailto:robenrietjerutten@gmail.com}{welcome} corrections and
suggestions for improvements.
\href{https://robrutten.nl}{My website} serves the
\href{https://robrutten.nl/rrweb/rjr-pubs/2021LingAstRep...2R.pdf}
{current version} with a
\href{https://robrutten.nl/Compendium_solar.html}{summary here}.
ADS serves the
\href{https://ui.adsabs.harvard.edu/abs/2021arXiv210302369R/abstract}
{latest arXived version}.

\mbox{}\\
\noindent
\phantomsection\label{sec:howto} 
{\em \normalsize How-to}\\
\noindent
ArXiv also serves my \href{https://arxiv.org/e-print/2103.02369}
{latex source files}.
Uncompress with \eg\ {\tt tar xvf 2103.02369}. 
You are welcome to my latex tricks in the {\tt .tex} files. 
They contain explanatory comments (to myself) with
Stackexchange referrals.
My bibtex automation (one command generates all references) is
\href{https://robrutten.nl/Recipes_publications.html}{described here}.
I have no manual for it but my webposted
\href{https://robrutten.nl/rrweb/rjr-ads/scripts/dircontent.html}
{download scripts} and
\href{https://robrutten.nl/rrweb/rjr-pubstuff/lar-2/lar.bst}{bibtex
bst style} are also self-commenting.

\end{footnotesize}

\newpage  
\fancyhead[RE,RO]{\nouppercase{References}}%
\renewcommand{\baselinestretch}{1.00}\small
\pagecolor{blue!8!white}  
\phantomsection\label{sec:references}   
\bibliographystyle{lar}  
\addcontentsline{toc}{section}{References}
\label{sec:refs}
\interlinepenalty=10000
\bibliography{rjrfiles,adsfiles} 

\label{sec:end}
\end{document}